%% file: 0_Like_top.tex
\documentclass[english,british,traditabstract,longauth]{aa}
\usepackage{fixltx2e}
\usepackage[T1]{fontenc}
\usepackage[utf8]{inputenc}
\usepackage{color}
\usepackage{babel}
\usepackage{url}
\usepackage{amsmath}
\usepackage{amssymb}
\usepackage{graphicx}
\usepackage{morefloats}
\graphicspath{{./figures/}{.figures/cspec/} } 

\usepackage[breaklinks,colorlinks,citecolor=blue,backref=section]{hyperref}
\makeatletter

\providecommand{\tabularnewline}{\\}
\pdfoutput=1
\usepackage[nonamebreak,authoryear]{natbib}
\bibpunct{(}{)}{;}{a}{}{,}
\usepackage{txfonts}
\usepackage{color}
\usepackage[mathscr]{eucal}
\usepackage{array}
\input{macros.tex}

\newcommand{\camspec}{\texttt{CamSpec}}
\newcommand{\camspecs}{\texttt{CamSpec}\ }
\newcommand{\plik}{\texttt{Plik}} 
\newcommand{\pliks}{\texttt{Plik}\ }
\newcommand{\commander}{\texttt{Commander}} 

\newcommand{\smica}{\texttt{SMICA}} 

\newcommand{\nilc}{\texttt{NILC}} 

\newcommand{\sevem}{\texttt{SEVEM}}

\newcommand{\FRB}[1]{\textbf{$\triangleleft$ #1 $\triangleright$}}

\newcommand{\Planck}{\textit{Planck}}
\newcommand{\Plancks}{\textit{Planck\/}\ }
\newcommand{\WMAP}{\textit{WMAP}}

\newcommand{\COBEs}{\textit{COBE\ }}

\def\GHz{\ifmmode $GHz$\else \,GHz\fi}
\def\ghz{\ifmmode $GHz$\else \,GHz\fi}
\def\MHz{\ifmmode $MHz$\else \,MHz\fi}
\def\Hz{\ifmmode $Hz$\else \,Hz\fi}

\def\muK{\ifmmode \,\mu$K$\else \,$\mu$\hbox{K}\fi}
\def\microK{\ifmmode \,\mu$K$\else \,$\mu$\hbox{K}\fi}
\def\muW{\ifmmode \,\mu$W$\else \,$\mu$\hbox{W}\fi}
\def\kms{\ifmmode $\,km\,s$^{-1}\else \,km\,s$^{-1}$\fi}
\def\kmsmpc{\ifmmode $\,km\,s$^{-1}\,$Mpc$^{-1}\else \,km\,s$^{-1}$Mpc$^{-1}$\fi}

\newcommand\ltsima{$\; \buildrel < \over \sim \;$}
\newcommand\simlt{\lower.5ex\hbox{\ltsima}}
\newcommand\gtsima{$\; \buildrel > \over \sim \;$}
\newcommand\simgt{\lower.5ex\hbox{\gtsima}}

\newbox\tablebox    \newdimen\tablewidth
\def\leaderfil{\leaders\hbox to 5pt{\hss.\hss}\hfil}
\def\endPlancktable{\tablewidth=\columnwidth 
    $$\hss\copy\tablebox\hss$$
    \vskip-\lastskip\vskip -2pt}

\def\tablenote#1 #2\par{\begingroup \parindent=0.8em
    \abovedisplayshortskip=0pt\belowdisplayshortskip=0pt
    \noindent
    $$\hss\vbox{\hsize\tablewidth \hangindent=\parindent \hangafter=1 \noindent
    \hbox to \parindent{$^#1$\hss}\strut#2\strut\par}\hss$$
    \endgroup}
\def\doubleline{\vskip 3pt\hrule \vskip 1.5pt \hrule \vskip 5pt}

\makeatother

\graphicspath{{figures/}{figures-arxiv/}}

\begin{document}
\global\long\def\pmb#1{\setbox0=\hbox{#1}
}
 \global\long\def\ltsima{$\frac{\;\buildrel<}{\sim\;}$}
 \global\long\def\gtsima{$\frac{\;\buildrel>}{\sim\;}$}
 \global\long\def\simlt{\lower.5ex\hbox{\ltsima}}
 \global\long\def\simgt{\lower.5ex\hbox{\gtsima}}
 \global\long\def\muk{$(\mu{\rm K})^{2}$ }
 \global\long\def\etal{{\it et al.}}
 \global\long\def\etals{{\it et al. }}
 \def\p 2Y{\;_2Y} \def\m 2Y{\;_{-2}Y} \global\long\def\beglet{ \addtocounter{equation}{1}
}
 \global\long\def\endlet{ \setcounter{equation}{\value{parentequation}}}

\global\long\def\adj{^{\dagger}}
\global\long\def\inv{^{-1} }

\global\long\def\nmc{\ensuremath{n_{{\rm MC}}}}
 \global\long\def\nmodes{\ensuremath{n_{{\rm modes}}}}
 \global\long\def\Bmean{\ensuremath{B_{{\rm mean}}}}
 \global\long\def\Wmean{\ensuremath{W_{{\rm mean}}}}
 \global\long\def\lmax{\ensuremath{\ell_{{\rm max}}}}
 \global\long\def\lmin{\ensuremath{\ell_{{\rm min}}}}
 \global\long\def\fixme#1{{\bf {\it #1}}}
\global\long\def\FRB#1{{\bf \textcolor{red}{\textbf{\ensuremath{\triangleleft}#1 \ensuremath{\triangleright}}}}}

\selectlanguage{english}%
\global\long\def\matthreethree#1#2#3#4#5#6#7#8#9{\left( \begin{array}{ccc}
\!#1\!  &  \!#2\!  &  \!#3\!\!\\
\!#4\!  &  \!#5\!  &  \!#6\!\!\\
\!#7\!  &  \!#8\!  &  \!#9\!\!%
\end{array}%
\right)}

\selectlanguage{british}%
\titlerunning{CMB power spectra \& likelihood} \authorrunning{\Planck\ collaboration}

\input{AuthorList_P08_Likelihood_authors_and_institutes.tex}

\title{\Planck~2013 results. XV. CMB power spectra and likelihood}

\abstract{\tiny
This paper presents the \Planck\  likelihood, a complete statistical description of the two-point correlation function of the CMB temperature fluctuations that accounts for all known relevant uncertainties, both instrumental and astrophysical in nature. We use this likelihood to derive our best estimate of the 
CMB angular power spectrum from \Planck\ over three decades in multipole moment, $\ell$, covering $2\le\ell\le2500$. The main source of error at $\ell \lesssim 1500$ is cosmic variance. Uncertainties in small-scale foreground modelling and instrumental noise dominate the error budget at higher $\ell$s. For $\ell < 50$, our likelihood exploits all \Planck\ frequency channels from 30 to 353~GHz, separating the cosmological CMB signal from diffuse Galactic foregrounds through a physically motivated Bayesian component separation technique. At $\ell \ge 50$, we employ a 
correlated Gaussian likelihood approximation based on a fine-grained set of angular cross-spectra derived from multiple detector combinations between the 100, 143, and 217~GHz frequency channels, marginalizing over power spectrum foreground templates. We validate our 
likelihood through an extensive suite of consistency tests, and assess the impact of residual foreground and instrumental uncertainties on the final 
cosmological parameters. 
We find good internal agreement among the high-$\ell$ cross-spectra with residuals below a few $\mu\textrm{K}^2$ at $\ell\lesssim 1000$, in agreement with estimated calibration uncertainties. We compare our results with foreground-cleaned CMB maps derived from all \Planck\ frequencies, as well as with cross-spectra derived from the 70~GHz \Planck\ map, and find broad agreement in terms of spectrum residuals and cosmological parameters. We further show that the best-fit \LCDM\ cosmology is in excellent agreement with preliminary \Planck\ $EE$ and $TE$ polarisation spectra. We find that the standard \LCDM\ cosmology is  well constrained by \Planck\ from the measurements at $\ell \lesssim 1500$. One specific example is the spectral index of scalar perturbations, for which we report a $5.4\,\sigma$ deviation from scale invariance, $n_{\mathrm s}\ne1$.  Increasing the multipole range beyond $\ell \simeq 1500$ does not increase our accuracy for the \LCDM\ parameters, but instead allows us to study extensions beyond the standard 
model. We find no indication of significant departures from the \LCDM\ framework. Finally, we report a tension between the \Planck\ best-fit $\Lambda$CDM model 
and the low-$\ell$ spectrum in the form of a power deficit of 5--10\% at $\ell\lesssim40$, with a statistical significance of 2.5--3$\,\sigma$.  Without a theoretically
motivated model for this power deficit, we do not elaborate further on its cosmological implications, but note that this is our most puzzling finding in an otherwise remarkably consistent 
dataset.}

\keywords{Cosmology: cosmic background radiation -- Surveys -- Methods: data
analysis}

\maketitle

\clearpage
\input 1_Introduction

\input 2_High-ell-likelihood

\input 3_Foreground-emission-model

\input 4_Combined-cross-spectra

\input 5_Reference-results-high-ell

\input 6_Assessment-accuracy-high-ell

\input 7_Consistency-checks

\input 8_Low-multipoles-likelihood

\input 9_Planck-CMB-spectrum-likelihood

\input A_10_Discussion-and-Conclusions


\begin{acknowledgements}
The development of \Planck\ has been supported by: ESA; CNES and CNRS/INSU-IN2P3-INP (France); ASI, CNR, and INAF (Italy); NASA and DoE (USA); STFC and UKSA (UK); CSIC, MICINN, JA and RES (Spain); Tekes, AoF and CSC (Finland); DLR and MPG (Germany); CSA (Canada); DTU Space (Denmark); SER/SSO (Switzerland); RCN (Norway); SFI (Ireland); FCT/MCTES (Portugal); and PRACE (EU). 

A description of the \Plancks Collaboration and a list of its members
with the technical or scientific activities they have been involved
into, can be found at \foreignlanguage{english}{\url{http://www.rssd.esa.int/index.php?project=PLANCK&page=PlanckCollaboration}}. 

We acknowledge the use of the CLASS Boltzmann code \citep{2011arXiv1104.2932L} and the Monte Python package \citep{2013JCAP...02..001A} in earlier stages of this work. 
The likelihood code and some of the validation work was built on the library pmclib from the CosmoPMC package \citep{2011arXiv1101.0950K}.

This research used resources of the IN2P3 Computer Center (http://cc.in2p3.fr) as well as of the Planck-HFI data processing center infrastructures hosted at the Institut d'Astrophysique de Paris (France) and financially supported by CNES.
\end{acknowledgements}

\bibliographystyle{aa}
\bibliography{Planck_bib,Like_bib}

\appendix


\section{High-$\ell$ likelihood details}
\input App-sec2

\input App-Sky-masks

\input App-Chance-correlations

\section{Validity tests}

\input App-sec7

\input App-Dust-353GHz

\bigskip{}
\bigskip{}
\clearpage
\raggedright
\end{document}

%% file: macros.tex
\newcommand{\WP}{WP}

\newcommand{\planckonly}{\planck}

\newcommand{\As}{A_{\rm s}}

\newcommand{\ns}{n_{\rm s}}

\newcommand{\zre}{z_{\text{re}}}
\newcommand{\yhe}{Y_{\text{P}}}

\newcommand{\rstar}{r_{\ast}}

\newcommand{\fixme}[1]{{\color{Red}{ FIXME: #1}}}

\providecommand{\planck}{\Planck}

\providecommand{\text}[1]{\rm{#1}}

\newcommand{\Mpc}{\text{Mpc}}

\providecommand{\muK}{\mu\rm{K}}

\newcommand{\lmax}{l_{\text{max}}}

\providecommand{\Omk}{\Omega_K}
\providecommand{\Oml}{\Omega_{\Lambda}}
\providecommand{\Omtot}{\Omega_{\mathrm{tot}}}
\providecommand{\Omb}{\Omega_{\mathrm{b}}}
\providecommand{\Omc}{\Omega_{\mathrm{c}}}
\providecommand{\Omm}{\Omega_{\mathrm{m}}}
\providecommand{\omb}{\omega_{\mathrm{b}}}
\providecommand{\omc}{\omega_{\mathrm{c}}}

\providecommand{\LCDM}{{$\rm{\Lambda CDM}$}}

\newcommand{\begm}{\begin{pmatrix}}
\newcommand{\enm}{\end{pmatrix}}

\newcommand\ba{\begin{eqnarray}}
\newcommand\ea{\end{eqnarray}}
\newcommand\bea{\begin{eqnarray}}
\newcommand\eea{\end{eqnarray}}

\newcommand\be{\begin{equation}}
\newcommand\ee{\end{equation}}



%% file: AuthorList_P08_Likelihood_authors_and_institutes.tex
\author{\small
Planck Collaboration:
P.~A.~R.~Ade\inst{89}
\and
N.~Aghanim\inst{62}
\and
C.~Armitage-Caplan\inst{94}
\and
M.~Arnaud\inst{76}
\and
M.~Ashdown\inst{73, 6}
\and
F.~Atrio-Barandela\inst{19}
\and
J.~Aumont\inst{62}
\and
C.~Baccigalupi\inst{88}
\and
A.~J.~Banday\inst{97, 10}
\and
R.~B.~Barreiro\inst{70}
% ...
}
\institute{\small
APC, AstroParticule et Cosmologie, Universit\'{e} Paris Diderot, CNRS/IN2P3, CEA/lrfu, Observatoire de Paris, Sorbonne Paris Cit\'{e}, 10, rue Alice Domon et L\'{e}onie Duquet, 75205 Paris Cedex 13, France\\
\and
Aalto University Mets\"{a}hovi Radio Observatory, Mets\"{a}hovintie 114, FIN-02540 Kylm\"{a}l\"{a}, Finland\\
\and
African Institute for Mathematical Sciences, 6-8 Melrose Road, Muizenberg, Cape Town, South Africa\\
\and
Agenzia Spaziale Italiana Science Data Center, c/o ESRIN, via Galileo Galilei, Frascati, Italy\\
\and
Agenzia Spaziale Italiana, Viale Liegi 26, Roma, Italy\\
\and
Astrophysics Group, Cavendish Laboratory, University of Cambridge, J J Thomson Avenue, Cambridge CB3 0HE, U.K.\\
\and
Astrophysics \& Cosmology Research Unit, School of Mathematics, Statistics \& Computer Science, University of KwaZulu-Natal, Westville Campus, Private Bag X54001, Durban 4000, South Africa\\
\and
Atacama Large Millimeter/submillimeter Array, ALMA Santiago Central Offices, Alonso de Cordova 3107, Vitacura, Casilla 763 0355, Santiago, Chile\\
\and
CITA, University of Toronto, 60 St. George St., Toronto, ON M5S 3H8, Canada\\
\and
CNRS, IRAP, 9 Av. colonel Roche, BP 44346, F-31028 Toulouse cedex 4, France\\
\and
California Institute of Technology, Pasadena, California, U.S.A.\\
\and
Centre for Theoretical Cosmology, DAMTP, University of Cambridge, Wilberforce Road, Cambridge CB3 0WA U.K.\\
\and
Centro de Estudios de F\'{i}sica del Cosmos de Arag\'{o}n (CEFCA), Plaza San Juan, 1, planta 2, E-44001, Teruel, Spain\\
\and
Computational Cosmology Center, Lawrence Berkeley National Laboratory, Berkeley, California, U.S.A.\\
\and
Consejo Superior de Investigaciones Cient\'{\i}ficas (CSIC), Madrid, Spain\\
\and
DSM/Irfu/SPP, CEA-Saclay, F-91191 Gif-sur-Yvette Cedex, France\\
\and
DTU Space, National Space Institute, Technical University of Denmark, Elektrovej 327, DK-2800 Kgs. Lyngby, Denmark\\
\and
D\'{e}partement de Physique Th\'{e}orique, Universit\'{e} de Gen\`{e}ve, 24, Quai E. Ansermet,1211 Gen\`{e}ve 4, Switzerland\\
\and
Departamento de F\'{\i}sica Fundamental, Facultad de Ciencias, Universidad de Salamanca, 37008 Salamanca, Spain\\
\and
Departamento de F\'{\i}sica, Universidad de Oviedo, Avda. Calvo Sotelo s/n, Oviedo, Spain\\
\and
Department of Astronomy and Astrophysics, University of Toronto, 50 Saint George Street, Toronto, Ontario, Canada\\
\and
Department of Astrophysics/IMAPP, Radboud University Nijmegen, P.O. Box 9010, 6500 GL Nijmegen, The Netherlands\\
\and
Department of Electrical Engineering and Computer Sciences, University of California, Berkeley, California, U.S.A.\\
\and
Department of Physics \& Astronomy, University of British Columbia, 6224 Agricultural Road, Vancouver, British Columbia, Canada\\
\and
Department of Physics and Astronomy, Dana and David Dornsife College of Letter, Arts and Sciences, University of Southern California, Los Angeles, CA 90089, U.S.A.\\
\and
Department of Physics and Astronomy, University College London, London WC1E 6BT, U.K.\\
\and
Department of Physics, Gustaf H\"{a}llstr\"{o}min katu 2a, University of Helsinki, Helsinki, Finland\\
\and
Department of Physics, Princeton University, Princeton, New Jersey, U.S.A.\\
\and
Department of Physics, University of California, Berkeley, California, U.S.A.\\
\and
Department of Physics, University of California, One Shields Avenue, Davis, California, U.S.A.\\
\and
Department of Physics, University of California, Santa Barbara, California, U.S.A.\\
\and
Department of Physics, University of Illinois at Urbana-Champaign, 1110 West Green Street, Urbana, Illinois, U.S.A.\\
\and
Dipartimento di Fisica e Astronomia G. Galilei, Universit\`{a} degli Studi di Padova, via Marzolo 8, 35131 Padova, Italy\\
\and
Dipartimento di Fisica e Scienze della Terra, Universit\`{a} di Ferrara, Via Saragat 1, 44122 Ferrara, Italy\\
\and
Dipartimento di Fisica, Universit\`{a} La Sapienza, P. le A. Moro 2, Roma, Italy\\
\and
Dipartimento di Fisica, Universit\`{a} degli Studi di Milano, Via Celoria, 16, Milano, Italy\\
\and
Dipartimento di Fisica, Universit\`{a} degli Studi di Trieste, via A. Valerio 2, Trieste, Italy\\
\and
Dipartimento di Fisica, Universit\`{a} di Roma Tor Vergata, Via della Ricerca Scientifica, 1, Roma, Italy\\
\and
Dipartimento di Matematica, Universit\`{a} di Roma Tor Vergata, Via della Ricerca Scientifica, 1, Roma, Italy\\
\and
Discovery Center, Niels Bohr Institute, Blegdamsvej 17, Copenhagen, Denmark\\
\and
Dpto. Astrof\'{i}sica, Universidad de La Laguna (ULL), E-38206 La Laguna, Tenerife, Spain\\
\and
European Southern Observatory, ESO Vitacura, Alonso de Cordova 3107, Vitacura, Casilla 19001, Santiago, Chile\\
\and
European Space Agency, ESAC, Planck Science Office, Camino bajo del Castillo, s/n, Urbanizaci\'{o}n Villafranca del Castillo, Villanueva de la Ca\~{n}ada, Madrid, Spain\\
\and
European Space Agency, ESTEC, Keplerlaan 1, 2201 AZ Noordwijk, The Netherlands\\
\and
Finnish Centre for Astronomy with ESO (FINCA), University of Turku, V\"{a}is\"{a}l\"{a}ntie 20, FIN-21500, Piikki\"{o}, Finland\\
\and
Haverford College Astronomy Department, 370 Lancaster Avenue, Haverford, Pennsylvania, U.S.A.\\
\and
Helsinki Institute of Physics, Gustaf H\"{a}llstr\"{o}min katu 2, University of Helsinki, Helsinki, Finland\\
\and
INAF - Osservatorio Astronomico di Padova, Vicolo dell'Osservatorio 5, Padova, Italy\\
\and
INAF - Osservatorio Astronomico di Roma, via di Frascati 33, Monte Porzio Catone, Italy\\
\and
INAF - Osservatorio Astronomico di Trieste, Via G.B. Tiepolo 11, Trieste, Italy\\
\and
INAF/IASF Bologna, Via Gobetti 101, Bologna, Italy\\
\and
INAF/IASF Milano, Via E. Bassini 15, Milano, Italy\\
\and
INFN, Sezione di Bologna, Via Irnerio 46, I-40126, Bologna, Italy\\
\and
INFN, Sezione di Roma 1, Universit\`{a} di Roma Sapienza, Piazzale Aldo Moro 2, 00185, Roma, Italy\\
\and
IPAG: Institut de Plan\'{e}tologie et d'Astrophysique de Grenoble, Universit\'{e} Joseph Fourier, Grenoble 1 / CNRS-INSU, UMR 5274, Grenoble, F-38041, France\\
\and
ISDC Data Centre for Astrophysics, University of Geneva, ch. d'Ecogia 16, Versoix, Switzerland\\
\and
IUCAA, Post Bag 4, Ganeshkhind, Pune University Campus, Pune 411 007, India\\
\and
Imperial College London, Astrophysics group, Blackett Laboratory, Prince Consort Road, London, SW7 2AZ, U.K.\\
\and
Infrared Processing and Analysis Center, California Institute of Technology, Pasadena, CA 91125, U.S.A.\\
\and
Institut N\'{e}el, CNRS, Universit\'{e} Joseph Fourier Grenoble I, 25 rue des Martyrs, Grenoble, France\\
\and
Institut Universitaire de France, 103, bd Saint-Michel, 75005, Paris, France\\
\and
Institut d'Astrophysique Spatiale, CNRS (UMR8617) Universit\'{e} Paris-Sud 11, B\^{a}timent 121, Orsay, France\\
\and
Institut d'Astrophysique de Paris, CNRS (UMR7095), 98 bis Boulevard Arago, F-75014, Paris, France\\
\and
Institute for Space Sciences, Bucharest-Magurale, Romania\\
\and
Institute of Astronomy and Astrophysics, Academia Sinica, Taipei, Taiwan\\
\and
Institute of Astronomy, University of Cambridge, Madingley Road, Cambridge CB3 0HA, U.K.\\
\and
Institute of Mathematics and Physics, Centre for Cosmology, Particle Physics and Phenomenology, Louvain University, Louvain-la-Neuve, Belgium\\
\and
Institute of Theoretical Astrophysics, University of Oslo, Blindern, Oslo, Norway\\
\and
Instituto de Astrof\'{\i}sica de Canarias, C/V\'{\i}a L\'{a}ctea s/n, La Laguna, Tenerife, Spain\\
\and
Instituto de F\'{\i}sica de Cantabria (CSIC-Universidad de Cantabria), Avda. de los Castros s/n, Santander, Spain\\
\and
Jet Propulsion Laboratory, California Institute of Technology, 4800 Oak Grove Drive, Pasadena, California, U.S.A.\\
\and
Jodrell Bank Centre for Astrophysics, Alan Turing Building, School of Physics and Astronomy, The University of Manchester, Oxford Road, Manchester, M13 9PL, U.K.\\
\and
Kavli Institute for Cosmology Cambridge, Madingley Road, Cambridge, CB3 0HA, U.K.\\
\and
LAL, Universit\'{e} Paris-Sud, CNRS/IN2P3, Orsay, France\\
\and
LERMA, CNRS, Observatoire de Paris, 61 Avenue de l'Observatoire, Paris, France\\
\and
Laboratoire AIM, IRFU/Service d'Astrophysique - CEA/DSM - CNRS - Universit\'{e} Paris Diderot, B\^{a}t. 709, CEA-Saclay, F-91191 Gif-sur-Yvette Cedex, France\\
\and
Laboratoire Traitement et Communication de l'Information, CNRS (UMR 5141) and T\'{e}l\'{e}com ParisTech, 46 rue Barrault F-75634 Paris Cedex 13, France\\
\and
Laboratoire de Physique Subatomique et de Cosmologie, Universit\'{e} Joseph Fourier Grenoble I, CNRS/IN2P3, Institut National Polytechnique de Grenoble, 53 rue des Martyrs, 38026 Grenoble cedex, France\\
\and
Laboratoire de Physique Th\'{e}orique, Universit\'{e} Paris-Sud 11 \& CNRS, B\^{a}timent 210, 91405 Orsay, France\\
\and
Lawrence Berkeley National Laboratory, Berkeley, California, U.S.A.\\
\and
Max-Planck-Institut f\"{u}r Astrophysik, Karl-Schwarzschild-Str. 1, 85741 Garching, Germany\\
\and
McGill Physics, Ernest Rutherford Physics Building, McGill University, 3600 rue University, Montr\'{e}al, QC, H3A 2T8, Canada\\
\and
MilliLab, VTT Technical Research Centre of Finland, Tietotie 3, Espoo, Finland\\
\and
Niels Bohr Institute, Blegdamsvej 17, Copenhagen, Denmark\\
\and
Observational Cosmology, Mail Stop 367-17, California Institute of Technology, Pasadena, CA, 91125, U.S.A.\\
\and
Optical Science Laboratory, University College London, Gower Street, London, U.K.\\
\and
SB-ITP-LPPC, EPFL, CH-1015, Lausanne, Switzerland\\
\and
SISSA, Astrophysics Sector, via Bonomea 265, 34136, Trieste, Italy\\
\and
School of Physics and Astronomy, Cardiff University, Queens Buildings, The Parade, Cardiff, CF24 3AA, U.K.\\
\and
School of Physics and Astronomy, University of Nottingham, Nottingham NG7 2RD, U.K.\\
\and
Space Sciences Laboratory, University of California, Berkeley, California, U.S.A.\\
\and
Special Astrophysical Observatory, Russian Academy of Sciences, Nizhnij Arkhyz, Zelenchukskiy region, Karachai-Cherkessian Republic, 369167, Russia\\
\and
Stanford University, Dept of Physics, Varian Physics Bldg, 382 Via Pueblo Mall, Stanford, California, U.S.A.\\
\and
Sub-Department of Astrophysics, University of Oxford, Keble Road, Oxford OX1 3RH, U.K.\\
\and
Theory Division, PH-TH, CERN, CH-1211, Geneva 23, Switzerland\\
\and
UPMC Univ Paris 06, UMR7095, 98 bis Boulevard Arago, F-75014, Paris, France\\
\and
Universit\'{e} de Toulouse, UPS-OMP, IRAP, F-31028 Toulouse cedex 4, France\\
% ...
}

%% file: 1_Introduction.tex
\section{Introduction}

This paper, one of a set associated with the 2013 release
of data from the \Planck\footnote{\Planck\ (\url{http://www.esa.int/Planck}) is a project of the European Space Agency (ESA) with instruments provided by two scientific consortia funded by ESA member states (in particular the lead countries France and Italy), with contributions from NASA (USA) and telescope reflectors provided by a collaboration between ESA and a scientific consortium led and funded by Denmark.}mission
\citep{planck2013-p01}, describes the CMB power spectra and the likelihood
that we derive from the \Planck\ data.

The power spectrum of the Cosmic Microwave Background (CMB) is a unique
signature of the underlying cosmological model \citep[e.g.,][]{spergel2003,hinshaw2012}.
It has been measured over the whole sky by \COBEs and \WMAP, and
over smaller regions by ground-based and sub-orbital experiments 
\citep[e.g.,][]{tristram/etal:2005,2006ApJ...647..823J,reichardt/etal:2009,fowler/etal:2010,das/etal:2011,keisler11,story/etal:prep,das/etal:prep}.
By mapping the whole sky to scales of a few arcminutes, \Plancks
now measures the power spectrum over an unprecedented range of scales
from a single experiment. To estimate cosmological parameters from
the power spectrum requires a likelihood function that propagates
uncertainties. 

In this paper we describe the power spectra obtained from the \Plancks
temperature data, as well as the associated likelihood function. Since
the probability distribution of the power spectrum is non-Gaussian at
large scales, we follow a hybrid approach to construct the likelihood
\citep{Efstathiou2004,Efstathiou2006}, using a Gibbs sampling based
approach at low multipoles, $\ell$, and a pseudo-C$_{\ell}$ technique
at high multipoles \citep{Hietal02}.

The high-$\ell$ part of the \Plancks likelihood is based on power
spectra estimated from each \Plancks detector in the frequency range
100 to 217\,GHz, allowing careful assessment of each detector's response
to the sky emission. We implement two independent likelihood methods.
The first, used in the distributed likelihood code, estimates the
power spectrum at every multipole, together with the associated covariance
matrix. The second takes a simplified form, binning the spectra, and
is used to explore the stability of the results with respect to different
instrumental and astrophysical systematic effects. The methods give
consistent results. 

Unresolved extragalactic foregrounds make a significant contribution
to the power spectra at high multipoles. We develop a model for these
foregrounds, designed to allow the \Plancks likelihood to be combined
with high resolution data from the Atacama Cosmology Telescope (ACT)
and the South Pole Telescope (SPT). We combine frequencies and model
unresolved foregrounds in a physical way, as in e.g.,
\cite{shirokoff10,dunkley10,reichardt12}, performing component
separation at small scales at the power spectrum level. On large
scales, $\ell<50$, Galactic contamination is more significant. We use
the \Plancks temperature maps in the range $30\le\nu\le353~\ghz$ to
separate Galactic foregrounds in the maps, and then estimate the full
probability distribution of the CMB power spectrum.

This paper is structured as follows. In Sect.~\ref{sec:High-multipoles-likelihood}
we describe the pseudo-C$_{\ell}$ likelihoods, and in Sect.~\ref{sub:Sky-model}
set up the foreground model. The power spectra and derived cosmological
parameters are presented in Sects.~\ref{sec:Combined-spectra}
and \ref{sec:Reference-model-results}, and an assessment of their
accuracy and robustness is made in Sects.~\ref{sec:HL-accuracy}
and~\ref{sec:consistency}. 

In Sect.~\ref{sec:Low-multipoles-Likelihood}
we describe the low-$\ell$ likelihood, and conclude by presenting
the complete \Plancks likelihood in Sect.~\ref{sec:Planck-Likelihood}.

%% file: 2_High-ell-likelihood.tex
\section{High-$\ell$ likelihoods\label{sec:High-multipoles-likelihood}}

The \Plancks maps consist of the order $5\times10^{7}$ pixels for
each detector, so a likelihood described directly at the
pixel level would be too time consuming. A significant compression of data can be achieved with
minimal information loss using pseudo-$C_{\ell}$ power spectra, even
in the case of incomplete sky coverage. Here we describe the form
of the likelihood function of the compressed data, given a sky signal
and instrumental model. 

Following \cite{HL08}, we assume a Gaussian form of the likelihood
based on pseudo-spectra that have been corrected to account for partial
sky masking. We use a `fine-grained' data description, computing spectra
of maps from individual detectors or detector sets. Table~\ref{tab:detsets}
describes the 13 maps used in the analysis, spanning 100 to 217~GHz.
We compute the spectra at these multiple frequencies to simultaneously
constrain the CMB and foreground contributions. We choose these frequencies
as a trade-off between adding further information, and adding further
complexity to the foreground model, which would be needed to including
the adjacent 70 and 353~GHz channels (see Sect.~\ref{sub:Sky-model}
for further discussion). In our baseline analysis the spectra are
computed at each multipole, together with an estimate of the full
covariance matrix with off-diagonal errors between different spectra
and multipoles. As in the \WMAP\ analysis, we use only cross-spectra
between detectors, alleviating the need to accurately model the mean
noise contribution.

\begin{table}[tmb] 
\begingroup 
\newdimen\tblskip \tblskip=5pt
\caption{Detectors used to make the maps for this analysis. Spider Web Bolometers (SWB)
are used individually; Polarised Sensitive Bolometer pairs (PSBs, denoted
a and b) are used in pairs, and we consider only the maps estimated
from two pairs of PSBs. The relevant effective beams, and their uncertainties,
are given in \cite{planck2013-p03c}. \label{tab:detsets}}
\vskip -6mm
\footnotesize 
\setbox\tablebox=\vbox{
\newdimen\digitwidth
\setbox0=\hbox{\rm 0}
\digitwidth=\wd0
\catcode`*=\active
\def*{\kern\digitwidth}
\newdimen\signwidth
\setbox0=\hbox{+}
\signwidth=\wd0
\catcode`!=\active
\def!{\kern\signwidth}
\newdimen\decimalwidth
\setbox0=\hbox{.}
\decimalwidth=\wd0
\catcode`@=\active
\def@{\kern\signwidth}
\halign{ \hbox to 1in{#\leaderfil}\tabskip=0em& 
    \hfil#\hfil\tabskip=1em& 
    \hfil#\hfil\tabskip=1em& 
    \hfil#\hfil& 
    \hfil#\hfil \tabskip=0pt\cr
\noalign{\doubleline}
\omit Set name\hfill  & Frequency  & Type  & Detectors   & FWHM\tablefootmark{a} \cr
\omit \hfill & {[}GHz{]}  &  &  & {[}arcmin{]}\cr
\noalign{\vskip 2pt\hrule\vskip 2pt}
100-ds0\dotfill  & 100  & PSB  & All 8 detectors  & 9.65\cr
\noalign{\vskip 4pt}
100-ds1\dotfill  & 100  & PSB  & 1a+1b + 4a+4b  & \cr
100-ds2\dotfill  & 100  & PSB  & 2a+2b + 3a+3b  & \cr
\noalign{\vskip 2pt\hrule\vskip 2pt}
143-ds0\dotfill  & 143  & MIX  & 11 detectors  & 7.25\cr
\noalign{\vskip 4pt}
143-ds1\dotfill  & 143  & PSB  & 1a+1b + 3a+3b  & \cr
143-ds2\dotfill  & 143  & PSB  & 2a+2b + 4a+4b  & \cr
143-ds3\dotfill  & 143  & SWB  & 143-5  & \cr
143-ds4\dotfill  & 143  & SWB  & 143-6  & \cr
143-ds5\dotfill  & 143  & SWB  & 143-7  & \cr
\noalign{\vskip 2pt\hrule\vskip 2pt}
217-ds0\dotfill  & 217  & MIX  & 12 detectors  & 4.99\cr
\noalign{\vskip 4pt}
217-ds1\dotfill  & 217  & PSB  & 5a+5b + 7a+7b  & \cr
217-ds2\dotfill  & 217  & PSB  & 6a+6b + 8a+8b  & \cr
217-ds3\dotfill  & 217  & SWB  & 217-1  & \cr
217-ds4\dotfill  & 217  & SWB  & 217-2  & \cr
217-ds5\dotfill  & 217  & SWB  & 217-3  & \cr
217-ds6\dotfill  & 217  & SWB  & 217-4  & \cr
\noalign{\vskip 2pt\hrule\vskip 2pt}
}}
\endPlancktable 
\endgroup
\end{table}

In this section, we begin with a reminder of the pseudo-spectrum
approach, and describe our baseline likelihood distribution, hereafter
referred to as the {\tt CamSpec} likelihood. We then show how a
compression of spectra within a given frequency can be achieved with
negligible loss of information.  We describe the signal and instrument
model, including detector noise properties, calibration, and beam
uncertainties.

Next, we describe an alternative, simpler, form of the likelihood,
hereafter referred to as \plik, based on binned power spectra with an
inverse-Wishart distribution. This does not require the
pre-computation of large covariance matrices, so changing the sky or
instrument modeling is straightforward. This simpler form of the
likelihood will be used to assess the robustness of our likelihood
methodology with respect to technical choices and astrophysical
foreground modeling.

In Sect.~\ref{sub:mapCheck} we also compare these likelihoods to
pseudo-spectra computed directly from CMB maps estimated by
multi-frequency component separation \citep{planck2013-p06}.

\subsection{The $\camspec$ likelihood}

We define $\tilde{T}_{\ell m}^{i}$ as the spherical harmonic coefficients
of the weighted temperature map of detector $i$. The pseudo-spectrum
at multipole $\ell$, for the detector pair $(i,j),$ is then given
by
\begin{equation}
\tilde{\mathbf{C}}_{\ell}^{ij}=\frac{1}{2\ell+1}\sum_{m}\tilde{\mathbf{T}}_{\ell m}^{i}\tilde{\mathbf{T}}_{\ell m}^{j\dagger},\label{C1}
\end{equation}
where the dagger, ${\dagger}$, denotes the Hermitian transpose. This
is related to the `deconvolved' spectrum, $\hat{C}^{T_{ij}}$, by a
coupling matrix,
\begin{equation}
\tilde{C}^{T_{ij}}=M_{ij}^{TT}\hat{C}^{T_{ij}}.\label{C2}
\end{equation}
For an isotropic signal on the sky, the ensemble average of these
deconvolved spectra are equal to the spectra of the theoretical models
(including CMB and isotropic unresolved foregrounds) that we wish
to test. For completeness, the coupling matrices are given explicitly
in Appendix~\ref{app:Power-Spectra}.

In the first method, \camspec, we form the deconvolved spectra
$\hat{C}_{\ell}$ without any prior smoothing of the pseudo-spectra
$\tilde{C}_{\ell}$.  Even for the largest sky masks used in our
analysis (see Sect.~\ref{sub:Sky-model}), the coupling matrices are
non-singular. The deconvolution requires the evaluation of $\sim
N_{\mathrm{map}}^{2}$ coupling matrices for a data set with
$N_{\mathrm{map}}$ sky maps, which takes a moderate, but not
excessive, amount of computer time.

A more challenging computational task is to compute the covariances
of the pseudo-spectra, i.e., ${\rm Cov}(\tilde{C}^{T_{ij}}\tilde{C}^{T_{pq}}).\label{C3}$
Here we need to compute $N_{{\rm map}}^{4}$ coupling matrices, and
the problem rapidly becomes computationally intractable even for relatively
low values of $N_{{\rm map}}$. For the moment we will assume that
these covariance matrices are available and describe their computation
in Appendix \ref{app:Covariance-matrix-of-combined-estimates}. We
will use the notation ${\bf \tilde{X}}={\rm Vec}({\bf \tilde{C}})\label{C4}$
to denote a column vector for which the index $p$ of a single element
$X_{p}$ denotes the map combination ($i,j$) and multipole $\ell$.
We denote the covariance matrix of this vector as 
\begin{equation}
\tilde{\mathcal{M}}=({\bf \tilde{X}}-\langle{\bf \tilde{X}}\rangle)({\bf \tilde{X}}-\langle{\bf \tilde{X}}\rangle)^{T}.\label{C5}
\end{equation}

As explained later,
the deconvolved detector set cross-spectra given by Eq.~\ref{C2} can
be efficiently combined within a given frequency pair after a small
effective recalibration, taking into account their respective
isotropised beam transfer function and noise levels (see
Appendix~\ref{app:Combining-intra-frequency} for the detailed
procedure). Covariance estimates of these combined spectra can be
deduced from those of the detector set cross-spectra.
The covariance matrix is computed for a fixed fiducial model, and we
approximate the likelihood as a Gaussian, described in
Appendix~\ref{app:The-fiducial-gaussian}.  The likelihood thus takes
the form $p={\rm e}^{-S}$ with
\begin{center}
$S=\frac{1}{2}\left({\bf \hat{X}}-{\bf X}\right)^T\hat{{\bf \mathcal{M}}}^{-1}\left({\bf \hat{X}}-{\bf X}\right).$
\par\end{center}

For the current analysis we include the following (deconvolved)
spectrum combinations,
\begin{equation}
{\bf {\hat{X}}}=(\hat{C}_{\ell}^{100\times100},\hat{C}_{\ell}^{143\times143},\hat{C}_{\ell}^{217\times217},\hat{C}_{\ell}^{143\times217}),\label{CSL1}
\end{equation}
coupled to a parametric model of the CMB and foreground power
spectra. The multipole ranges we select depend on frequency, as
described in Sect.~\ref{sec:Reference-model-results}. We do not
include the $100\times143$ and $100\times217$ spectra since these
spectra carry little additional information about the primary CMB
anisotropies, but would require us to solve for additional unresolved
foreground parameters. This tradeoff of information versus complexity
was also considered for the use of the 70~GHz and 353~GHz data, which
we choose not to include except for cross-checks.

\begin{figure*}
\begin{centering}
\includegraphics[bb=0bp 0bp 1039bp 288bp,clip,width=1\textwidth]{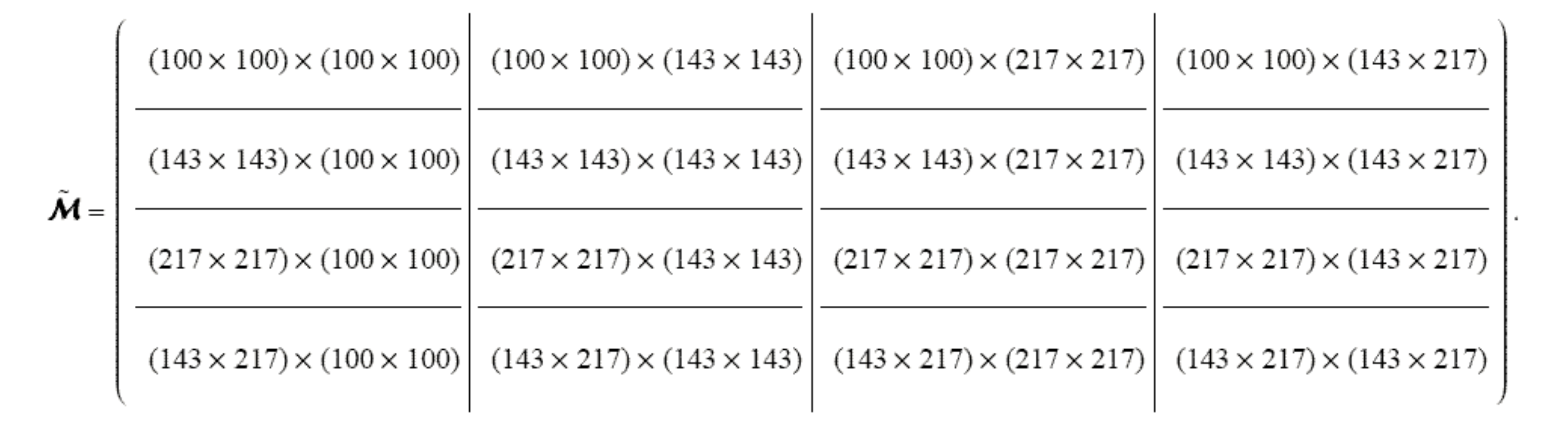} 
\par\end{centering}

\caption{The covariance matrix blocks used in the likelihood, accounting for
the correlations between cross-spectra estimated from the 100, 143,
and 217\,GHz channels. \emph{\label{fig:likecoveq}}}
\end{figure*}

The fiducial covariance matrix is composed of the blocks shown in
Fig.~\ref{fig:likecoveq}. The off-diagonal blocks in this matrix
accurately account for the correlations between the power spectra at
different frequencies.

This description would be sufficient for perfectly known calibrations
and beam transfer functions of each detector sets' cross-spectra. \cite{planck2013-p03c}
describes in detail these uncertainties, and shows
that for each detector set pair, $(i,j)$, the effective beam transfer
function can be expressed as
\begin{align}
B^{ij}(\ell) & =\Bmean^{ij}(\ell)\exp\left(\sum_{k=1}^{n_{\mathrm{modes}}}g_{k}^{ij}E_{k}^{ij}(\ell)\right),
\label{eq:detsets-spectra-beams}
\end{align}
described further in Appendix~\ref{app:Det-beam-errors},
with $n_{\mathrm{modes}}$ beam error eigenmodes $E_{k}^{ij}(\ell)$, and their
covariance matrix. These modes are then combined into generalised
beam eigenmodes corresponding to the  spectra ${\bf \hat{X}}$.
The associated
covariance matrix is used to construct a Gaussian posterior distribution
of the eigenmodes, which allows marginalization over the uncertainties.

Finally, in the construction of the covariance matrix, one needs to
accurately specify the contribution of the instrumental noise. 
Even if there is no bias on the spectra due to instrumental noise, having removed auto-spectra, the latter dominates
the covariance matrix on small scales. Fortunately, 
the \Planck\ scanning strategy at the ring level allows us to make estimates of the noise
pseudo-spectra from half-ring difference maps \citep[see~][]{planck2013-p06}.
These half-ring difference maps, together with the knowledge of the
noise variance per pixel for each detector set, can be used to derive
the noise contribution to the covariance matrix with good accuracy
(see Appendix~\ref{app:Detector-noise-model} for details).

\subsection{The \plik\ likelihood\label{sub:Plikdef}}

We now describe the alternative form of the likelihood, inspired by \cite{smica}, used for
cross-checks and robustness tests. We start from the full-sky exact
likelihood for a Gaussian signal, which for $N_{\textrm{map}}$ detector maps is given by
\[
p(\mathrm{maps}|\theta)\propto\exp-\bigl\{\sum_{\ell}(2\ell+1)\mathcal{K}\bigl(\hat{C}_{\ell},\,\mathbf{C}_{\ell}(\theta)\bigr)\bigr\},
\]
where $\theta$ is a vector containing the parameters
of the signal model, and $\hat{C}_{\ell}$ are the empirical
angular spectra.  $\mathcal{K}\bigl(\mathbf{A},\,\mathbf{B}\bigr)$
denotes the Kullback divergence between two $n$-variate zero-mean
Gaussian distributions with covariance matrices $\mathbf{A}$ and
$\mathbf{B}$, and is given by
\[
\mathcal{K}\bigl(\mathbf{A},\,\mathbf{B}\bigr)=\frac{1}{2}\Bigl[\mathrm{tr}\bigl(\mathbf{A}\mathbf{B}\inv\bigr)-\log\det\bigl(\mathbf{A}\mathbf{B}\inv\bigr)-n\Bigr].
\]

As already noted, a sky cut introduces off-diagonal couplings between
different multipoles. In this method we bin the power spectra in such
a way that these off-diagonal terms of the covariance are
negligible. This is adequate to model sources with slowly varying
spectra, such as foregrounds, and the CMB anisotropies for standard
cosmologies. In this case, the likelihood takes the form
\begin{equation}
p(\mathrm{maps}|\theta)\propto\exp-\mathcal{L}(\theta),\quad\mathrm{with\quad\mathcal{L}(\theta)=\sum_{q=1}^{Q}n_{q}\,\mathcal{K}\bigl(\hat{C}_{q},\,\mathbf{C}_{q}\bigr),}
\end{equation}
where the angular spectra $\hat{C}_{\ell}$ for each cross-frequency
spectrum have been averaged into $Q$ spectral bins using spectral
windows $w_{q}(\ell)$ ($q=1,\ldots,Q$), with 
\[
\hat{C_{q}}=\sum_{\ell}w_{q}(\ell)\,{}_{\ell}\hat{C_{\ell}},\qquad\mathbf{C}_{q}=\sum_{\ell}w_{q}(\ell)\,\mathbf{C}_{\ell}.
\]
Here $w_{q}(\ell)$ denotes the window function for the $q$-th bin,
and the same symbol, $\mathbf{C}$, is used to denote binned or unbinned
spectra. The effective number of modes in the $q$-th bin is 
\[
n_{q}=f_{\mathrm{sky}}\cdot\frac{(\sum_{\ell}w_{q}(\ell)^{2})^{2}}{\sum_{\ell}w_{q}(\ell)^{4}/(2\ell+1)}.
\]
We adopt a spectral binning defined by 
\[
w_{q}(\ell)=\begin{cases}
\frac{\ell(\ell+1)(2\ell+1)}{\sum_{\ell_{\mathrm{min}}^{q}}^{\ell_{\mathrm{max}}^{q}}\ell(\ell+1)(2\ell+1)} & \ell_{\mathrm{min}}^{q}\leq\ell\leq\ell_{\mathrm{max}}^{q},\\
0 & \text{otherwise.}
\end{cases}
\]
The $\pliks$ bin width is $\Delta\ell=9$ from $\ell=100$
to $\ell=1503$, then $\Delta\ell=17$ to $\ell=2013$, and finally
$\Delta\ell=33$ to $\ell_{\textrm{max}}=2508$. This ensures that correlations
between any two bins are smaller than 10\,\%.

While this binned likelihood approximation does not fully capture all
couplings between different multipoles, it has a notable advantage in
computational speed, and it agrees well with the primary
likelihood. It is therefore very well suited for performing an extensive
suite of robustness tests, as many more parameters can be considered
in a short time. Further, instrumental effects can be investigated quickly to assess
the agreement between pairs of detectors within a frequency channel,
such as individual detector calibrations and beam errors.

A specific example is the impact of beam uncertainty parameters on the
likelihood. This can be investigated by re-expressing the model
covariance matrices as
\begin{equation}
\mathbf{C}_{\ell}=\mathbf{B}_{\ell}(\gamma)\,\mathbf{C}_{\ell}(\theta)\,\mathbf{B}_{\ell}(\gamma)^{\mathrm{T}},\label{eq:plik-beam-equation}
\end{equation}
where $\mathbf{C}_{\ell}(\theta)$ is the model covariance including
both signal and noise, and ${\bf B}_{\ell}(\gamma)$ is a diagonal
matrix encoding the beam (and calibration) errors with elements given
by \footnote{From Eqs. \ref{eq:detsets-spectra-beams}, \ref{eq:plik-beam-equation}, and \ref{eq:detsets-auto-beams}, we have $\delta_{k}^{i}=g_{k}^{ii}/2$
at first order.}
\begin{equation}
B_{\ell}^{i}(\gamma)={\rm exp} \left(\sum_{k=1}^{\nmodes}\delta_{k}^{i}\, E_{k}^{ii}(\ell)\right).
\label{eq:detsets-auto-beams}
\end{equation}
Here, $E_{k}^{ii}(\ell)$ are the eigenmodes of the (auto-)spectra,
similar to Eq.~\ref{eq:detsets-spectra-beams}.  Note that
Eq.~\ref{eq:plik-beam-equation} does not contain the \emph{mean} beam
transfer function, since it is already included in the empirical
spectra. Thus, using Eq.~\ref{eq:plik-beam-equation} \plik\ approximates
the cross-spectrum beam errors as the harmonic mean of the
corresponding auto-spectrum beam errors, under the assumption that
$\mathbf{B}_{\ell}$ is diagonal between detectors. This approximate factorisation
is intrinsically linked to the assumed Kullback shape of the \pliks
likelihood, and is later demonstrated to work well for both
simulations and data.

The \pliks\ likelihood method also provides a direct estimate of the
detector noise power spectra as it can include the empirical
auto-spectra, and we find that these noise estimates are in good
agreement with the noise spectra used to construct the \camspecs
likelihood covariance matrix. The method can also produce a binned CMB
power spectrum independent of the underlying cosmological model,
providing a direct quality assessment of the foreground model
parametrisation. In practice, we proceed in two steps. First, we
jointly estimate the noise together with all other parameters using
both auto and cross-spectra. Then we fix the noise estimates, and use
the fiducial Gaussian approximation to explore the remaining free
parameters excluding the auto-spectra, optionally including only
specific data combinations.

%% file: 3_Foreground-emission-model.tex
\section{Foreground emission model and sky masks\label{sub:Sky-model}}

\subsection{Sky masks\label{sub:Sky-masks}}

The Galactic emission varies strongly in both complexity and strength
across the sky. It is therefore necessary to find a balance between
maximizing the sky coverage to reduce statistical uncertainties, and
establishing a simple yet efficient foreground model. In this paper,
we threshold the $353$\,\GHz\ temperature map to define a basic set of
diffuse Galactic masks (shown in Appendix\,\ref{App:Sky-masks}), which
form a sequence of increasing sky fraction, to minimize the
contribution from diffuse dust emission. The sky fractions retained by
these masks is summarized in Table~\ref{tab:mask_fsky}.

\begin{table}[tmb]
\begingroup
\newdimen\tblskip \tblskip=5pt
\caption{Area of sky retained by combining diffuse foreground and point source masks, once apodised.\label{tab:mask_fsky}}
\vskip -6mm
\footnotesize
\setbox\tablebox=\vbox{ 
\newdimen\digitwidth
\setbox0=\hbox{\rm 0}
\digitwidth=\wd0
\catcode`*=\active
\def*{\kern\digitwidth}
\newdimen\signwidth
\setbox0=\hbox{+}
\signwidth=\wd0
\catcode`!=\active
\def!{\kern\signwidth}
\newdimen\decimalwidth
\setbox0=\hbox{.}
\decimalwidth=\wd0
\catcode`@=\active
\def@{\kern\signwidth}
\halign{ \hbox to 1in{#\leaderfil}\tabskip=0em& 
    \hfil#\hfil\tabskip=1em& 
    \hfil#\hfil\tabskip=1em& 
    \hfil#\hfil& 
    \hfil#\hfil \tabskip=0pt\cr
\noalign{\doubleline}
\omit Mask\hfill  & Sky fraction  & Sky area  \cr
\omit \hfill & {[}\%{]}  &{[}deg$^2${]} \cr
\noalign{\vskip 2pt\hrule\vskip 2pt}
CL31& 30.71& 12\,668\cr
CL39& 39.32& 16\,223\cr
CL49& 48.77& 20\,121\cr
\noalign{\vskip 2pt\hrule\vskip 2pt}
}}
\endPlancktable
\endgroup
\end{table}
\begin{figure}
\begin{centering}
\includegraphics[bb=0bp 0bp 483bp 243bp,clip,width=1\columnwidth]{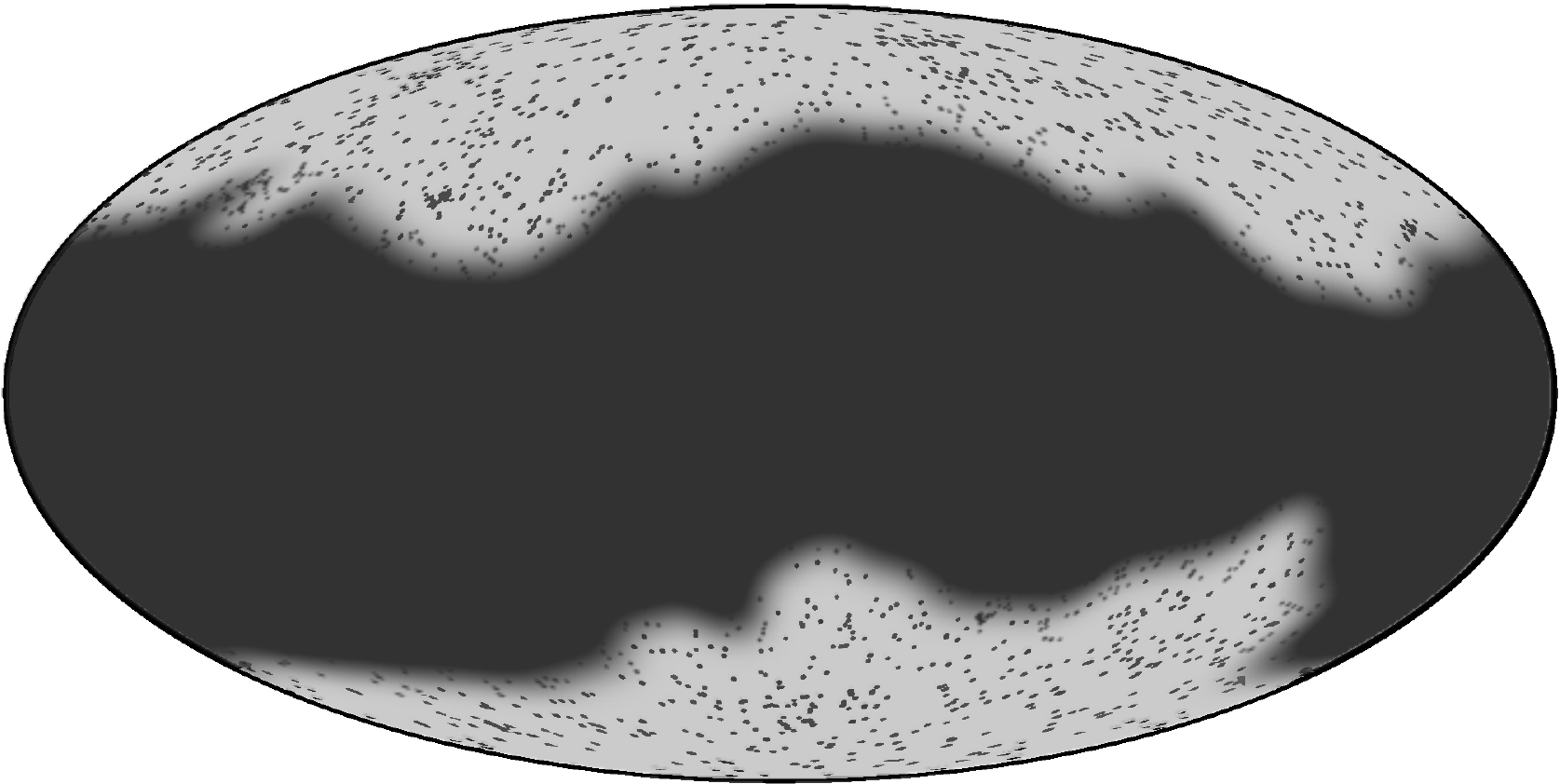}
\includegraphics[bb=0bp 0bp 483bp 243bp,clip,width=1\columnwidth]{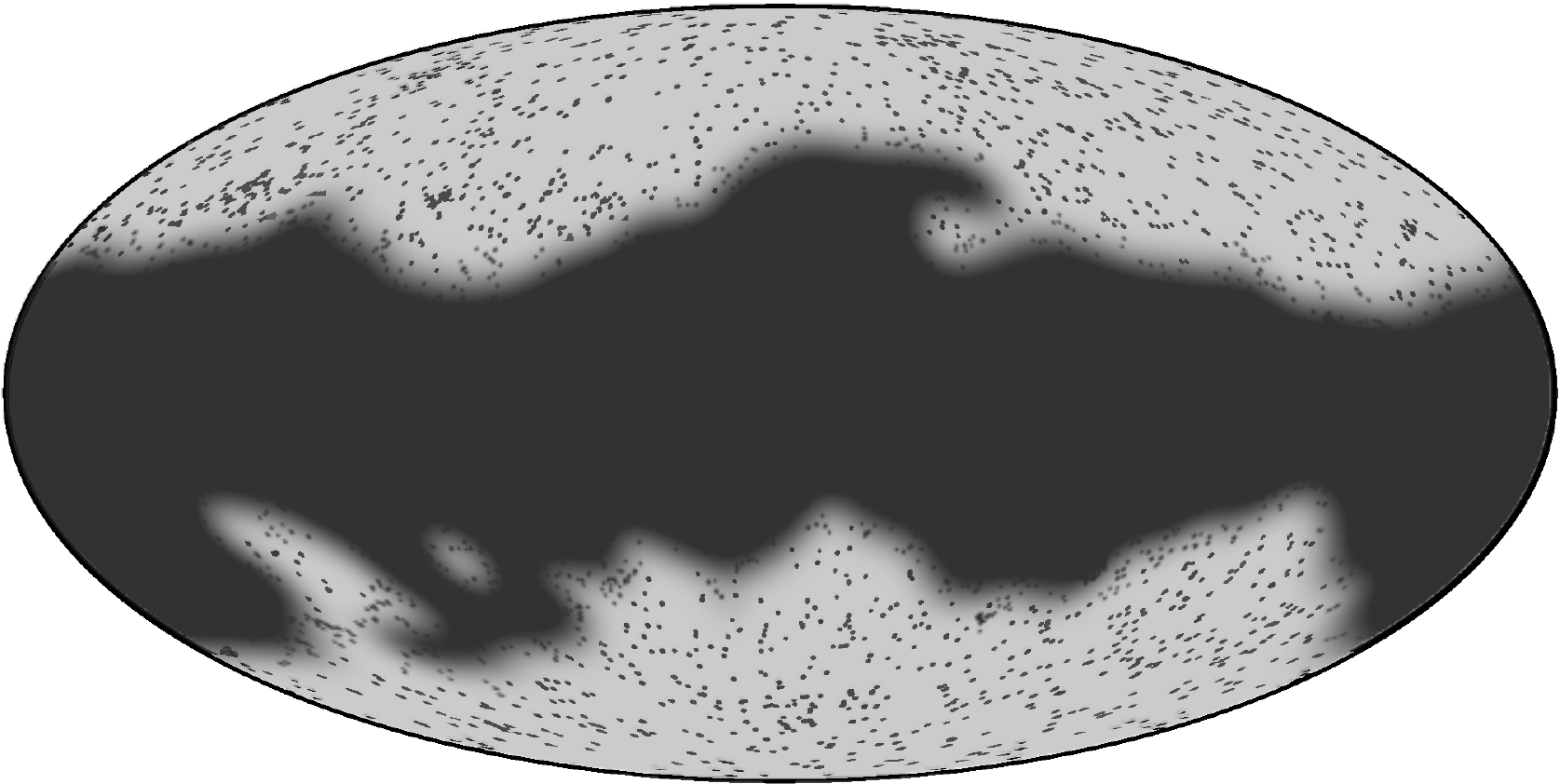} 
\includegraphics[bb=0bp 0bp 483bp 243bp,clip,width=1\columnwidth]{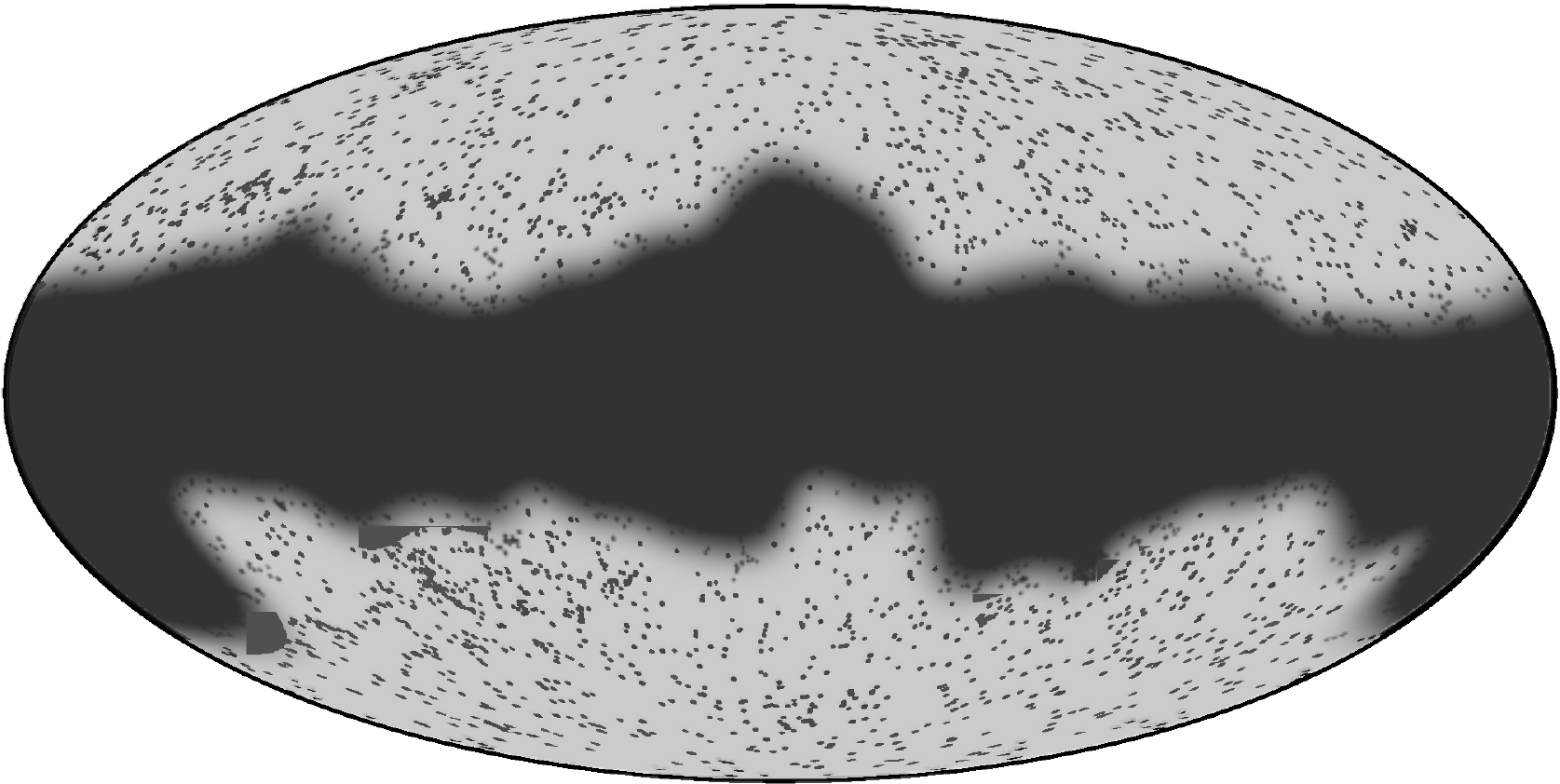}
\caption{The set of masks (CL31, CL39, CL49) used for the likelihood analyses. \label{fig:CLmasks} }
\end{centering}
\end{figure}

For \Planck, we need to estimate the covariance matrices to percent
level precision. For temperature spectra, and in the absence of point
source holes, this precision can be achieved with sharp, non-apodised
Galactic masks \citep{Efstathiou2004}. However, the inclusion of point
source holes introduces non-negligible low-$\ell$ power leakage,
which in turn can generate errors of a few percent in the covariance
matrices. We reduce this leakage by apodising the diffuse Galactic
masks (see Appendix~\ref{App:Sky-masks} for details). 
The point source mask is based on the union of the point sources
detected between $100$ and $353$\,GHz, and is also apodized.  
The point source flux cut is not critical, since the
amplitudes of the Poisson contributions of unresolved sources are
allowed to vary over a wide range in the likelihood analysis. Thus, we
do not impose tight priors from source counts and other CMB
experiments on the Poisson amplitudes. A set of the combined Galactic and point source masks, referred to as `CLx', where `x' is the percentage of sky retained, are shown in Fig.~\ref{fig:CLmasks}.

\subsection{Galactic emission\label{sub:Galactic-emission}}

\begin{figure}
\centering{}\includegraphics[bb=40bp 70bp 600bp 510bp,clip,width=1\columnwidth]{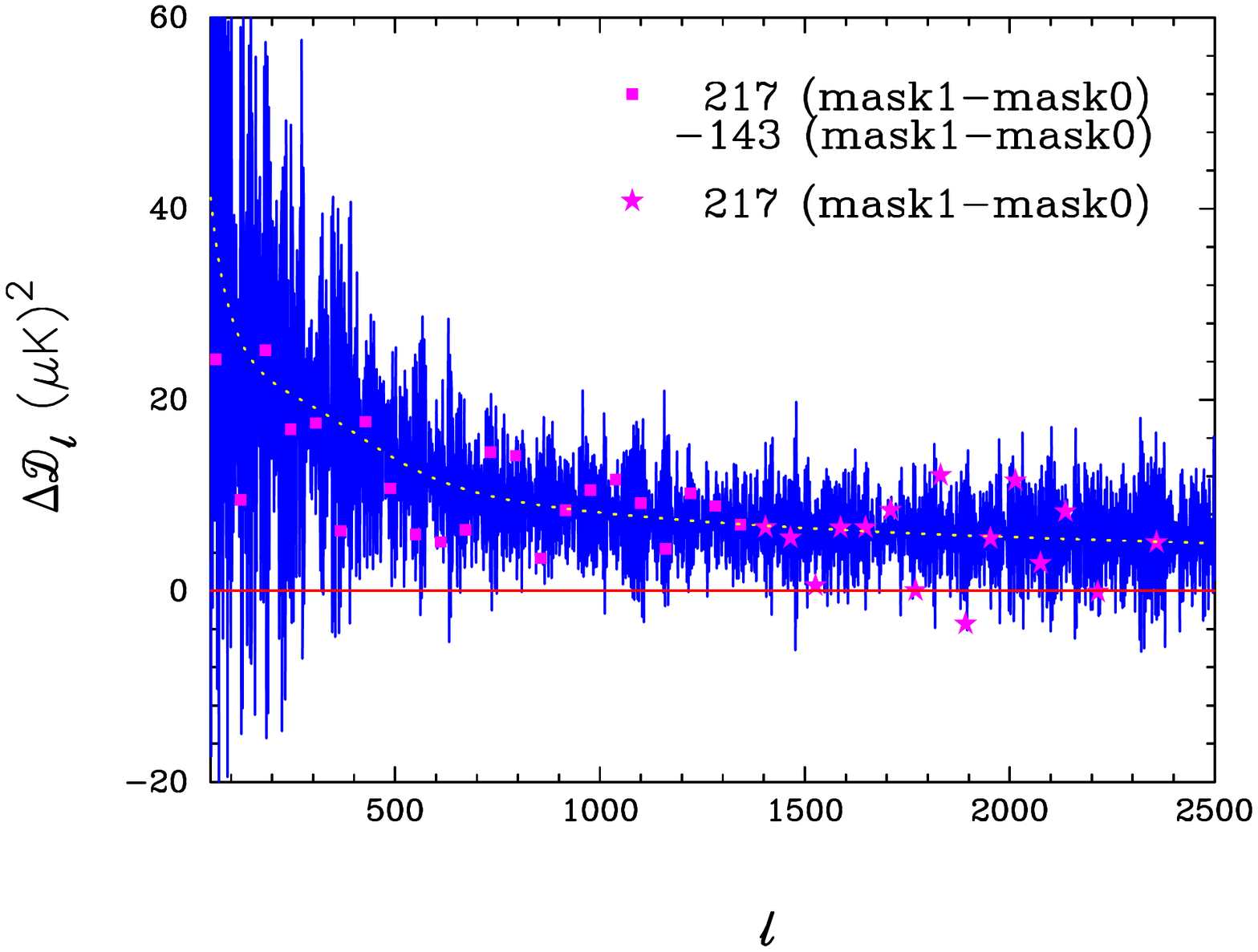}
\caption{Differences between power spectra evaluated from masks1 and mask0, showing
  the presence of Galactic dust. For $\ell\le1400$ the spectra show
  the 217$-$143 ``double-differenced'' power spectrum, rescaled to
  correct for dust emission at $143$\,GHz. For $\ell>1400$ the $217$
  mask differenced power spectrum is plotted. The blue
  line shows the 857\,GHz mask-differenced power spectrum scaled to
  $217$\,GHz as described in the text, fit by the dotted
  line. \label{fig:doublediff}}
\end{figure}

The contamination from diffuse Galactic emission at low to
intermediate multipoles can be reduced to low levels compared to CMB
anisotropies by a suitable choice of masking. However, even with
conservative masking, the remaining Galactic emission at high
multipoles is non-negligible compared to other unresolved components,
such as the Cosmic Infrared Background (CIB) anisotropies at 143 and
217\,GHz. A clear way of demonstrating this is by differencing the
power spectra computed with different masks, thereby highlighting the
differences between the isotropic and non-isotropic unresolved
components. Figure~\ref{fig:doublediff} shows (up to $\ell \le 1400$)
the 217\,GHz power spectrum difference for the mask1 and mask0 masks\footnote{
These are the combination of the non-apodised Galactic masks G35 and G22 with 
the apodised point source mask PSA82.}, minus
the corresponding difference for the $143$\,GHz frequency channel. Any
isotropic contribution to the power spectrum (CMB, unresolved
extragalactic sources, etc.)  will cancel in such a double difference,
leaving a non-isotropic signal of Galactic origin, free of the CMB induced cosmic variance scatter. 
Above $\ell>1400$, Fig.~\ref{fig:doublediff} shows the mask differenced 217~GHz power
spectrum, as the instrumental noise becomes significant at
$\ell\gtrsim1400$ for the 143~GHz channel.

In the same figure, these difference spectra are compared to the
unbinned mask-differenced 857\,GHz power spectrum, scaled to 217\,GHz
adopting a multiplicative factor\footnote{The scaling coefficient for
  the 143\,GHz spectrum is $(3.14\times10^{-5})^{2}$, derived from the
  7-parameter fitting function of Eq.~\ref{N3}.} of
$(9.93\times10^{-5})^{2}$; the dotted line shows a smooth fit to the
unbinned spectrum.  The agreement between this prediction and the
actual dust emission at $217$\,GHz is excellent, and this demonstrates
conclusively the existence of a small-scale dust emission component
with an amplitude of $\sim5-15\,\mu\textrm{K}^{2}$ at 217\,GHz if mask1 is used.

For cosmological parameter analysis this small-scale dust component
must be taken into account, and several approaches may be considered:
\begin{enumerate}
\item Fit to a template shape, e.g., as shown by the dotted line in Fig.~\ref{fig:doublediff}.
\item Reduce the amplitude by further masking of the sky.
\item Attempt a component separation by using higher frequencies.
\end{enumerate}
The main disadvantage of the third approach is a potential
signal-to-noise penalty, depending on which frequencies are used, as
well as confusion with other unresolved foregrounds. This is
particularly problematic with regards to the CIB, which has a spectrum
very similar to that of Galactic dust. In the following we therefore
adopt the two former solutions. 

\begin{figure}
\centering{}\includegraphics[bb=40bp 70bp 600bp
  510bp,clip,width=0.5\textwidth]{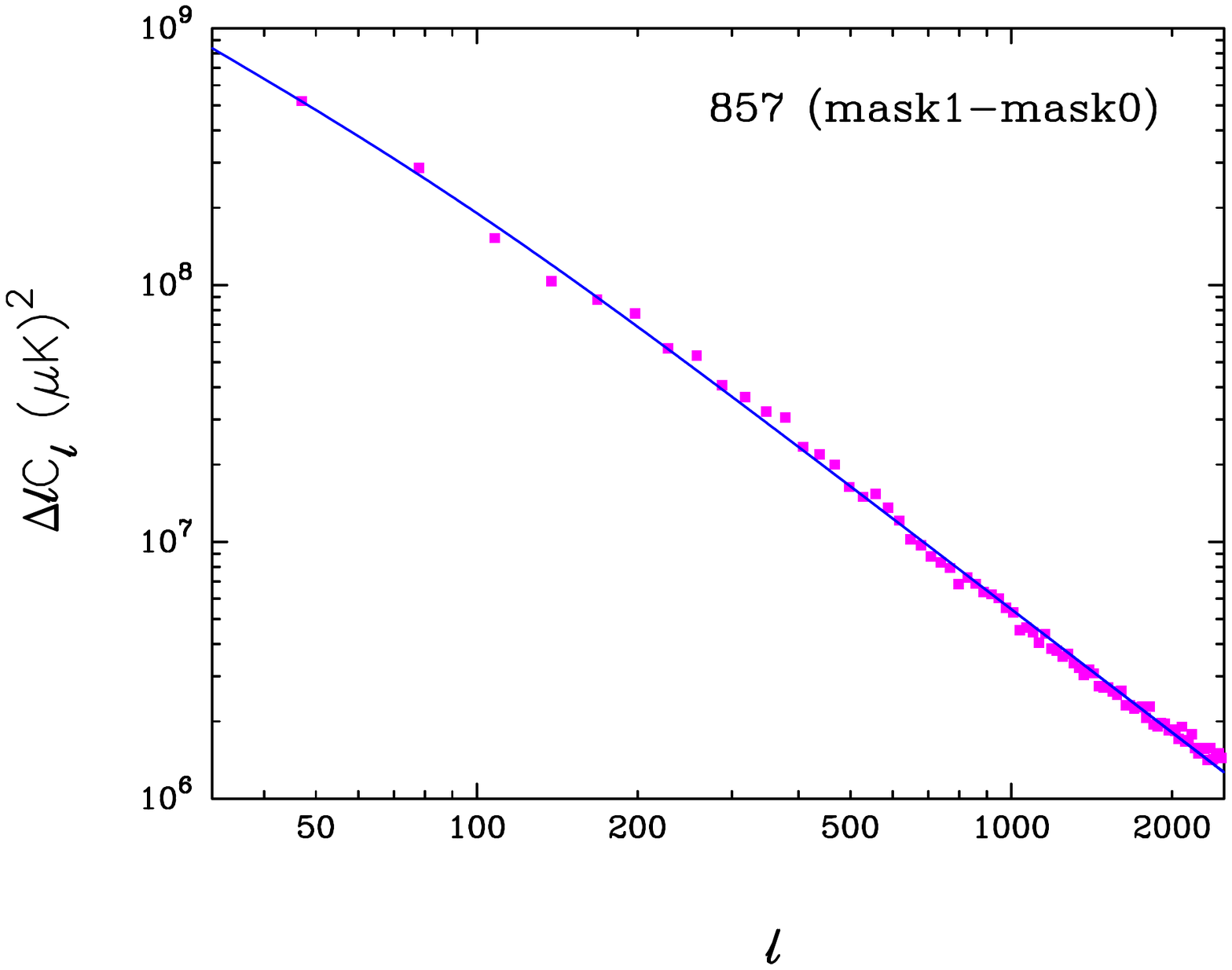}\caption{\label{857log}857\,GHz
  mask-differenced power spectrum (points), interpreted
as Galactic dust emission. The solid line shows the best-fit model
defined by Eq.~\ref{SSD1}.}
\end{figure}

It is important to understand the nature of the small scale dust
emission, and, as far as possible, to disentangle this emission from
the CIB contribution at the HFI cosmological frequencies. We use the
$857$\,GHz power spectrum for this purpose, noting that the dust
emission at 857~GHz is so intense that this particular map provides an
effectively noise-free dust emission map. In Fig.~\ref{857log} we
again show the 857\,GHz mask power spectrum difference, but this time
plotted on a log-log scale. The solid line shows the corresponding
best-fit model defined by
\begin{equation}
{\cal D}_{\ell}=\frac{A\,(100/\ell)^{\alpha}}{[1+(\ell/\ell_{c})^{2}]^{\gamma/2}},\label{SSD1}
\end{equation}
with $A=5.729\times10^{8}\,\mu\textrm{K}^2$, $\alpha=0.169$,
$\ell_{c}=905$, and $\gamma=0.427$.  At high multipoles this fit
asymptotically approaches $C_{\ell}\propto\ell^{-2.6}$, which is
compatible with previous knowledge about diffuse Galactic emission,
i.e., a power-law behaviour with an index close to $-3$ extending to
high multipoles \citep[see e.g.,][]{MDetal07}.

\begin{figure}[b]
\begin{centering}
\includegraphics[bb=40bp 70bp 600bp 570bp,clip,width=1\columnwidth]{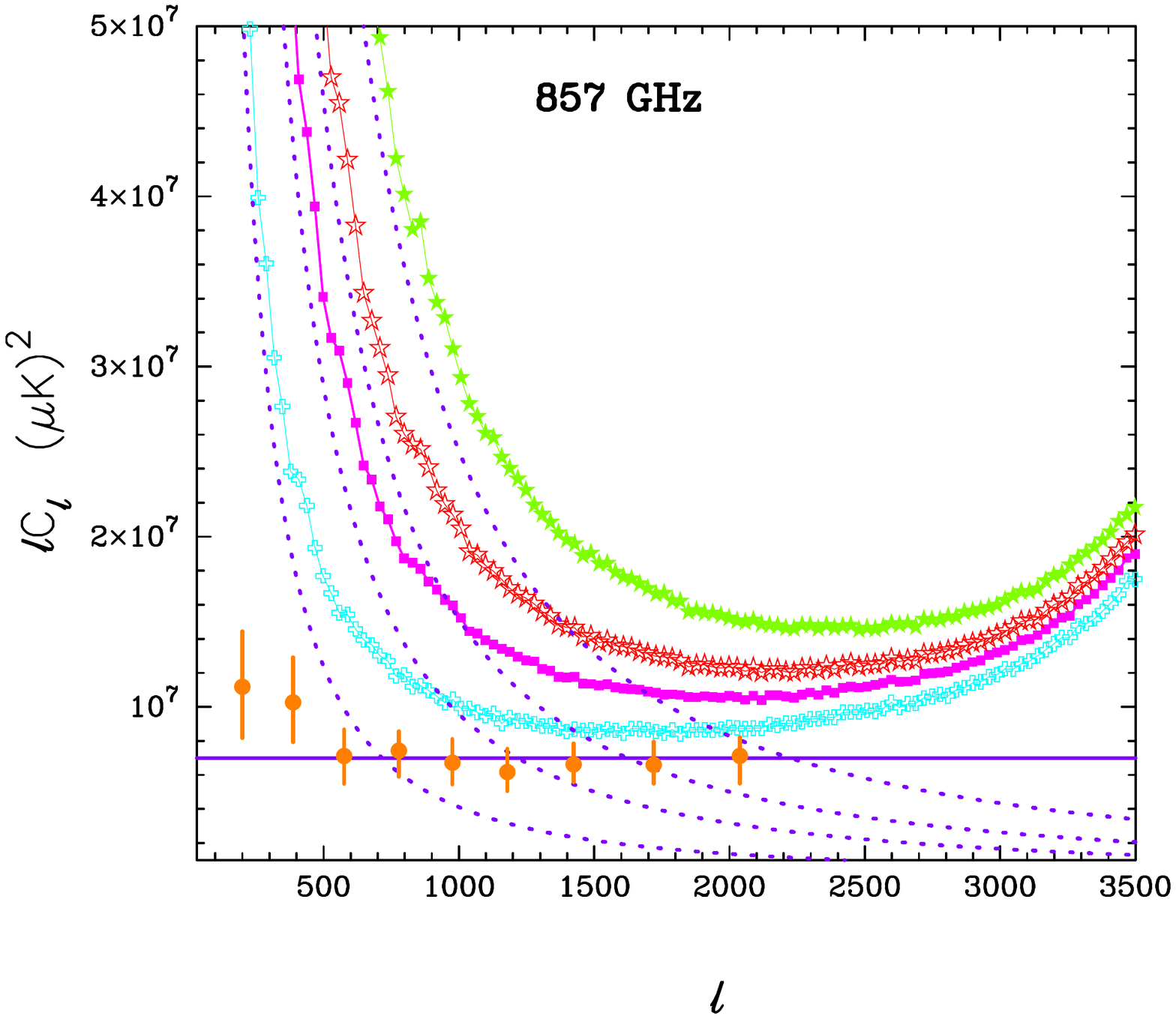} 
\par\end{centering}
\begin{centering}
\includegraphics[bb=40bp 70bp 600bp 570bp,clip,width=1\columnwidth]{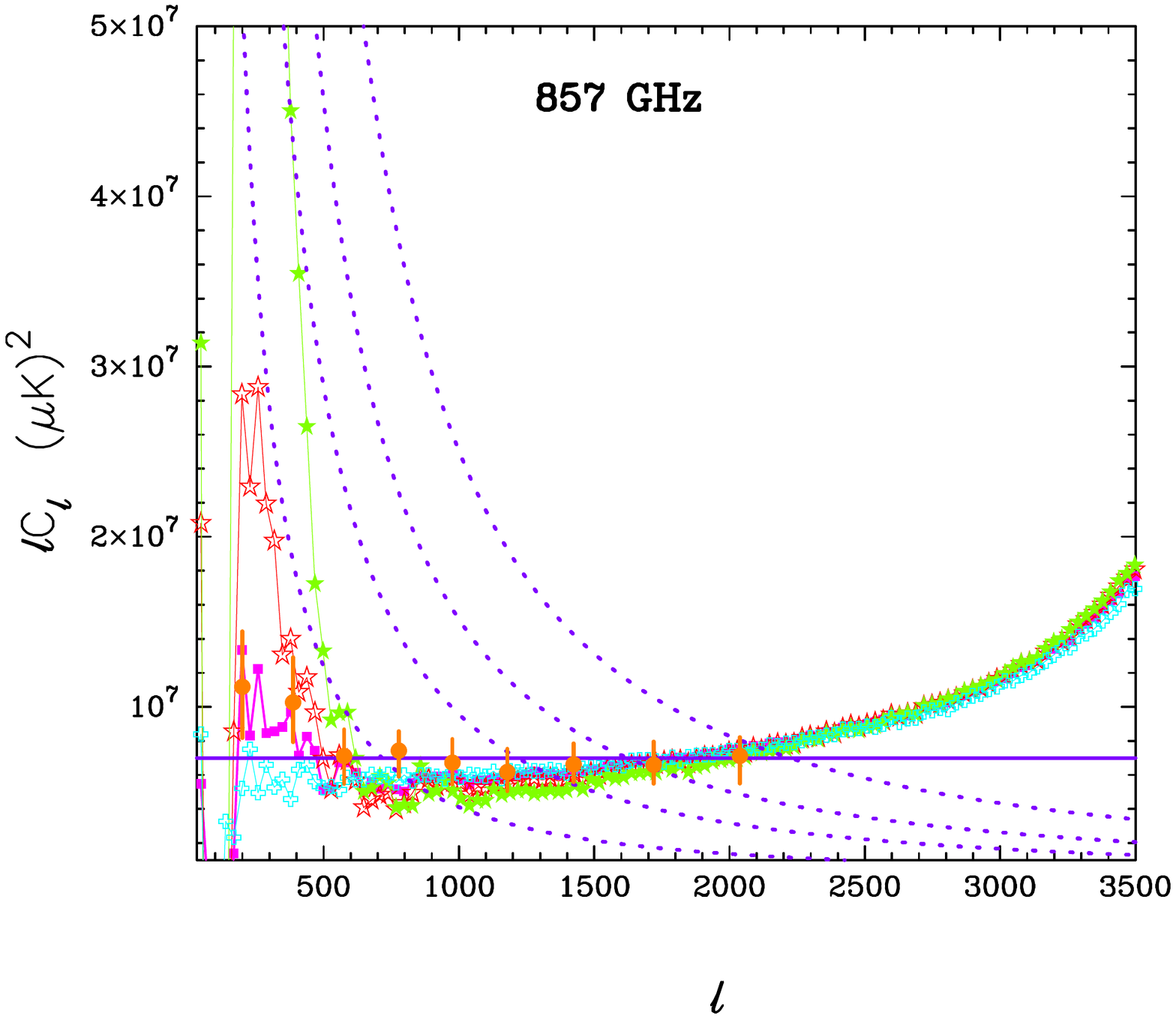} 
\par\end{centering}
\caption{\label{857fit} {\emph{Top}}: 857\,GHz power spectra for
  the four different masks defined in the text. The dotted lines show
  the best-fit model defined by Eq.~\ref{SSD1} fit to $\ell\le500$,
  capturing the Galactic dust. An estimate of the CIB power spectrum
is shown in orange points \cite{PlanckCIB}. {\emph{Bottom}}: Power spectra after subtracting the Galactic dust model. }
\end{figure}

The upper panel in Fig.~\ref{857fit} shows the 857\,GHz spectra for
the four Galactic masks (G22, G35, G45 through to G56) with the point-source mask
applied. They are compared to the 857\,GHz CIB power spectrum from
\cite{PlanckCIB}, which for $\ell>500$ can be described approximately
as $\ell\, C_{\ell}\approx7\times10^{6}\,\mu{\rm K}^{2}$. The best
fit models of Eq.~\ref{SSD1} are also shown, fitted to $\ell\le500$,
where we expect diffuse emission to be dominant. The lower panel of
Fig.~\ref{857fit} shows the same power spectra after subtracting the
best-fit dust model. After subtracting the Galactic dust component,
the recovered power spectra are consistent with the CIB measured in
\citet{PlanckCIB} for all masks. The excess at high multipoles may be
due to a combination of aliasing of large scale power through the
point source masks at $\ell\gtrsim3000$, Galactic point sources, and
uncertainties in the 857\,GHz beams.

The model explains by construction the `double-difference' plot shown
in Fig.~\ref{fig:doublediff}. Specifically, this emission is
consistent with a cirrus-like power spectrum,
$C_{\ell}\propto\ell^{-2.6}$, extrapolated to high
multipoles. Furthermore, the results of Fig.~\ref{857fit} demonstrate
that over a wide area of sky, we can understand the 857\,GHz power
spectrum in terms of a `universal' cirrus spectrum together with an
isotropic CIB component. These results provide strong evidence that an
extragalactic CIB component dominates over the diffuse Galactic
emission at multipoles $\ell\gtrsim500$ over the full range of HFI
frequencies outside the CL31 mask.

We take a different approach for the Galactic dust correction with 
the \plik\ likelihood.  Rather than correcting the empirical spectra during 
a pre-processing step, the \plik\ likelihood implements an explicit
one-parameter model that describes the dust contribution to the
cross-spectrum between detectors $i$ and $j$,
\begin{equation}
C_{\ell}^{\mathrm{Dust}}(i,\, j)=A^{\mathrm{Dust}}\, F(\nu_{i},\nu_{0})F(\nu_{j,}\nu_{0})\,\left(\frac{\ell}{500}\right)^{-\gamma_{d}}\, g_{i}^{\mathrm{Dust}}g_{j}^{\mathrm{Dust}}.\label{eq:DustModel}
\end{equation}
Here
\begin{equation}
F(\nu_{,}\nu_{0})=\frac{\nu^{\beta_{d}}B(T_{d},\nu)}{\frac{\partial B(T_{\mathrm{CMB}},\nu)}{\partial T}}\,/\,\frac{\nu^{\beta_{d}}B(T_{d},\nu_{0})}{\frac{\partial B(T_{\mathrm{CMB}},\nu_{0})}{\partial T}},\label{eq:DustModel-1}
\end{equation}
where the dust amplitude, $A^{\mathrm{Dust}}$, is measured in units of
$\mu\mathrm{{K}}^{2}$, $\nu_{i}$ is the reference frequency for map
$i$, $\nu_{0}$ is a reference frequency which is taken to be 143\,GHz,
$B(T,\nu)$ is the emission law of a blackbody with temperature $T$,
and the dust color-correction terms, $g_{i}^{\mathrm{Dust}}$, are
computed by integrating the dust spectrum within the spectral band of
each detector (set). We fix the frequency and angular scaling
parameters to $\gamma_{d}=2.6$, $\beta_{d}=1.6$ and
$T_{d}=18\mathrm{K}$.

\subsection{Poisson power from unresolved point sources\label{sub:Poisson-point-sources}}

Unresolved galaxies contribute both shot noise and clustered power to
the \Plancks maps. The Poisson contribution leads to a scale
independent tem, $C_{\ell}={\rm constant}$. We model this power with a
single amplitude parameter for each auto-spectrum ($A_{100}^{{\rm
    PS}}$, $A_{143}^{{\rm PS}}$, and $A_{217}^{{\rm PS}}$) and a cross
correlation coefficient for each cross spectrum
($A_{143\times217}^{{\rm PS}} =  r_{{\rm CIB}}\sqrt(A_{143}^{{\rm CIB}}A_{217}^{{\rm CIB}})$).
These quantities are not of primary interest for cosmological results,
so to avoid modelling error we do not separate the power into that
sourced by ``dusty\textquotedbl{} or ``radio\textquotedbl{} galaxies
(i.e., with increasing or decreasing brightness with frequency, respectively)
as is done in the analysis of the ACT and SPT power spectra \citep{dunkley10,reichardt12,dunkley/etal:prep}.
We also make no assumptions about their coherence between frequencies.

The Poisson power can be related to the flux density $dN/dS$ via
\begin{equation}
C_{\ell}=\frac{1}{4\pi}\int d\hat{n}\int_{0}^{S_{{\rm cut}}(\hat{n})}dS\, S^{2}\frac{dN}{dS}\label{eq:exact_poisson},
\end{equation}
where we have explicitly introduced the \Plancks flux cut $S_{{\rm
    cut}}(\hat{n})$.  Since \Plancks utilizes a constant
signal-to-noise cut, and the \Plancks noise varies significantly
across the sky, this flux cut has a spatial dependence. Although this does not
alter the shape of the Poisson term, extra care must be taken when
comparing results%
\footnote{One must also account for the fact that these numbers correspond to
the amplitude for a suitably averaged spectral band, which is approximately
that of the map, and is described in detail in \cite{planck2013-p03d}. %
} with models of $dN/dS$. In Sect.~\ref{sub:FGModelCheck}, we explore the consistency between
the Poisson power recovered from the \Plancks power spectrum analysis
and predictions from source count measurements.

\subsection{Clustered power from unresolved point sources\label{sub:Clustered-IR-sources}}

Unresolved galaxies also contribute power because they trace
large-scale structures. The mean flux from the radio
   galaxies is much smaller than that from the dusty galaxies, so only the
dusty galaxies contribute a significant clustering term \citep{millea12}. The CIB
clustering has been studied extensively, starting with
\cite{bond86,bond91}. Further theoretical investigation
\citep{scott99,haiman00} was stimulated by the detection of the
infrared background in the {\it COBE} data \citep{puget96,fixsen98},
and the detection of bright ``sub-millimeter'' galaxies in SCUBA data
\citep{hughes98}. Subsequently, the clustering has been detected at
160\,microns \citep{lagache07}, at 250, 350 and 500\,microns by the
Balloon-borne Large Aperture Submillimeter Telescope
\citep[BLAST,][]{viero09,hajian11} and at 217\,GHz by SPT and ACT
\citep{hall10,dunkley10}.  Recent \Plancks measurements of the Cosmic
Infrared Background \citep{PlanckCIB} have extended the measurements
at 217\,GHz, 353\,GHz, and 545\,GHz to larger scales, and recent
\textit{Herschel} measurements \citep{amblard11} have improved on the
BLAST measurements and extended them to smaller angular scales.

Rather than attempt to establish a physical model of the CIB, we adopt
in this analysis a phenomenological model that captures the CIB
uncertainties for both Planck and high-$\ell$ experiments.  Our
baseline CIB model is a power-law spectrum with a free spectral index,
${\cal D}_{\ell}^{{\rm CIB}}\propto\ell^{\gamma_{{\rm }}^{\mathrm{CIB}}},\label{CIB2}$
with an amplitude at each frequency, $A_{143}^{{\rm CIB}}$ and
$A_{217}^{{\rm CIB}}$, and a cross-correlation between frequencies,
$A_{143\times217}^{{\rm CIB}} = r_{{\rm CIB}}\sqrt(A_{143}^{{\rm
    CIB}}A_{217}^{{\rm CIB}})$. We assume that the CIB clustering
power at 100\,GHz is negligible.

\subsection{Unresolved Sunyaev-Zeldovich effects\label{sub:Unresolved-Sunyaev-Zeldovich-eff}}

Based on analysis of ACT and SPT data, the thermal Sunyaev-Zeldovich
(tSZ) contribution is expected to contribute approximately ${\cal
  D}_{\ell=3000}^{{\rm tSZ}}\sim9\,\mu{\rm K}^{2}$ at 100\,GHz and
${\cal D}_{\ell=3000}^{{\rm tSZ}}\sim4\,\mu{\rm K}^{2}$ at 143\,GHz
\citep{reichardt12,dunkley/etal:prep,sievers/etal:prep}.  The kinetic
Sunyaev-Zeldovich (kSZ) effect is expected to have a similar, or
smaller, contribution, with ${\cal D}_{\ell=3000}^{{\rm
    kSZ}}\lesssim7\,\mu{\rm K}^{2}$.  In addition, theoretical
arguments \citep{reichardt12,Aetal12b} suggest that there should be a
tSZ\,x\,CIB correlation that should contribute about the same order of
magnitude as the kSZ term at 143\,GHz. 

For \Planck, all of these SZ contributions are small in comparison to
other unresolved foregrounds and are therefore poorly constrained by
\Planck\ data alone. Nevertheless, to eliminate biases in cosmological
parameters \citep{millea12,zahn05}, we model their contributions, with
appropriate constraints from higher resolution CMB experiments, using
three templates.

\begin{figure}
\begin{centering}
\includegraphics[bb=30bp 60bp 520bp 490bp,clip,width=1\columnwidth]{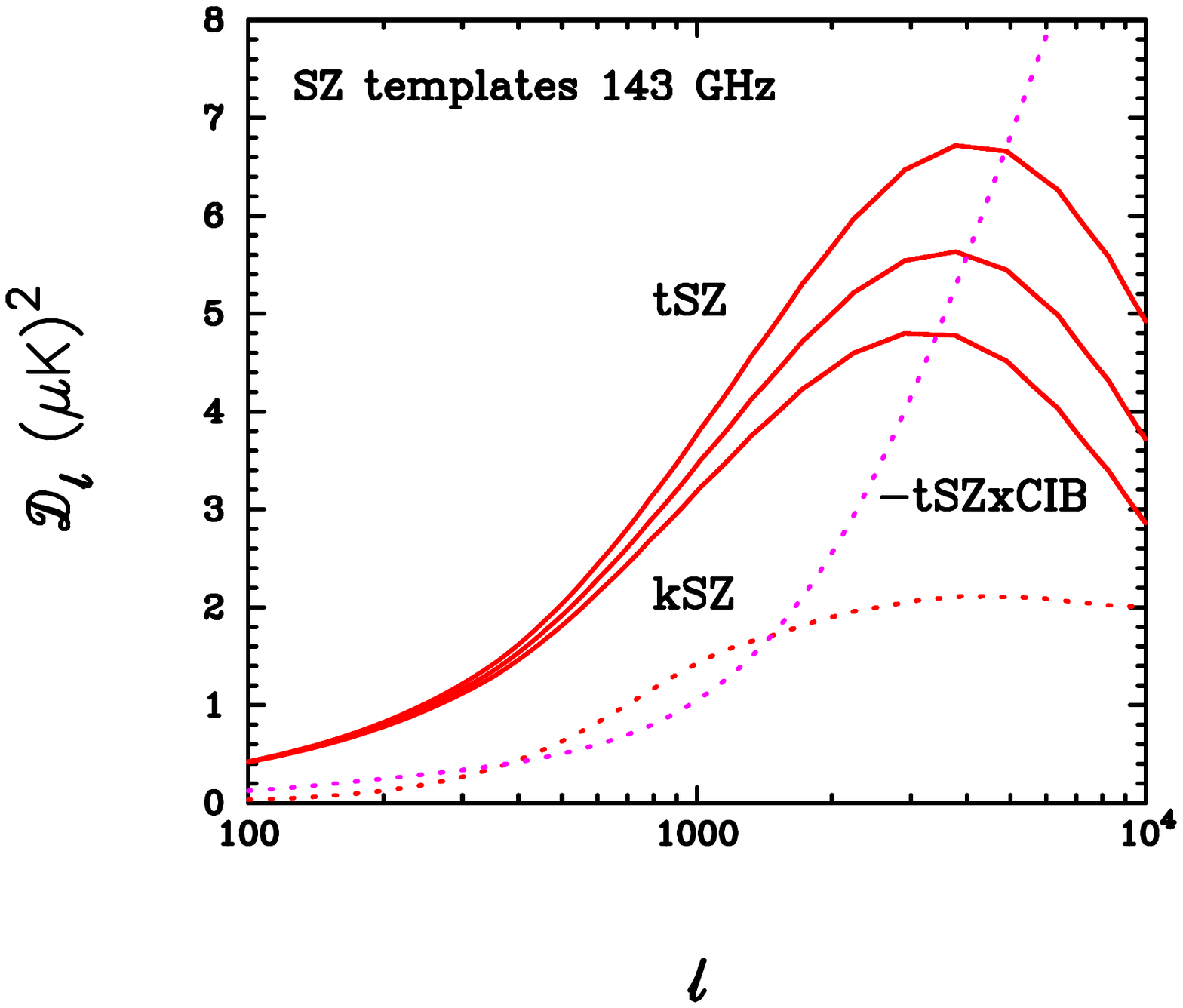}\caption{SZ templates at 143\,GHz computed for a normalization of $\sigma_{8}=0.8$.
The tSZ templates are from the model of \cite{EM012} for three values
of the evolution parameter $\epsilon$, $\epsilon=0$ (top), $\epsilon=0.25$
(middle) and $\epsilon=0.5$ (lower). The kinetic SZ template is from
\cite{TBO11}. The tSZ\,x\,CIB cross correlation ($143\times143$\,GHz)
is from the \cite{Aetal12b} template with parameters described in
the text, and is negative for $143\times143$\,GHz.\label{fig:SZtemplate} 
}

\par\end{centering}

\end{figure}

First, for the thermal SZ effect we adopt the family of templates described by
\cite{EM012}.  These are based on the \cite{KS02} model, but use the
`universal' X-ray electron pressure profile, $P_e$, of
\cite{Arnetal10} extrapolated to high redshift via
\begin{equation}
P_{e}(z)\propto[(1-\Omega_{\Lambda})(1+z)^{3}+\Omega_{\Lambda}]^{4/3-\epsilon/2}.\label{SZ2}
\end{equation}
Here, $\epsilon$ describes departures from self-similar evolution, and
a value of $\epsilon=0.5$, which is adopted as the default for
parameter estimation purposes, provides a good match to the results
from recent hydrodynamical numerical simulations incorporating
feedback processes
\citep{Batetal10,Batetal11}. Figure~\ref{fig:SZtemplate} shows the tSZ
templates for three values of $\epsilon$; the template shape is not
particularly sensitive to $\epsilon$. We treat the (dimensionless)
normalization of the tSZ template at $143$\,GHz as an adjustable
parameter,
\begin{equation}
{\cal D}_{\ell}^{{\rm tSZ}}=A_{143}^{{\rm tSZ}}{\cal D}_{\ell}^{{\rm template}}.\label{SZ3}
\end{equation}
This parameter fixes the amplitude at $100$\,GHz via the frequency
dependence of the tSZ effect,
\begin{equation}
C^{{\rm tSZ}}\propto\left(x\frac{e^{x}+1}{e^{x}-1}-4\right)^{2},\quad x=\frac{h\nu}{kT}.\label{SZ4}
\end{equation}
We neglect the tSZ at 217~GHz.

Second, for the kinetic SZ effect we adopt the template described by \cite{TBO11},
and as in Eq.~\ref{SZ3} we treat the dimensionless amplitude
of the template, $A^{{\rm kSZ}}$, as a free parameter, 
\begin{equation}
{\cal D}_{\ell}^{{\rm kSZ}}=A^{{\rm kSZ}}{\cal D}_{\ell}^{{\rm template}}.\label{SZ5}
\end{equation}

Third and finally, for the cross-correlation between the thermal SZ
component and the CIB we adopt the template described by \cite{Aetal12a}.
In this case, the amplitude is parameterised in terms of a single correlation coefficient,
\begin{equation}
\left.\begin{array}{ccl}
{\cal D}_{\ell}^{{\rm tSZxCIB}}=-2\xi\sqrt{{\cal D}_{3000}^{{\rm tSZ143}}{\cal D}_{3000}^{{\rm CIB143}}}{\cal D}_{\ell}^{{\rm template}}\quad(143\times143),\\
{\cal D}_{\ell}^{{\rm tSZxCIB}}=-\xi\sqrt{{\cal D}_{3000}^{{\rm tSZ143}}{\cal D}_{3000}^{{\rm CIB217}}}{\cal D}_{\ell}^{{\rm template}}\quad(143\times217).
\end{array}\;\right\} 
\label{SZ6}
\end{equation}
These templates are plotted in Fig.~\ref{fig:SZtemplate}, normalized
to $\sigma_{8}=0.8$ and with $\xi=1.0$ using a fiducial CIB amplitude.
Note that with these parameters, the tSZ\,x\,CIB
cross-spectrum approximately cancels the kSZ spectrum at $143$\,GHz.

As seen in Fig.~\ref{fig:SZtemplate}, the SZ contributions are at the
level of a few $\mu\textrm{K}^2$, which, although small, must be taken
into account to assess inter-frequency residuals. However, one can see
that these templates have similar shapes at multipoles $\lesssim2000$,
and therefore they cannot be disentangled using \Planck\ data
alone. On the other hand, higher resolution experiments can break this
degeneracy, and as shown in \cite{planck2013-p11}, the combination of
\Planck, ACT, and SPT, better constrains the amplitude of the thermal SZ
effect. The ACT and SPT data at $150$\,GHz can be fitted to high
(sub-$\mu$K$^2$) accuracy \textit{without} kSZ and tSZ\,x\,CIB
templates, yet we expect a kSZ contribution of at least the amplitude
shown in Fig.~\ref{fig:SZtemplate}, and larger if we account for
patchy reionization \citep[see, e.g.,][and references therein]{Knox03} and references
therein). This implies a cancellation of the kSZ and tSZ\,x\,CIB
contributions at $150$\,GHz \citep{Aetal12b}, as discussed in greater
detail in \cite{planck2013-p11}.

%% file: 4_Combined-cross-spectra.tex
\section{Combined cross-spectra and consistency checks\label{sec:Combined-spectra}}

\begin{figure*}[h]
\begin{centering}
\includegraphics[bb=40bp 40bp 600bp 600bp,clip,width=0.33\textwidth]{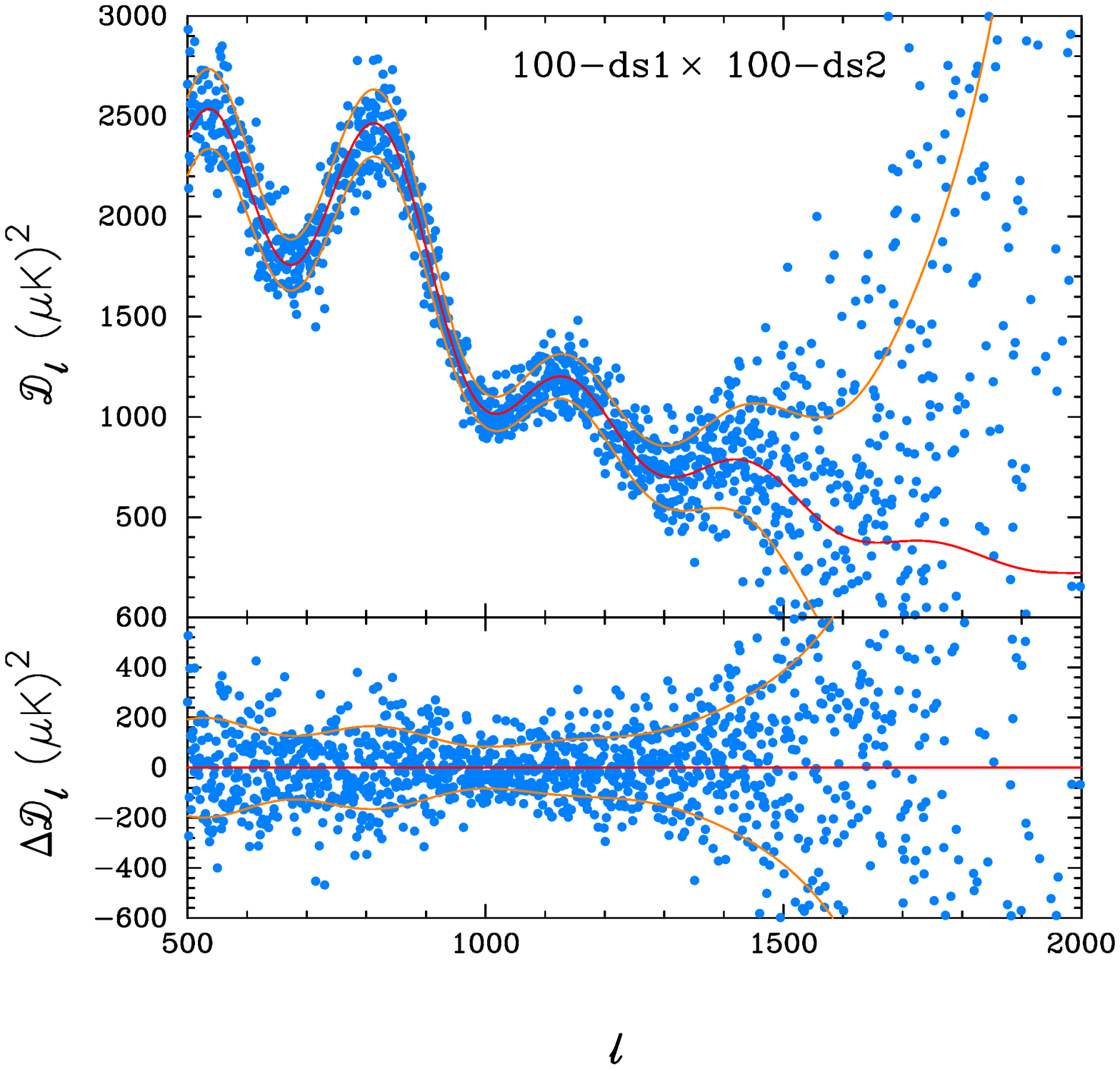}\includegraphics[bb=40bp 40bp 600bp 600bp,clip,width=0.33\textwidth]{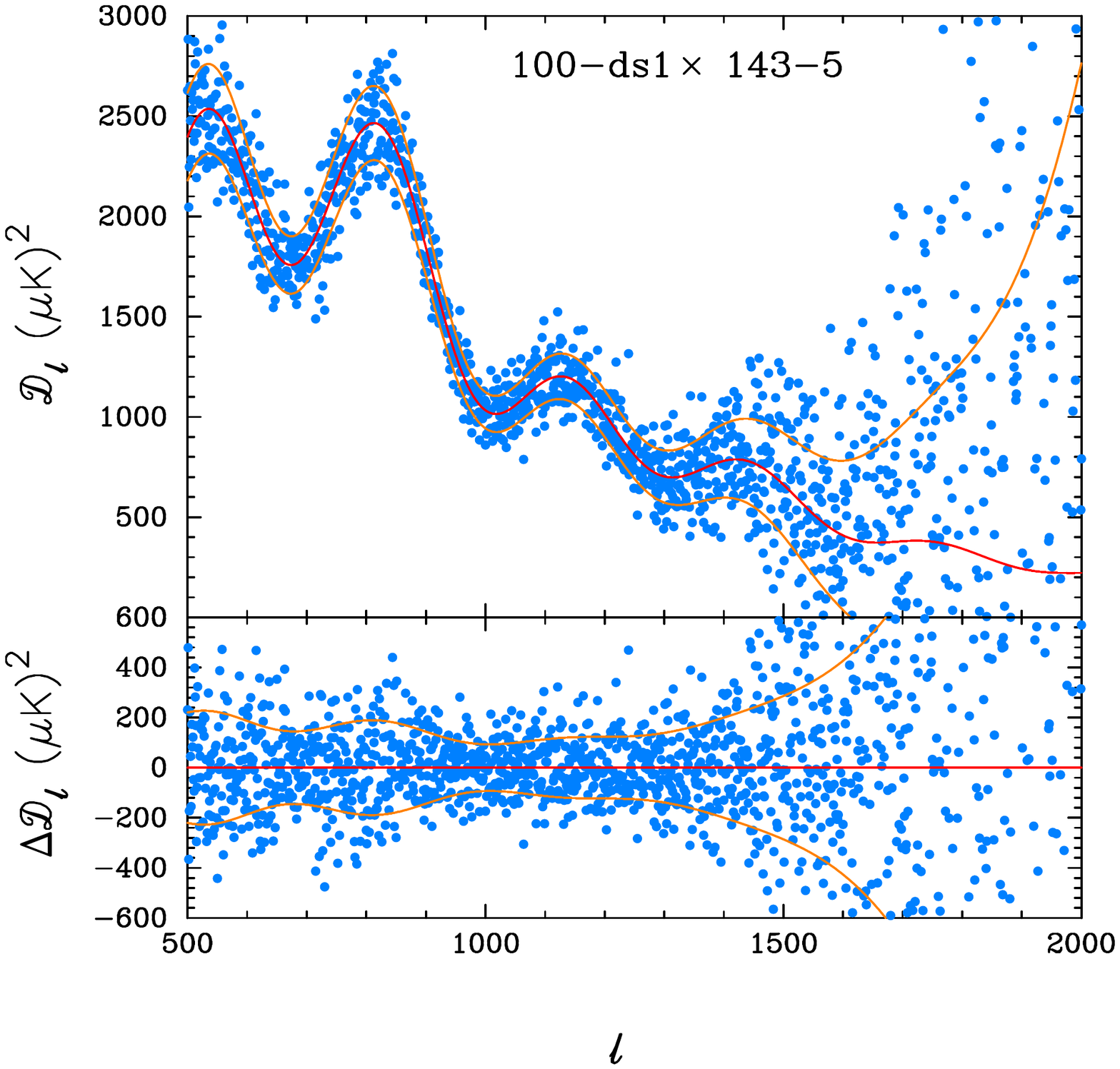}\includegraphics[bb=40bp 40bp 600bp 600bp,clip,width=0.33\textwidth]{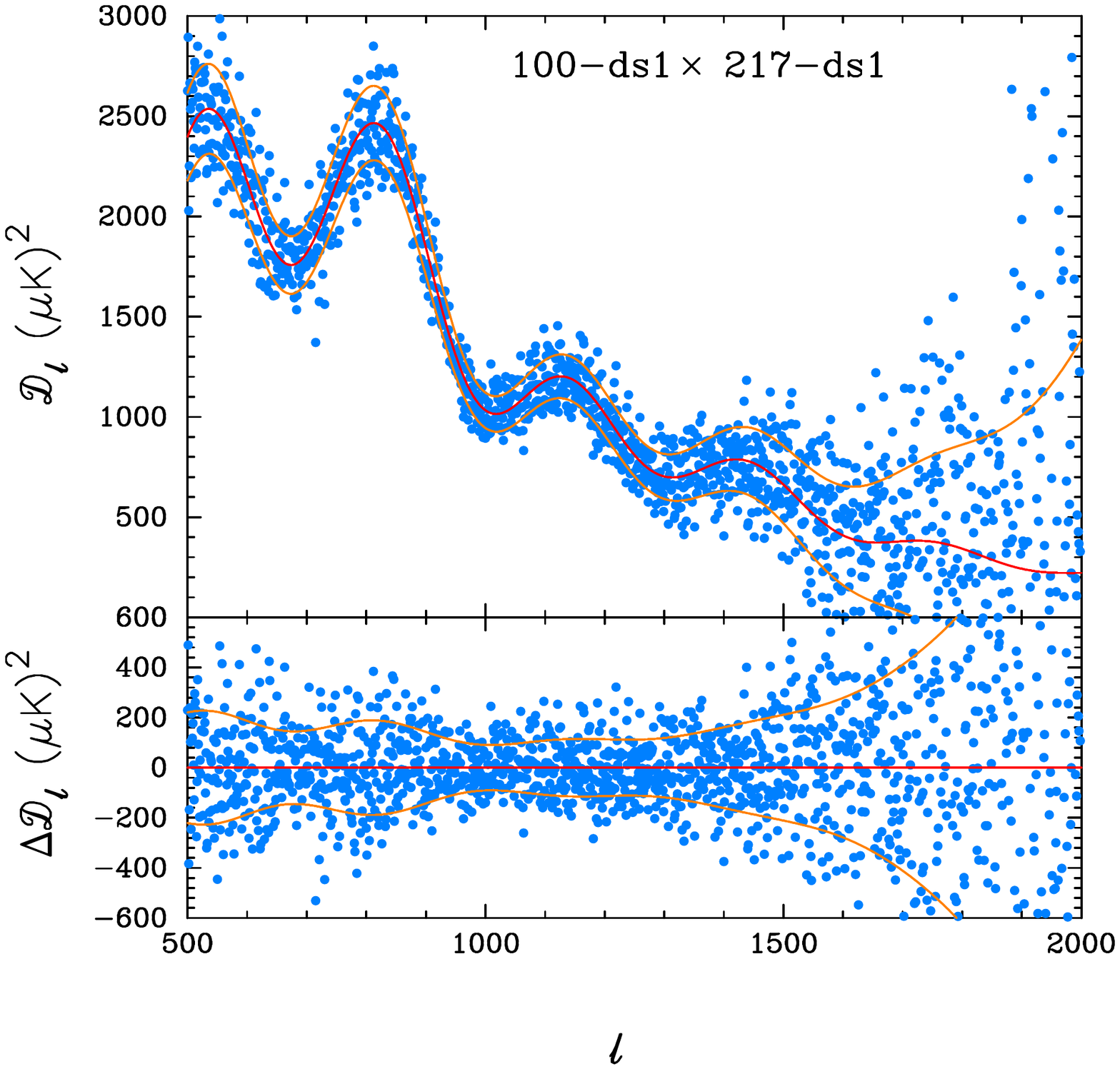} 
\par\end{centering}

\begin{centering}
\includegraphics[bb=40bp 40bp 600bp 600bp,clip,width=0.33\textwidth]{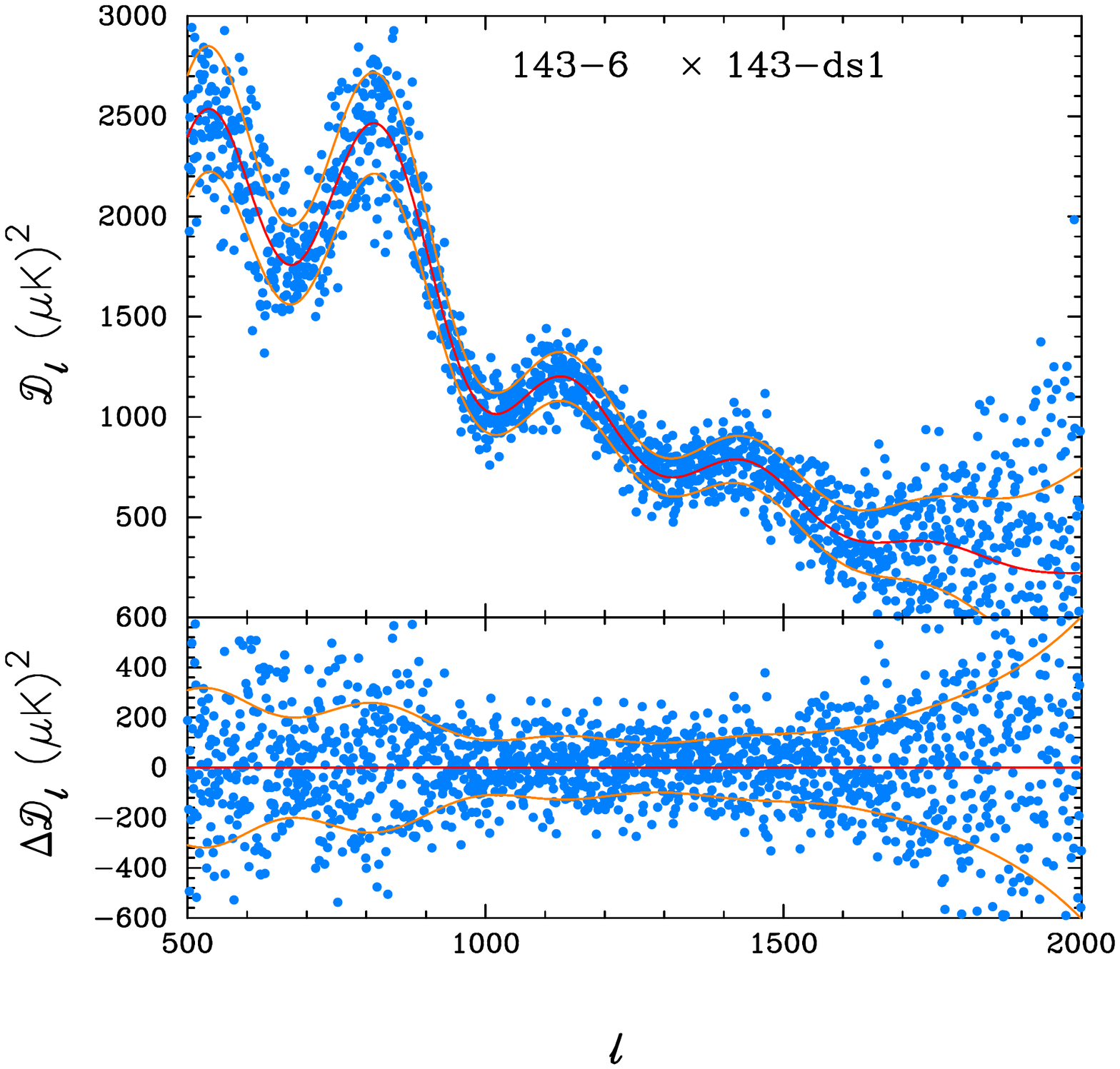}\includegraphics[bb=40bp 40bp 600bp 600bp,clip,width=0.33\textwidth]{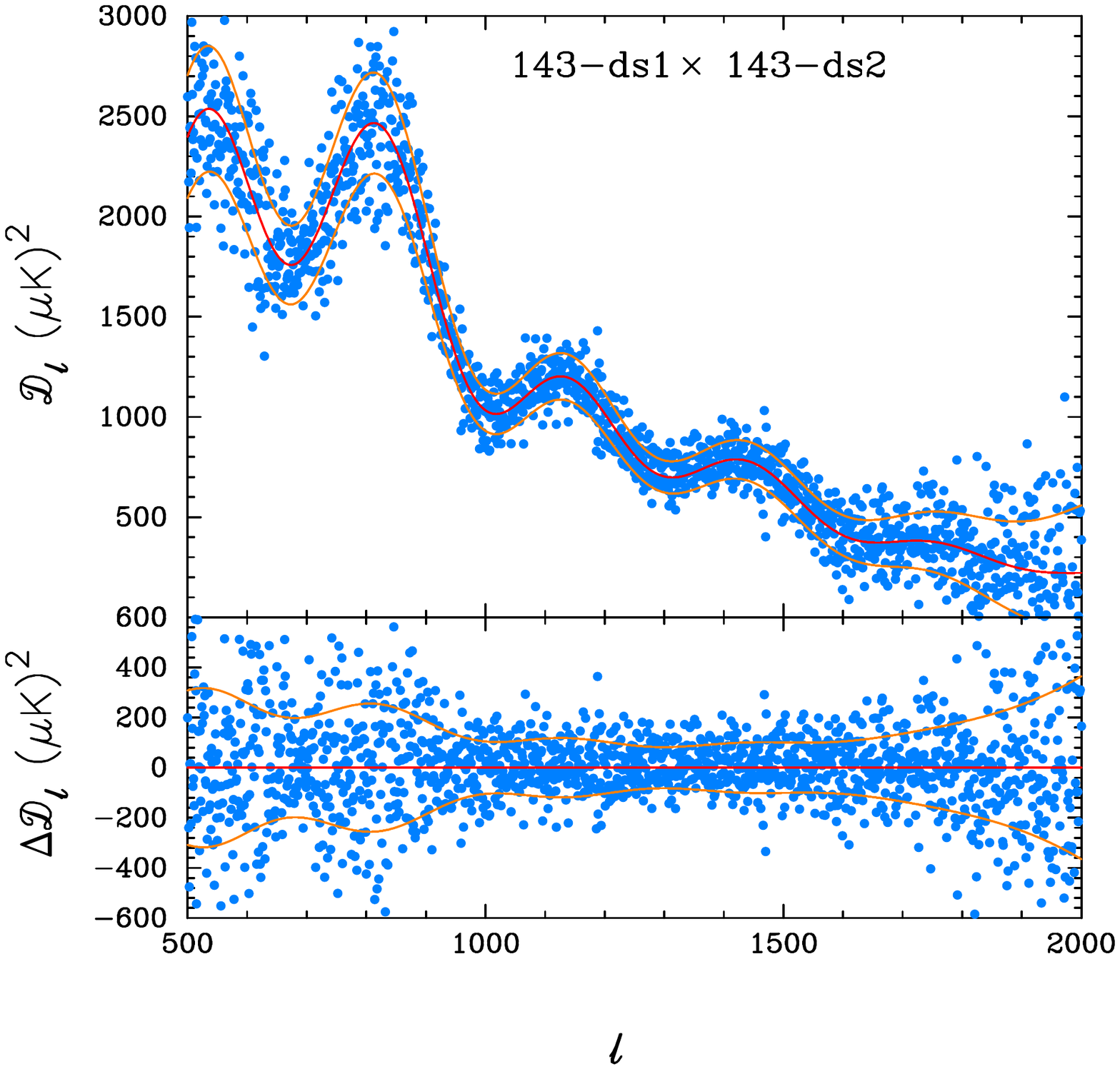}\includegraphics[bb=40bp 40bp 600bp 600bp,clip,width=0.33\textwidth]{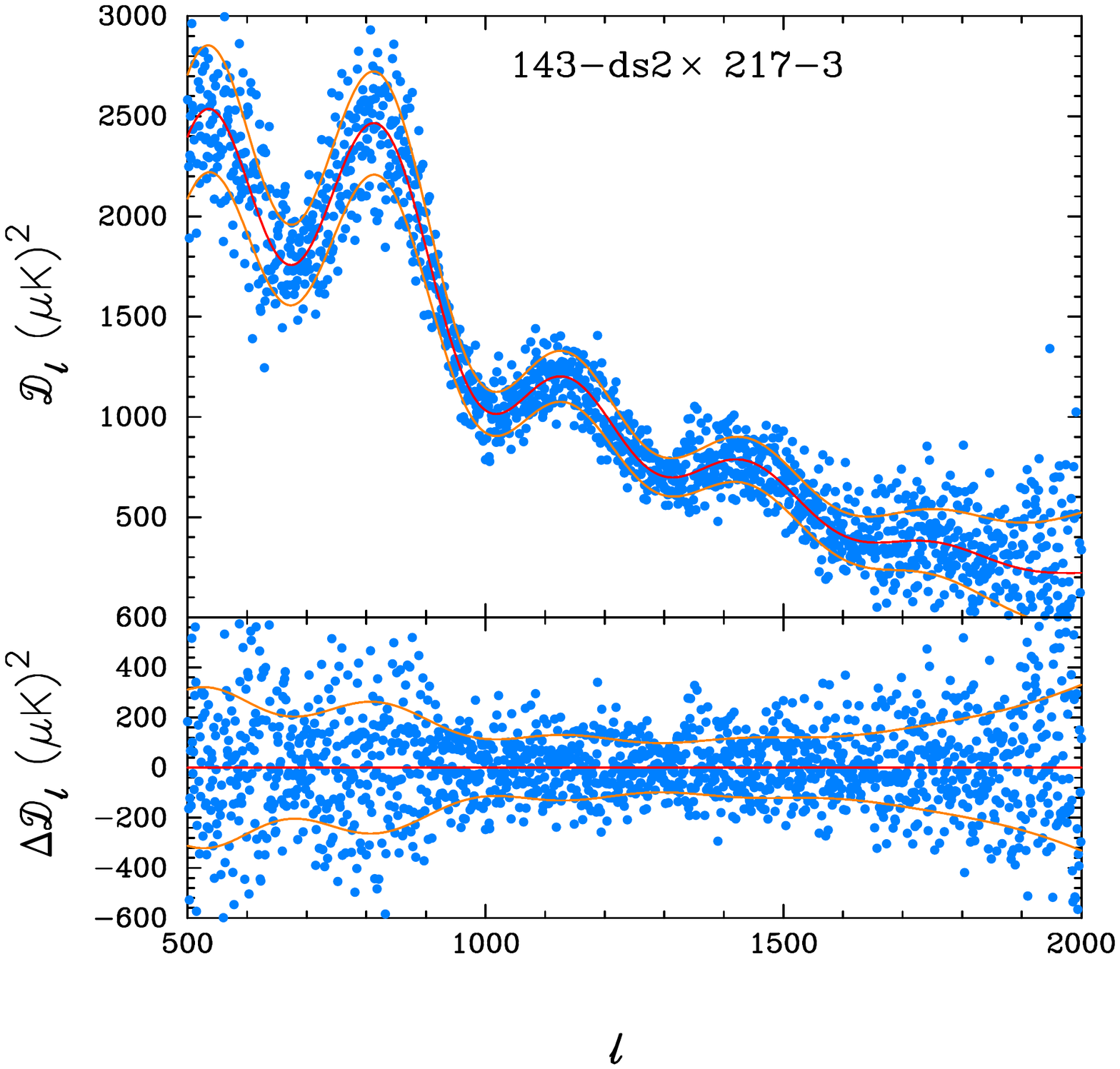} 
\par\end{centering}

\begin{centering}
\includegraphics[bb=40bp 40bp 600bp 600bp,clip,width=0.33\textwidth]{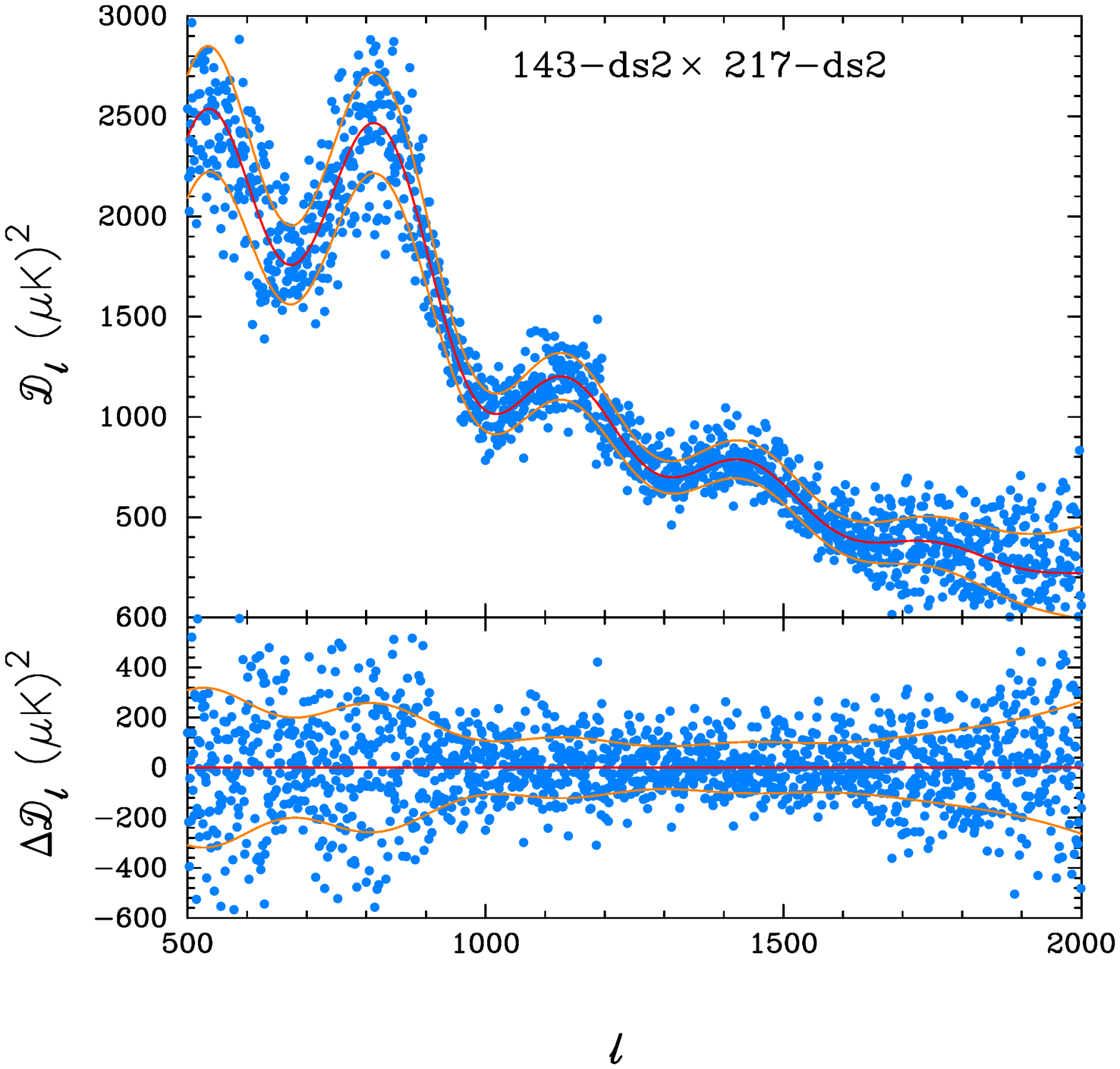}\includegraphics[bb=40bp 40bp 600bp 600bp,clip,width=0.33\textwidth]{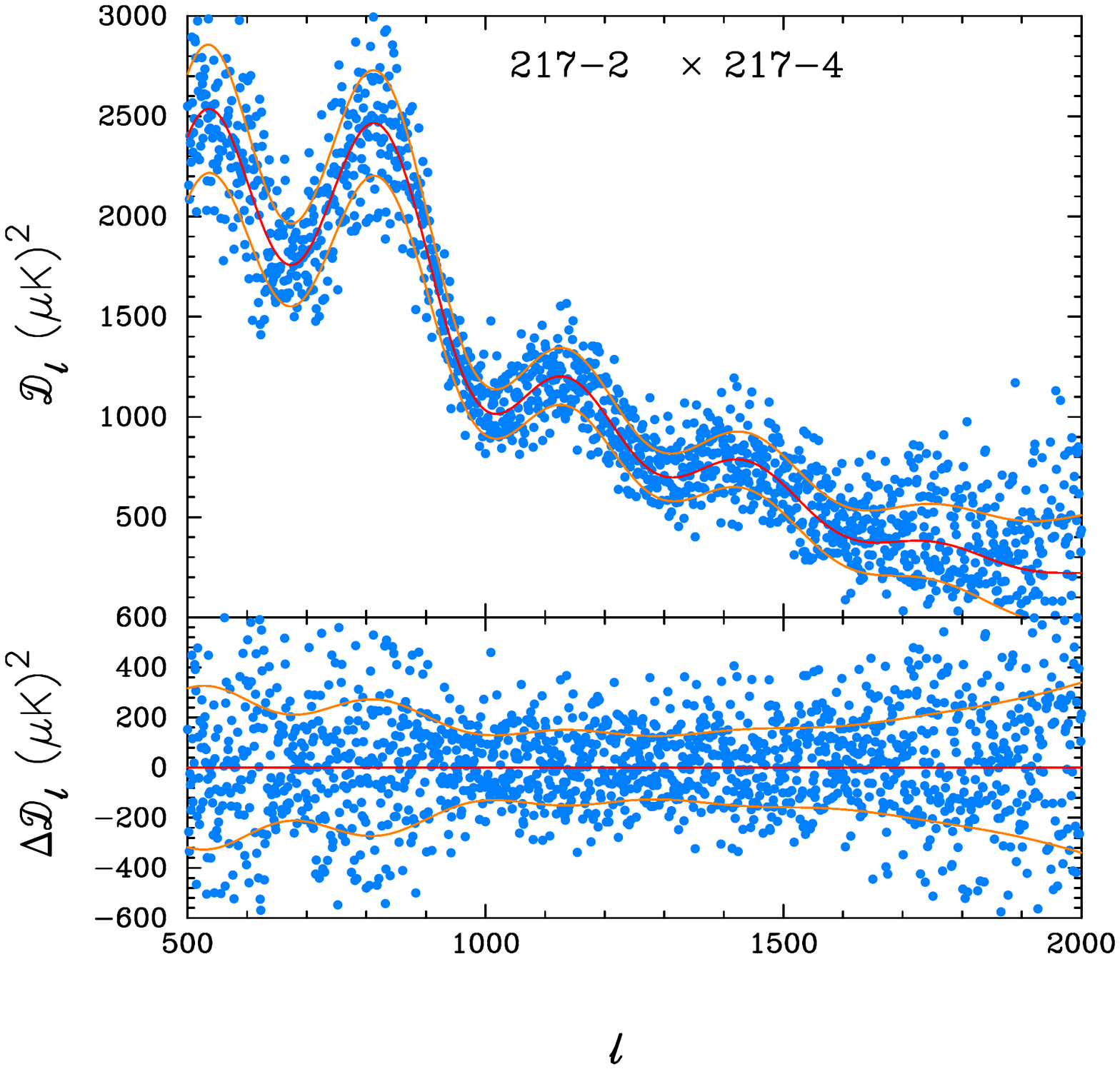}\includegraphics[bb=40bp 40bp 600bp 600bp,clip,width=0.33\textwidth]{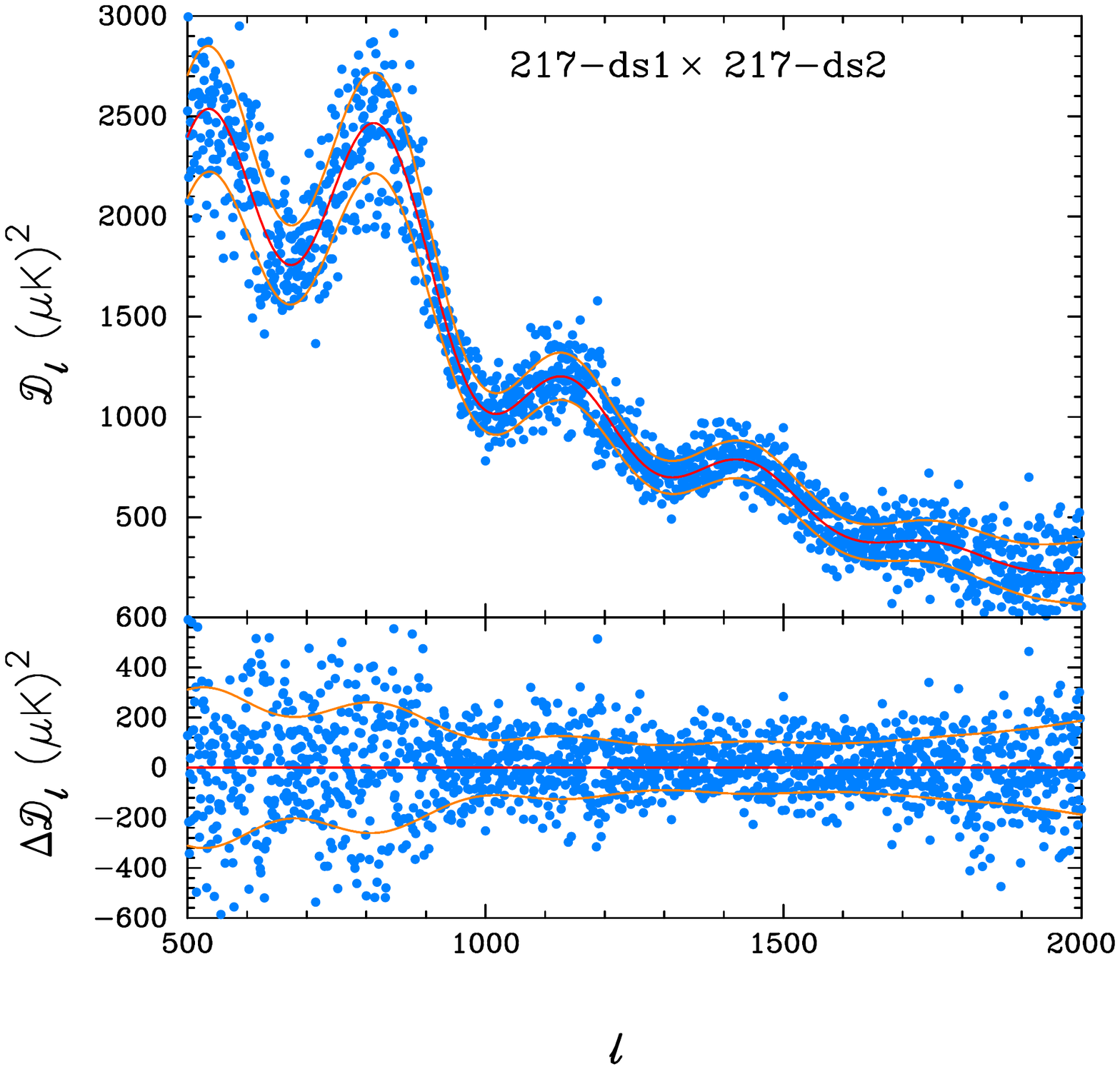} 
\par\end{centering}

\caption{A selection of cross spectra from individual detectors, with the best
fit unresolved foreground model subtracted. The best-fit six parameter
$\Lambda$CDM model is shown, with $\pm1\sigma$ errors determined from the
diagonal components of the analytic covariance matrices. The lower
panel in each plot shows the residuals with respect to the model.}

\label{fig:spec_examples} 
\end{figure*}

The large number of cross spectra in a detector-by-detector power
spectrum analysis allows for a number of internal consistency checks
of the data. Within a frequency band, we expect to see exactly the
same sky signals (primordial CMB, Galactic, and extra-galactic foregrounds),
and so any intra-frequency residuals will reflect instrumental systematics, for example beam errors, `gain' fluctuations, and band-pass mismatch.
In contrast, inter-frequency residuals are harder to analyse because
the sky signals vary with frequency. An accurate model of the unresolved
foregrounds is therefore required to assess inter-frequency residuals.
Furthermore, as we will show below, the scatter caused by chance CMB--foreground
cross-correlations can dominate the inter-frequency residuals. For
a precision experiment such as \Planck, where the power spectra are
expected to be signal dominated over a wide multipole range, intra-
and inter-frequency residuals provide a powerful way of assessing
possible systematic errors. It is essential that contributions of
systematic errors to both types of residual are small enough that
they have negligible impact on cosmological parameter analysis.

Figure~\ref{fig:spec_examples} shows a selection of temperature
cross-spectra and estimates of the analytic covariance matrices,
together with the best-fit cosmological model described in
Sect.~\ref{sec:Reference-model-results}. Unresolved foregrounds have
been subtracted using the best-fit foreground parameters of the model
described in Sect.~\ref{sub:Sky-model}.  The scatter varies
substantially between cross-spectra, reflecting differences in the
instrument noise and effective resolution of different detector
combinations. The analytic error model summarized in
Appendix~\ref{app:PCL-Covariance-Matrices} is indicated, modified by
the non-white noise correction.
This model provides an excellent description of the scatter seen in
the data, over the full multipole range shown in plots, with an
accuracy of a few percent or better.

\subsection{Intra-frequency residuals \label{sub:Intra-frequency-residuals}}

In this section we analyse the intra-frequency residuals at $143$ and
$217$\,GHz. There are $N_{{\rm spec}}=10$ cross-spectra at $143$\,GHz
and $15$ cross-spectra at $217$\,GHz\footnote{there is only one cross-spectrum at $100$\ghz}. 
At each frequency, we solve for
multiplicative (`effective' calibration) coefficients, $y_{i}$, that
minimise
\begin{equation}
\chi^{2}=\sum_{\ell}\sum_{ij,j>i}(y_{i}y_{j}\hat{C}_{\ell}^{ij}-\langle\hat{C}_{\ell}\rangle)^{2},\label{IF1a}
\end{equation}
where 
\begin{equation}
\langle\hat{C}_{\ell}\rangle=\frac{1}{N_{{\rm spec}}}\sum_{ij,j>i}y_{i}y_{j}\hat{C}_{\ell}^{ij},\label{IF1b}
\end{equation}
subject to the constraint that $y_{1}=1$ (where $i=1$
corresponds to detector 5 at $143$\,GHz and detector 1 at
$217$\,GHz). Note that the power spectra in Eq.~\ref{IF1a} and
\ref{IF1b} are corrected for beam transfer functions. To minimise the
possible impact of beam errors and noise, we restrict the sum in
Eq.~\ref{IF1a} to the multipole range $50\le\ell\le500$ where the
spectra are signal dominated. Numerical values for the calibration
coefficients are given in
Table~\ref{tab:Map-calibration-coefficients.}, using mask CL31. The calibration factors are insensitive to
the choice of mask or multipole range.

\begin{table}
\begingroup
\newdimen\tblskip \tblskip=5pt
\caption{Map calibration coefficients.}
\label{tab:Map-calibration-coefficients.}
\nointerlineskip
\footnotesize
\setbox\tablebox=\vbox{ %
\newdimen\digitwidth %
\setbox0=\hbox{\rm 0}
\digitwidth=\wd0
\catcode`*=\active
\def*{\kern\digitwidth}
\newdimen\signwidth
\setbox0=\hbox{+}
\signwidth=\wd0
\catcode`!=\active
\def!{\kern\signwidth}
\halign{
\hfil#\hfil\tabskip=0.2cm&
\hfil#\hfil\tabskip=0.5cm&
\hfil#\hfil\tabskip=0.2cm&
\hfil#\hfil\tabskip=0pt\cr
\noalign{\doubleline}
map& $y_{i}$& map& $y_{i}$\cr
\noalign{\vskip 3pt\hrule\vskip 3pt}
143-5**& 1.0000& 217-1**& 1.0000\cr
143-6**& 0.9988& 217-2**& 0.9992\cr
143-7**& 0.9980& 217-3**& 0.9981\cr
          -&          -& 217-4**& 0.9985\cr
143-ds1& 0.9990& 217-ds1& 0.9982\cr
143-ds2& 0.9994& 217-ds2& 0.9975\cr
\noalign{\vskip 3pt\hrule\vskip 3pt}}}
\endPlancktable
\endgroup
\end{table}

The results of Table~\ref{tab:Map-calibration-coefficients.} show that
effective calibration factors of $\sim$0.2\,\% are quite typical for
HFI maps, in the $100-217$\,GHz frequency range. These recalibrations 
are of the order of magnitude of the statistical errors of the calibrations on 
dipole \citep[see][Table 2]{planck2013-p03f}. Note that the data
are corrected for individual bolometer time transfer functions (TTFs;
\citealp{planck2013-p03c}). For each detector, the TTF model is tuned
to minimise survey differences and by construction normalised to unity
at the spin frequency of the satellite ($0.01666$\,Hz) to preserve the
dipole calibration. The consistency of intra-frequency power spectrum residuals
therefore provides a test of the consistency of the TTFs in addition to
the beam transfer functions.

\begin{figure}[h!]
\begin{centering}
\includegraphics[bb=40bp 40bp 600bp 600bp,clip,width=1\columnwidth]{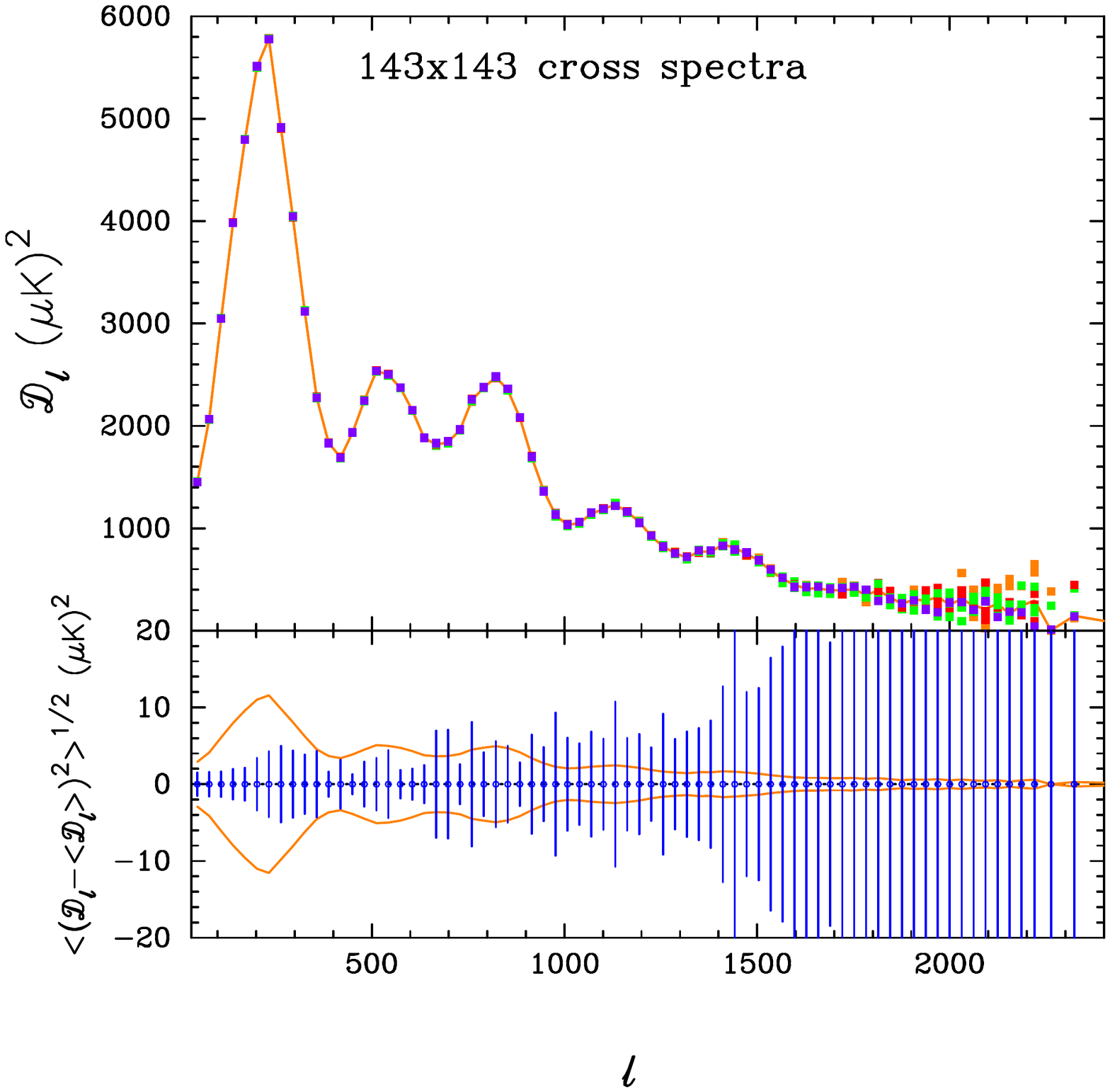} 
\par\end{centering}
\centering{\includegraphics[bb=40bp 40bp 600bp
  600bp,clip,width=1\columnwidth]{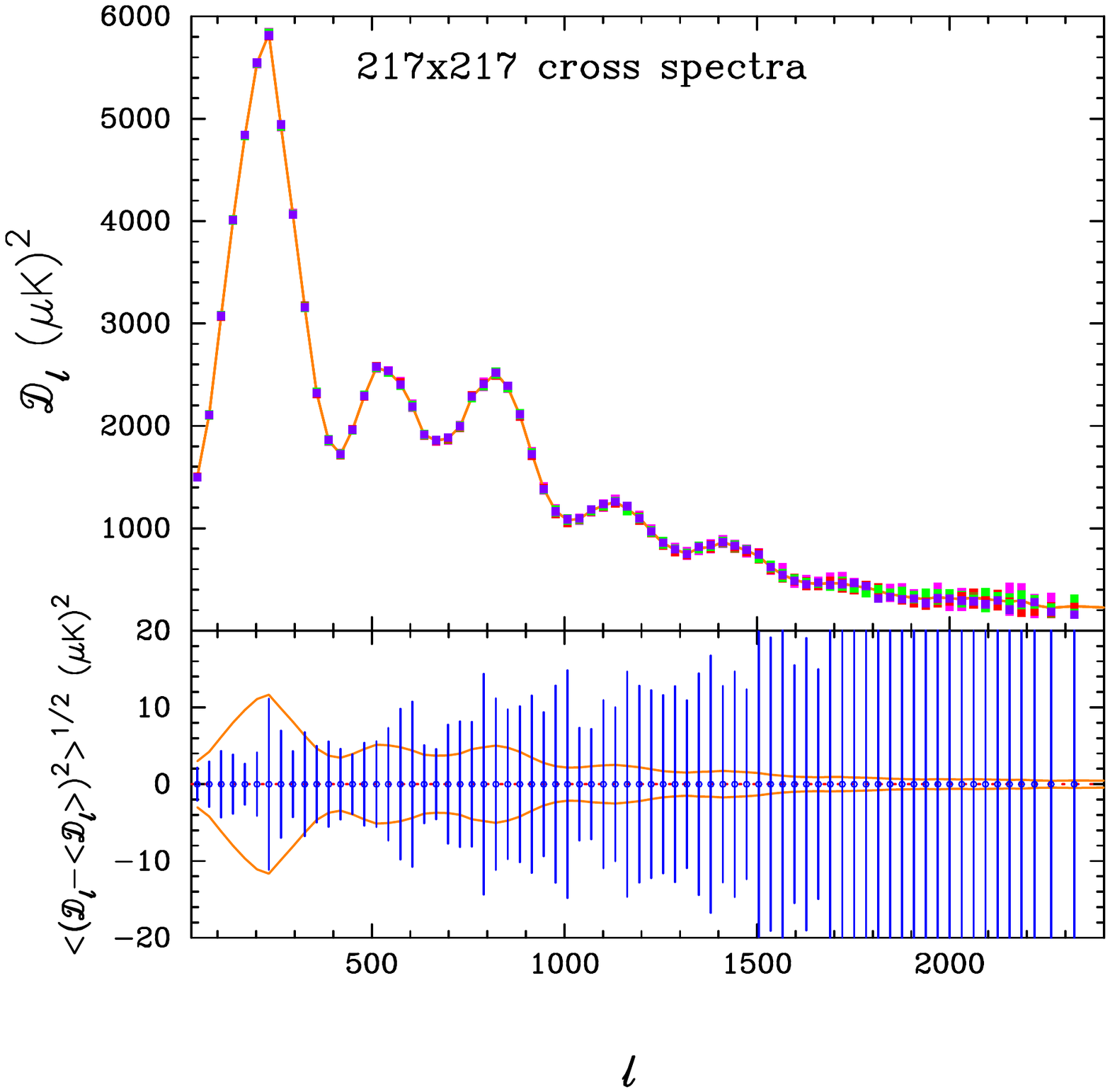}}
  \caption{Cross spectra
  corrected for the beam and effective calibration, together with the
  mean power spectrum.  {\it Top}: The 10 cross spectra at
  143\,GHz. {\it Bottom}: The 15 cross spectra at 217\,GHz, with
  SWB$\times$SWB spectra (magenta), SWB$\times$ds1 (red),
  SWB$\times$ds2 (green), ds1$\times$ds2 (purple).  The power spectra
  are distinguishable only at
  high multipoles where the data become noise dominated. The lower
  panels show the dispersion of the cross spectra around the mean,
  together with a $\pm0.2\%$ calibration
  error. \label{fig:intrafrequency}}
\end{figure}

Figure~\ref{fig:intrafrequency} shows the remarkable consistency
of the power spectra at each frequency. The upper panels 
show the spectra corrected for the beam and effective calibration,
together with the mean cross spectra. The lower panels
show the dispersion around the mean. In the signal dominated regime
the cross spectra show an RMS dispersion of a few $\mu{\rm K}^2$
(in bands of width $\Delta\ell=31$), i.e., the band-averaged
spectra are consistent to an accuracy of $\sim$0.1--0.2\,\%. This
excess scatter has negligible impact on cosmological parameter analysis.

\begin{figure}[b]
\begin{centering}
\includegraphics[bb=40bp 40bp 600bp 600bp,clip,width=1\columnwidth]{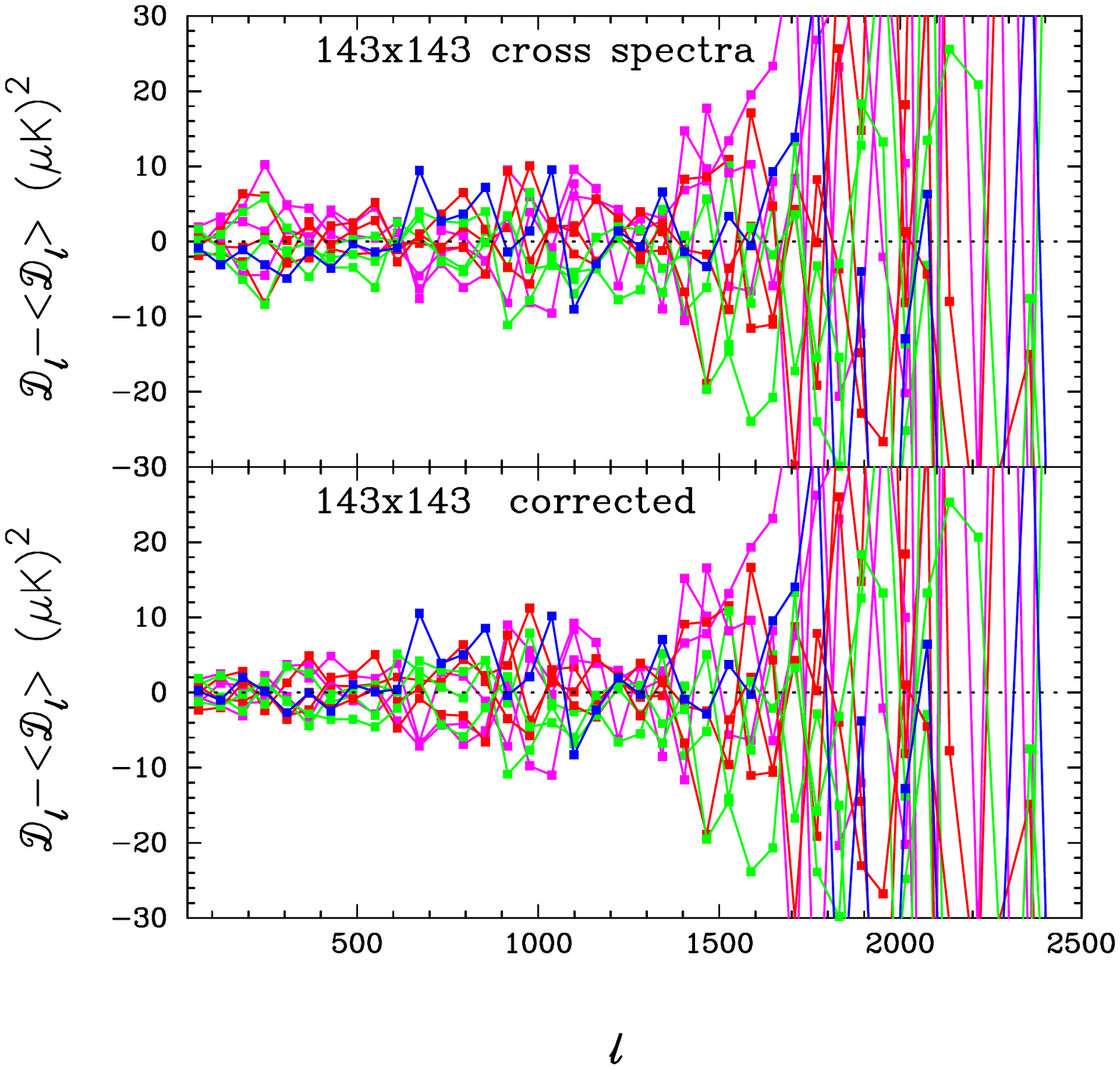}
\par\end{centering}

\begin{centering}
\includegraphics[bb=40bp 40bp 600bp 600bp,clip,width=1\columnwidth]{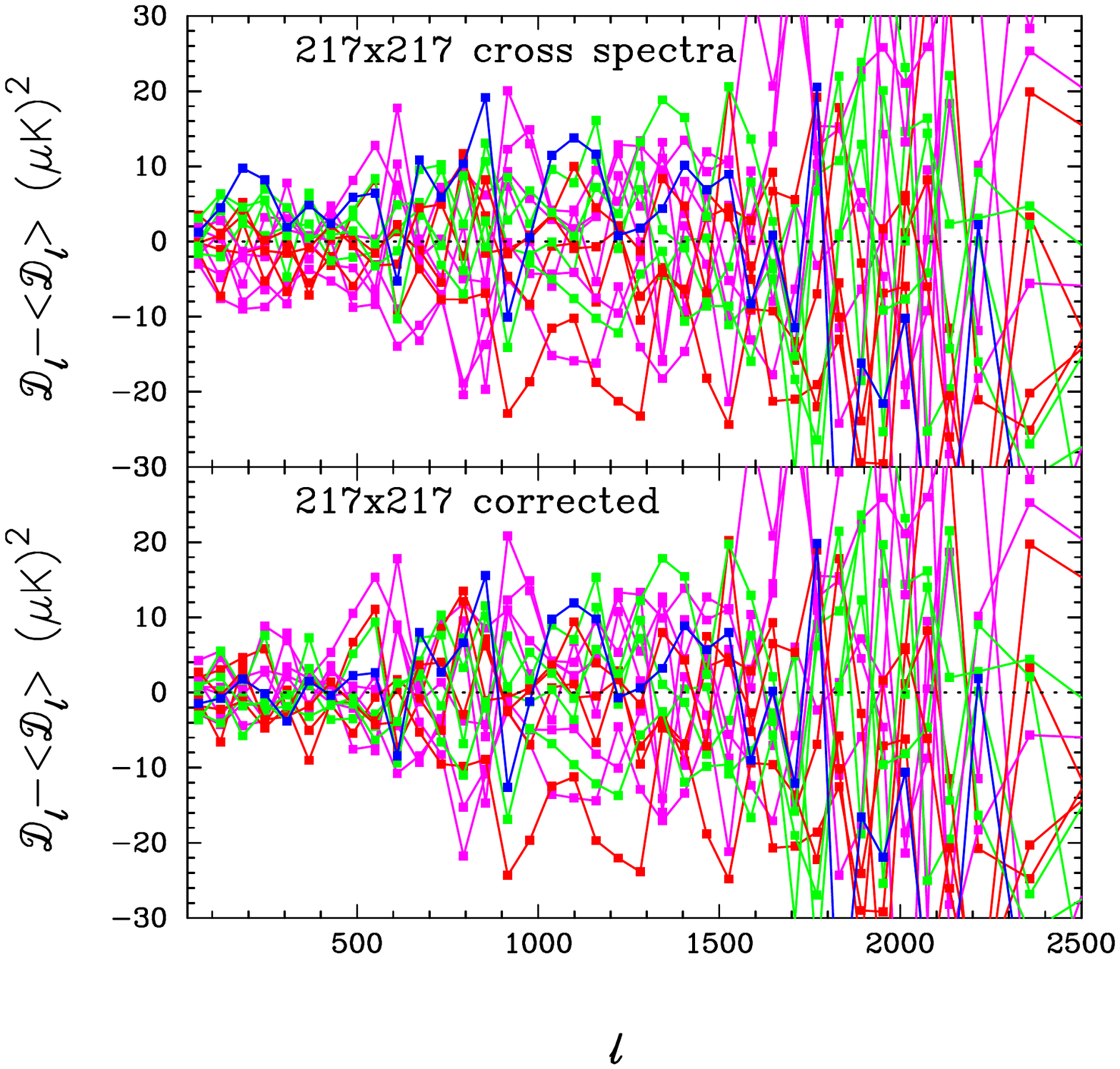}
\par\end{centering}

\centering{}\caption{Cross spectra for the 143\,GHz ({\it top}) and 217\,GHz ({\it bottom}) channels, as in Fig.~\ref{fig:intrafrequency},
before correction for multiplicative intra-frequency calibration coefficients
(above), and after correction (below). \label{fig:143-217consistency} }
\end{figure}

The residuals of the cross spectra in band averages of width $\Delta\ell\sim61$
are shown in Fig.~\ref{fig:143-217consistency}, before and after
correction for the effective intra-frequency calibrations. The reduction
in scatter after correction is evident at $\ell\lesssim 500$, and the residual scatter is
consistent with instrument noise and beam errors. At 217\,GHz, beam
errors dominate over noise at multipoles $\lesssim1000$. There is
no evidence that the excess scatter is caused by a small number of
`anomalous' detectors.

\subsection{Inter-frequency residuals}

The results of the previous section show that the intra-frequency
cross-spectra between detector/detector sets are consistent to within
a few $\mu\textrm{K}^2$ at multipoles $\ell\lesssim1000$. In a likelihood analysis,
there is therefore little loss of information in compressing the power
spectra for each distinct frequency combination, 
as opposed to retaining the spectra for each map pair. This
compression greatly reduces the size of the data vector and its covariance
matrix, and speeds up the likelihood computation at high
multipoles. In this section we inter-compare the residuals of these
compressed power spectra. 
\begin{figure}
\begin{centering}
\includegraphics[bb=30bp 30bp 600bp 600bp,clip,width=1\columnwidth]{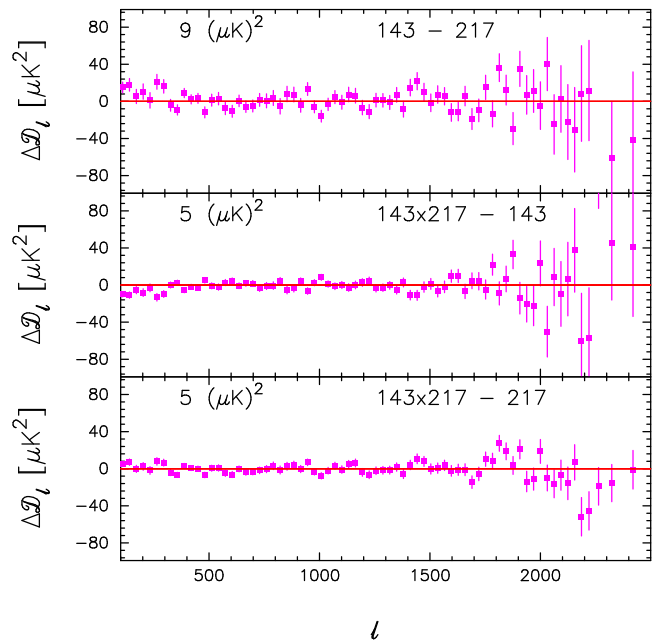} 
\par\end{centering}

\caption{Differences between the $143\times143$, $143\times217$, and
$217\times217$ cross spectra (plotted in bins of width $\delta\ell\approx31$).
The best-fit model for unresolved foregrounds has been subtracted
from each spectrum. 
The numbers list the dispersions over the multipole range $800\le\ell\le1500$.
\label{fig:interband_residues} }
\end{figure}

One might naïvely expect that with accurate foreground modelling, the
inter-frequency residuals in the signal dominated regime should be
reduced to levels comparable to those seen in the intra-frequency
comparisons described in the previous section. This is incorrect.
Figure~\ref{fig:interband_residues} shows power spectrum differences
between the cosmologically significant spectra for \Planck\ at high
multipoles ($143\times143$, $143\times217$, $217\times217$). In this
figure, which is independent of the cosmological model, the best-fit
unresolved foreground model has been subtracted from each spectrum,
and relative calibration factors have been applied.
Residual beam, calibration and unresolved foreground errors would show
up in this figure as large-scale smooth residuals.

In fact, we see small-scale residuals at multipoles $\ell\lesssim800$
which are considerably larger than expected from instrumental noise.
This excess scatter arises from chance CMB--foreground
cross-correlations.  Even if the foreground contamination is much
smaller than the CMB, chance cross-correlations can produce scatter in
the inter-frequency power spectra that is large in the signal
dominated regime. We demonstrate in Appendix~
\ref{app:ChanceCorrealtions} that the observed scatter can be
predicted quantitatively.

At high enough multipoles, instrument noise, beam errors, and errors
in foreground modelling dominate the inter-frequency residuals. A
complete analysis of inter-frequency residuals therefore requires the
full likelihood machinery and MCMC analysis to determine foreground,
beam and calibration parameters. We will therefore revisit the
inter-frequency residuals in the following sections.

%% file: 5_Reference-results-high-ell.tex
\section{Reference results of the high-$\ell$ likelihood \label{sec:Reference-model-results}}

\begin{figure*}
\begin{centering}
\includegraphics[clip,width=1\textwidth]{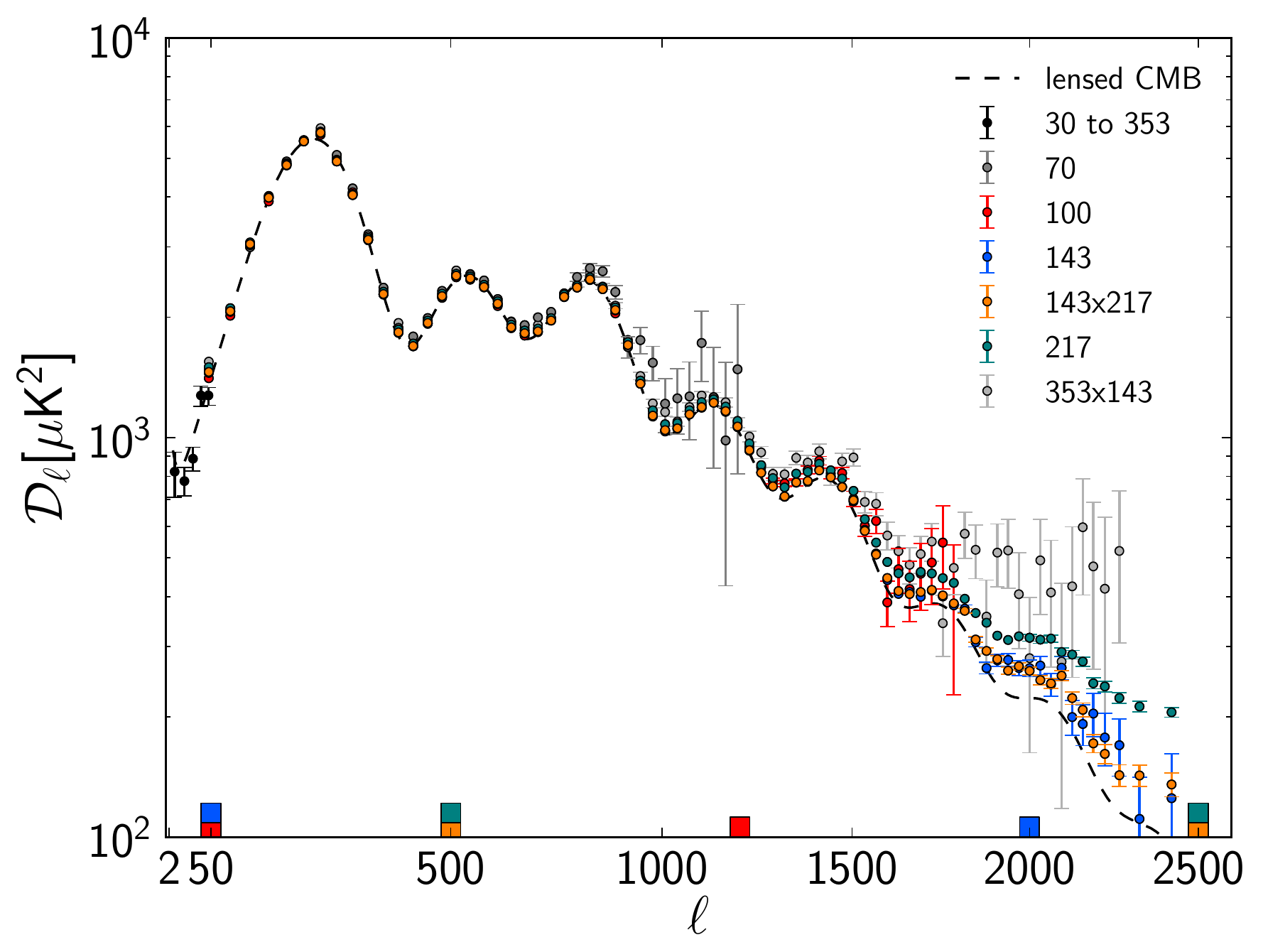} 
\par\end{centering}
\caption{\Planck\ power spectra and data selection. The coloured tick marks
indicate the $\ell$-range of the four cross-spectra included in \camspec\
(and computed with the same mask, see Table\,\ref{tab:nu-mask-l-ref}).
Although not used, the 70\,GHz and 143 x 353\,GHz spectra demonstrate
the consistency of the data. The dashed line indicates
the best-fit \Planck\ spectrum.\textcolor{red}{{} \label{fig:PS_CMB+nu}}}
\end{figure*}

In this section we study the high-$\ell$ \Plancks likelihood, and
present the power spectrum and parameters derived from the baseline
likelihood for the basic six-parameter $\Lambda$CDM model.
In order to break the well-known degeneracy between the optical depth,
$\tau$, and the scalar index of scalar perturbations,
$n_{\textrm{s}}$, we adopt a Gaussian prior on $\tau$ \citep[inspired from WMAP7
data, i.e., $0.088 \pm 0.015$, see ][]{komatsu2010} instead of the
low-$\ell$ likelihood at $\ell<50$.  We return to the global \Plancks
results after introducing the low-$\ell$ likelihood.

We choose separate masks for each frequency map to minimise Galactic
foreground emission. First, since the HFI data are signal-dominated at
$\lesssim500$, and diffuse Galactic emission is low at $100$\,GHz
outside the CL49 mask, there is little to be gained from analyzing the
dustier $217$\,GHz with the same region; it contains no new
information about the primordial CMB. At higher multipoles,
$\ell\gtrsim500$, we use the CL31 mask, optimizing the sky coverage
while ensuring a low amplitude of small-scale Galactic emission
relative to the isotropic unresolved foregrounds and the primordial
CMB. In addition, we tune the multipole range for each frequency to
mitigate Galactic foreground contamination and beam errors.

The choices of masks and angular ranges used in the high-$\ell$
likelihood are summarised in Table~\ref{tab:nu-mask-l-ref}, together
with basic $\chi^2$ statistics with respect to the minimal
$\Lambda$CDM model per cross-spectrum and combined.  The 100\,GHz
cross-spectrum is computed over the largest sky fraction, a total of
49\% of the sky, and measures the largest scales. On the other hand,
it has lowest resolution, and it is therefore only used for
$\ell\le1200$. The 143\,GHz cross-spectrum has higher resolution, and
is used for $\ell\le2000$. Finally, the 217\,GHz cross-spectrum has
the highest resolution, but also the most Galactic dust contamination,
and is therefore evaluated from only 31\% of the sky, but including an
angular range of $500\le \ell\le2500$.

\begin{table}[tmb] 
\begingroup 
\newdimen\tblskip \tblskip=5pt
\caption{Overview of of cross-spectra, multipole ranges and masks
  used in the \Plancks\ high-$\ell$ likelihood.
  Reduced $\chi^2$s with respect to the best-fit minimal
  $\Lambda$CDM model are given in the fourth column, and the
  corresponding probability-to-exceed in the fifth column. \label{tab:nu-mask-l-ref}}
\vskip -6mm
\footnotesize 
\setbox\tablebox=\vbox{ %
\newdimen\digitwidth 
\setbox0=\hbox{\rm 0}
\digitwidth=\wd0
\catcode`*=\active
\def*{\kern\digitwidth}
\newdimen\signwidth
\setbox0=\hbox{+}
\signwidth=\wd0
\catcode`!=\active
\def!{\kern\signwidth}
\newdimen\decimalwidth
\setbox0=\hbox{.}
\decimalwidth=\wd0
\catcode`@=\active
\def@{\kern\signwidth}
\halign{ \hbox to 1in{#\leaderfil}\tabskip=0em& 
    \hfil#\hfil\tabskip=0.5em& 
    \hfil#\hfil\tabskip=0.5em& 
    \hfil#\hfil\tabskip=0.5em& 
    \hfil#\hfil\cr 
\noalign{\doubleline}
\omit Spectrum  & Multipole range & Mask  &
$\chi^2_{\Lambda\textrm{CDM}}/\nu_{\textrm{dof}}$  & PTE \cr
\noalign{\vskip 2pt\hrule\vskip 2pt}
$100\times100$  & *50 -- 1200 & CL49  & 1.01 & 0.40 \cr
$143\times143$  & *50 -- 2000 & CL31  & 0.96 & 0.84 \cr
$143\times217$  & 500 -- 2500 & CL31  & 1.04 & 0.10 \cr
$217\times217$  & 500 -- 2500 & CL31  & 0.96 & 0.90 \cr
\noalign{\vskip 4pt}
Combined& *50 -- 2500 & CL31/49 & 1.04 & 0.08 \cr
\noalign{\vskip 2pt\hrule\vskip 2pt}
}}
\endPlancktable 
\endgroup
\end{table}

Given these masks and angular ranges, we compute the angular power
spectra and covariance matrices, and construct the \camspecs
likelihood.  The angular power spectra for each frequency combination
are shown in Fig.~\ref{fig:PS_CMB+nu}, and compared to spectra derived
from the 70~GHz and 353~GHz \Planck\ maps.

\begin{table*}[h]
\caption{Overview of cosmological parameters used in this analysis, including symbols, the baseline values if fixed for the standard \LCDM\ model, and their definition (see text for further details).
The top block lists the estimated parameters, with (uniform) prior ranges priors given in square
brackets. The lower block lists derived parameters.\label{tab:cosmo-params}}
\begin{centering}
\medskip{}
\par\end{centering}
\centering{}\input{Table-params_cosmo_def.tex} 
\end{table*}

\begin{table*}[h]
\caption{Overview of parameters describing astrophysical foreground modeling, instrumental calibration and beam uncertainties, including symbols, definitions, and prior ranges (see text for further details). Square brackets denote
hard priors, parentheses indicate Gaussian priors. The `Likelihood'
column indicates whether a parameter is used by the \camspec\ (C) and/or
\plik\ (P) likelihood. Note that the beam eigenmode amplitudes 
require a correlation
matrix to fully describe their joint prior, and that all but $\beta^1_1$
are marginalized over internally rather than sampled explicitly.\label{tab:fg-params}}
\begin{centering}
\medskip{}
\par\end{centering}
\centering{}\input{Table-params_foregrounds_def.tex}
\end{table*}

\begin{figure}[t]
\begin{centering}
\includegraphics[bb=50bp 50bp 400bp 600bp,clip,width=1\columnwidth]{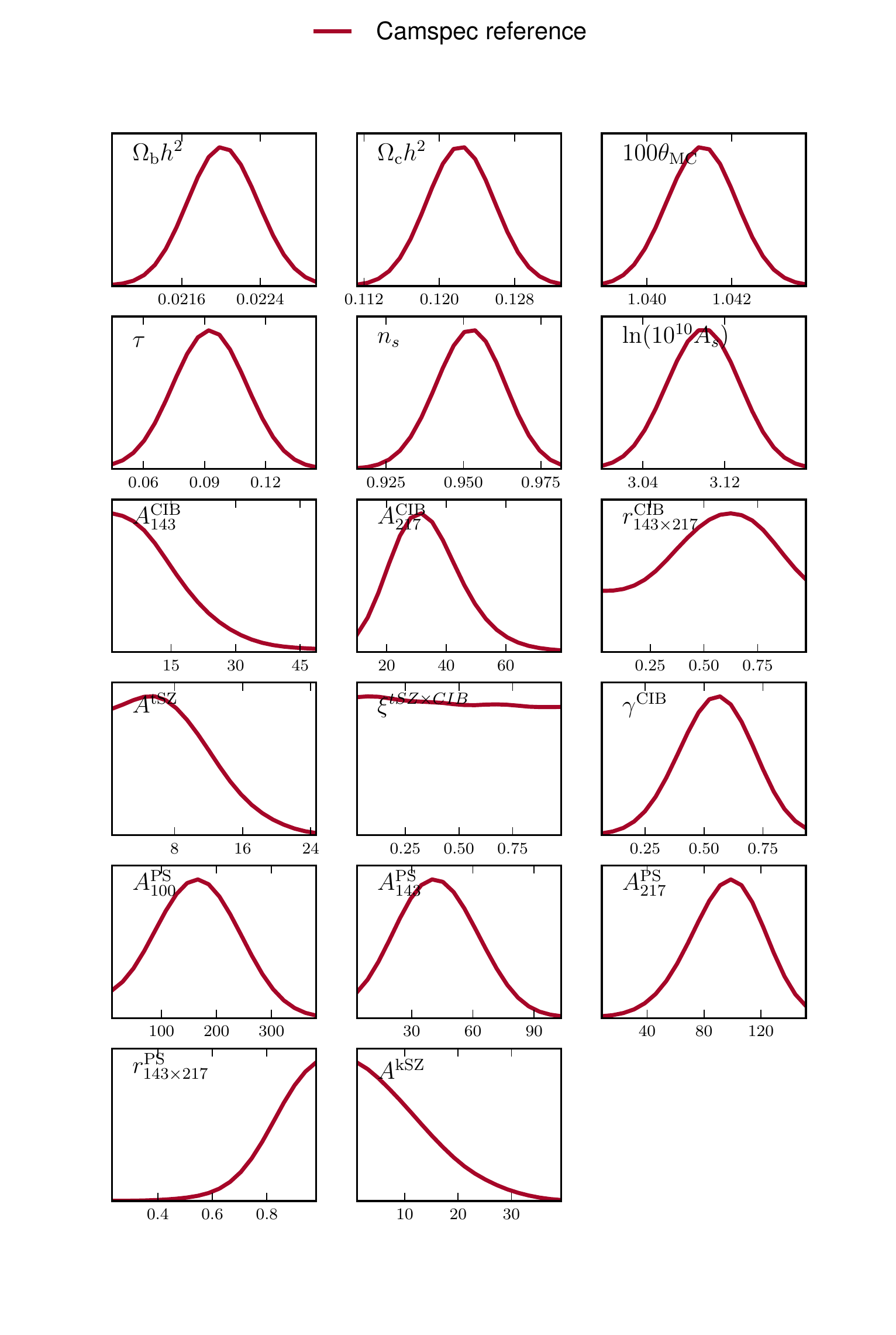} 
\par\end{centering}
\caption{Marginal posterior distributions for the six cosmological (top two rows) and eleven nuisance
parameters (lower four rows) estimated with the \camspec\ likelihood. \textcolor{red}{\label{fig:RefHL_Post}}}
\end{figure}

We use the likelihood to estimate six $\Lambda$CDM cosmological parameters,
together with a set of 14 nuisance parameters (11 foreground parameters,
two relative calibration parameters, and one beam error parameter\footnote{The calibration parameters $c_{100}$ and 
$c_{217}$ are relative to the $143\times143$\ghz cross-spectrum, whose calibration is held fixed. Only the first beam error
eigenmode of the $100\times100$~GHz cross-spectrum is explored, all other eigenmodes being internally marginalised over}, 
described in Sect.~\ref{sub:Sky-model}. Tables\,\ref{tab:cosmo-params}
and \ref{tab:fg-params} summarize these parameters and the associated priors\footnote{We use the approximation $\theta_{\mathrm{MC}}$ to the acoustic
  scale $\theta_{\star}$ (the ratio of the comoving size of the
  horizon at the time of recombination, $r_{S}$, to the angular
  diameter distance at which we observe the fluctuations, $D_{A}$) which was
  introduced by \cite{1996ApJ...471..542H}. $\theta_{\mathrm{MC}}$ is commonly used, e.g.,
  in {\tt CosmoMC}, to speed up calculations; see also
  \cite{2002PhRvD..66f3007K} for further details. }.  Apart from the
beam eigenmode amplitude and calibration factors, we adopt uniform
priors. To map out the corresponding posterior distributions we use
the methods described in \cite{planck2013-p11}, and the resulting marginal
distributions are shown in Fig.~\ref{fig:RefHL_Post}. Note that on the parameters $A^{\mathrm{tSZ}}$, $A^{\mathrm{kSZ}}$ and $A^{\mathrm{CIB}}_{143}$ we are using larger prior ranges as compared to 
\cite{planck2013-p11}. 

Figure~\ref{fig:RefHL_Post} shows the strong constraining power of the \Planck\ data, but also highlights some of the deficiencies of a
`\Plancks-alone' analysis. The thermal SZ amplitude provides a good
example; the distribution is broad, and the `best fit' value is
excluded by the ACT and SPT high resolution CMB experiments
\citep{reichardt12}.
For the CIB amplitudes, the upper bound on e.g., $A_{143}^{\mathrm{CIB}}$ is significantly weaker than the ACT and SPT constraints. To accurately estimate
the foreground parameters at the $\lesssim\mu{\rm K}^{2}$
level, we need to supplement the \Planck\ power spectra with 
temperature data from  ACT and SPT, as described in \cite{planck2013-p11}.
The fiducial model and foreground parameters used in the \camspec\ likelihood are therefore 
derived from a joint \Planck+ACT+SPT analysis and is \textit{not} based on the parameters listed
in Table~\ref{tab:params-ref}. In the rest of this section, we will
use the parameters of Table~\ref{tab:params-ref} to discuss
inter-frequency residuals.

\begin{figure*}
\begin{centering}
\includegraphics[bb=40bp 110bp 590bp 581bp,clip,width=0.93\columnwidth]{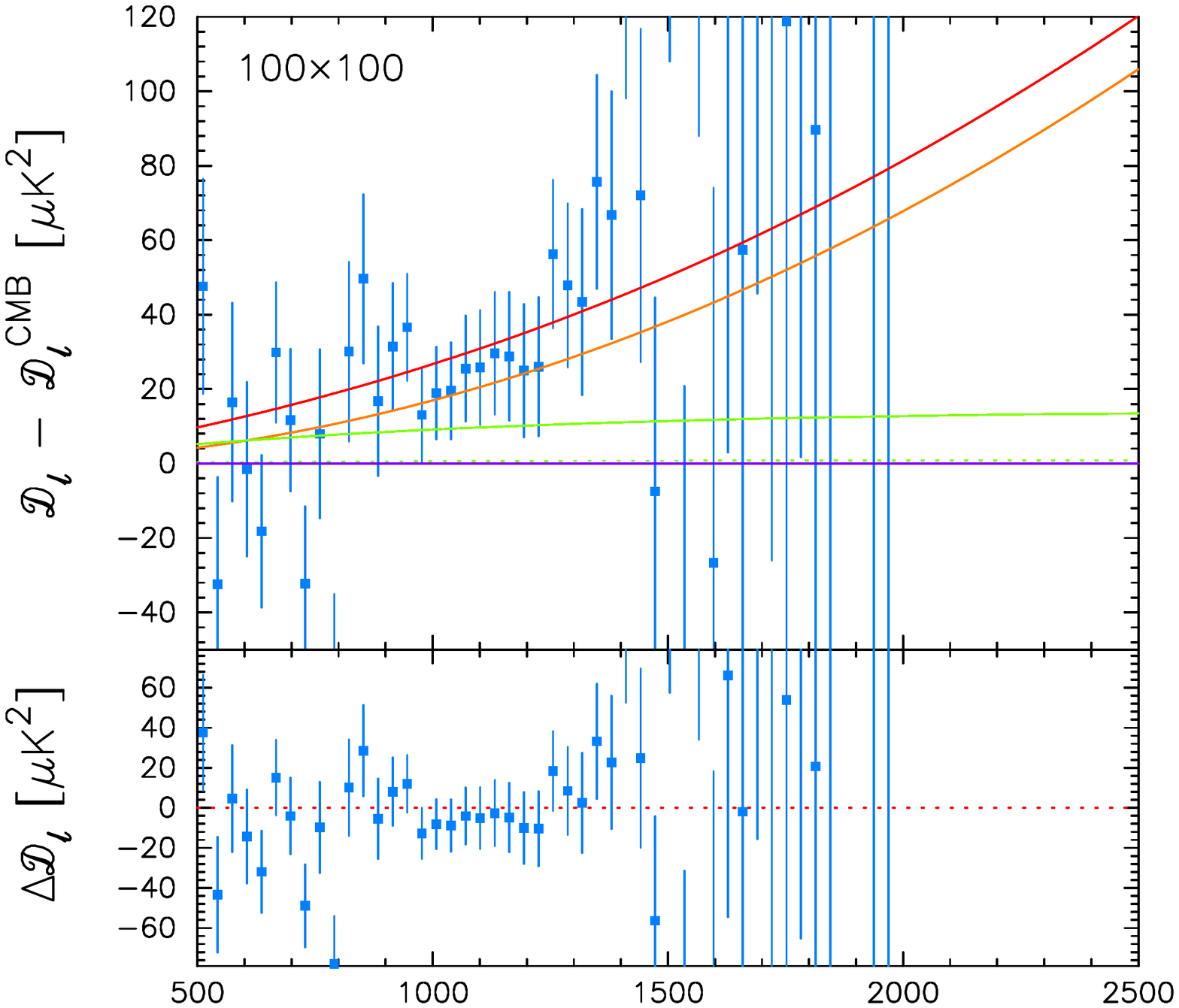}\includegraphics[bb=40bp 110bp 590bp 581bp,clip,width=0.93\columnwidth]{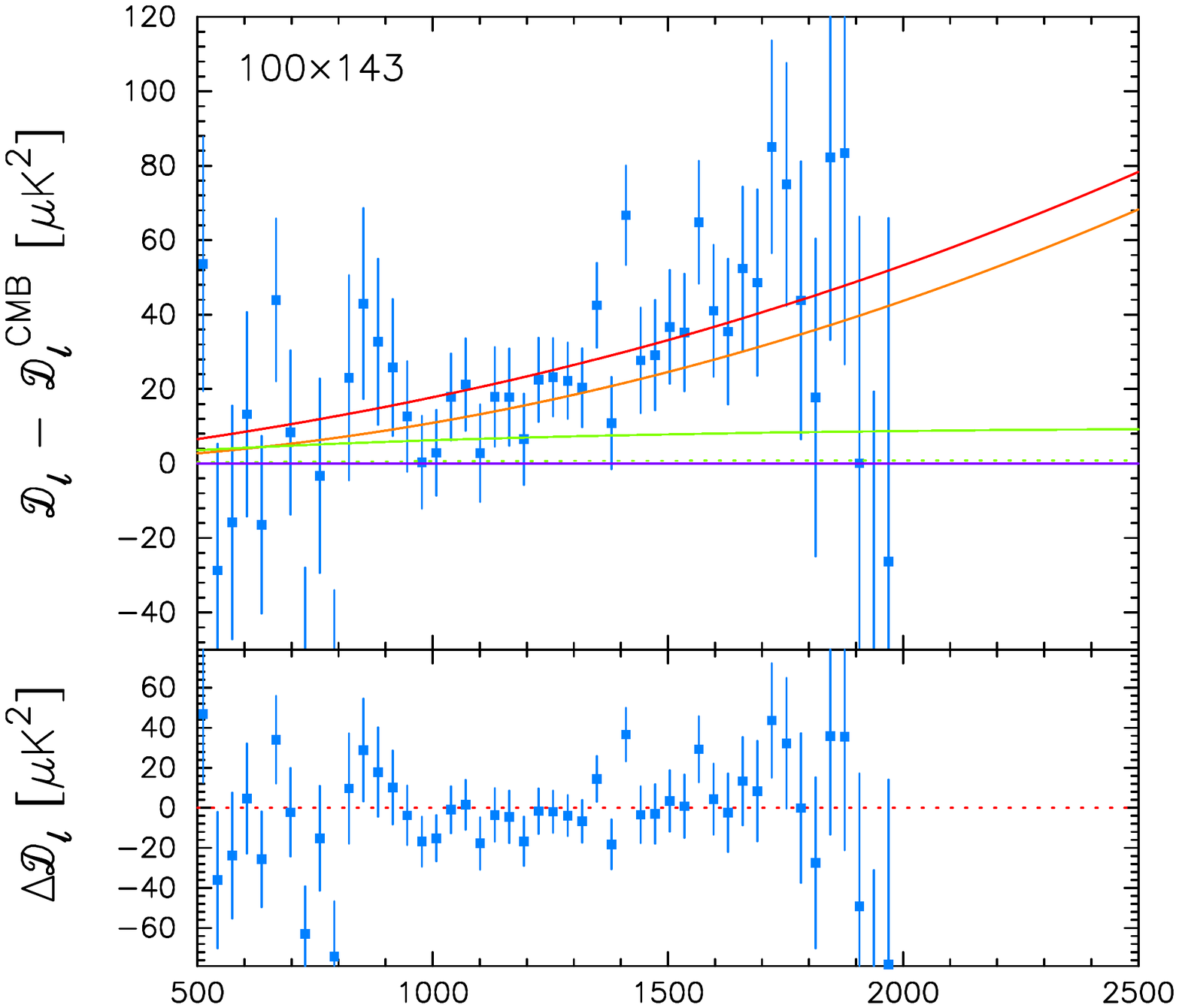} 
\includegraphics[bb=40bp 110bp 590bp 581bp,clip,width=0.93\columnwidth]{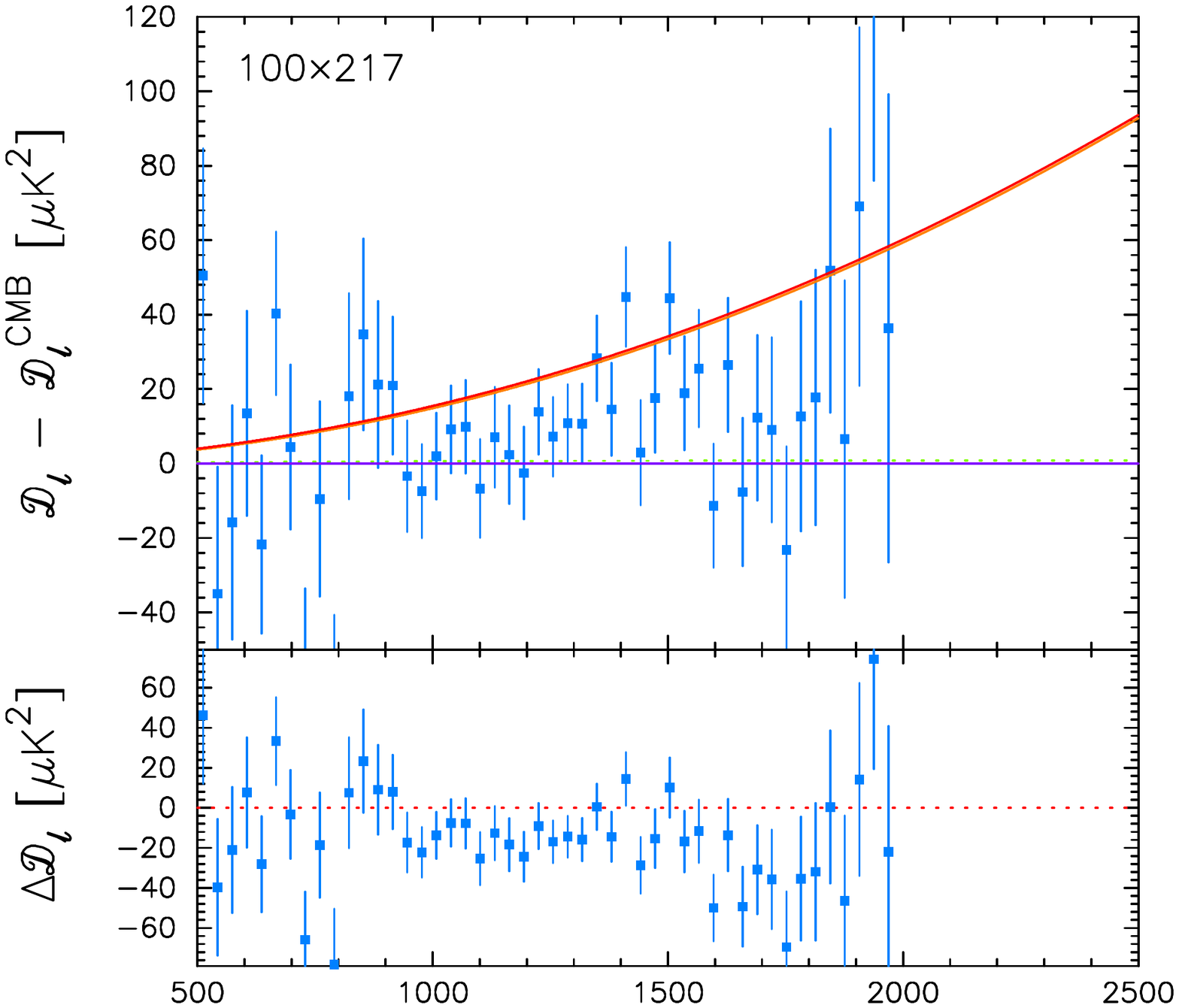}\includegraphics[bb=40bp 110bp 590bp 581bp,clip,width=0.93\columnwidth]{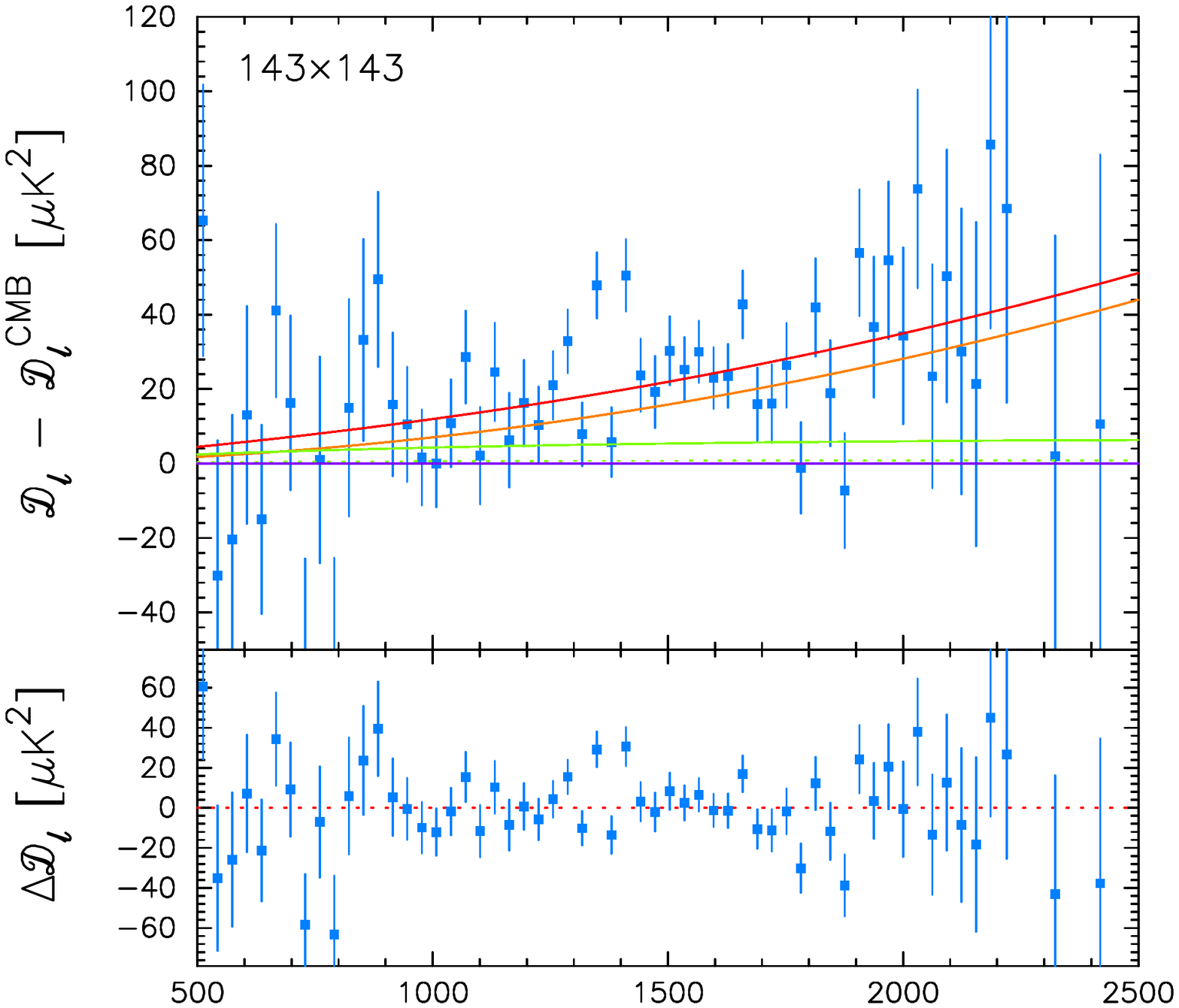} 
\includegraphics[bb=40bp 70bp 590bp 581bp,clip,width=0.93\columnwidth]{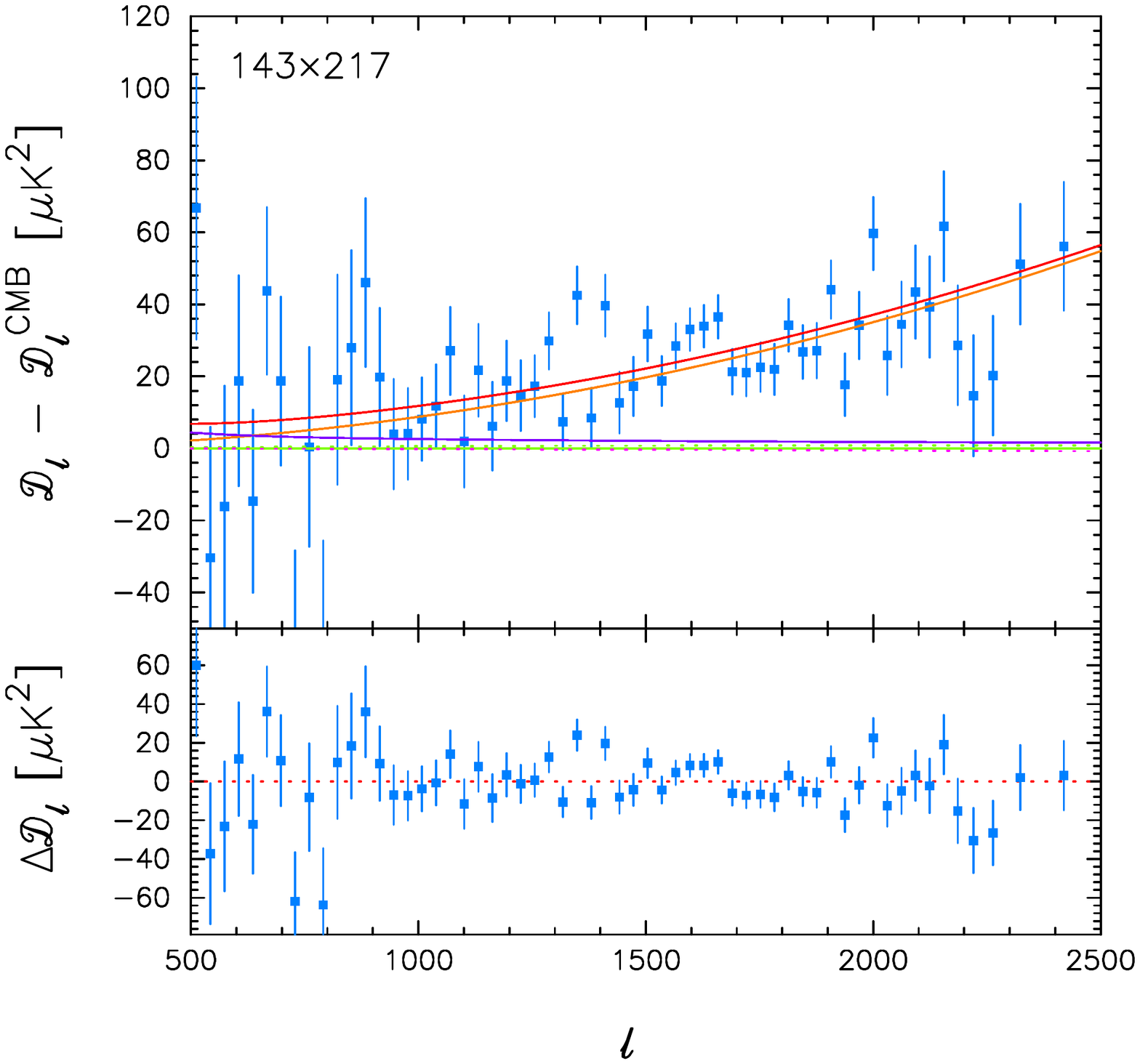}\includegraphics[bb=40bp 70bp 590bp 581bp,clip,width=0.93\columnwidth]{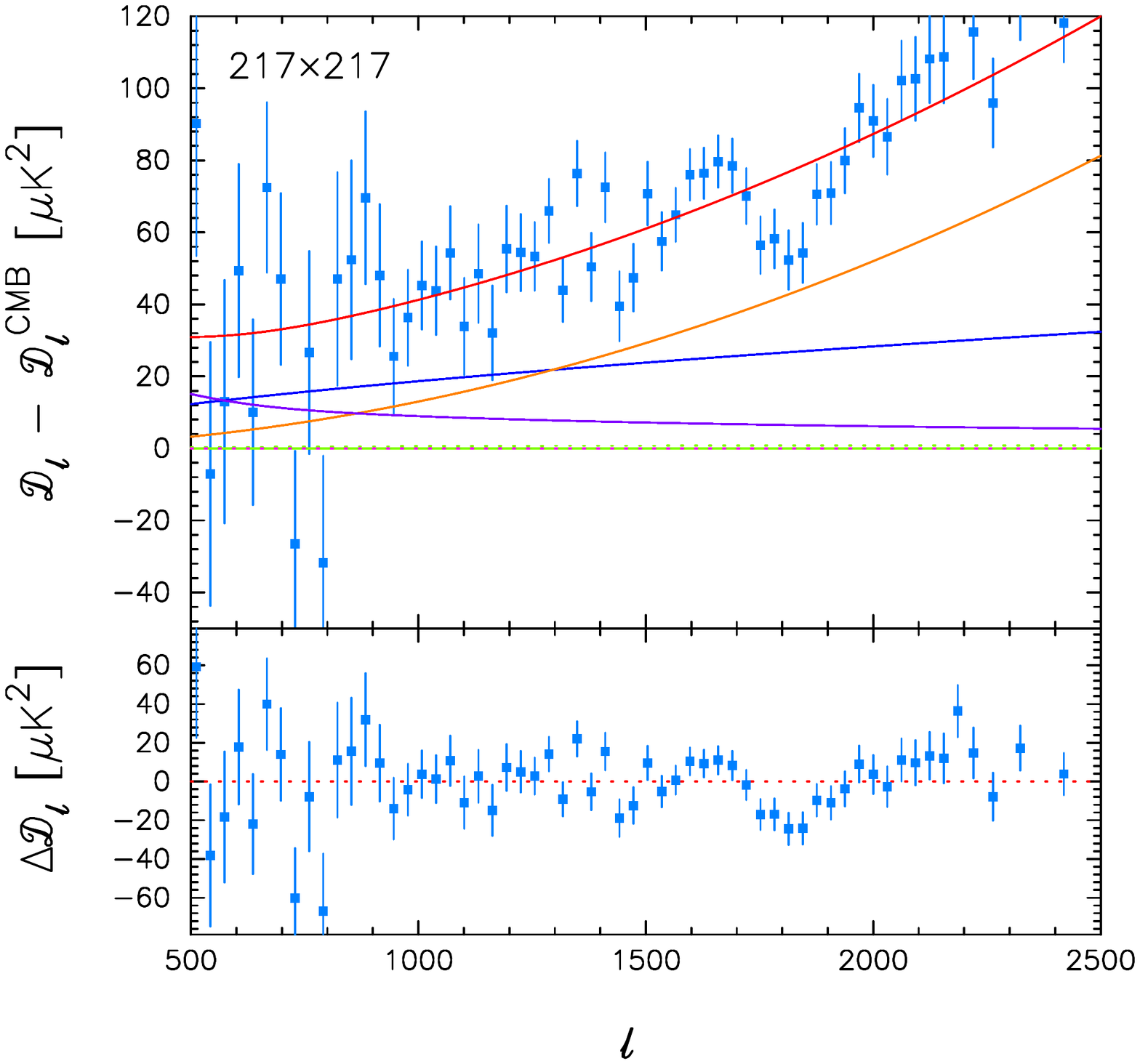}
\par\end{centering}
\centering{}\caption{Foreground model over the full range of HFI cosmological frequency
combinations. The upper panel in each plot shows the residual between
the measured power spectrum and the `best-fit' primary CMB power spectrum, i.e., the unresolved foreground residual for each frequency
combination. The lower panels show the residuals after removing the
best-fit foreground model. The lines in the upper panels show the
various foreground components. Major foreground components are shown
by the solid lines, colour coded as follows: total foreground spectrum
(red); Poisson point sources (orange); CIB (blue); thermal SZ (green).
Minor foreground components are shown by the dotted lines: kinetic SZ (green); tSZ\,X\,CIB cross correlation (purple).
The $100\times143$ and $100\times217$\,GHz spectra are
not used in the \camspec\ likelihood. Here we have assumed $r_{100\times143}^{PS}=1$
and $r_{100\times217}^{PS}=1$. \label{fig:ps_nu_FG_ref}}
\end{figure*}

Figure~\ref{fig:ps_nu_FG_ref} shows the foreground residuals and total
residuals after removing the best-fit foreground model for all spectra
(including the $100\times143$ and $100\times217$ spectra, which are
not used in the \camspecs likelihood). The first point to note here is
that \Plancks has a limited ability to disentangle foregrounds. While
the \Plancks data constrain the Poisson point source amplitudes at
each frequency, as well as the CIB amplitude at $217$\,GHz (which
dominates over the Poisson point source amplitude over much of the
multipole range), they have only marginal sensitivity to the tSZ
amplitude in the $100\times100$ spectrum, though the thermal SZ is
strongly degenerate with the Poisson point source amplitude. The
remaining foreground parameters are highly degenerate.  For \Plancks
alone, these minor foreground contributions combine to absorb
inter-frequency residuals.

\begin{figure*}[h!]
\begin{centering}
\includegraphics[bb=35bp 30bp 595bp 580bp,clip,width=1\columnwidth]{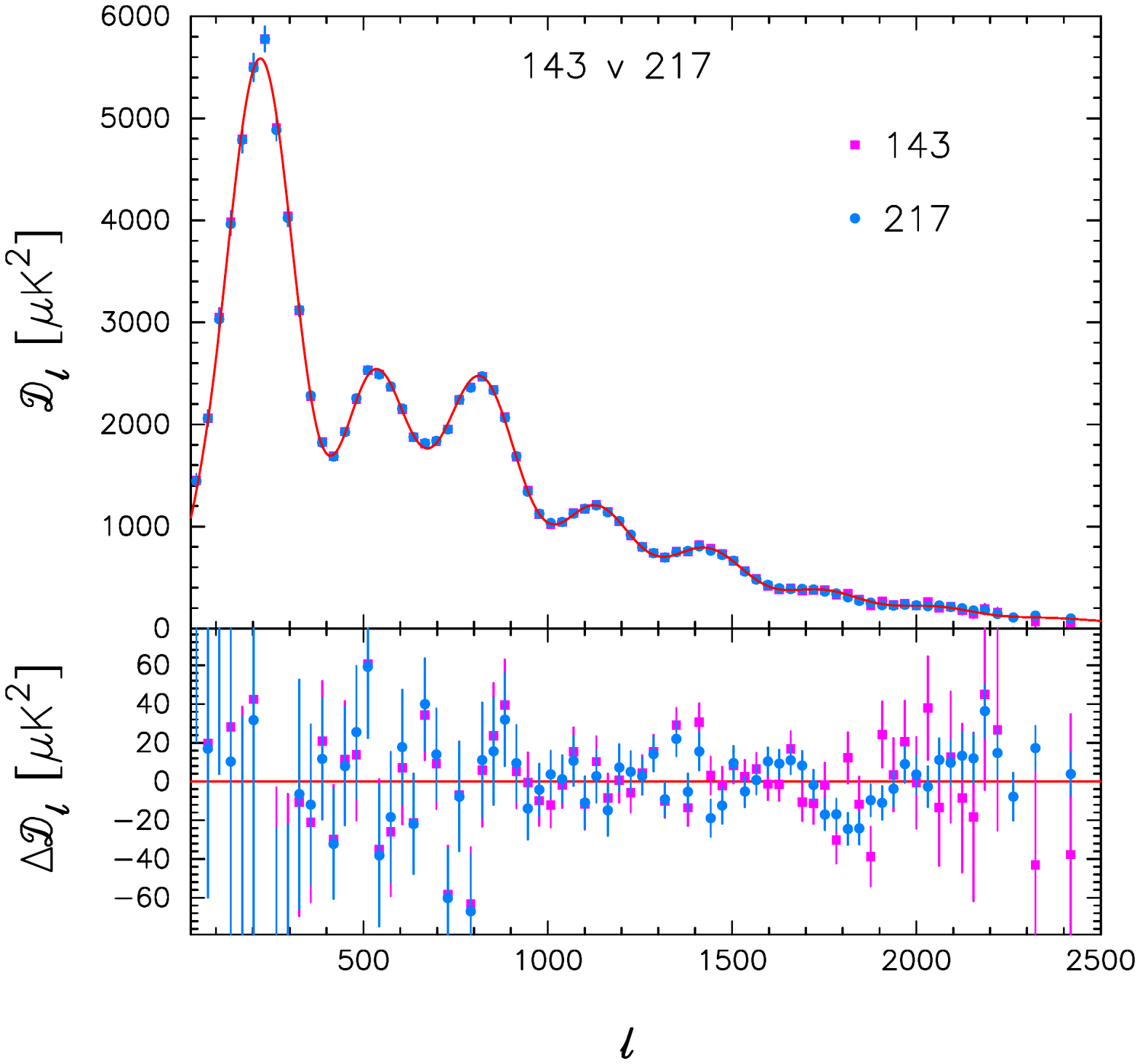}\includegraphics[bb=35bp 30bp 595bp 580bp,clip,width=1\columnwidth]{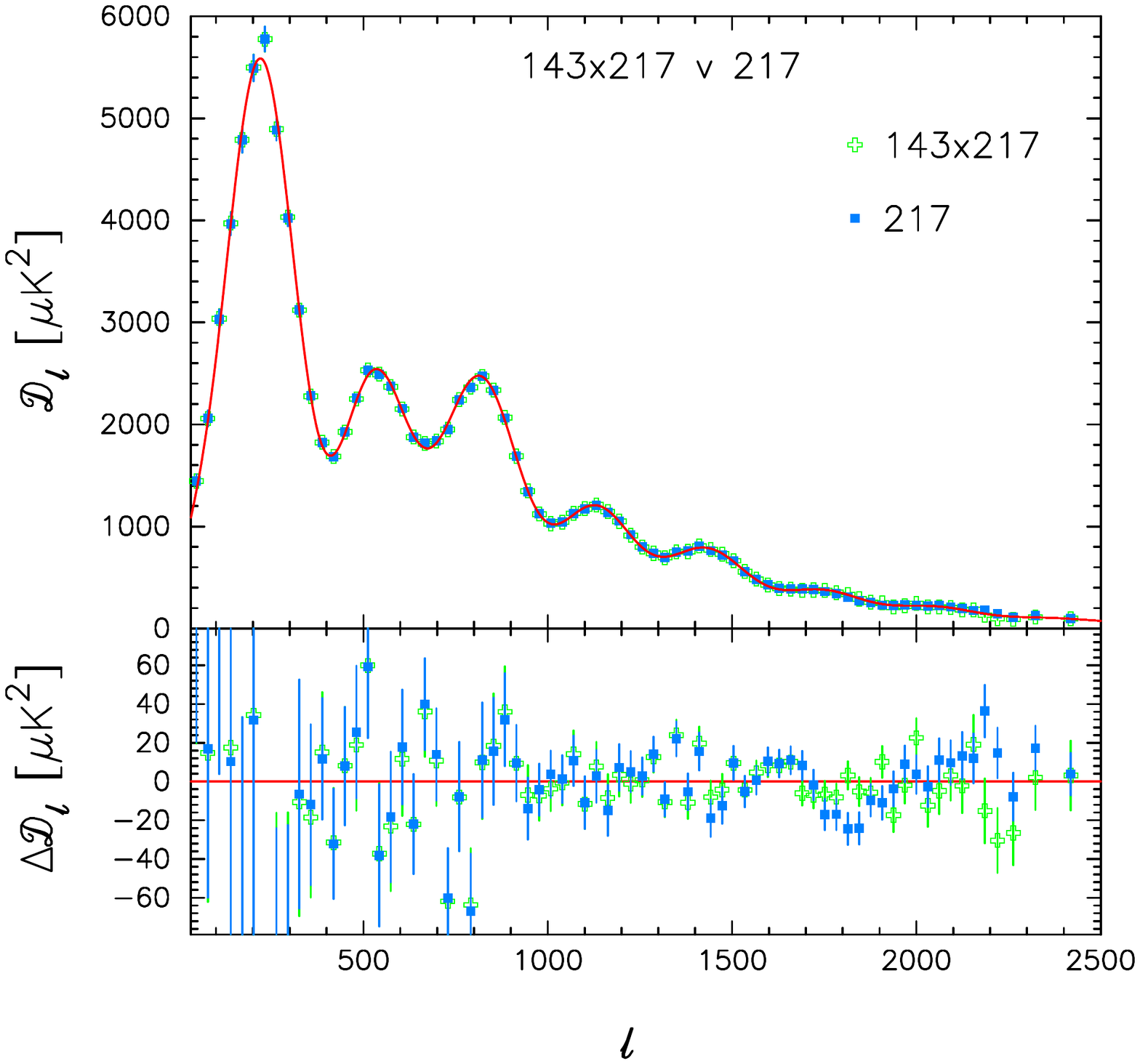} 
\includegraphics[bb=35bp 30bp 595bp 580bp,clip,width=1\columnwidth]{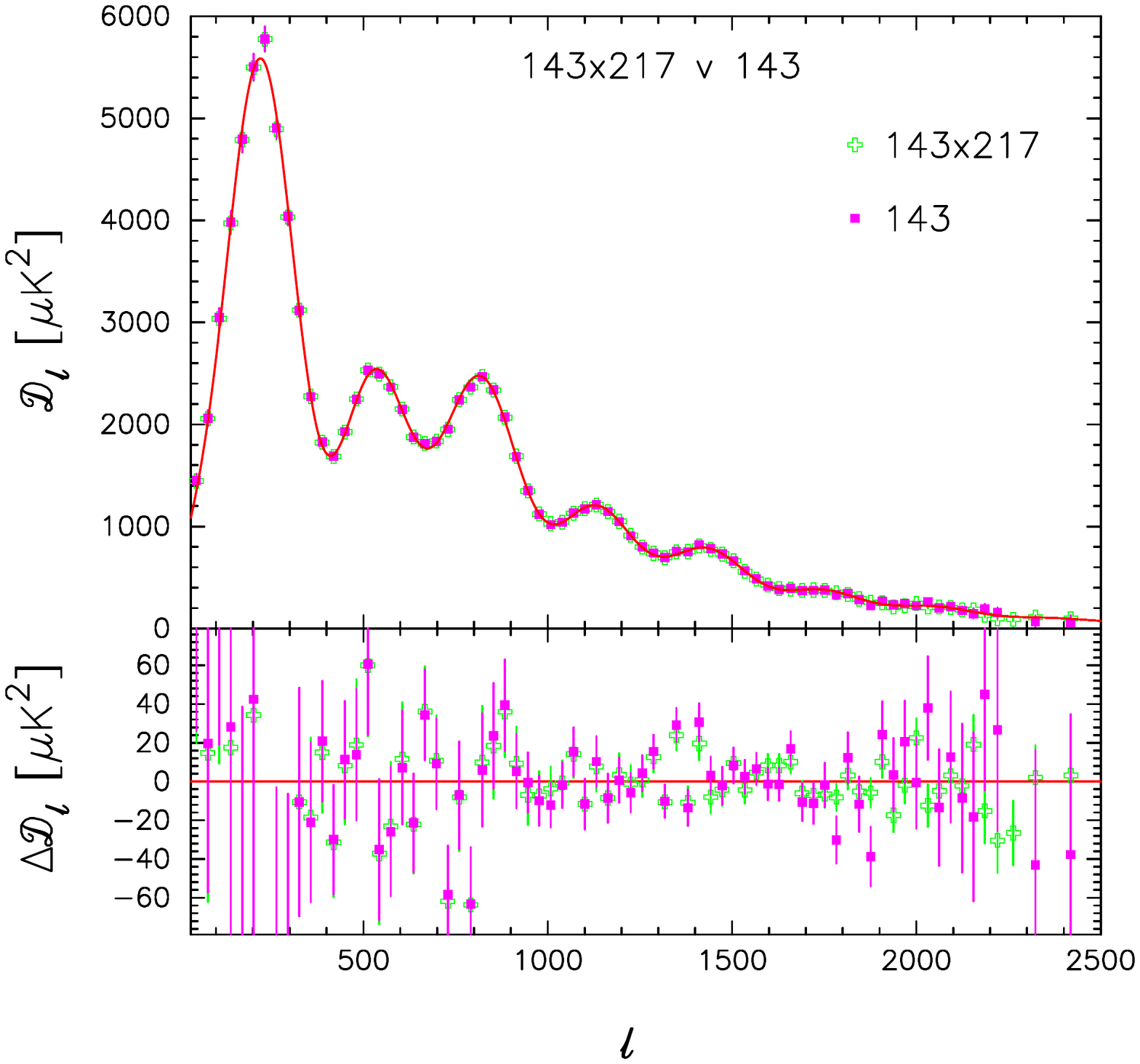}\includegraphics[bb=35bp 30bp 595bp 580bp,clip,width=1\columnwidth]{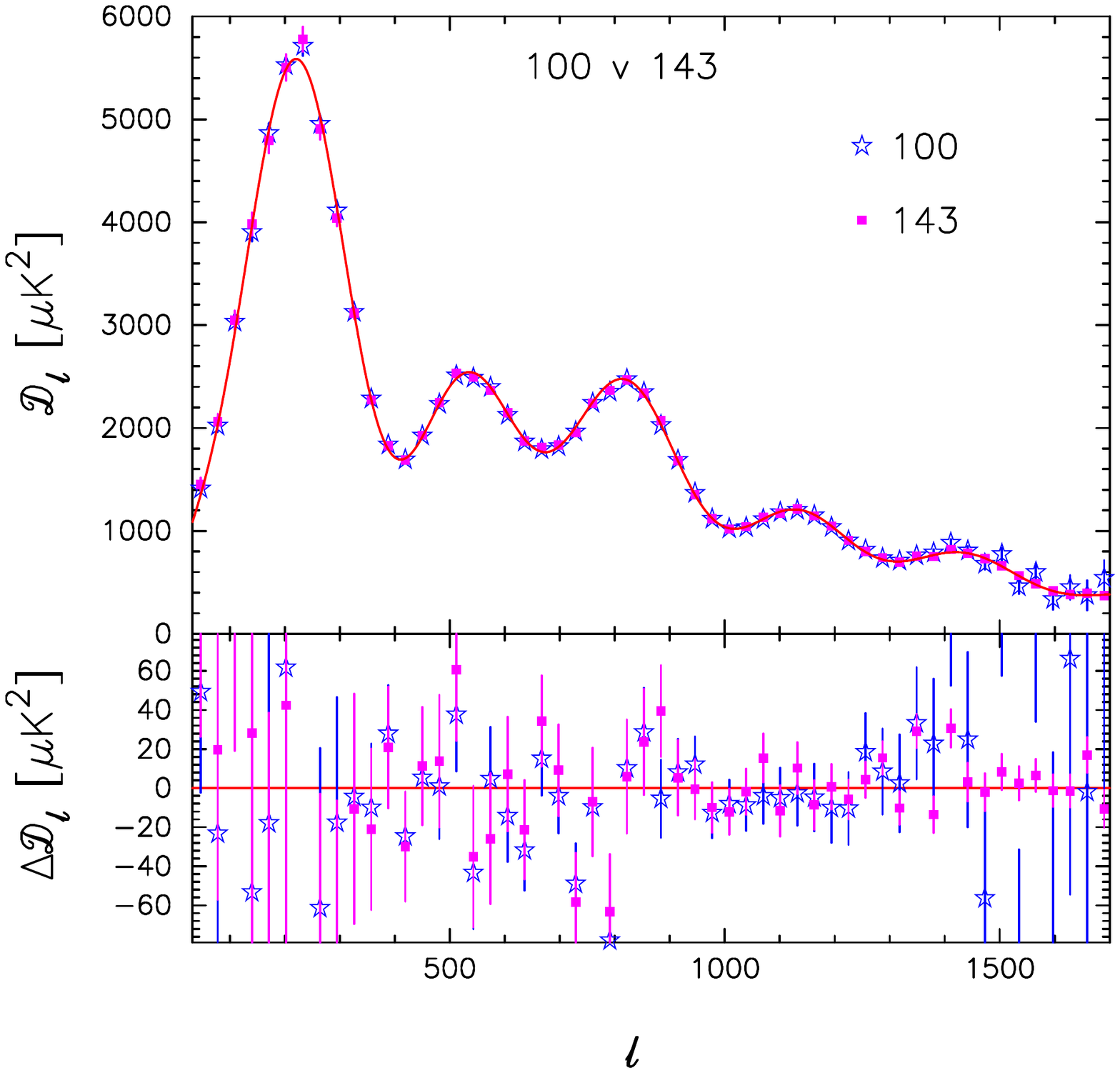}
\par\end{centering}
\caption{Comparison of pairs of foreground subtracted cross spectra, demonstrating
consistency of the residuals with respect to the best-fit theoretical
model. The red line in each of the upper panels shows the theoretical
six parameter $\Lambda$CDM spectrum for the best-fit parameters listed
in Table~\ref{tab:params-ref}. The lower panels show the residuals
with respect to this spectrum, together with error bars computed from
the diagonal components of the covariance matrices of the band averages.
The points here are band-averaged in bins of width $\Delta\ell\sim31$.
\label{fig:ps_nu_CMB_ref}}
\end{figure*}

Pairs of spectra are compared in Fig.~\ref{fig:ps_nu_CMB_ref},
averaged over bands of width $\Delta\ell=31$ below $\ell \lesssim
2000$ and wider bands above 2000. (The error bars show the diagonals of
the covariance matrices of these averages, but it is important to note
that the points are highly correlated even with bin widths as large as
these.)
This comparison shows that each of the spectra used in the
\textit{\camspecs} likelihood is consistent with the best-fit
theoretical spectrum to high accuracy. In fact, each spectrum can be
used to form a likelihood, and each gives a reduced $\chi^{2}$ close
to unity (see Table~\ref{tab:nu-mask-l-ref}).  Thus, the six parameter
$\Lambda$CDM model provides an excellent fit to the
\Planck\ high-$\ell$ power spectra at all frequencies between 100 and
217\,GHz.

\begin{figure}[h]
\centering{}\includegraphics[bb=30bp 40bp 750bp
  595bp,clip,width=1\columnwidth]{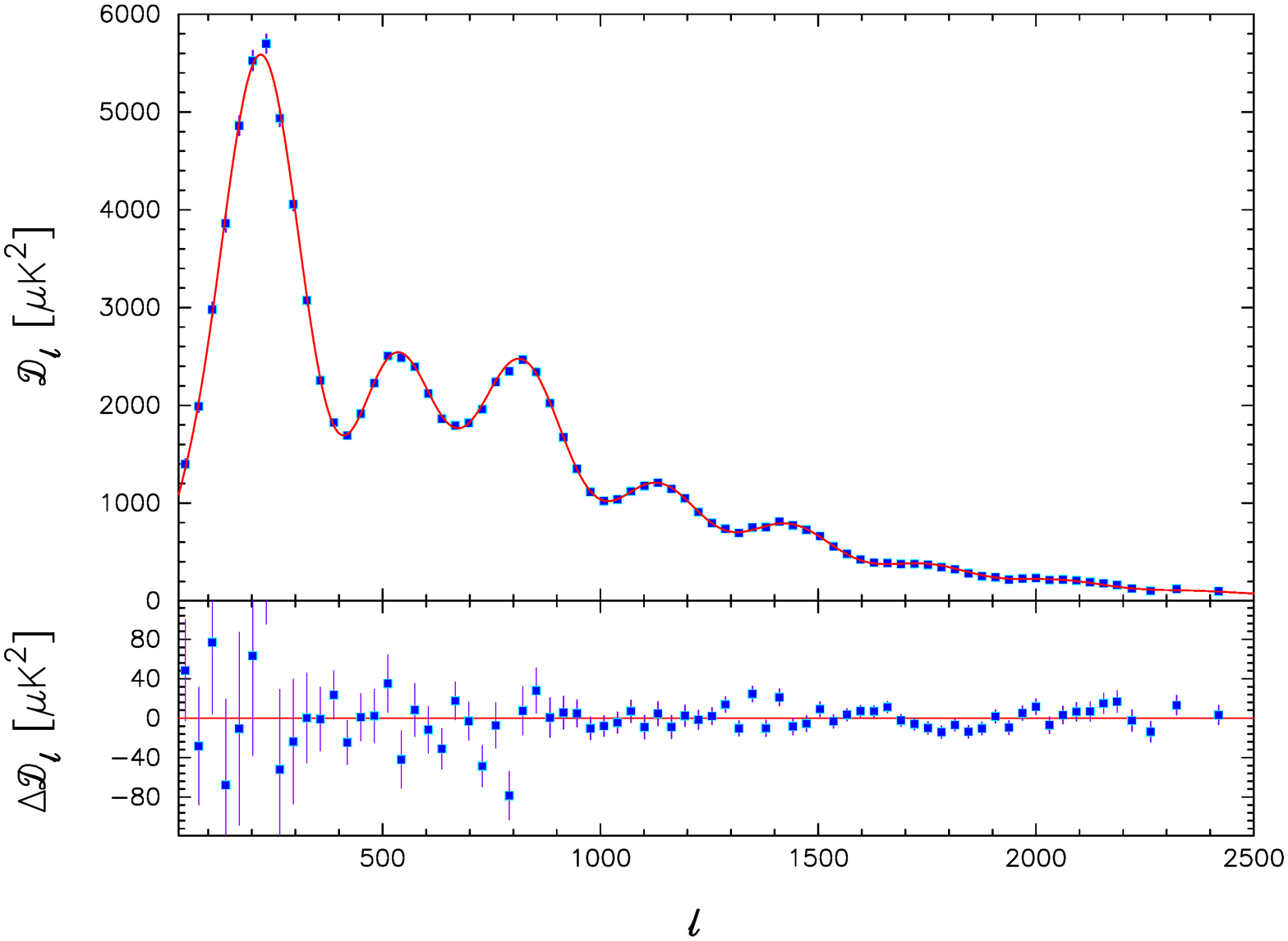}\caption{\emph{Top:}
  \Plancks maximum-likelihood (primary)
CMB spectrum compared 
with the best-fit six parameter $\Lambda$CDM spectrum. \emph{Bottom:}
Power spectrum residuals with respect to the $\Lambda$CDM model.  The
error bars are computed from the diagonal elements of the
band-averaged covariance matrix, as given by Eq.~\ref{CSL4}, including
contributions from foreground and beam transfer function
errors.\label{fig:ps_CMB_ref}}
\end{figure}

Figure~\ref{fig:ps_CMB_ref} shows our maximum likelihood primary CMB
spectrum,
together with the best-fit theoretical spectrum. The residuals with
respect to the model are shown in the lower panel. The error bars are
computed from the diagonal components of the band-averaged covariance
matrix.
The binning scheme is the same as in Fig.~\ref{fig:ps_nu_CMB_ref}.

\begin{figure*}
\begin{centering}
\includegraphics[bb=34bp 63bp 741bp 582bp,clip,width=1\columnwidth]{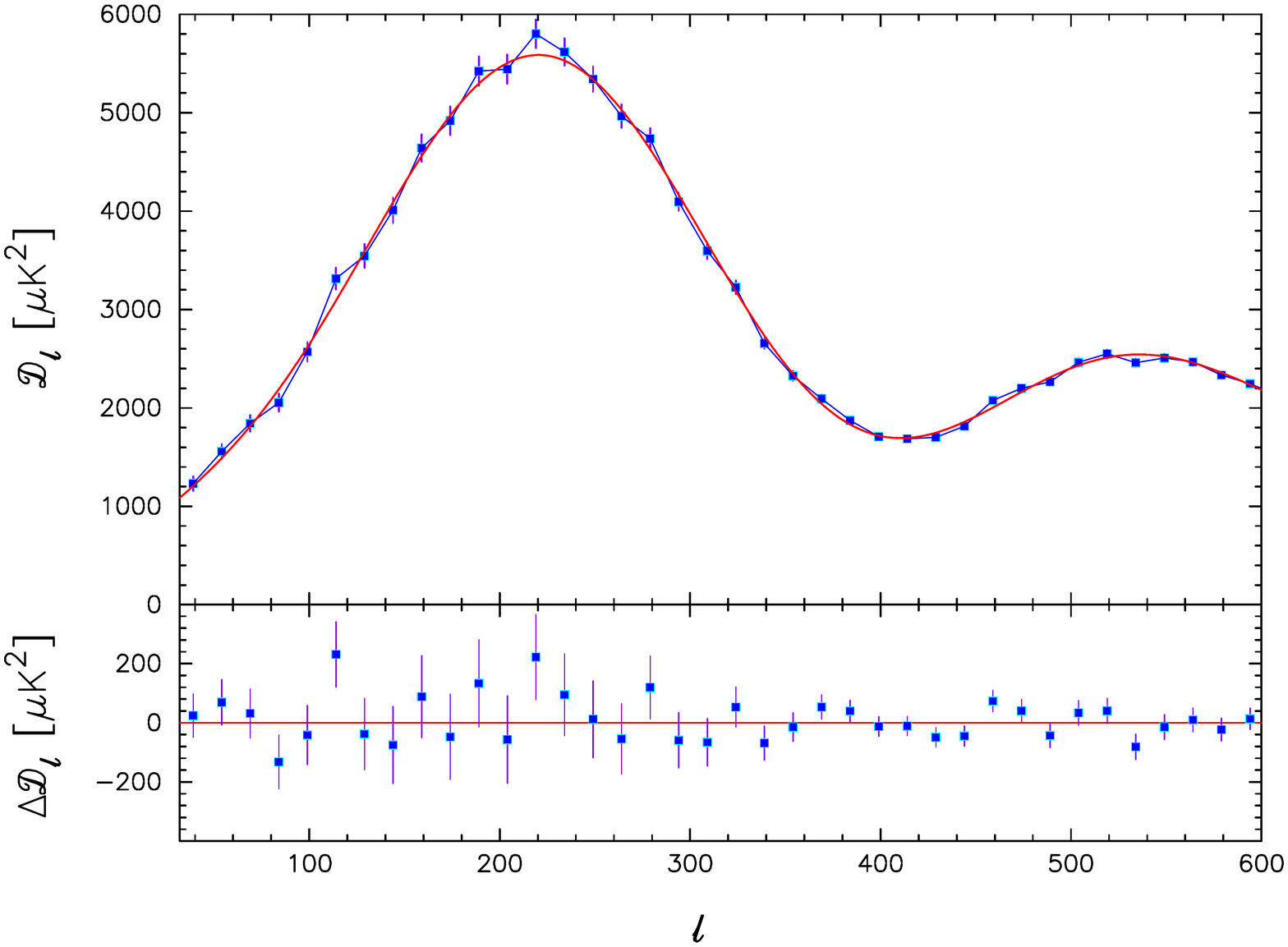}\includegraphics[bb=34bp 63bp 741bp 582bp,clip,width=1\columnwidth]{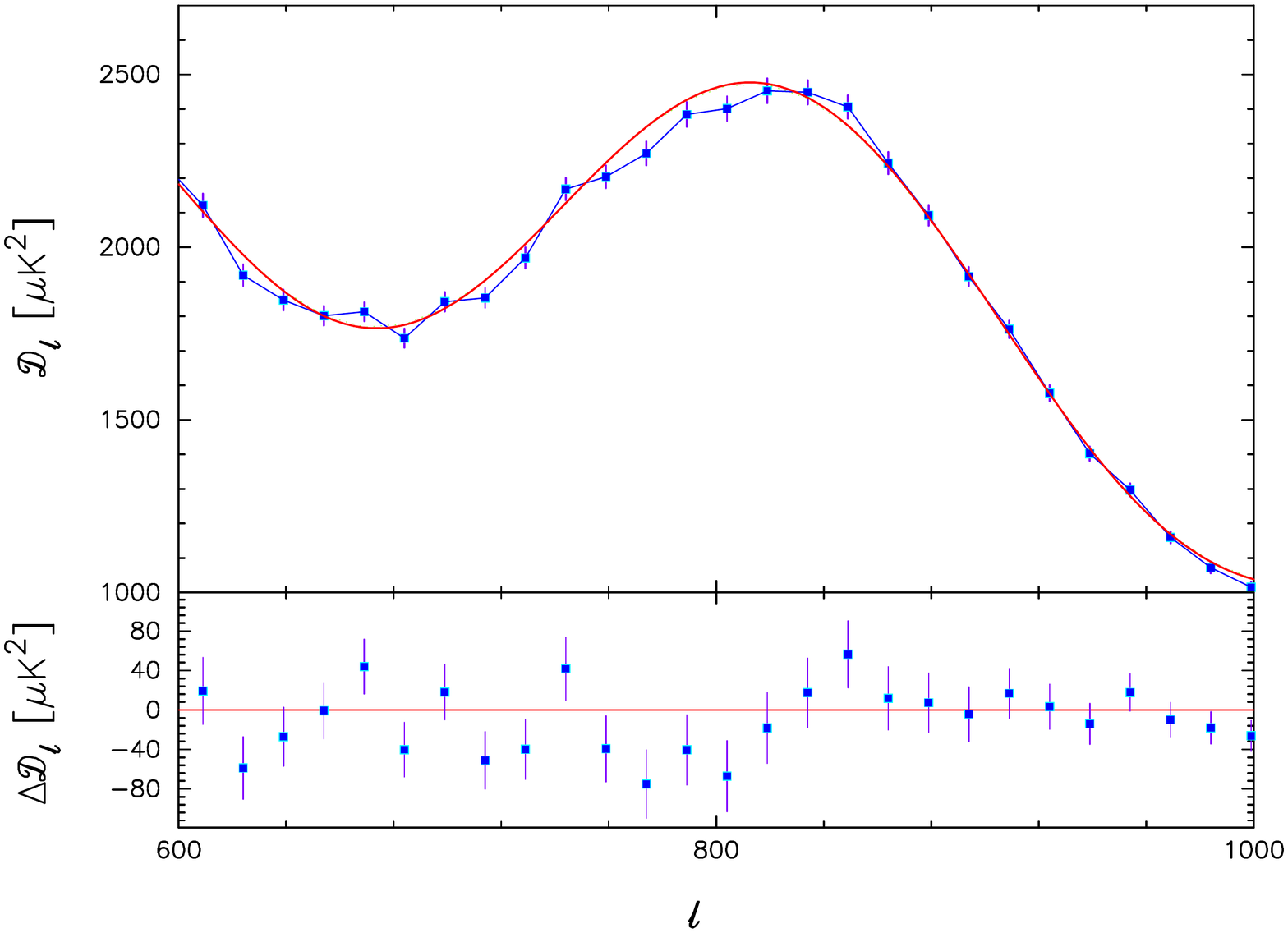} 
\par\end{centering}

\begin{centering}
\includegraphics[bb=34bp 63bp 741bp 582bp,clip,width=1\columnwidth]{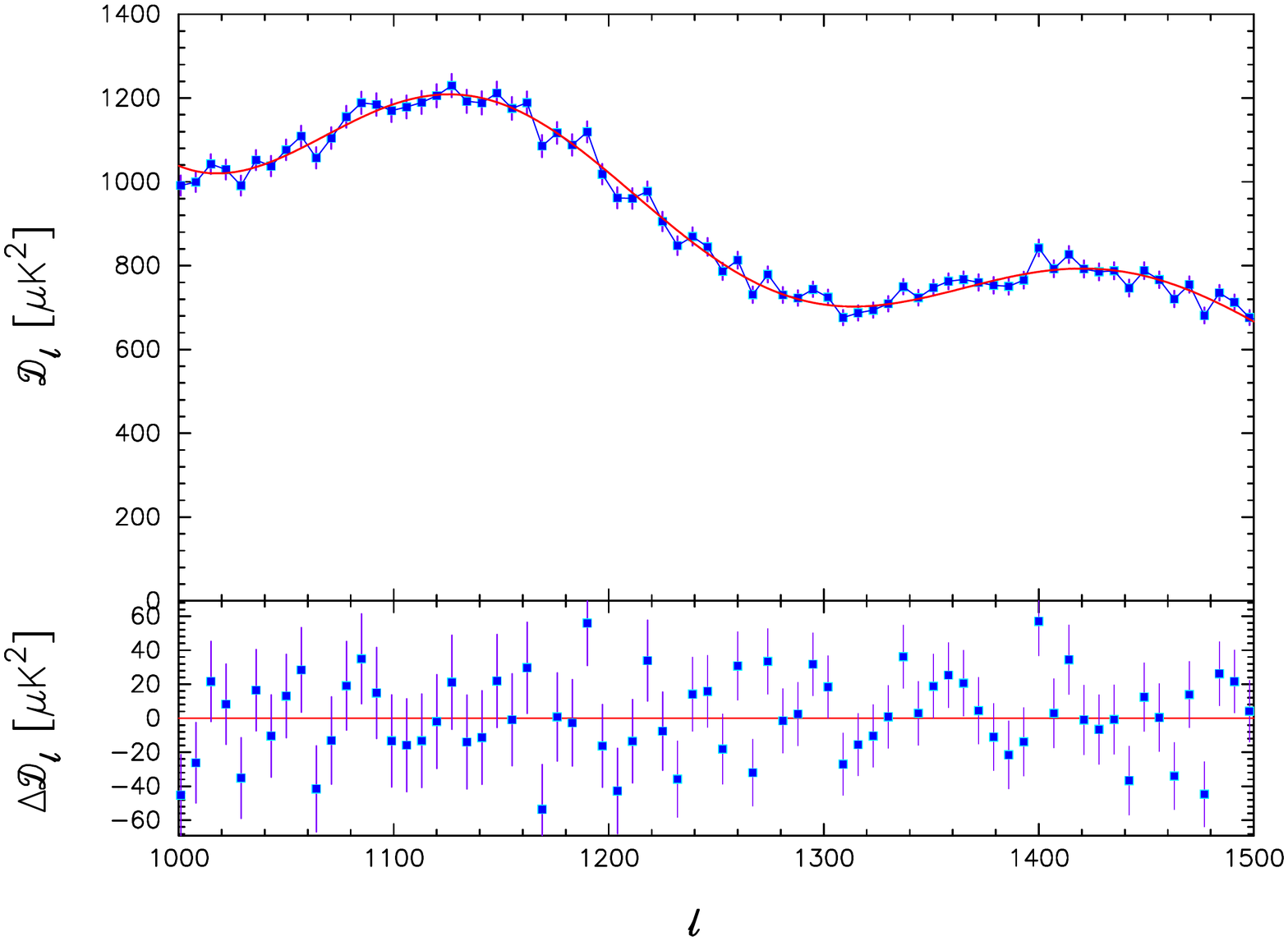}\includegraphics[bb=34bp 63bp 741bp 582bp,clip,width=1\columnwidth]{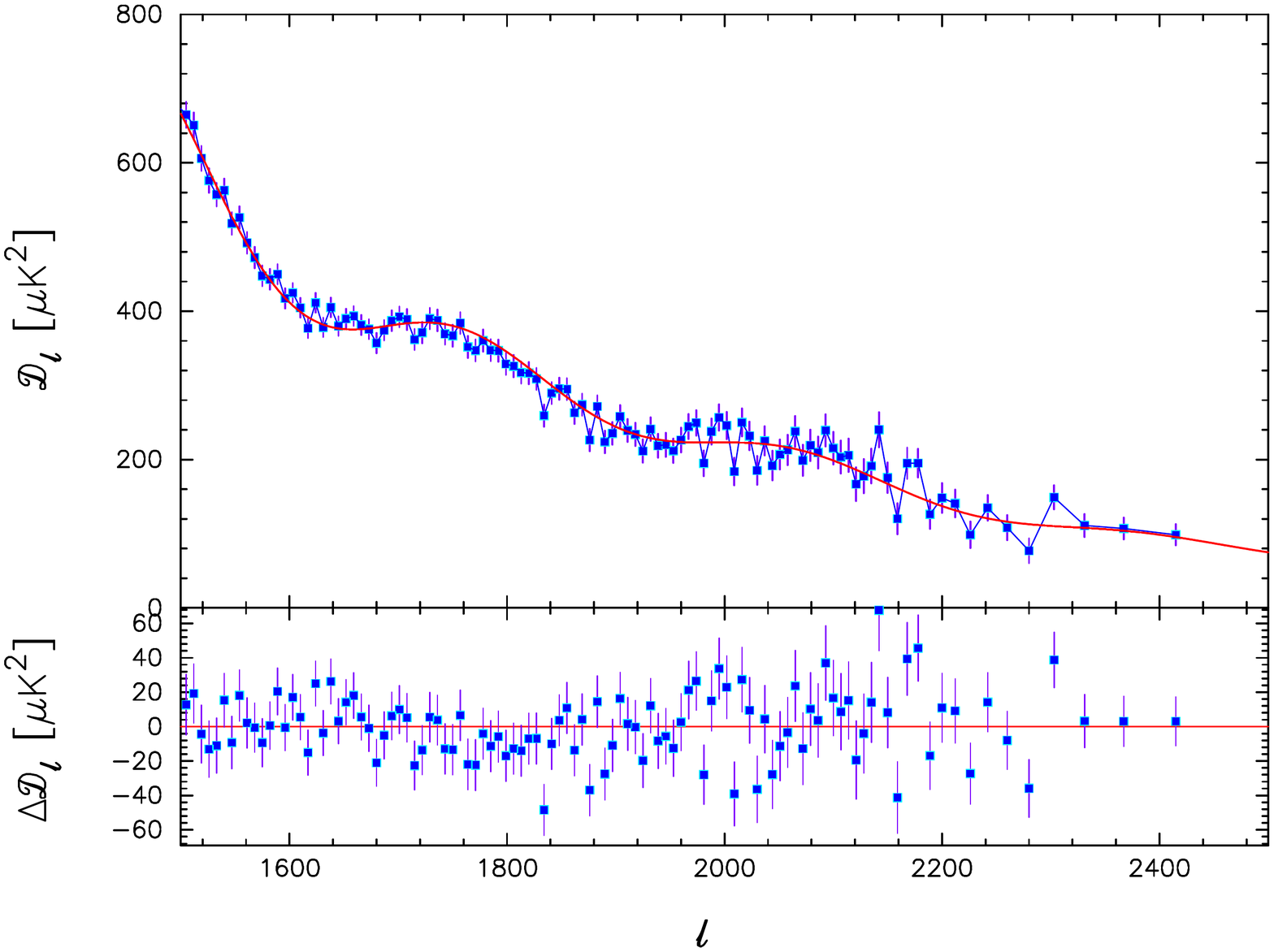}
\par\end{centering}

\centering{}\caption{Zoom-in of regions of the \Plancks primary CMB
  power spectrum using finer bin widths ($\Delta\ell=15$ for
  $\ell<1000$ and $\Delta\ell=7$ for $1000\le\ell\le2200$. In the
  upper panels, the red lines show the best-fit $\Lambda$CDM spectrum,
  and the blue lines join the \Plancks data points. Error bars are
  computed as in Fig.~\ref{fig:ps_CMB_ref}.
\label{fig:ps_CMB_ref-zoom}}
\end{figure*}

Finally, in Fig.~\ref{fig:ps_CMB_ref-zoom} we zoom in on this spectrum
in four multipole ranges using finer binning. The correlated
fluctuations seen in this figure are mask-induced, and perfectly
compatible with the six parameter $\Lambda$CDM model. Features such as
the `bite' missing from the third peak at $\ell\sim800$ and the
oscillatory features in the range $1300\lesssim\ell\lesssim1500$ are
in excellent agreement with what we expect from our covariance
matrices and from simulations; see
Appendix~\ref{app:Covariance-matrix-of-combined-estimates} for a
few specific examples.

%% file: Table-params_cosmo_def.tex

\begingroup 
\newdimen\tblskip \tblskip=5pt
\vskip -6mm
\footnotesize 
\setbox\tablebox=\vbox{ %
\newdimen\digitwidth 
\setbox0=\hbox{\rm 0}
\digitwidth=\wd0
\catcode`*=\active
\def*{\kern\digitwidth}
\newdimen\signwidth
\setbox0=\hbox{+}
\signwidth=\wd0
\catcode`!=\active
\def!{\kern\signwidth}
\halign{\hbox to 2.7cm{#\leaderfil}\tabskip=0.4cm&  \hfil#\hfil\tabskip=0.6cm& \hfil#\hfil\tabskip=0.6cm&  #\hfil\tabskip=0pt\cr
\noalign{\doubleline}
\omit\hfil Parameter\hfil&\omit\hfil Prior range\hfil&\omit\hfil Baseline\hfil&\omit\hfil Definition\hfil\cr
\noalign{\vskip 3pt\hrule\vskip 3pt}
$\omb \equiv \Omb h^2$& $[!0.005, !*0.1*]$& \dots& Baryon density today\cr
$\omc \equiv \Omc h^2$& $[!0.001, !*0.99]$& \dots& Cold dark matter density today\cr
$100\theta_{\mathrm{MC}}$ & $[!0.5**, !10.0*]$& \dots& $100\,{\times}$ approximation to $\rstar/D_{\rm A}$ (used in CosmoMC)\cr
$\tau                $&   $(*0.088\pm 0.015*)$& \dots& Thomson scattering optical depth due to reionization\cr
$\Omk            $&  $[-0.3**, !*0.3*]$& 0& Curvature parameter today with $\Omtot= 1 - \Omk$\cr
$\yhe                $& $[!0.1**, !*0.5*]$& BBN& Fraction of baryonic mass in helium\cr
$\ns           $& $[!0.9**, !*1.1*]$& \dots& Scalar spectrum power-law index ($k_0 = 0.05~\Mpc^{-1}$)\cr
$\ln(10^{10}\As) $& $[!2.7**, !*4.0*]$& \dots& Log power of the primordial curvature perturbations ($k_0 = 0.05~\Mpc^{-1}$)\cr
\noalign{\vskip 3pt\hrule\vskip 3pt}
$\Oml      $&     & \dots& Dark energy density divided by the critical density today\cr
Age                 &  & \dots& Age of the Universe today (in Gyr)\cr
$\Omm     $&  & \dots& Matter density (inc.\ massive neutrinos) today divided by the critical density\cr
$\zre                $&   & \dots& Redshift at which Universe is half reionized\cr
$H_0                 $&[!20**,*!100*]& \dots& Current expansion rate in $\rm{km}\, \rm{s}^{-1}\Mpc^{-1}$\cr
\noalign{\vskip 3pt\hrule\vskip 3pt}}}
\endPlancktable 
\endgroup

%% file: Table-params_foregrounds_def.tex
%
\begingroup 
\newdimen\tblskip \tblskip=5pt
\vskip -3mm
\footnotesize 
\setbox\tablebox=\vbox{
\newdimen\digitwidth
\setbox0=\hbox{\rm 0}
\digitwidth=\wd0
\catcode`*=\active
\def*{\kern\digitwidth}
\newdimen\signwidth
\setbox0=\hbox{+}
\signwidth=\wd0
\catcode`!=\active
\def!{\kern\signwidth}
\halign{\hbox to 2cm{#\leaderfil}\tabskip=0.2cm&  #\hfil\tabskip=0.2cm&
 \hfil#\hfil\tabskip=0.2cm&  #\hfil\tabskip=0pt\cr
\noalign{\doubleline}
Parameter& Prior range& Likelihood&Definition\cr
\noalign{\vskip 3pt\hrule\vskip 3pt}
$A^{\mathrm{PS}}_{100}$&$[0,360]$&C&Contribution of Poisson point-source power to $\mathcal{D}^{100\times 100}_{3000}$ for \planck\ (in $\mu\mathrm{K}^2$)\cr
&$[0,400]$&P&\cr
$A^{\mathrm{PS}}_{143}$&$[0,270]$&C&As for $A^{\mathrm{PS}}_{100}$ but at $143\,$GHz\cr
&$[0,400]$&P&\cr
$A^{\mathrm{PS}}_{217}$&$[0,450]$&C&As for $A^{\mathrm{PS}}_{100}$ but at $217\,$GHz\cr
&$[0,400]$&P&\cr
$r^{\mathrm{PS}}_{143\times 217} $&$[0,1]$&C,P&Point-source correlation coefficient for \planck\ between $143$ and $217\,$GHz\cr
$A^{\mathrm{CIB}}_{143}$&$[0,50]$&C,P&Contribution of CIB power to $\mathcal{D}^{143\times 143}_{3000}$ at the \planck\ CMB frequency for $143\,$GHz (in $\mu\mathrm{K}^2$)\cr
$A^{\mathrm{CIB}}_{217}$&$[0,80]$&C&As for $A^{\mathrm{CIB}}_{143}$ but for $217\,$GHz\cr
&$[0,120]$&P&\cr
$r^{\mathrm{CIB}}_{143\times 217}$&$[0,1]$&C,P&CIB correlation coefficient between $143$ and $217\,$GHz\cr
$\gamma^{\mathrm{CIB}}$&$[-2,2]\,(0.7\pm0.2)$&C&Spectral index of the CIB angular power spectrum ($\mathcal{D}_\ell \propto \ell^{\gamma^{\mathrm{CIB}}}$)\cr
&$[-5,+5]$&P&\cr
$A^{\mathrm{tSZ}}$&$[0,50]$&C,P&Contribution of tSZ to $\mathcal{D}_{3000}^{143\times 143}$ at $143\,$GHz (in $\mu\mathrm{K}^2$)\cr
$A^{\mathrm{kSZ}}$&$[0,50]$&C,P&Contribution of kSZ to $\mathcal{D}_{3000}$ (in $\mu\mathrm{K}^2$)\cr
$\xi^{\mbox{\scriptsize{tSZ$\times$CIB}}}$&$[0,1]$&C,P&Correlation coefficient between the CIB and tSZ (see text)\cr
$A^{\mathrm{Dust}}$&$[0,0.001]$&P&Amplitude of Galactic dust power (in $\mu\mathrm{K}^2$)\cr
\noalign{\vskip 3pt\hrule\vskip 3pt}
$c_{100}$&$[0.98,1.02]$&C&Relative power spectrum calibration for \planck\ between $100\,$GHz and $143\,$GHz\cr
&$(1.0006\pm0.0004)$&&\cr
$c_{217}$&$[0.95,1.05]$&C&Relative power spectrum calibration for \planck\ between $217\,$GHz and $143\,$GHz\cr
&$(0.9966\pm0.0015)$&&\cr
$\beta^i_j$&$(0\pm1)$&C&Amplitude of the $j-$th beam eigenmode ($j=1$--5) for the $i-$th cross-spectrum ($i=1$--4)\cr 
\noalign{\vskip 3pt\hrule\vskip 3pt}
$\delta^0_j$&$[-3,+3]$&P&Amplitude of the calibration eigenmode for the $i-$th detector (set) ($i=1$--13)\cr 
$\delta^i_j$&$[-3,+3]$&P&Amplitude of the $j-$th beam eigenmode ($j=1$--5) for the $i-$th detector(set) ($i=1$--13)\cr 
\noalign{\vskip 3pt\hrule\vskip 3pt}}}
\endPlancktable 
\endgroup

%% file: 6_Assessment-accuracy-high-ell.tex
\section{Accuracy assessment of the high-$\ell$ likelihoods\label{sec:HL-accuracy}}

In this section we compare the power spectra and likelihoods derived
using our two independent methods, and test the likelihoods
using full \Planck\ simulations.

\subsection{Comparison of the $\pliks$ and $\camspecs$ likelihoods \label{sub:plik versus camspec on data}}

To allow a consistent comparison between the \camspec\ and
\plik\ likelihoods, we use the same frequency cross-spectra for both
codes in the following, i.e., we discard the $100\times143$ and
$100\times217$\,$\ghz$ frequency combinations from the default
\plik\ likelihood. To achieve this, we modify the \plik\ likelihood to use
the fiducial Gaussian approximation instead of the Kullback
divergence. On the other hand, while we use the same multipole
coverage, $100\leq\ell\leq2500$, and only one mask (CL39) for 
all cross-spectra for \plik, we still
use multipole ranges and masks as defined in
Table~\ref{tab:nu-mask-l-ref} for \camspec. In addition, we perform one
\plik\ analysis with the CL49 mask, which matches the \camspec\ mask at
$100$\ghz.

\begin{figure*}[h]
\begin{centering}
\includegraphics[width=1\textwidth]{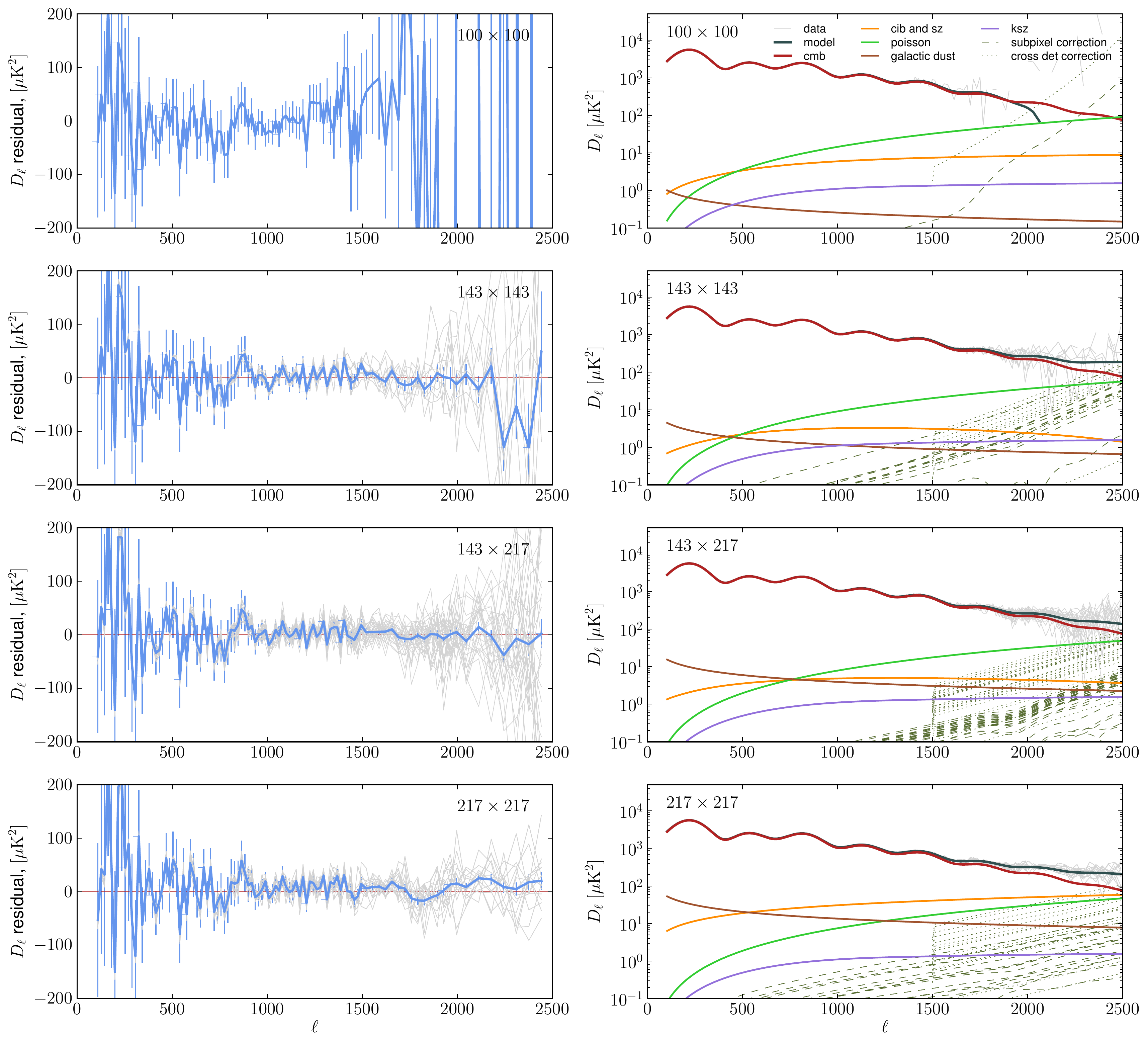}
\par\end{centering}
\caption{\emph{Left}: Residuals between the \planck\ power spectrum derived with \plik\ using 'validation' settings (described in the text) and the best-fit model. The light grey lines show residuals
for individual detector pairs within each frequency combination. The blue lines show the inverse covariance weighted averages of the individual
residuals, together with their errors computed from the covariance
matrix. \emph{Right}: Decomposition of the total best-fitting model power
spectra into CMB, combined thermal SZ and CIB, unresolved point sources, 
kinetic SZ, and Galactic dust.}
\label{fig:plamspec-residuals-models-per-frequencies}
\end{figure*}

The left column of
Fig.~\ref{fig:plamspec-residuals-models-per-frequencies} shows the
differences between the \plik\ power spectra, adopting the above
validation settings, and the corresponding best-fit model. The right
column shows the total spectra decomposed into cosmological and
foreground components. The residuals agree with those in
Fig.~\ref{fig:ps_nu_FG_ref}, and do not show any evidence of features,
except for some excess power in the $217\times217$\,GHz spectra at
small scales, where foreground modelling has the highest impact. At
scales $\ell\lesssim1500$, the residuals are coherent between cross
spectra as they are computed with the same Galactic mask, and the
residuals are dominated by cosmic variance. At smaller scales
($\ell\gtrsim1500$) the residuals are dominated by noise and become
uncorrelated.

\begin{figure}[htbp]
\centering{}\includegraphics[bb=10bp 10bp 545bp
  410bp,clip,width=1\columnwidth]{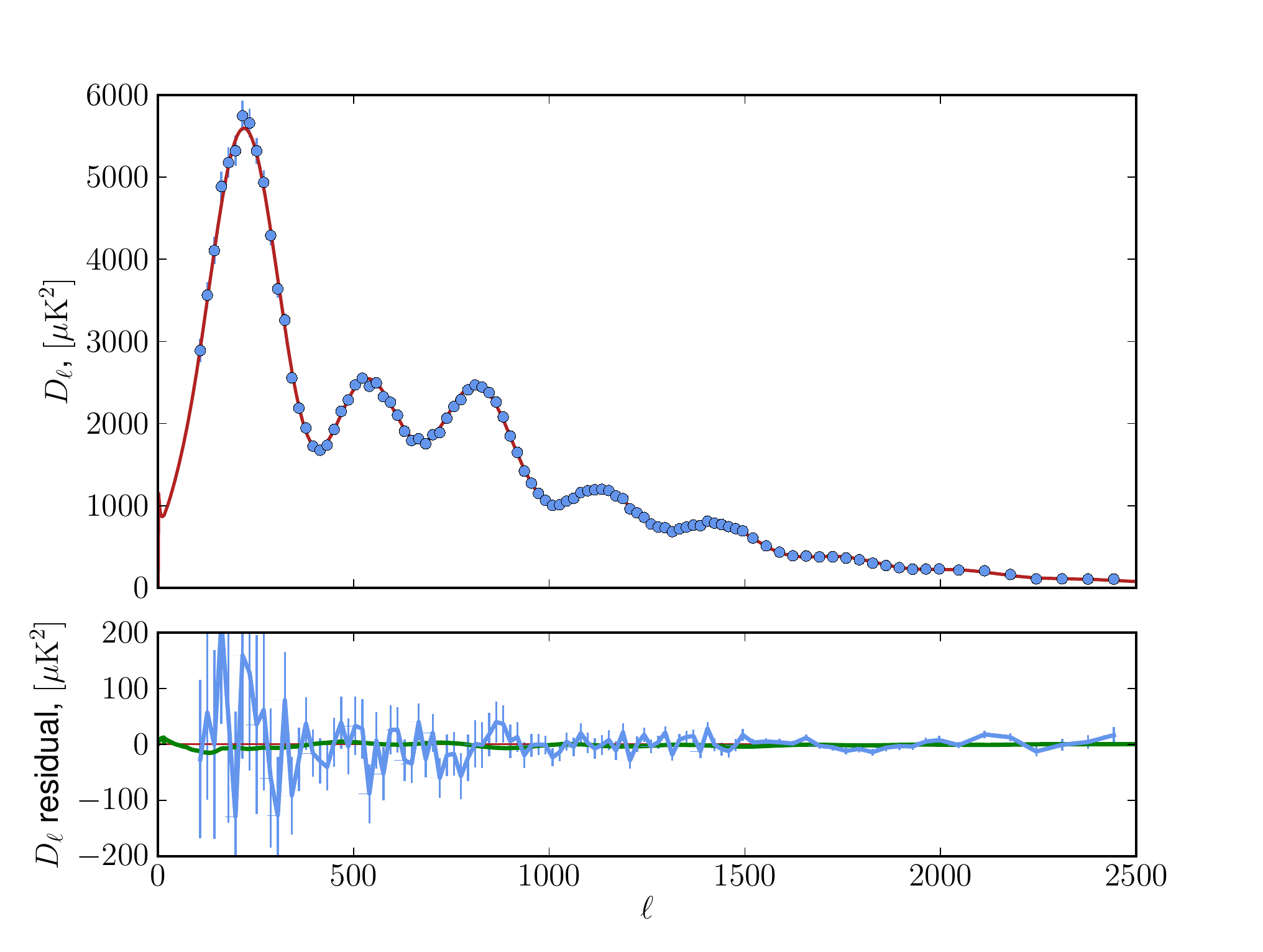}\caption{\emph{Top:}
  CMB power spectrum estimated using the \plik\ likelihood, after subtracting the best-fit foreground model. 
\emph{Bottom:} Residual difference (blue line) between
the \plik\ spectrum and the best-fit $\Lambda$CDM model. The solid green
line shows the difference between the best-fit models derived from the
\pliks and \camspec\ likelihoods. 
\label{fig:ps_CMB_ref-plik}}
\end{figure}

In Fig.~\ref{fig:ps_CMB_ref-plik} we show the CMB power spectrum
recovered by \plik\, estimated by removing the best-fit foreground
amplitudes from each cross-spectrum and computing their optimally
weighted average, and the corresponding difference with respect to the
best-fit $\Lambda$CDM model. The large scatter at low multipoles is
expected due to cosmic variance. The residual scatter at higher
multipoles is at the $\pm20\,\mu\mathrm{K}^{2}$ level, demonstrating
the good fit provided by the sum of the $\Lambda$CDM and foreground
models. These CMB residuals can be compared to the $\camspecs$
inverse-covariance weighted CMB residuals shown in
Fig.~\ref{fig:ps_CMB_ref}, which are of the same order of
magnitude. Thus, the \Planck\ likelihood fit to the $\Lambda$CDM model
is robust with respect to the detailed shape of the likelihood, as
quantified in terms of power spectrum residuals.

\begin{figure}[b]
\centering{\includegraphics[bb=70bp 30bp 600bp 400bp,clip,width=1\columnwidth]{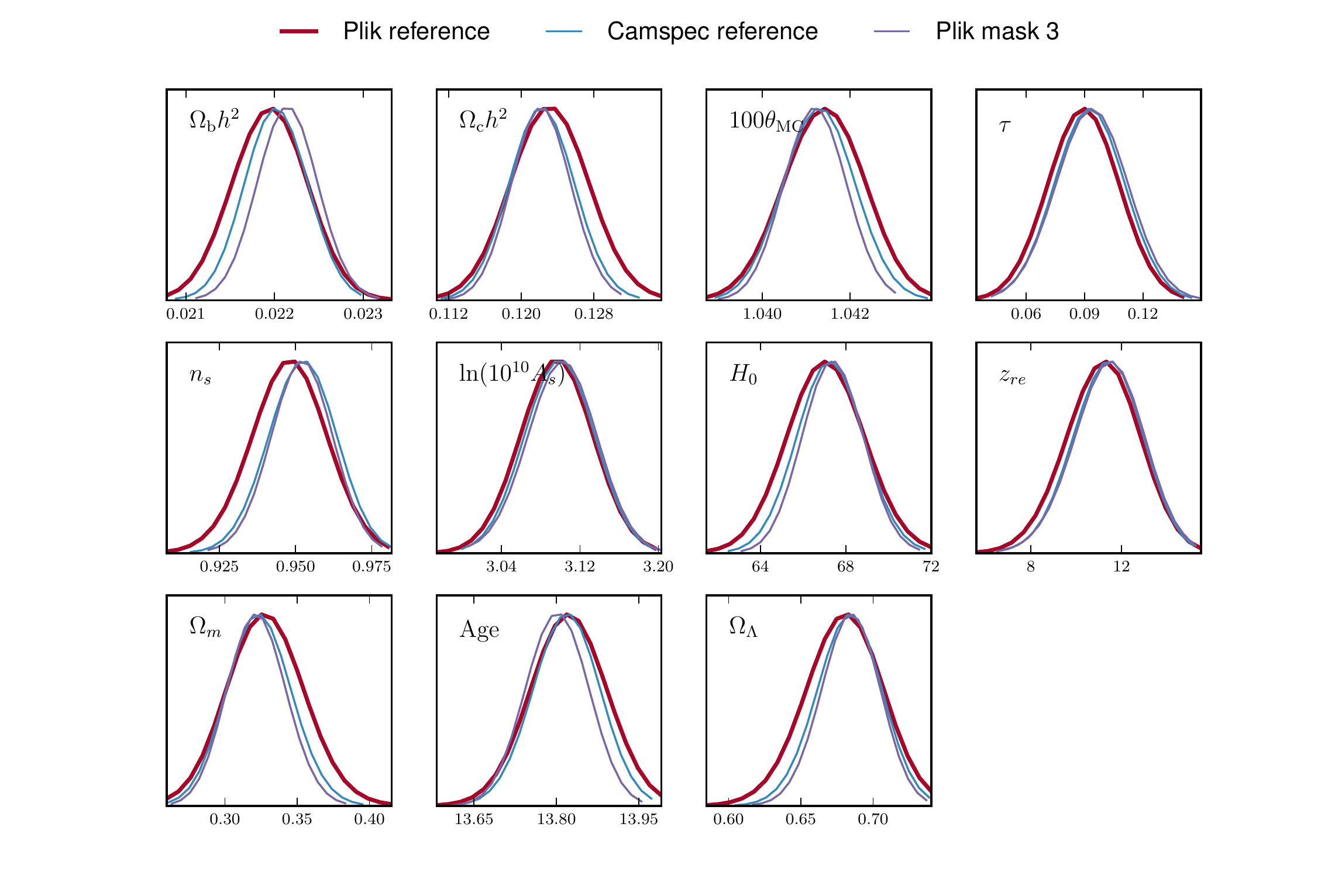}}
\caption{Comparison of cosmological parameters estimated from the
  \camspec\ (blue) and \plik\ (red for mask C48; purple for mask C58)
  likelihoods. All parameters agree to better than $0.2\sigma$ when
  C58 is used for \plik, matching the sky area used by \camspec\ at
  100~GHz.
\label{fig:posteriors_covspec+plik}}
\end{figure}

\begin{figure}[htbp]
\begin{centering}
\includegraphics[bb=75bp 30bp 600bp 405bp,clip,width=1\columnwidth]{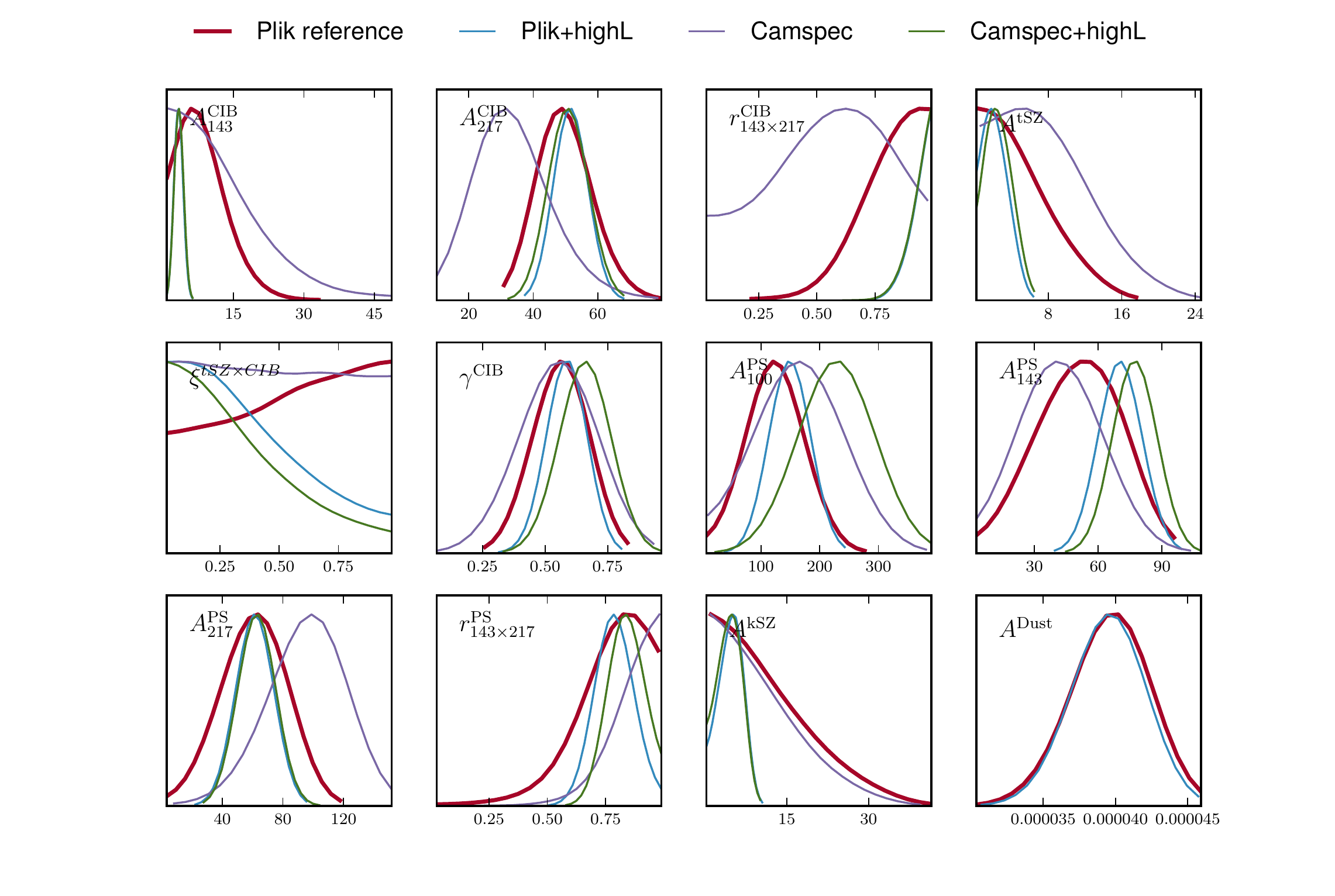}
\par\end{centering}
\caption{Comparison of foreground parameters estimated with the
  \camspec\ and \plik\ likelihoods. The red (purple) lines show the
  \camspec\ (\plik) distributions using only \planck\ data, and the green
  (blue) lines show the
  \camspec\ (\plik) results when additionally including ACT and SPT. 
\label{fig:fg_posteriors_camspec_plik}}
\end{figure}

The $\Lambda$CDM parameter constraints derived from the two
likelihoods are shown in Fig.~\ref{fig:posteriors_covspec+plik}, while
Fig.~\ref{fig:fg_posteriors_camspec_plik} shows the foreground
parameters. For the cosmological parameters, the agreement between the
two likelihoods is excellent: when the CL49 mask is adopted for the
\plik\ likelihood, which is also used by \camspec\ at
100~GHz, all parameters agree to $0.2\sigma$ in terms of maximum
posterior values. We also see that the widths of the distributions are quite
similar, with the \plik\ ones slightly broader
than the \camspec\ ones. Significantly larger differences are
seen for the foreground parameters.

These effects can be understood
as follows: we use the \camspecs likelihood with Galactic mask CL49 for the $100\times100$~$\ghz$
spectra, to minimize the cosmic variance in the low to intermediate
multipole range ($\ell\lesssim1200)$, taking advantage of the low
level of Galactic emission in this channel. At higher multipoles we
use the more conservative Galactic mask CL31 for both the $143$
and $217$~$\ghz$ channels, at the price of a higher variance. However,
in the specific case of the $\Lambda$CDM model considered here, most
of the constraints on cosmological parameters come from relatively
modest multipoles, $\ell<1500$, rather than from the damping tail. 
This explains why when we repeat the \pliks analysis, enlarging the sky area from Galactic mask CL39 to CL49, we find parameter distributions in good agreement with those of \camspec. However,
the detailed choice of masks, as well as multipole range cuts, does
affect the foreground parameters. This is investigated further in Sect.~\ref{sec:consistency}. 

Figure~\ref{fig:fg_posteriors_camspec_plik} shows the foreground
parameters estimated from both \pliks and \camspec. We consider the case for \Planck\ data alone, and with the inclusion of data from ACT and SPT.
We also impose, for \camspec, a Gaussian prior of $0.7\pm0.2$ on the CIB slope parameter,
$\gamma^{{\rm CIB}}$. We find that the upper bounds
on the CIB amplitude at $143$~$\ghz$, ${\rm A_{143}^{CIB}}$, and on
the SZ amplitudes (both thermal and kinetic, ${\rm A^{tSZ}}$ and
${\rm A^{kSZ}}$) are in good agreement using  \Plancks data alone,
but we see differences ($\sim1.5\sigma$) in the CIB and Poisson
amplitudes at $217$~GHz (${\rm A_{217}^{CIB}}$, ${\rm A_{217}^{PS}}$)
as well as a difference in the CIB correlation coefficient, ${\rm r_{143\times217}^{CIB}}$.

We understand this effect in the following way: \camspecs uses a more
limited multipole range and a more conservative mask at $217$~$\ghz$
than \plik, in order to minimize the Galactic emission in this
channel.  This enhances the degeneracy between the foreground
parameters at 217~GHz, and enhances the sensitivity to possible
deviations of the CIB power spectrum from the pure power law assumed
here. This is artificially enhanced by normalizing the components at
$\ell=3000$, which is more suitable for high resolution experiments
than for \Planck. Despite the disagreement of the precise
decomposition into the different physical components, the sum of the
foreground contributions at $217$~GHz in \pliks and \camspecs is in
good agreement. When ACT and SPT data are added, these problems are
largely alleviated, as shown in
Fig.~\ref{fig:fg_posteriors_camspec_plik}.

\begin{figure*}
\begin{centering}
\includegraphics[bb=130bp 290bp 1390bp 1400bp,clip,width=1\textwidth]{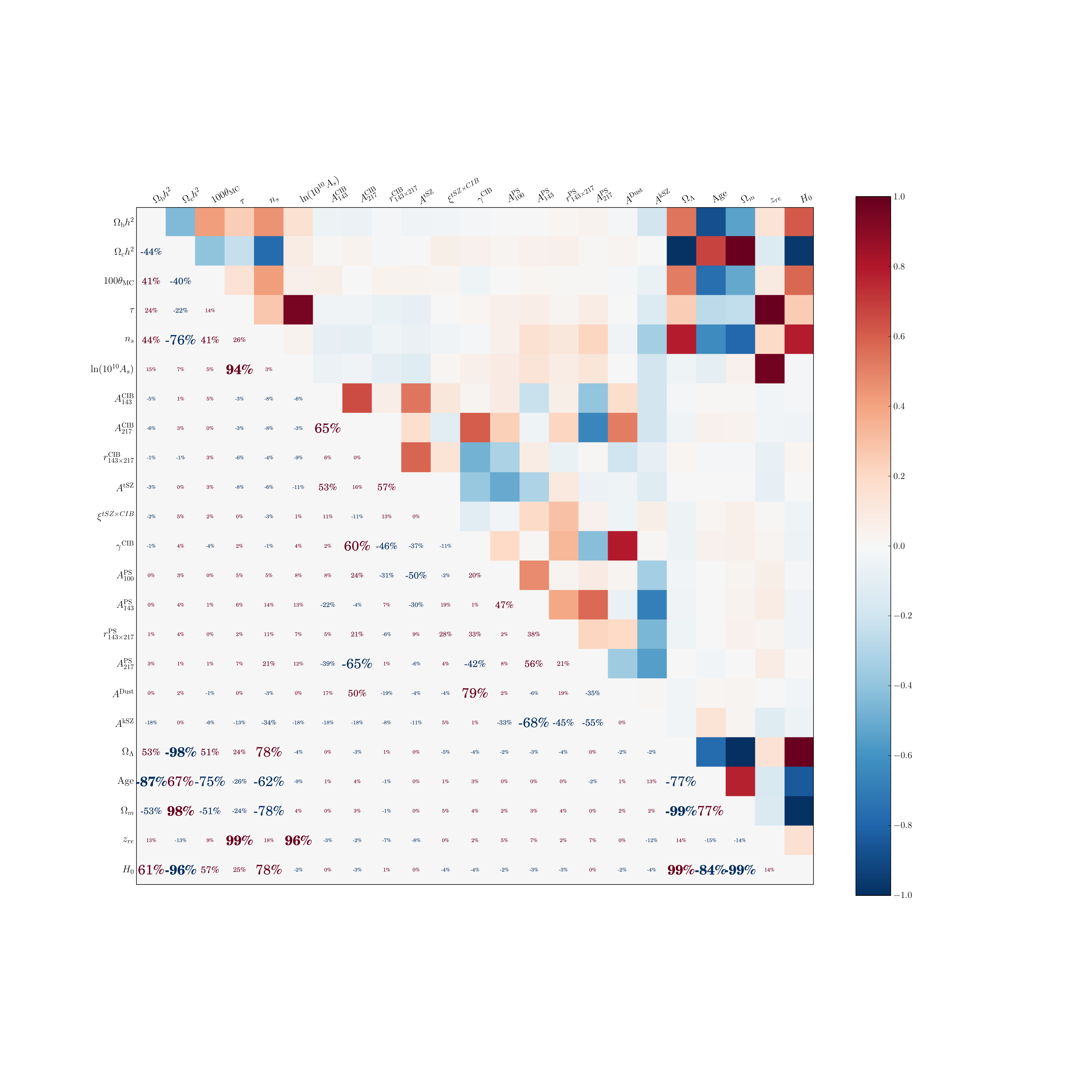}
\par\end{centering}
\caption{Correlation matrix between all the cosmological (top block), 
foreground (middle block), and derived (bottom block) parameters, 
estimated using the \plik\ likelihood. 
\label{fig:par_cor_PLik}}
\end{figure*}

Are these differences important for cosmology? To address
this question, we examine the covariance between cosmological and
foreground parameters. Figure~\ref{fig:par_cor_PLik} shows the 
correlation matrix for all the estimated parameters. The basic $\Lambda$CDM
parameters have well-known correlations: the scalar spectral index,
$n_{\mathrm s}$, is anti-correlated with both the amplitude $A_{\mathrm{s}}$ 
and the
dark matter density $\Omega_{\mathrm{c}}h^{2}$, which is itself correlated
with the amplitude. 
Within the foreground parameters, there are strong correlations between the
Galactic dust amplitude, the CIB amplitudes, and the SZ amplitude,
as well as between point source amplitudes at different frequencies.
These correlations result from the conservative foreground model adopted
here, where all amplitudes of the CIB and Poisson contributions are
left free to vary in each frequency pair; this choice results in small
residuals in the fits, at the price of partial degeneracies between
the foreground parameters. 

Despite this conservative foreground model, there is a small
correlation between cosmological and foreground parameters, the strongest
effect being a $34\%$ anti-correlation between the kinetic SZ amplitude
and the scalar spectral index. The kinetic SZ power spectrum amplitude
is positive, so marginalizing over it affects the peak value of $n_{\mathrm s}$
despite the fact that $A^{\mathrm{kSZ}}$ is not significantly different
from zero. The addition of smaller scale data from ACT and SPT helps to
break this degeneracy. In addition, as can be seen in Fig.~\ref{fig:fg_posteriors_camspec_plik},
the posterior distributions of \camspecs and \pliks for $A^{\mathrm {kSZ}}$
are in good agreement, showing the stability of the cosmological
parameters to the likelihood method. 

In Sect.~\ref{sec:consistency}, we will further explore the 
stability of the cosmological and foreground distributions to technical choices made in the likelihood and data selections.

\subsection{Comparison to simulations\label{sub:Comparison-to-simulations}}

It is important to demonstrate the precision and accuracy with which cosmological
parameters, and foreground parameters to a lesser extent, can be recovered
on realistically simulated data. Here we compare the posterior distributions
of cosmological and foreground parameters, together with calibration
and beam error parameters, inferred using the \plik\ likelihood, with the input
values of a set of simulations, referred to as `Full Focal Plane' (FFP6). 
The signals in these
simulations are based on the `Planck Sky Model' \citep{delabrouille2012}
which includes a detailed model of the astrophysical emission,
both Galactic and extragalactic, at the \Plancks frequencies. The simulations
also reproduce in detail the main instrumental systematic effects
of \Planck, including correlated timeline noise, instrumental pointing,
flags, anisotropic detector beams, and spectral bandpasses. One thousand
CMB and noise realisations were generated using the same
foreground emission, and a hundred realisations were performed at
the level of different detector sets. These simulations are described further in \citet{planck2013-p28}.

\begin{figure}
\begin{centering}
\includegraphics[bb=10bp 10bp 640bp 560bp,clip,width=1\columnwidth]{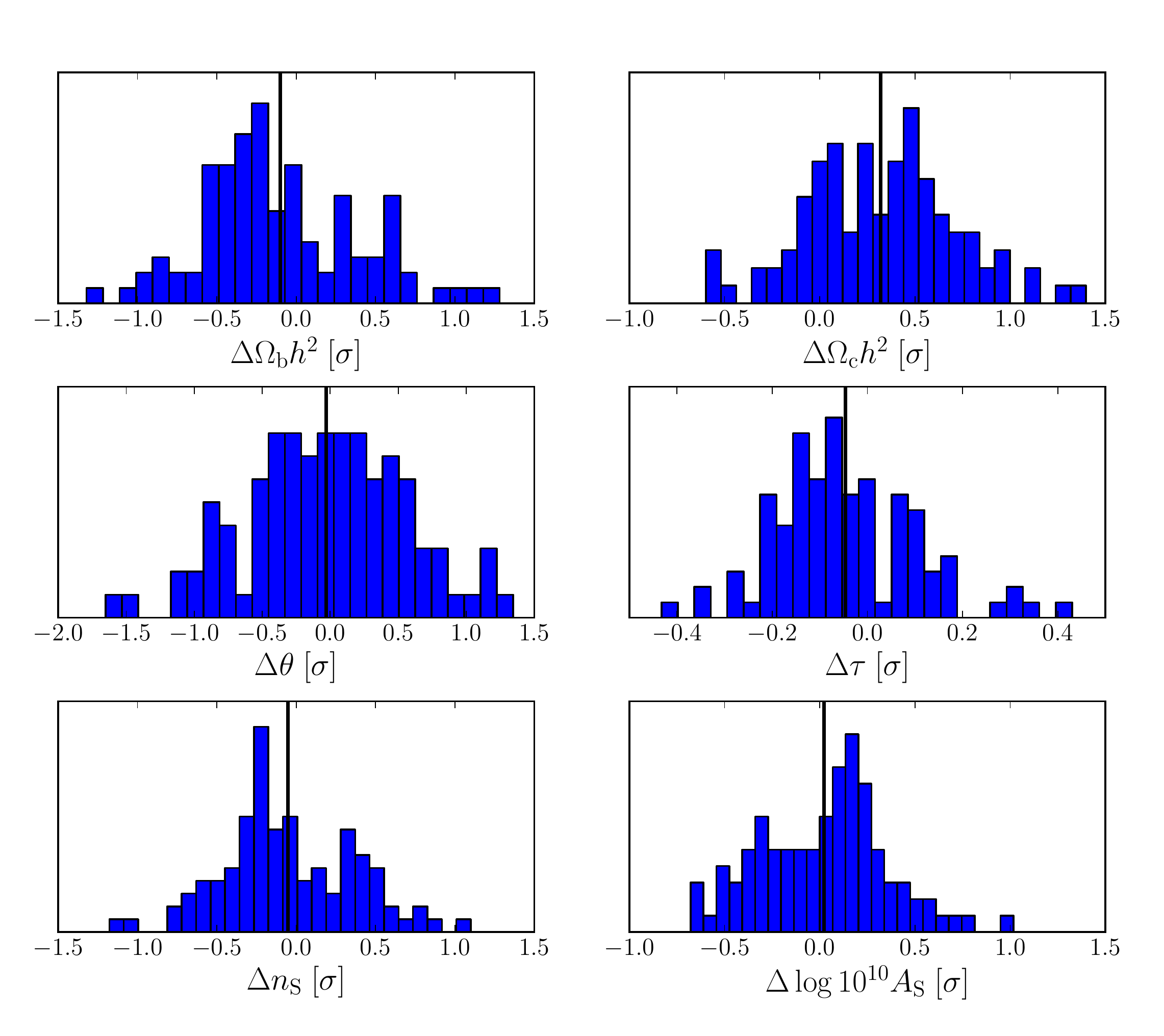}
\par\end{centering}
\caption{Distribution of the difference between parameter mean (as estimated by \pliks assuming foreground spectra are known) and the input
values in units of the posterior standard deviation. 
This demonstrates the absence of methodological bias at the level of the intrinsic dispersion 
between realizations.  
\label{fig:FFP6x100_PLik_pars}}
\end{figure}

In order to test for any methodological bias, we estimate cosmological 
parameters from 100 FFP6 simulations using the \plik\ likelihood. They consist of random realisations of CMB anisotropies and
noise, superimposed with a single realisation of a  frequency-dependent
foreground template. We assume the
foreground power spectra to be known exactly, and include their
additional power as a constant component in our model. We do not
estimate the foreground parameters in this case. 
This goal of this test is to demonstrate the reliability of our analysis pipeline and 
explore the effects of the
noise and CMB-foreground chance correlations on parameter estimation.

To compare our results to the simulation inputs, 
we also estimate cosmological parameters from the 100 CMB realisations, without including noise or foregrounds. 
In the likelihood evaluations we down-weight the
high-$\ell$ part of the power spectrum as if there were noise at the level of
the noise simulations. A direct comparison of
the derived parameters from this CMB-only analysis and the full simulation (that contains noise and foreground emission) allows us to remove the scatter introduced by cosmic variance.

The result of the analysis is shown in
Fig.~\textcolor{red}{\,\ref{fig:FFP6x100_PLik_pars}}, where we plot
the distribution of the difference of the mean estimated parameters, between the CMB-only simulation and the noisy CMB simulation with foregrounds, in
units of the standard deviation of each individual distribution. An unbiased pipeline will give a difference consistent with zero. Averaged over 100 simulations, only the bias on $\Omega_{\mathrm c}h^2$  is statistically significant, 
but is $<0.3\sigma$. An interesting point to note from these histograms is that the
distributions are rather wide, mostly around $0.5\sigma$, while by
construction, we only studied the effect of noise and CMB-foreground
chance correlations in this test.

\begin{figure}
\begin{centering}
\includegraphics[bb=45bp 30bp 400bp 420bp,width=1\columnwidth]{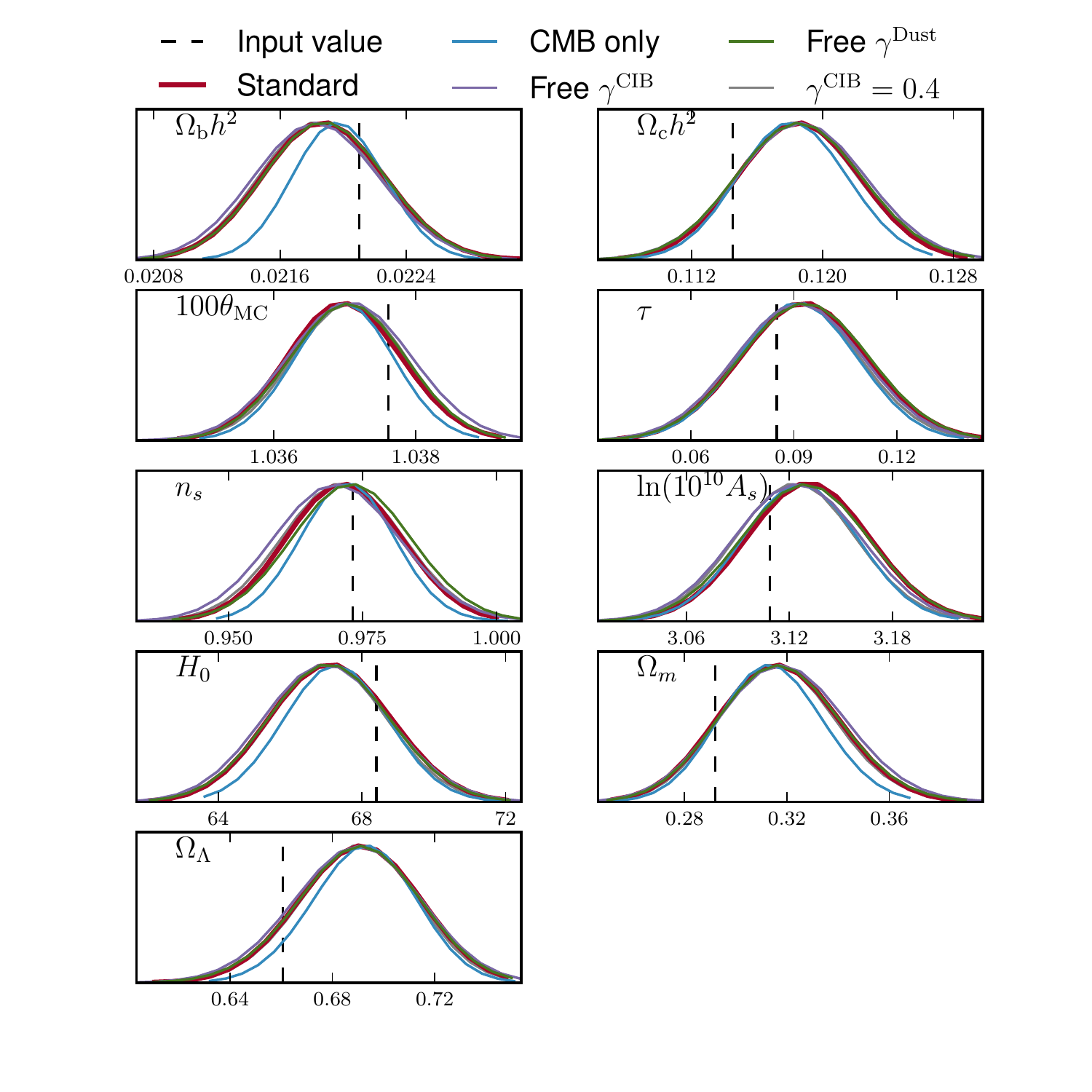}
\par\end{centering}
\caption{Posterior marginal distributions of cosmological parameters obtained by \pliks on a single FFP6 realisation, 
under different assumptions on the foreground model. Red thick lines correspond to the
foreground model parametrisation described in Sect.~\ref{sub:Sky-model} with fixed $\gamma^{\rm CIB}$. Purple lines (resp. green, grey) show
the effect of leaving $\gamma^{\rm CIB}$ free (resp. letting $\gamma^{\rm Dust}$ free, or fixing $\gamma^{CIB}$ to $0.4$, $2\sigma$ away from the peak posterior value). Finally, the blue lines  correspond to the analysis made on CMB-only simulations. The input values of the parameters of the simulation are shown as dashed vertical lines.
\label{fig:ffp6_marg}}
\end{figure}

The FFP6 simulations can also be used to assess the sensitivity to
foreground modelling errors. In Fig.~\ref{fig:ffp6_marg}, we compare the posterior marginal
distributions of  cosmological parameters, estimated with different assumptions about the foreground model. The blue lines, 
which correspond to the analysis of CMB-only simulations (but accounting for the noise covariance in the likelihood) correspond
to the idealised case where the foregrounds play no role, and where the noise-induced variance has been averaged.
The red, thick lines show results obtained when marginalising over the parameters of the model of foregrounds that is applied to the \Planck\ data (see Sect.~\ref{sub:Sky-model}), with a fixed value of the CIB spectral index $\gamma^{\rm CIB}$. 
Purple and grey lines show respectively the effect of leaving $\gamma^{\rm CIB}$ free when marginalising, or fixing it to the displaced value of $0.4$, more than $2\sigma$ away from the peak posterior. The green lines show the effect of leaving $\gamma^{\rm Dust}$ (spectral index of the Galactic dust emission) free in the marginalisation. 

The distributions are all in reasonable agreement with the input parameters of the simulation. In addition, we see that varying assumptions on the parameters of the foreground model (red, green, purple, and yellow lines) have negligible impact on the recovered cosmological parameters. Finally, the broadening of the posteriors between the CMB-only exploration and the full case (including noise random realisations and foregrounds) is expected, as the latter includes all the sources of variance, including CMB-foreground and CMB-noise chance correlations.  

It is worth noting that the FFP6 foreground simulations, based on extrapolations of existing observations, cannot be described by the simple foreground model used in the likelihood analyses. The upper bounds on biases introduced by a possible mismatch between the simulated foreground templates and the model used in the analysis, inferred from Fig.~\ref{fig:FFP6x100_PLik_pars}, should be representative of the \Planck\ data analysis. The negligible impact of the various assumptions made on the foreground model parameters of Fig.~\ref{fig:ffp6_marg} confirms that the cosmological parameter estimations are robust to details of the foreground model.

%% file: 7_Consistency-checks.tex
\section{Consistency checks\label{sec:consistency} }

In this section we investigate the stability of the distributions
of cosmological and foreground parameters. The technical choices made
in constructing the high-$\ell$ likelihood fall into three broad
categories. The first category covers internal parameter choices that
leave the data selection unchanged. This includes choices such as
the binning strategy, marginalizing or not over calibration and beam
errors, and the description of the noise model. The second category
includes variations in the data selection, such as the multipole range
used, and the choice of masks. The final category accounts for variation
in the foreground model. We perform a suite of tests to investigate
the impact of these choices on parameters. We use the \pliks likelihood,
and all tests are compared to the baseline \pliks spectra. Most
of the results can be summarized by `whisker plots' which compare
the main properties of the posterior distribution of the cosmological
and foreground parameters. More detailed results are reported 
in Appendix~\ref{App:Checks}.

In this section we also compare our estimated cosmological parameters to those
derived from spectra computed from the LFI 70\,GHz channel. We additionally 
check the consistency of parameters with results obtained using
the power spectrum of CMB maps derived by component separation methods
\citep[described in][]{planck2013-p06} that use \Plancks data at all frequencies. This battery of tests demonstrates the stability of the inferred cosmological
parameters.

A final test is to compare the predicted polarisation spectrum 
of the best fitting $\Lambda$CDM model 
with spectra measured from \Planck. 
As discussed in \cite{planck2013-p01} and
\cite{planck2013-p03}, the \Plancks
polarisation data 
is not yet used in our cosmological analysis, as further tests 
must be performed,
but the current results 
increase our confidence in the robustness of the high-$\ell$ temperature
likelihood.

\begin{figure*}[t]
\begin{centering}
\includegraphics[bb=100bp 85bp 1100bp 842bp,clip,width=1\textwidth]{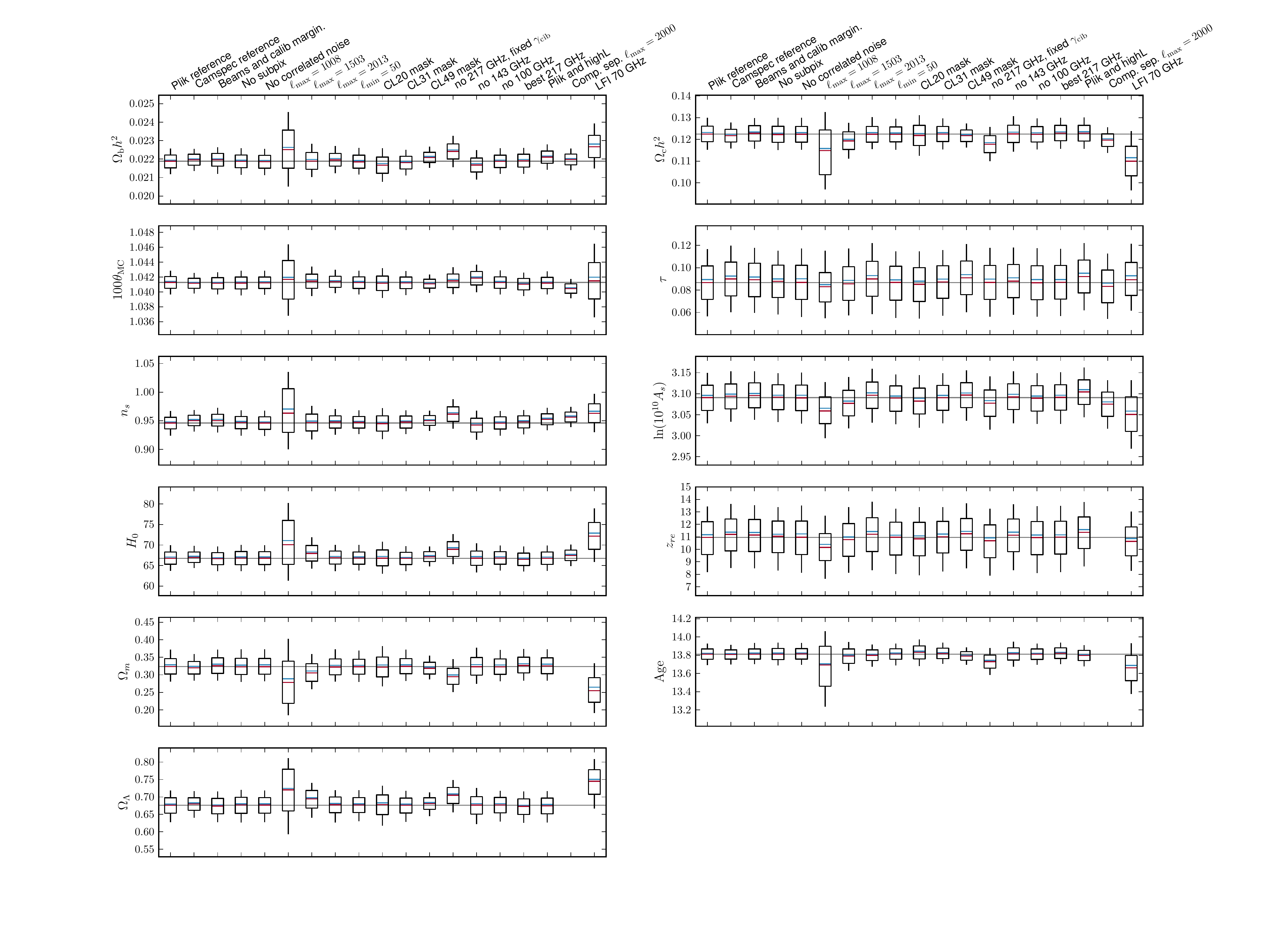}
\par\end{centering}
\centering{}
\caption{
Comparison of the distributions of cosmological parameters in the
reference case (left) with a set of validation test cases. The red
line indicates the median and blue the mean, computed from the posterior
histograms. The box shows the 68\% confidence interval; the outer
line the 95\% interval.
\label{fig:whiskers_cosmo}}
\end{figure*}

\begin{figure*}[t]
\begin{centering}
\includegraphics[bb=107bp 85bp 1083bp 841bp,clip,width=1\textwidth]{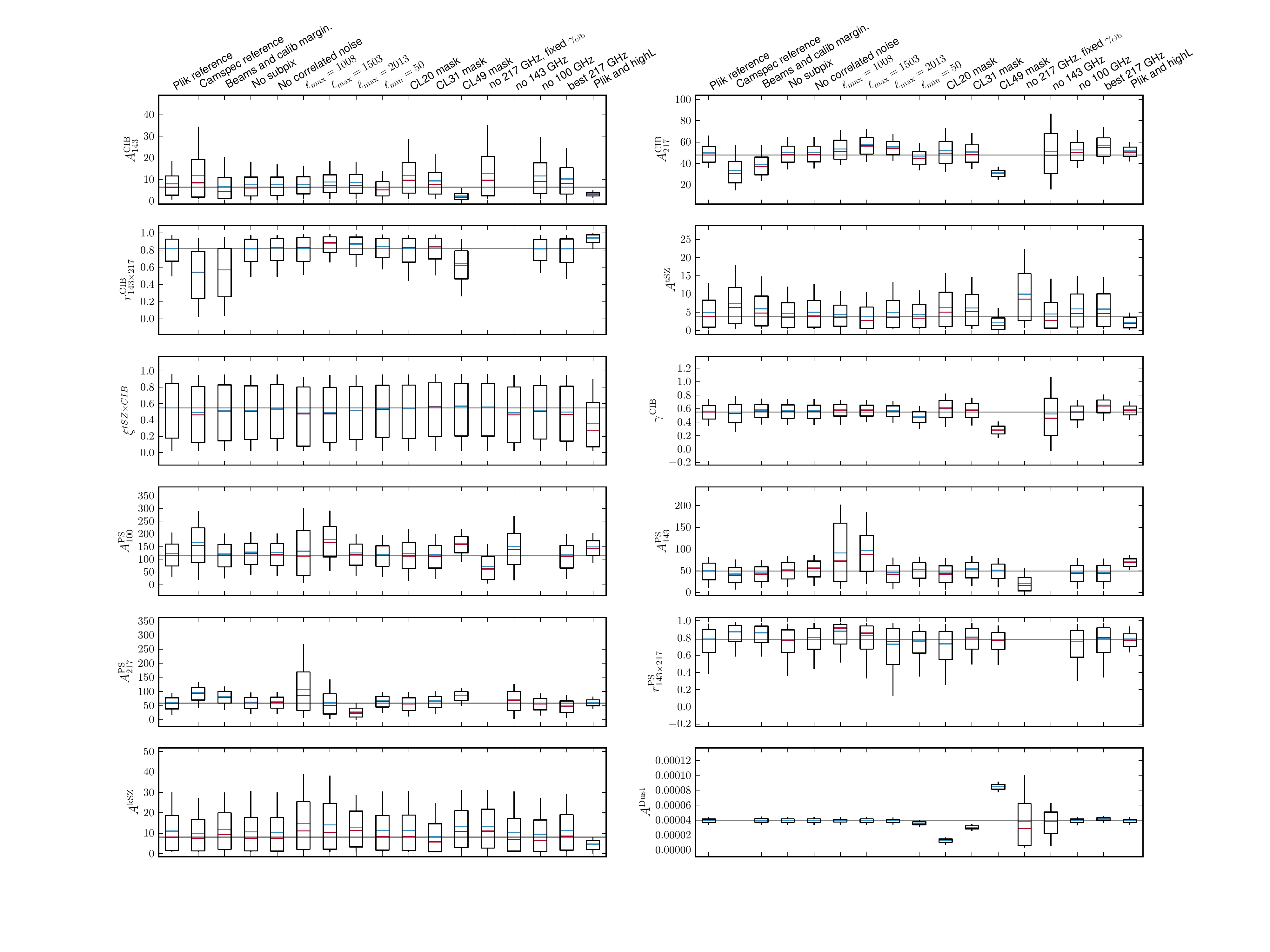}
\par\end{centering}
\centering{}
\caption{
Comparison of the distributions of foreground model parameters, as
in Figure \ref{fig:whiskers_cosmo}.
\label{fig:whiskers_FG}
}
\end{figure*}

\subsection{Impact of technical choices for fixed data selection \label{sub:Stability-against-technical-changes}}

Here we consider three changes: (1) fixing the inter-frequency calibration
and beam errors to the best-fit values, rather than marginalizing,
(2) including sub-pixel effects, and (3) including noise correlation
between detectors. Figure~\ref{fig:whiskers_cosmo} shows the corresponding
impact on parameters. 

The effect of fixing the calibration and beam errors is negligible
on most cosmological parameters within the six parameter $\Lambda$CDM model, with the exception of $n_{\mathrm s}$ where
we see a $0.16\sigma$ shift. There is a bigger effect
on the foreground parameters due to their partial degeneracy, and
their subdominant contribution to the total power. There is only a
small correlation between the cosmological parameters and the calibration
coefficients, so marginalizing or fixing their value has little impact
on the cosmology.

\begin{figure*}[t]
\begin{centering}
\includegraphics[width=1\columnwidth]{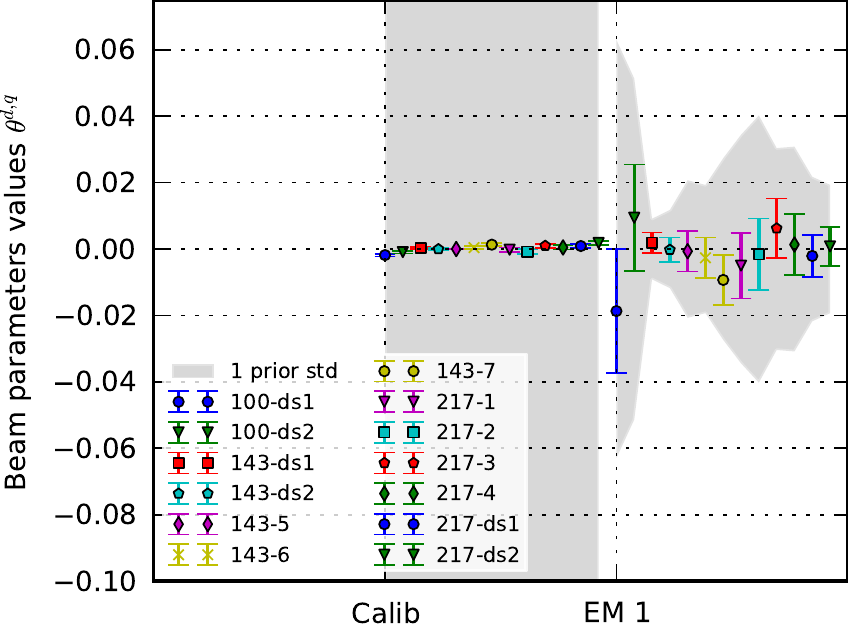}{\includegraphics[width=1\columnwidth]{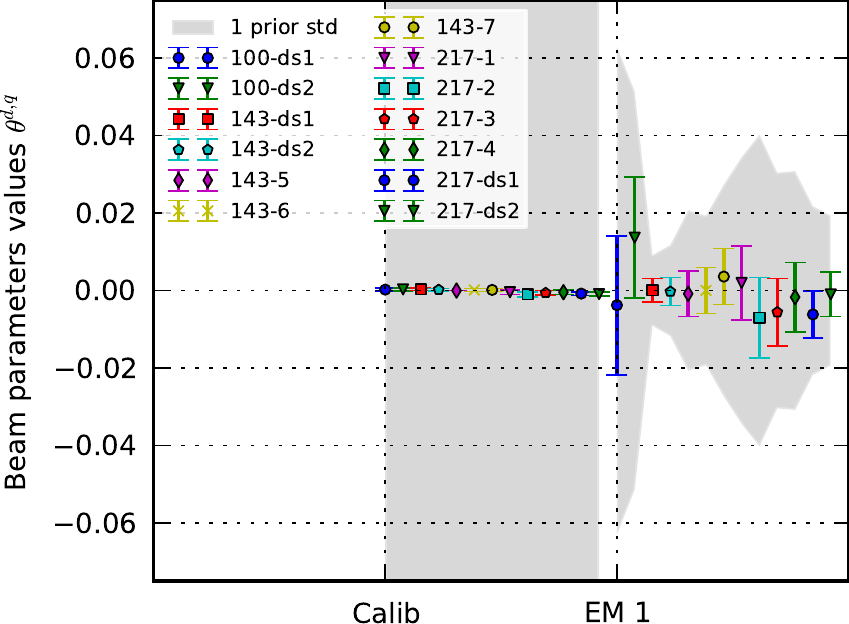}}
\par\end{centering}
\centering{}
\caption{
Estimated calibration and beam eigenmode parameters, compared to the priors, 
using (left) 100
to 217\,GHz data, and (right) a single `FFP6' simulation where there was no calibration or beam errors. The shaded area shows the width of the prior imposed on the first
beam eigenvalue.
\label{fig: calibeams-peak-values}
}
\end{figure*}

The calibration coefficients are, however, strongly correlated with each other, in particular at $217\,\ghz$,
since they are also significantly correlated with e.g., the CIB amplitudes
in the $217\times217$ and $217\times143\,\ghz$ spectra. This is
important to keep in mind when comparing the  calibration
estimates obtained here with those obtained from the CMB dipole in the
HFI data processing paper \citep{planck2013-p03}. Nevertheless, as can be seen
in Fig.~\ref{fig: calibeams-peak-values}, the peak
posterior values of the relative calibration coefficients are found
to differ from $1$ at most at the \emph{few parts per thousand }for
all the 13 detectors sets involved, in agreement with the estimates
of the calibration accuracy of the maps \citep{planck2013-p03}, although
a wide flat prior has been applied on these coefficients. The same
test applied on simulations with no beam or calibration errors shows how well
this test is passed. This confirms that the deviations found at the
0.1\% are significantly detected, and it is important to
show that these deviations have little impact on the cosmology.

The estimated values of the beam errors do not imply that extra 
beam corrections are required, with the possible
exception of the $100-\mathrm{ds}1$ and 143 -7 detectors (although
the corresponding marginal distribution is far from Gaussian). A comparison
of the prior and posterior distributions suggests that we have conservatively estimated the uncertainties from the beam determination. We keep this conservative approach in \camspec, in which we marginalise analytically over all beam eigenmodes
except for the dominant 100\,GHz mode, $\beta^1_1$, which we sample directly.

\begin{figure}
\begin{centering}
\includegraphics[bb=55bp 45bp 535bp 530bp,clip,width=1\columnwidth]{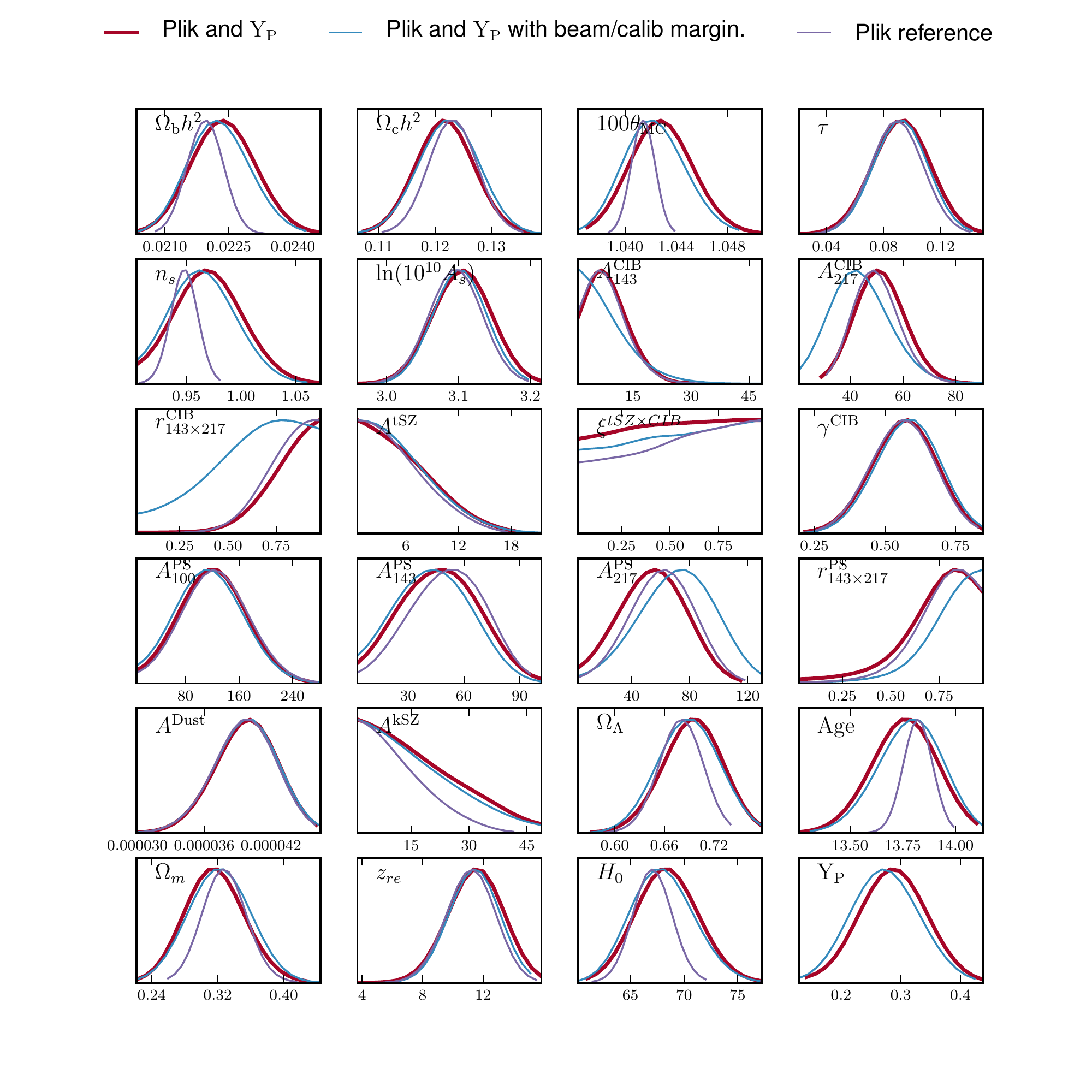}
\par\end{centering}
\caption{Posterior distributions for cosmological parameters, varying
the Helium abundance ${\rm Y_{P}}$ (red).
We show the effect of marginalizing over calibration
and beam errors (blue), and the results obtained using the reference
settings (purple), where ${\rm Y_{P}}$ is constrained by BBN.
\label{fig:Plik-helium}}
\end{figure}

As an extended test, we investigate the effect of possible errors in the beam transfer function when the helium abundance $Y_{\mathrm P}$ is also allowed to vary freely (i.e., without imposing constraints from Big Bang Nucleosynthesis), as it has a larger effect on the small scale spectrum. Varying this parameter leads to a substantial broadening of the posterior distributions for $\Omega_{{\rm b}}h^{2}$, $\theta_{{\rm MC}}$, and $n_{\mathrm s}$, as can be seen in Fig.~\ref{fig:Plik-helium}. We confirm that marginalizing over calibration and beam errors
has a small impact on all cosmological parameters, including
$Y_{\mathrm P}$. On the other hand, it has a somewhat 
larger impact on some of the foreground parameters (see Appendix~\ref{App:Checks}). 

We then investigate the impact of two subdominant
effects: the ``sub-pixel effect'' and the possible presence of a
correlation in the noise between detectors or detector sets. The sub-pixel
effect has a convolutive effect on the power spectra that is 
similar to gravitational lensing of the CMB, but is purely a result
of the \Plancks scanning strategy and the map-making
procedure \citep{planck2013-p03c}. The scanning strategy on rings
with very low nutation levels results in the centroid of the samples
being slightly shifted from the pixel centres; however, the map-making
algorithm assigns the mean value of samples in the pixel to the position at the centre of the pixel. This has a non-diagonal effect on the power
spectra, but the correction can be computed given the estimated
power spectra for a given data selection, and recast into an additive, fixed
component of the model covariance matrix. 

The possible noise correlation between detectors may appear due to
factors such as common residual thermal fluctuations, electronic chain
noise, or cosmic ray showers. To build a model of this correlated
component, we compute the cross-power spectra between detectors of
difference maps that are free of signal. This procedure should capture
all correlations on timescales shorter than half a ring's observation. 
These estimates are noisy, so we compute an average amplitude
of the correlation for $\ell\geq1000$. 

We find that the impact on parameters of both these effects is negligible, with less then a $0.1\sigma$ effect.

\subsection{Impact of data selection \label{sub:Impact-of-data-selection}}

Here we consider three changes: (1) varying the angular range used
in the likelihood, (2) varying the Galactic mask, and (3) discarding individual
frequency channels.
These are expected to result in changes in the parameter distributions
due to the fact that we are changing the input data. 

We first vary the maximum and minimum multipole. Using $\ell_{\mathrm{max}}=1008$
gives parameter distributions of similar width to those obtained by
\WMAP. We find that all basic $\Lambda$CDM cosmological parameters
have converged by $\ell_{\mathrm{max}}\simeq1500$, since no parameters are
specifically sensitive to the damping tail. The convergence of the
posteriors on foreground parameters is, as expected, slower when increasing
$\ell_{\mathrm{max}}$, as most of them are dominant at small scales. The Galactic
dust normalization decreases with $\ell_{\mathrm{max}}$ due to its correlation
with the CIB components. Similarly, changing $\ell_{\mathrm{min}}$ from $100$
to $50$ has a negligible effect on cosmological parameters, and mostly
affects the determination of the Galactic dust amplitude, which decreases
for $\ell_{\mathrm{min}}$ as it is better measured on large scales; its
correlations with the CIB clustered and Poisson contributions explain
the slight variations in the corresponding parameters ${\rm A^{CIB}_{217}}$,
$\gamma^{\rm CIB}$, and ${\rm A^{PS}_{217}}$. 

We then investigate the impact of varying the Galactic mask, from
the most conservative (CL20) to the least conservative (CL49), for
fixed multipole range $100\leq\ell\leq2508$. As expected, the errors
decrease as the sky fraction increases. From CL20 to CL39,
cosmological and foreground parameters are stable. The foreground
parameters change significantly however when we use CL49 for all
channels, showing that our foreground model is unable to properly
fit the data: the clearest sign of this failure is the unphysically
low value of the CIB spectral index ($\gamma^{\rm CIB}$), indicating
that our CIB component determination is getting contaminated by a
dust-like component with a steeper angular power spectrum than the
CIB, but shallower than that of our (single) Galactic dust component.
At low Galactic latitudes, the presence of compact Galactic sources
leads to a flatter angular power spectrum than that of (high-latitude)
diffuse thermal dust, and this likely effects the CIB determination.
This justifies the choice made in $\camspecs$to use a conservative
masking strategy (CL31) for the $143$ and $217$\ghz\ channels.

Next, we remove one frequency channel at a time. Results change by
less than $0.5\sigma$ \emph{except when removing the $217$\ghz\ channel}.
This removes a large part of the information and amounts to retaining
only 21 cross-spectra out of the 78.

\subsection{Testing the Foreground Model\label{sub:FGModelCheck}}

In this section we describe various tests of the foreground model used in the likelihood, checking the validity of our model for the extragalactic sources.  Further tests are also reported in Appendix B of \citet{planck2013-p11}.

\subsubsection{Poisson power from extragalactic sources}\label{sub:poisson-counts}

\begin{figure*}
\begin{centering}
\includegraphics[bb=53bp 18bp 498bp 325bp,clip,width=0.33\textwidth]{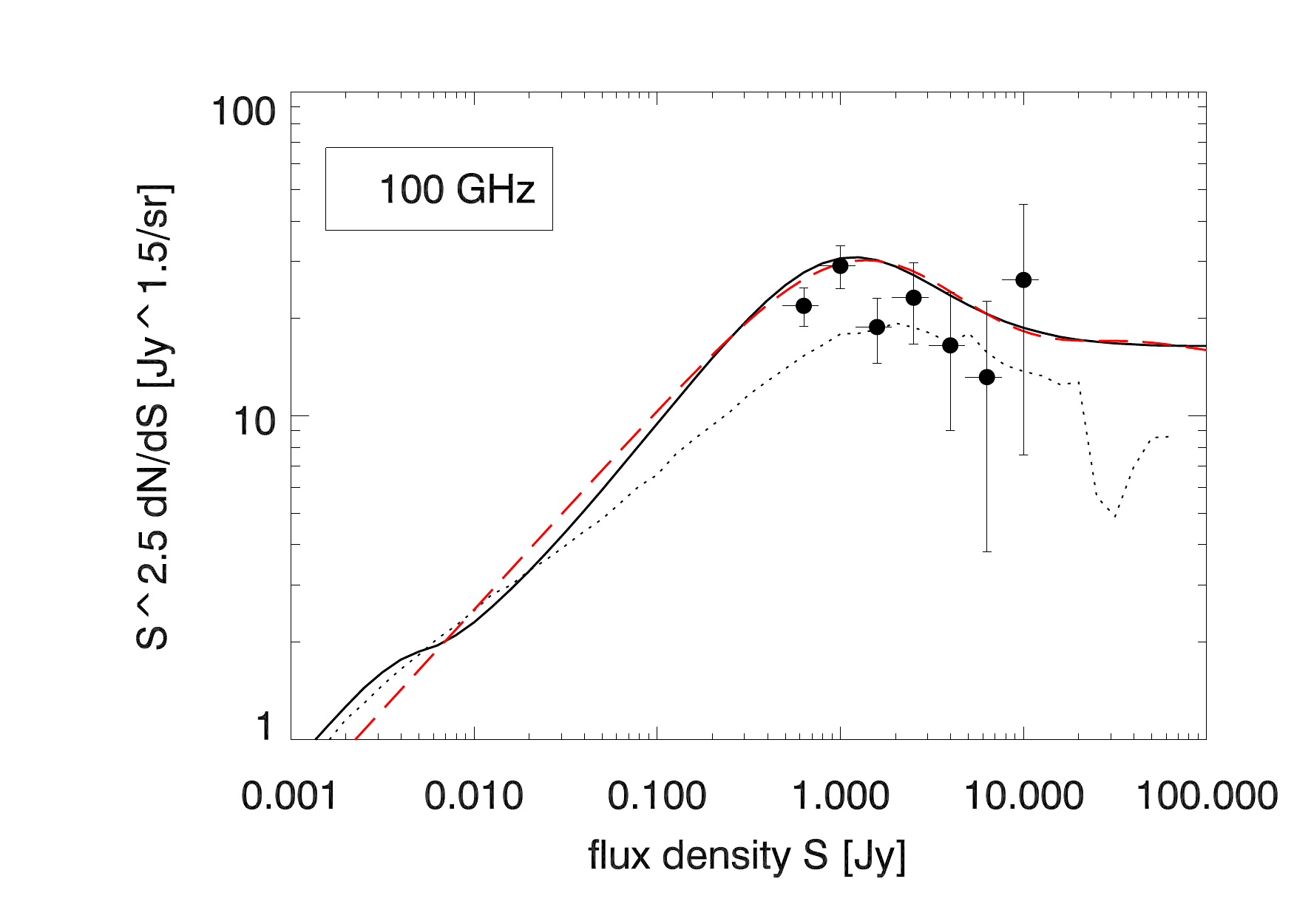} 
\includegraphics[bb=53bp 18bp 498bp 325bp,clip,width=0.33\textwidth]{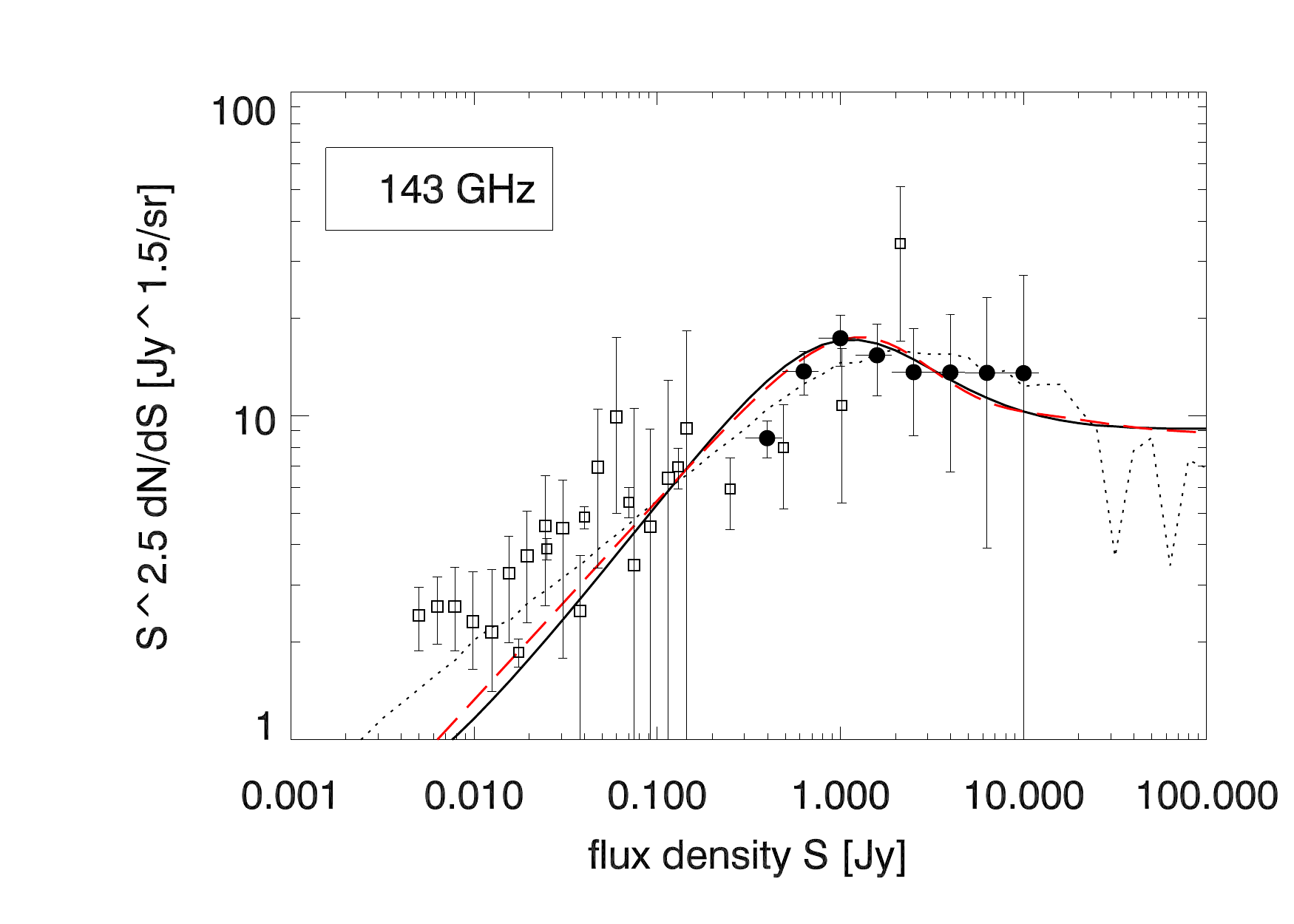}
\includegraphics[bb=53bp 18bp 498bp 325bp,clip,width=0.33\textwidth]{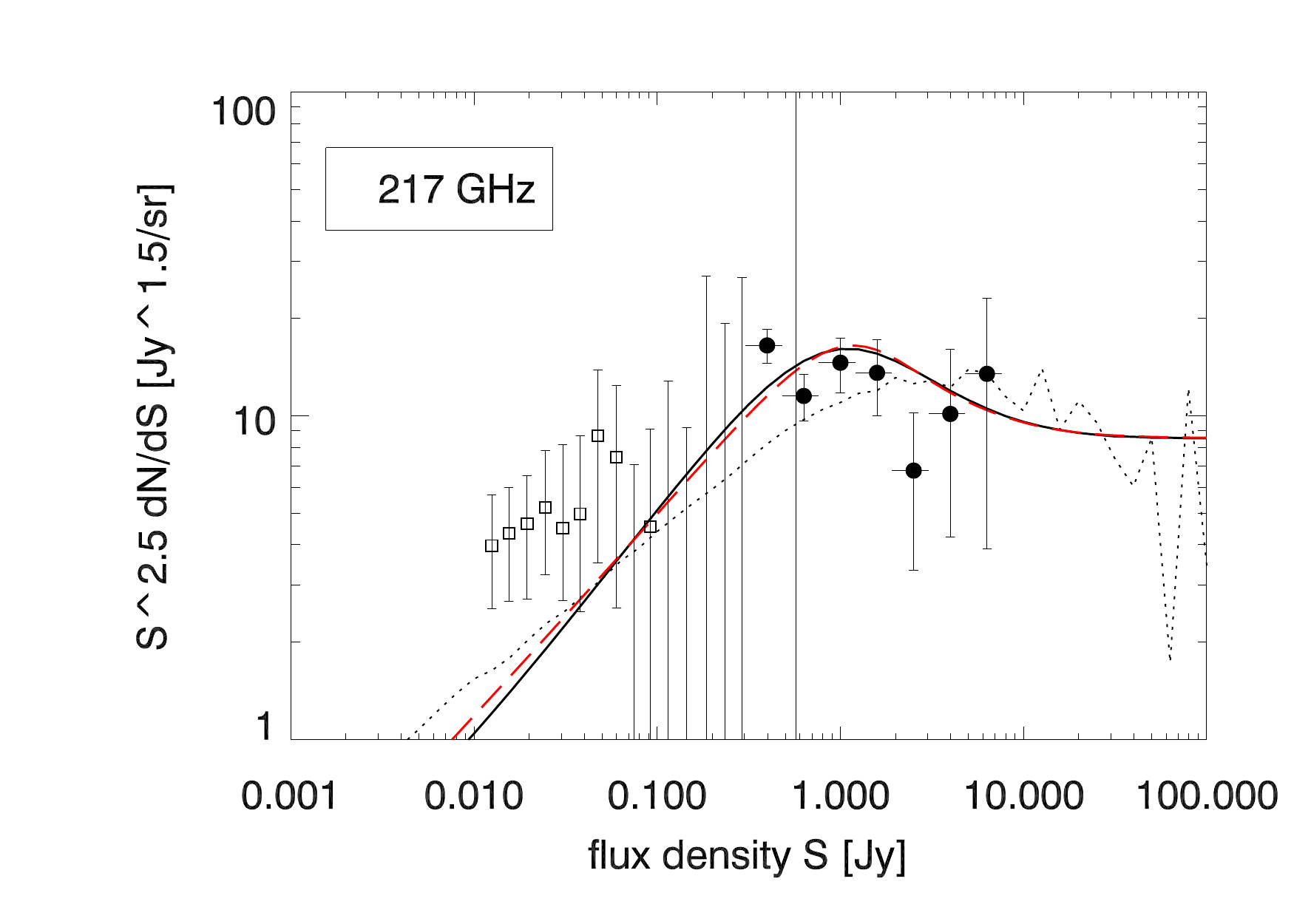} 
\par\end{centering}
\caption{
  Number counts from \Planck\  \citep[filled circles, ][]{planck2012-VII}, ACT, and SPT (open squares) as described
  in the text, from 100 GHz to 217 GHz. The models from \citet[solid line]{dezotti05}  and \citet[dots]{tucci11} are overplotted. The analytical fit from Eq.~\ref{PS1-1} and Table~\ref{tab:Parameters-for-pointsources} is shown dashed,
  and shows a similar behaviour to the \cite{dezotti05} model.
\label{fig:pscounts}}
\end{figure*}

Here we check that the Poisson power estimated in the likelihood, which comes from sources below \Planck's detection threshold, is
consistent with the level expected given number counts of detected galaxies. 
Figure~\ref{fig:pscounts} shows source counts from \Planck\ \citep{planck2012-VII}
 at $100$, $143$, and $217\;{\rm GHz}$
derived from the Planck Early Release Compact Source Catalogue
\citep{PlanckECSC,PlanckPS}.  At $143$ and $217$\,GHz we also show the
source counts from SPT as reported in \cite{vieira10} at $150$ and
$220$~GHz, and from ACT \citep{Metal11} at $150$~GHz. 
The models of \cite{dezotti05} and \cite{tucci11}
are also shown and are  discussed in \cite{planck2012-VII}.

\cite{planck2012-VII} use spectral information to separate the sources
into `synchrotron' and `dusty' sources, and show that the counts at
100 -- 217 GHz are dominated by synchrotron sources at flux densities
above $\sim 400~{\rm mJy}$. \cite{vieira10} performed a similar
separation. The counts at $150$~GHz are dominated by
synchrotron sources at flux densities $S>10~{\rm mJy}$, but dusty
galaxies contribute roughly equally at $220$~GHz 
at flux densities $\lesssim30\,{\rm mJy}$
\citep{vieira10,hall10}. The ACT counts have not been separated
according to spectral type, but should be dominated by radio sources
at these flux densities.

Figure \ref{fig:pscounts} show models fit to the counts using the function
\begin{equation}
S^{5/2}\frac{dN}{dS}= \frac{A S^{5/2} }{(S/S_0)^{\alpha}+(S/S_0)^{\beta}} + B \left(1-\exp{\left(\frac{S}{S_1}\right)}\right),
\label{PS1-1}
\end{equation}
where $A$, $S_0$, $\alpha$, $\beta$, $B$, and $S_1$ are free parameters. The best-fit values of these
parameters are given in
Table~\ref{tab:Parameters-for-pointsources}.
\begin{table}[tmb] 
\begingroup 
\newdimen\tblskip \tblskip=5pt
\caption{Parameters for point source model of Eq.~\ref{PS1-1}, fitting detected source counts shown in Fig.~\ref{fig:pscounts}.
\label{tab:Parameters-for-pointsources}}
\vskip -6mm
\footnotesize 
\setbox\tablebox=\vbox{ %
\newdimen\digitwidth 
\setbox0=\hbox{\rm 0}
\digitwidth=\wd0
\catcode`*=\active
\def*{\kern\digitwidth}
\newdimen\signwidth
\setbox0=\hbox{+}
\signwidth=\wd0
\catcode`!=\active
\def!{\kern\signwidth}
\newdimen\decimalwidth
\setbox0=\hbox{.}
\decimalwidth=\wd0
\catcode`@=\active
\def@{\kern\signwidth}
\halign{ \hbox to 1in{#\leaderfil}\tabskip=0em& 
    \hfil#\hfil\tabskip=1em& 
    \hfil#\hfil\tabskip=1em& 
    \hfil#\hfil \tabskip=0pt\cr
\noalign{\doubleline}
\omit Parameter\hfill  & 100~GHz  & 143~GHz  & 217~GHz  \cr
\noalign{\vskip 2pt\hrule\vskip 2pt}
$A$      & $18.24$ & $8.38$ & $8.58$\cr
$S_{0}$   & *$1.58$ & $1.65$ & $1.48$\cr
$\alpha$ & *$1.88$ & $1.89$ & $1.90$\cr
$\beta$  & *$3.35$ & $3.78$ & $4.10$\cr
$B$      & $14.91$ & $8.73$ & $8.53$\cr
$S_{1}$   & $14.91$ & $5.17$ & $1.78$\cr
\noalign{\vskip 2pt\hrule\vskip 2pt}
}}
\endPlancktable 
\endgroup
\end{table}

Given this model, and given the approximate flux cut applied to the \Planck\ maps, the expected contribution of radio sources to the \Planck\ power spectra, at flux densities smaller than $400$, $350$, and $225 \pm 50$~mJy at 100, 143, and
217~GHz, are $8.47\pm 1$, $6.05\pm 0.8$, and $3.10\pm 0.7$~Jy$^2$/sr.
The contribution of unresolved infrared galaxies to the power spectra is not negligible. They are expected to dominate at $217$\ghz, even if they are subdominant
in the \Planck\ counts. Indeed, faint
IR galaxies create a ``bump'' in the $S^{5/2}{\rm d}N/{\rm d}S$ distribution,
 below the detection limit of ACT or SPT. This bump is
 seen at higher frequencies, e.g., with the {\it Herschel} SPIRE instrument 
\citep[see][for details]{planck2012-VII}.

This bump of infrared galaxies has not been measured at frequencies of $217$~GHz and below.
However, measurements with the AzTEC telescope at 1.1mm \cite[270 GHz,][]{scott12} can be used to extrapolate the counts down to 217 GHz. This
leads to a predicted peak in the number counts ($S^{5/2}{\rm d}N/{\rm d}S$) around $1.4$~mJy at a
level of $190\ {\rm Jy}^{1.5}/{\rm sr}$, somewhat higher than the values in
\cite{hall10}.  The corresponding contribution of infrared galaxies to the 
power spectrum is estimated in \cite{planck2011-6.6} to be $16\ {\rm Jy}^2/{\rm sr}$ at $217$\ghz\, but with significant uncertainty.

Summing the expected contributions from radio and IR galaxies, we estimate the following values
for $D_{3000}$ for \Planck:  $200$, $75$, and $120\ \muK^2$ at $100$, $143$, and $217$\ghz\
respectively. These predictions are much less certain at $217$\ghz\ due 
to the absence of infrared galaxy counts at this frequency.

\subsubsection{Clustered power in CIB fluctuations}

Here we check the consistency of the estimated clustered power in CIB fluctuations. As already noted, from the \camspec\ and \plik\ likelihoods we find only an upper limit on the clustered CIB power at $143$~GHz. With \camspec\ we detect clustered power at 217~GHz with $A^{\mathrm{CIB}}_{217}= 32 \pm 10 \,\muK^2$, and Poisson power with $A^{\mathrm{PS}}_{217}=92\pm22\,\muK^2$. This Poisson power is dominated by the CIB fluctuations. For the \plik\ likelihood, we have $A^{\mathrm{CIB}}_{217}= 49\pm7\,\muK^2$ and $A^{\mathrm{PS}}_{217}=58\pm19\,\muK^2$. The sum of the CIB power at 217~GHz, and at pivot scale $\ell=3000$, is in the range $\approx 105-125\,\muK^2$.\footnote{Note that \Planck\ does not measure directly the CIB power at the pivot scale $\ell=3000$, hence these extrapolated values are sensitive to possible shape mismatch of the clustered CIB fluctuation power spectra at lower multipoles.}

We compare this level to the measurements by the ACT and SPT experiments, which probe higher angular resolution. Fitting a common model to the ACT power spectra from \citet{das/etal:prep}, and the SPT spectra from \citet{keisler11,reichardt12}, the analysis in \citet{dunkley/etal:prep} finds $A^{\mathrm{CIB}}_{219.6}=54\pm16\,\muK^2$ for the CIB clustered component (and $A^{\mathrm{PS, CIB}}_{219.6} = 78\pm12\,\muK^2$ for the CIB Poisson component) at effective frequency 219.6~GHz. For SPT, the clustered level is $A^{\mathrm{CIB}}_{219.6}=59\pm12\,\muK^2$  (and Poisson $A^{\mathrm{PS, CIB}}_{219.6}=69\pm10\,\muK^2$), also at an effective frequency of $219.6$~GHz. This is consistent with the SPT analysis in \cite{reichardt12}. These are estimated assuming $\gamma^{\mathrm{CIB}}=0.8$ and $r^{\mathrm{CIB}}_{143\times217}=1$, and that the CIB emission can be modeled with frequency as a modified blackbody, following \citet{Aetal12a}. 

The total CIB signal seen by \Planck, extrapolated to $\ell=3000$ scales, is therefore consistent with the ACT and SPT observations, but given the limited angular range of \Planck, the clustered and Poisson part are degenerate. This motivates us to include the ACT and SPT data in many of our cosmological analyses.

When combining \Planck, ACT and SPT data together, using the same foreground model (except for Poisson power which depend on the respective flux cuts of the experiments), both \camspec\ and \plik\ give $A^{\mathrm{CIB}}_{217}= 50 \pm 5\,\muK^2$, $A^{\mathrm{PS}}_{217}=60\pm 10\,\muK^2$, $A^{\mathrm{CIB}}_{143}= 32 \pm 8\,\muK^2$, $A^{\mathrm{PS}}_{143}=75\pm 8\,\muK^2$, and 
$A^{\mathrm{PS}}_{100}=220\pm 53\,\muK^2$. These estimates of the Poisson power are in good agreement with the predictions given in Sect.~\ref{sub:poisson-counts} for the $100$ and $143$\ghz\ channels. In the latter, radio sources below the Planck flux cuts dominate the Poisson power, which can be reliably estimated from existing source counts measurements.

We also consider modifying our model for the clustered part of the CIB. There have been a wealth of CIB models \citep[e.g.,][]{knox01,amblard07,hall10,penin11a,addison12}, most assuming that the dust is a biased tracer of the dark matter distribution, but differing in their parametrization of the dust emissivity and its evolution, and their treatment of the
dark matter power spectrum. 
Recent papers \citep{PlanckCIB,addison12} have shown that the addition of \Planck\ CIB measurements, when combined
with other small scale probes including SPT, ACT, BLAST, and {\it Herschel}, rule out models that assume the underlying dark matter power spectrum is linear. 

We therefore test a set of models that try to simultaneously fit the non-linear spectrum with one or more template spectra. We consider fixing the scale dependence to a power law, either $\ell^{0.8}$  or $\ell^{0.6}$, or extending the power law model to have a running of the index. The $\ell^{0.8}$ model has been used in \citet{shirokoff10,reichardt12,Aetal12a,dunkley/etal:prep}, while the $\ell^{0.6}$ more closely matches the CIB model of \cite{addison12}. 
We find that a simple power law does not fit both the \Plancks and high-$\ell$ data sufficiently well, but that allowing the additional freedom of a running spectrum opens up the parameter space too much, with little improvement in goodness of fit, motivating our use of the varying $\gamma^{{\rm CIB}}$ model.

We also test a simple model using just a linear theory dark matter 
power spectrum, assuming that the non-linear power can be absorbed into the Poisson term. This results in an estimate of the Poisson level at $217$~GHz that is inconsistent with ACT and SPT, so we do not use this model.  For all these models we test the effect on cosmological parameters, using the Hubble constant as a test case, and find the effect on parameters to be small. This is also investigated in \citet{planck2013-p11}.

\subsection{Consistency of the \Planck\ 70\,GHz data\label{sub:LFIcheck}}

Figure\,\ref{fig:PS_CMB+nu} shows individual frequency spectra
from 70 to 353\,GHz. We only use data
from the 100 to 217\,GHz channels to form the high-$\ell$ likelihoods, but 
here we compare cosmological parameters
derived from the \Plancks 70\,GHz channel alone. Maps at 70\,GHz are easier
to characterize than the higher frequency channels in terms of instrumental
and foreground properties, but the resolution and sensitivity are lower. 
The 70\,GHz noise properties are in general
well described by a simple three parameter model involving $1/f$
and white noise contributions \citep{planck2013-p02}. The 70\,GHz channel
also has the least diffuse foreground emission \citep{planck2013-p06},
and the extragalactic source contribution is dominated by radio galaxies
whose emission is well known at these frequencies \citep{planck2013-p05}.
We adopt a Galactic plane cut leaving $\sim 70\%$ of the sky for the analysis (CS70) to which we add a point source mask optimized for 70\,GHz. 
In Appendix\,\ref{App:70GHz-Cosmo} we describe how cosmological 
parameters are estimated from the 70\,GHz channel, which are summarised in Fig.\,\ref{fig:whiskers_cosmo}. Accounting for the lower sensitivity and
angular resolution at 70\,GHz, which translate into a narrower multipole range ($\ell < 1200$), the parameter distributions are consistent with the reference values.

\subsection{Consistency with power spectra of CMB maps obtained by component
separation methods \label{sub:mapCheck}}

\begin{figure}[h]
\begin{centering}
\includegraphics[width=1\columnwidth]{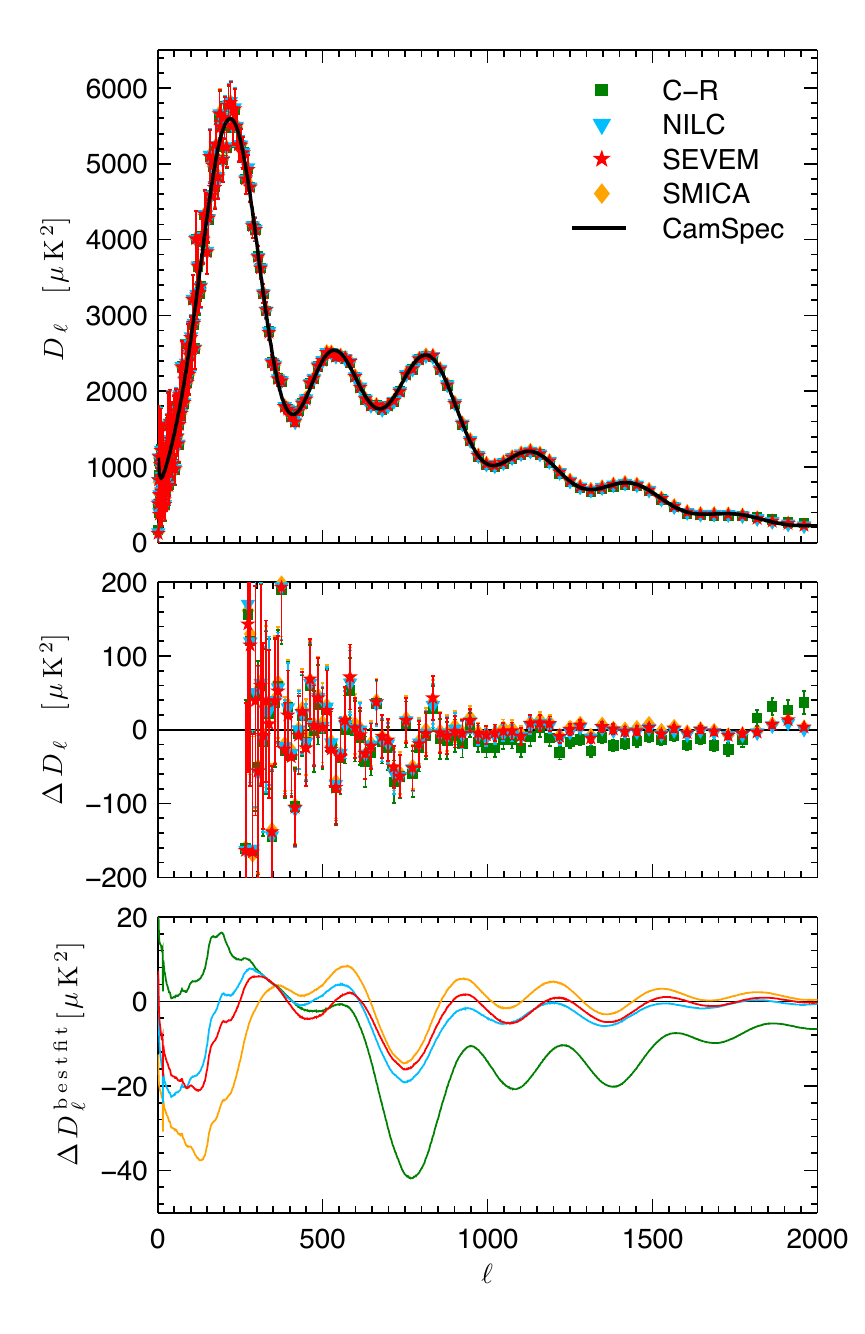}
\par\end{centering}
\caption{\emph{Top:} CMB power spectrum, with best-fitting foreground model
removed, compared to the \camspecs best fit model. \emph{Middle:}
the residuals with respect to this best-fit model. \emph{Bottom:}
residuals of the best-fit models from the map-based likelihoods, with
respect to the \camspecs best fit model. 
\label{fig:cl_CMBmaps_dx9}}
\end{figure}

The likelihoods we consider in this paper account for component separation
by modeling the multi-frequency data at the power spectrum level, to 
fully exploit the signal at the smallest
scales probed by \Planck. We can compare the results to those derived
from an alternative approach, measuring the power spectrum of CMB
maps estimated from component separation techniques. Here we present
results obtained with four CMB maps, derived using methods referred
to as Commander-Ruler, SMICA, NILC, and SEVEM, described in detailed
in the accompanying paper \citep{planck2013-p06}. We compare
their angular power spectra and cosmological parameters with those
from \camspec. To estimate the power spectra we use the XFaster method,
an approximation to the iterative, maximum likelihood, quadratic band
power estimator based on a diagonal approximation to the quadratic
Fisher matrix estimator \citep{rocha2009,rocha2010b}. The noise
bias is estimated using difference maps, as described in \citet{planck2013-p06}.
The resulting spectra are shown in Fig.\,\ref{fig:cl_CMBmaps_dx9},
and agree well out to scales $\ell_{\mathrm{max}}=2000$, even though the agreement
is less striking for the Commander-Ruler small-scale spectrum. 

\begin{figure}[h]
\centering{}\includegraphics[width=1\columnwidth]{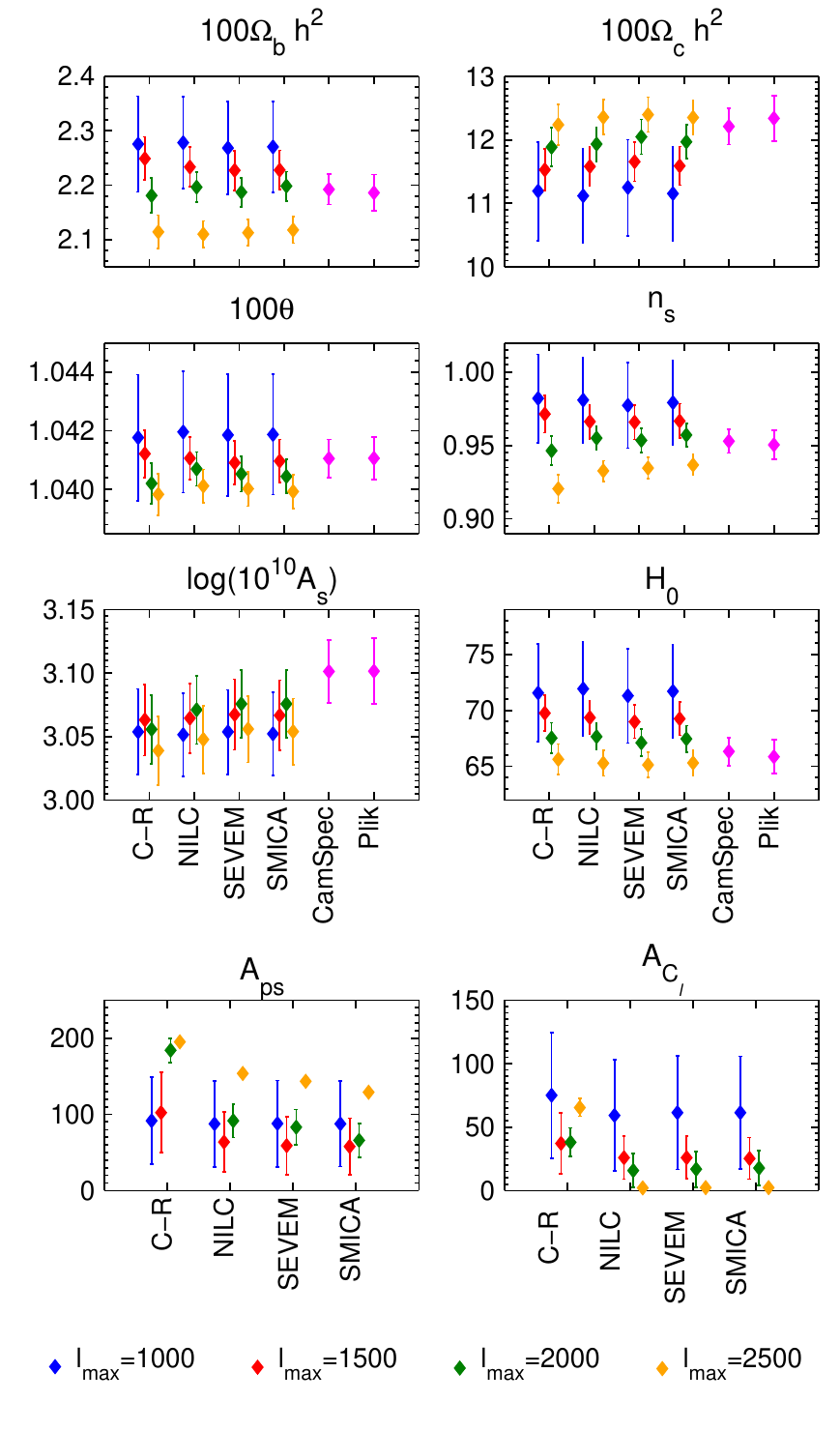}
\caption{
The variation of cosmological and foreground
parameters estimated for the four CMB maps as function of $\ell_{\mathrm{max}}=1000,\ 1500,\,2000,\,2500$, compared to those obtained
with the \camspecs and \pliks likelihoods. Parameters are consistent to $1\sigma$ for  $\ell_{\mathrm{max}} \le 2000$
\label{fig:par-summ-ddx9-2}}
\end{figure}

We then estimate cosmological parameters
using a Gaussian correlated likelihood derived from these band-powers.
To model the residual extragalactic foregrounds, we consider two nuisance
parameters: $A_{\mathrm{ps}}$, the amplitude of a Poisson component, 
and $A_{C_{\ell}}$, the amplitude of a clustered component, scaling a term with shape $D_{\ell} \propto \ell^{0.8}$.
Figure\,\ref{fig:par-summ-ddx9-2}
compares the parameters obtained as a function of $\ell_{\mathrm{max}}$ for
each method, compared to the values from the high-$\ell$
likelihoods. The results are consistent to 1$\sigma$ or better for $\ell_{\mathrm{max}}=1500$
and $\ell_{\mathrm{max}}=2000$.
Despite adopting a simple two-parameter model
for the extragalactic foregrounds, the likelihood using a CMB map as input data 
works reasonably well, and may be further exploited with analysis of simulations, improved 
extragalactic foreground modeling, and the development of an error model.

\subsection{Consistency with high-$\ell$ polarisation\label{sub:Consistency-Polar-}}

\begin{figure*}
\begin{centering}
\includegraphics[bb=9bp 10bp 503bp 425bp,clip,width=1\columnwidth]{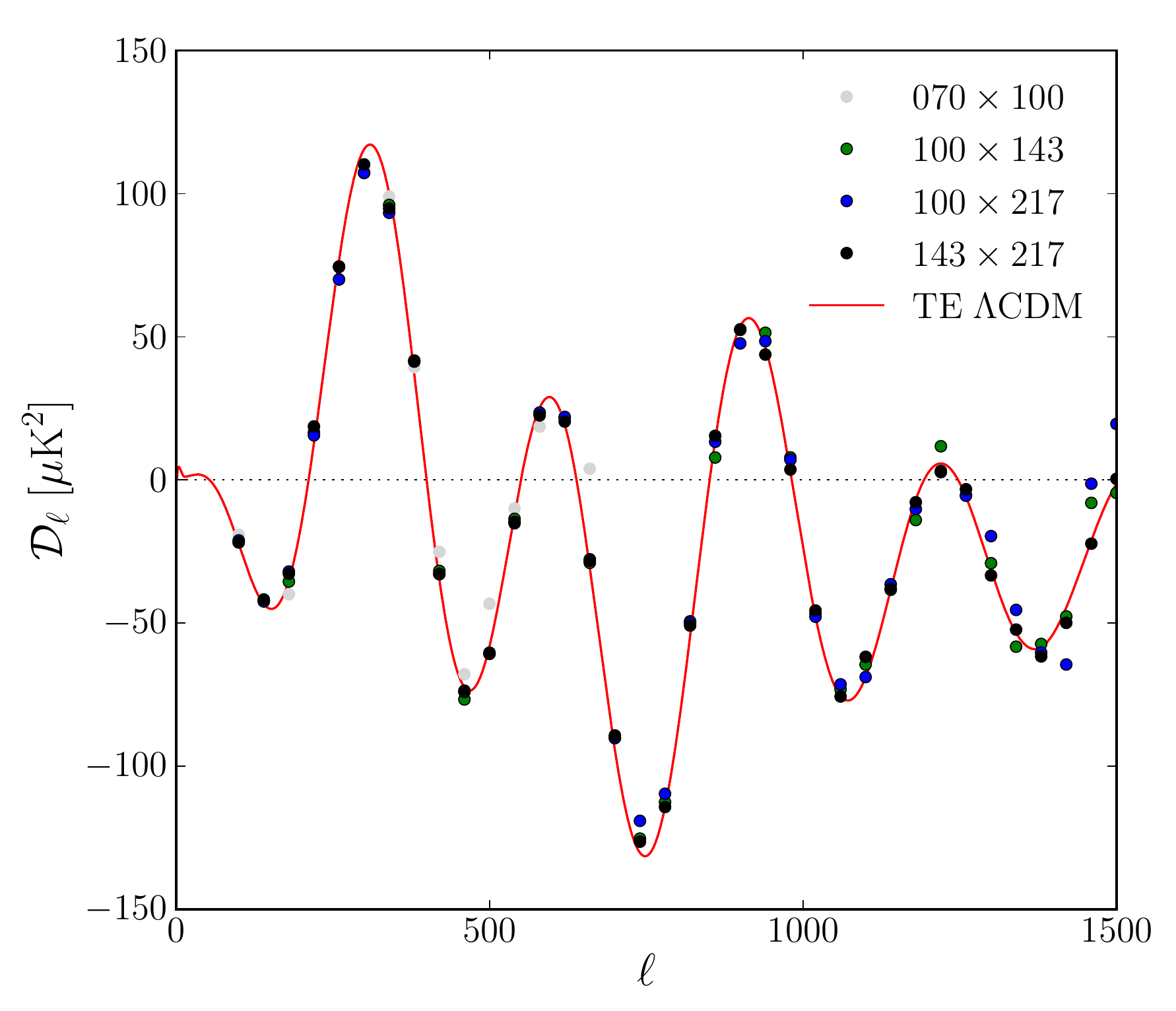}
\includegraphics[bb=9bp 10bp 503bp 425bp,clip,width=1\columnwidth]{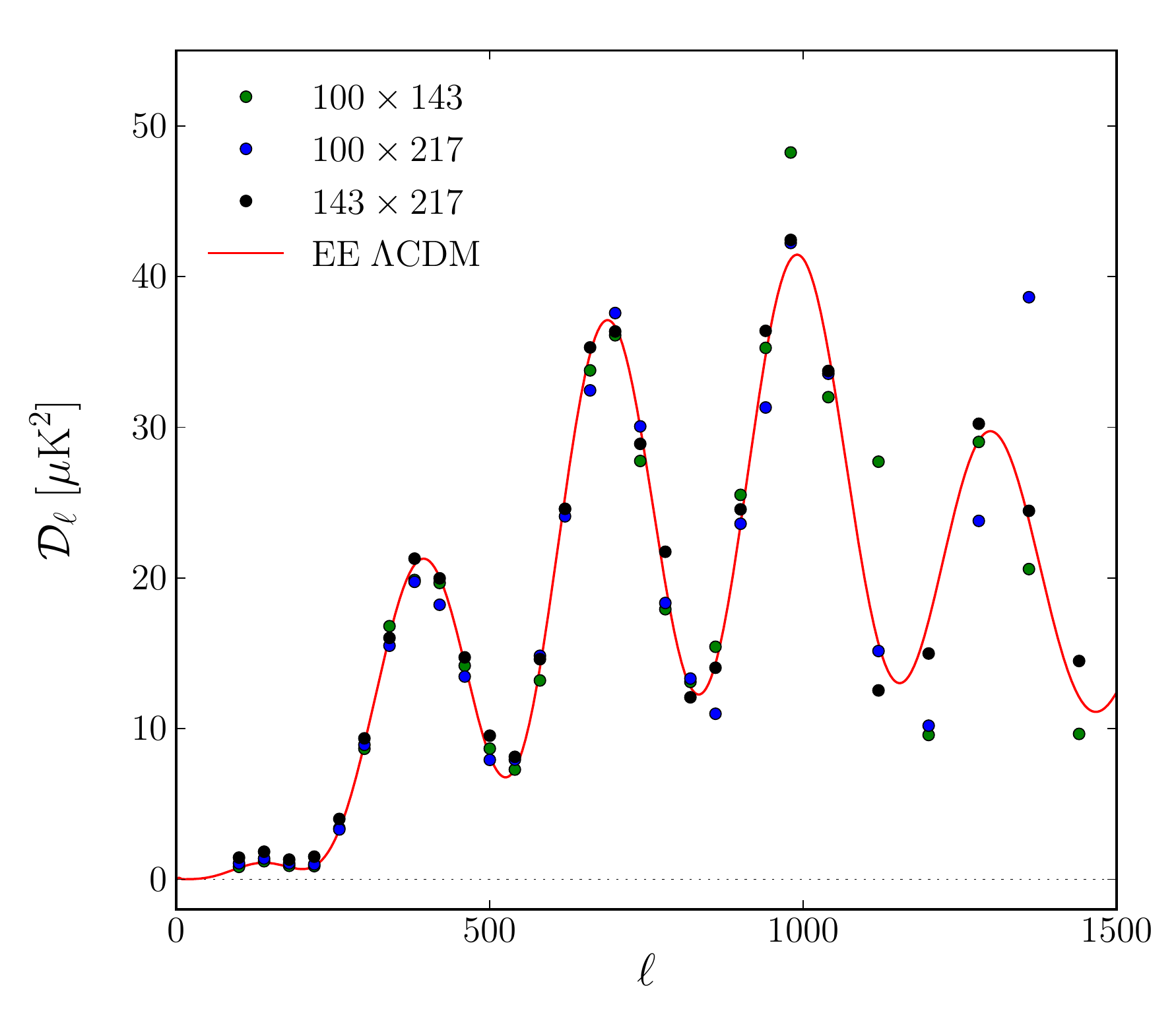}
\par\end{centering}
\caption{\Planck\ TE  and EE polarisation spectra computed as described in
the text, together with the polarisation spectra predicted from the six-parameter
$\Lambda$CDM model, fit only to the \Planck\ temperature data.
\label{fig:cl_pol_consistency}}
\end{figure*}

Our final consistency test is illustrated in Fig.\,\ref{fig:cl_pol_consistency}, showing the polarisation power spectra derived from the \Planck\ data.
Both the TE and EE cross-spectra are shown, in bins of  width of $\Delta \ell = 40$.
These spectra are computed by performing a uniformly weighted average
of all detector sets combinations at $70\times 100$, $100\times143$,
$100\times217$, and $143\times217$~GHz. We use the temperature beam
window functions for beam deconvolution.
For the analysis, we applied CL39 to the temperature maps, and discarded
$60\,\%$ of the polarisation data (i.e., $f_{\mathrm{sky}}=0.4$). Other than
masking, no efforts have been made to subtract foreground contributions
or take into account instrumental effects such as leakage. Despite the
substantial masking applied, we see evidence for residual contributions
of non-cosmological origin. Besides demonstrating the potential of
\Planck\ to deliver high quality polarisation maps and spectra \citep[with
the limitations explained in][]{planck2013-p03}, the figure demonstrates
the high level of consistency of these polarisation spectra between
themselves, and with the prediction from the model fit using just the temperature
spectrum shown in Fig.\,\ref{fig:ps_CMB_ref}. 

As discussed in \cite{planck2013-p03} and \cite{planck2013-p02}, at present,
the HFI and LFI polarisation spectra at low multipoles are affected
by systematic errors that cause biases which will need to be accurately
modeled or removed for the next \Planck\ release. However, these
systematics rapidly become unimportant at higher multipoles.
While not yet fit for cosmological parameter analysis, the consistency
at the level of a few $\muK$ of these \Planck\ polarisation spectra
adds to our confidence in the analysis of temperature
data.
It shows that within the $\Lambda$CDM framework, the cosmological 
parameters estimated  from \Planck\ temperature data are not 
strongly affected by the uncertainties in the modelling of 
unresolved foregrounds.

%% file: 8_Low-multipoles-likelihood.tex
\section{Low-$\ell$ likelihood\label{sec:Low-multipoles-Likelihood}}

At low multipoles ($\ell\lesssim50$), the distribution of the
estimated $C_{\ell}$s is not well approximated by a Gaussian due to
the limited degrees of freedom per $\ell$
\citep[e.g.,][]{Efstathiou2004}. However, both the CMB signal,
$\mathbf{s}$, and instrumental noise, $\mathbf{n}$, are individually
nearly Gaussian distributed at the map level, provided that foreground
emission and instrumental systematics effects are negligible
\citep[e.g.,][]{planck2013-p09,planck2013-p09a}, and the actually
observed map, ${\bf m}={\bf s}+{\bf n}$, is therefore also nearly
Gaussian distributed. Under this assumption, the CMB power spectrum
likelihood is given by
\begin{equation}
\mathcal{L}(C_{\ell}) = P(\mathbf{m}|C_{\ell})=\frac{1}{2\pi^{n/2}|\mathbf{M}|^{1/2}}\exp\left(-\frac{1}{2}\mathbf{m}^t\,\mathbf{M}^{-1}\mathbf{m}\right),\label{pbLike}
\end{equation}
where $n$ is the number of observed pixels,
$\mathbf{M}(C_{\ell})=\mathbf{C}(C_{\ell})+\mathbf{N}$ is the data
covariance matrix, and $\mathbf{C}$ and $\mathbf{N}$ are the CMB and
noise covariance matrices, respectively.

In the general case, the data vector $\mathbf{m}$ includes both
temperature ($T$) and linear polarisation ($Q$, $U$) Stokes parameter
maps. Pixels exhibiting high foreground contamination are removed by
masking, such that the data vector is restricted to the subset of
valid pixels, ${\bf
  m}=(T_{i_{1}},T_{i_{2}},...,T_{n_{T}},Q_{j_{1}},Q_{j_{2}},...Q_{n_{P}},U_{j_{1}},U_{j_{2}},...U_{n_{P}})$.
The corresponding rows and columns are removed from $\mathbf{M}$,
effectively corresponding to marginalizing over the masked region of
the sky. Note that in general, $n_{T}\ne n_{P}$, and the sets of
indexes of temperature and polarisation measurements will be
different. We assume the same number of pixels in $Q$ and $U$,
although this is not a requirement.  

The signal covariance matrix can be written symbolically as
\begin{equation}
{\bf C}=\begin{pmatrix}\left<TT\right>_{(n_{T}\times n_{T})} & \left<TQ\right>_{(n_{T}\times n_{P})} & \left<TU\right>_{(n_{P}\times n_{P})}\\
\left<QT\right>_{(n_{P}\times n_{T})} & \left<QQ\right>_{(n_{P}\times n_{P})} & \left<QU\right>_{(n_{P}\times n_{P})}\\
\left<UT\right>_{(n_{P}\times n_{T})} & \left<UQ\right>_{(n_{P}\times n_{P})} & \left<UU\right>_{(n_{P}\times n_{P})}
\end{pmatrix},\label{eq:TQUcovar}
\end{equation}
where the signal correlations for the temperature component are
explicitly given by
\begin{equation}
\langle T_{i_{1}}T_{i_{2}}\rangle=\sum_{\ell=2}^{\ell_{{\rm max}}}\frac{{2\ell+1}}{{4\pi}}{\hat{C}}_{\ell}P_{\ell}(\theta_{i_{1}i_{2}})+{\mathbf{N}_{i_{1}i_{2}}}.\label{eq:ttcorr}
\end{equation}
Here $P_{\ell}$ are the Legendre polynomials, and
$\theta_{i_{1}i_{2}}$ is the angle between the centres of pixels
$i_{1}$ and $i_{2}$. Similar expressions are available for the
polarisation correlations \citep[e.g.,][]{Tegmark2001}. The effect of
the (azimuthally symmetric) instrumental beam, $b_{\ell}$, and pixel
window function, $w_{\ell}$, are encoded in ${\hat{C}}_{\ell} =
C_{\ell}^{{\rm th}} b_{\ell}^{2}w_{\ell}^{2}$.

The main problem with the likelihood expression given in
Eq.~\ref{pbLike} is its high computational cost. This is determined by
the matrix inversion and determinant evaluations, both of which scale
as $\mathcal{O}(N^3)$ with $N=n_T+2n_P$. In practice, this approach is
therefore limited to coarse pixelizations, $N_{\rm side}\le16$, which
reliably only supports multipoles below $\ell\lesssim30$. On the other
hand, the Gaussian approximation adopted by the high-$\ell$ likelihood
is not sufficiently accurate for the stringent requirements of
\Planck\ below $\ell\lesssim50$. In the next section, we therefore
describe a faster low-$\ell$ likelihood estimator, based on Gibbs/MCMC
sampling, which allows us to exploit the full range up to $\ell\le50$
with low computational cost, while additionally supporting physically
motivated foreground marginalization.

\citet{WMAP-3yrsPol} pointed out that the temperature and polarisation
parts of the likelihood can be separated and evaluated independently,
under the assumption of negligible noise in temperature and in the
temperature-polarisation cross correlations (i.e., the $TQ$ and $TU$
blocks of the pixel level noise covariance matrices).  Further
assuming vanishing primordial $B$ modes and $TB$ correlations, the $TE$
correlations can be accounted for by redefining the modified $Q$ and
$U$ maps as
\begin{align}
Q & \rightarrow Q-\frac{1}{2}\sum_{\ell=2}^{\ell\mathrm{max}}\frac{C_{\ell}^{TE}}{C_{\ell}^{TT}}\sum_{m=-\ell}^{\ell}a_{\ell m}^{T}\left(_{+2}Y_{\ell m}+_{-2}Y_{\ell m}^{*}\right)\label{eq:Qtransf}\\
U & \rightarrow U-\frac{i}{2}\sum_{\ell=2}^{\ell\mathrm{max}}\frac{C_{\ell}^{TE}}{C_{\ell}^{TT}}\sum_{m=-\ell}^{\ell}a_{\ell m}^{T}\left(_{+2}Y_{\ell m}-_{-2}Y_{\ell m}^{*}\right),\label{eq:Utransf}
\end{align}
where $_{\pm2}Y_{\ell m}$ are spin weighted spherical harmonics and
$a_{\ell m}^{T}$ are the harmonic coefficients of the signal in the
temperature map. One can show by direct substitution that these
modified $Q$ and $U$ maps are free of temperature correlations. The
polarisation likelihood can be then computed independently from the
temperature likelihood and, possibly, at lower resolution to save
computational expenses. We test this strategy in
Sect.~\ref{sec:low-ell-pol-like}, and adopt it for the current release
of the \Planck\ likelihood.

\subsection{Low-$\ell$ temperature likelihood }

\label{app:commander} 

As discussed above, we do not implement the likelihood expression
given in Eq.~\ref{pbLike} directly, due to its high computational cost
and limited flexibility with respect to foreground modelling.
Instead, we adopt the Gibbs sampling approach
\citep{Eriksen2004,Jewell2004,Wandelt2004}, as implemented by the
\commander\ code \citep{Eriksen2008a}, which allows both for
physically motivated component separation and accurate likelihood
estimation. A similar Gibbs sampling method was used to estimate the
low-$\ell$ temperature likelihood for {\it WMAP}
\citep{dunkley2009,larson2011}, although not simultaneously accounting
for component separation.

\subsubsection{Methodology}

We start by generalizing the above data model to include both
multi-frequency observations and a set of foreground signal terms,
\begin{equation}
\mathbf{d}_{\nu}=\mathbf{s}+\sum_{i}\mathbf{f}_{\nu}^{i}+\mathbf{n}_{\nu}.
\end{equation}
Here $\mathbf{d}_{\nu}$ denotes the observed sky map at frequency
$\nu$, and $\mathbf{f}_{\nu}^{i}$ denotes a specific foreground signal
component. As above, the CMB field is assumed to be a Gaussian random
field with power spectrum $C_{\ell}$, and the noise is assumed
Gaussian with covariance $\mathbf{N}_\nu$. The foreground model can be
adjusted as needed for a given data set, and a full description of the
model relevant for \Planck\ is presented in \citet{planck2013-p06}. In
short, this consists of a single low-frequency foreground component
(i.e., the sum of synchrotron, anomalous microwave emission, and
free-free emission), a carbon monoxide (CO) component, and a thermal
dust component, in addition to unknown monopole and dipole components
at each frequency.

Given this data model, we map out the full posterior distribution,
$P(\mathbf{s},\mathbf{f}^{i},C_{\ell}|\mathbf{d})$, using a Monte
Carlo sampling algorithm called Gibbs sampling. Directly drawing samples
from $P(\mathbf{s},\mathbf{f}^{i},C_{\ell}|\mathbf{d})$ is computationally
prohibitive, but this algorithm achieves the same by iteratively sampling
from each corresponding \emph{conditional} distribution, 
\begin{align*}
\mathbf{s} & \leftarrow P(\mathbf{s}|\mathbf{f},C_{\ell},\mathbf{d})\\
\mathbf{f} & \leftarrow P(\mathbf{f}|\mathbf{s},C_{\ell},\mathbf{d})\\
C_{\ell} & \leftarrow P(C_{\ell}|\mathbf{s},\mathbf{f}^{i},\mathbf{d}).
\end{align*}
It is straightforward to show that
$P(\mathbf{s}|\mathbf{f},C_{\ell},\mathbf{d})$ is a multivariate
Gaussian distribution, and
$P(C_{\ell}|\mathbf{s},\mathbf{f}^{i},\mathbf{d})$ is an inverse Gamma
distribution. The foreground distribution,
$P(\mathbf{f}|\mathbf{s},C_{\ell},\mathbf{d})$, does not have a closed
analytic form, but can easily be mapped out numerically
\citep{Eriksen2008a}. Thus, all three distributions are
associated with simple textbook sampling algorithms.

For CMB likelihood estimation, the crucial intermediate product from
the above sampling process is the ensemble of CMB sky samples,
$\mathbf{s}^{k}$.  Each individual sample corresponds to one possible
CMB realization consistent with the observed data. In the absence of
sky cuts, foreground contamination and instrumental noise, this map is
identical to the true sky. In that case, the likelihood as a function
of $C_{\ell}$ is determined by cosmic variance alone, and given by an
inverse gamma distribution,
\begin{equation}
\mathcal{L}^{k}(C_{\ell})\propto\frac{\sigma_{\ell,k}^{\frac{2\ell-1}{2}}}{C_{\ell}^{\frac{2\ell+1}{2}}}e^{-\frac{2\ell+1}{2}\frac{\sigma_{\ell},k}{C_{\ell}}}.\label{eq:exact_like}
\end{equation}
Here we have introduced the realization specific power spectrum,
$\sigma_{\ell,k}\equiv\frac{1}{2\ell+1}\sum_{\ell=-m}^{m}|a_{\ell
  m}^{k}|^{2}$, where $a_{\ell m}^{k}$ are the spherical harmonic
coefficients of $\mathbf{s}^{k}$. In the case of realistic data, we
need to marginalize over uncertainties due to sky cuts, foregrounds,
and instrumental noise. Hence, Eq. \ref{eq:exact_like} is replaced
by an average over all possible sampled CMB realizations,
\begin{equation}
\mathcal{L}(C_{\ell})\propto\sum_{k=1}^{N_{\textrm{samp}}}\mathcal{L}^{k}(C_{\ell}).
\end{equation}
This expression is known as the Blackwell-Rao estimator
\citep{Chu2005}, and is guaranteed to converge to the exact likelihood
as the number of samples, $N_{\textrm{samp}}$, increases. Note that
the normalization factor in this expression is unknown, but since the
likelihood function is only used to compare different models through
an effective likelihood ratio, this factor is irrelevant for actual
calculations.

\subsubsection{Data selection and preprocessing}\label{sec:commander_dataselection_prep}

As described in \citet{planck2013-p06}, we include
\Plancks\ frequencies between 30 and 353\,GHz in the low-$\ell$
likelihood. Each frequency map is downgraded from its native
resolution to a common resolution of 40\arcmin, and projected onto an
$N_{\textrm{side}}=256$ HEALPix grid. Uncorrelated Gaussian
regularization noise is added to each frequency channel map, with an
RMS proportional to the spatial mean of the instrumental noise of the
corresponding channel, $\left<\sigma_{\nu}\right>$, conserving
relative signal-to-noise between channels. The regularization noise
level at frequency $\nu$ is
$5\mu\textrm{K}\cdot\left<\sigma_{\nu}\right>/\left<\sigma_{\textrm{143GHz}}\right>$.
The purpose of this is to make the results insensitive to unmodelled
features at scales comparable to and beyond the smoothing scale of 40\arcmin, in addition to improve the convergence speed of the Gibbs
sampler. The resulting signal-to-noise is unity at $\ell\sim400$, and
the additional uncertainty due to the regularization noise is less
than $0.2\,\mu\textrm{K}^{2}$ below $\ell=50$, and less than
$1\,\mu\textrm{K}^{2}$ below $\ell=100$.

\begin{figure}
\includegraphics[width=88mm]{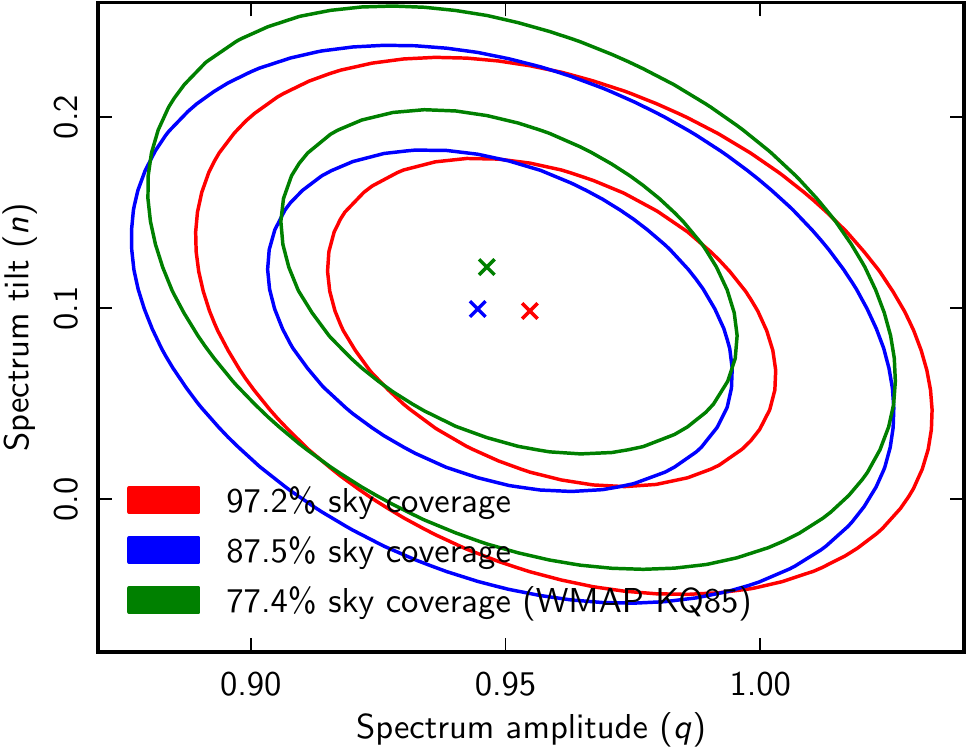} 
\caption{Two-parameter probability distribution for an amplitude-tilt
  model constrained by the low-$\ell$ \Planck\ likelihood using three
  different masks. Angular scales between $2\le\ell\le49$ are included
  in this fit.}
\label{fig:A_n_mask} 
\end{figure}

To study the stability of the low-$\ell$ likelihood with respect to
sky fraction, we constructed a suite of five different masks, covering
between 81 and 100\% of the sky, and for completeness we include the
\emph{WMAP} KQ85 mask, covering 75\% of the sky, as a sixth
case. These low-$\ell$ masks are distinct from those employed for the
high-$\ell$ likelihood, and are produced in a preliminary
\commander\ full-sky analysis in which we estimate individual
foreground components and residual $\chi^{2}$ values per pixel. These
maps are thresholded at various levels to produce a useful range of
sky fractions. 

For each mask, we fit a two-parameter amplitude and tilt power
spectrum model of the form
$C_{\ell}(q,n)=qC_{\ell}^{\textrm{fid}}\left(\ell/\ell_{0}\right)^{n}$,
using the low-$\ell$ likelihood between $\ell_{\textrm{min}}=2$ and
$\ell_{\textrm{max}}=49$, where $C_{\ell}^{\textrm{fid}}$ is the
best-fit \Planck\ $\Lambda$CDM spectrum,
and $\ell_0 = (\ell_{\textrm{min}}+\ell_{\textrm{max}})/2$. The
resulting distributions are shown in Fig.~\ref{fig:A_n_mask} for three
masks, covering 77.4 (A; WMAP KQ85), 87.5 (B) and 97.2\% (C) of the
sky, respectively. The internal agreement is excellent, with
parameters differing by less than $0.3\,\sigma$ between the very
aggressive mask A and the conservative mask C. While any of these
masks would establish an acceptable likelihood, we adopt Mask B as our
fiducial mask for two reasons. On the one hand, the parameter
uncertainties obtained with Mask B are only 4\% larger than those
obtained for the minimal Mask A, indicating that both nearly saturate
the cosmic variance limit. On the other hand, analysis of realistic
simulations indicate the presence of statistical significant map
residuals near the Galactic plane that are accepted by Mask A, but
rejected by Mask B \citep{planck2013-p06}. The latter therefore
represents a good compromise between rejecting foreground residuals
and maximizing statistical power.

We include 100\,000 Gibbs samples in the likelihood estimator,
ensuring excellent convergence characteristics for the Blackwell-Rao
estimator for $\ell<50$.

\subsection{Low-$\ell$ polarisation likelihood}\label{sec:low-ell-pol-like}

The present \Planck\ data release includes only temperature data.  In
this release, we therefore supplement the \Planck\ likelihood with the
9-year \emph{WMAP} polarisation
likelihood\footnote{http://lambda.gsfc.nasa.gov} derived from the {\it
  WMAP} polarisation maps at 33, 41, and 61 GHz (Ka, Q, and V bands)
\citep{WMAP-3yrsPol,BennettWMAP9}.  However, we introduce one
modification to this pixel-based likelihood code, replacing the
spherical harmonics coefficients of the temperature field, $a_{\ell
  m}^{T}$, in Eq.~\ref{eq:Qtransf} and \ref{eq:Utransf} with those
derived from the \Planck\ temperature map derived by \commander, for
which the Galactic plane has been replaced with a Gaussian constrained
realization. 

\begin{figure}
\begin{centering}
\includegraphics[width=1\columnwidth]{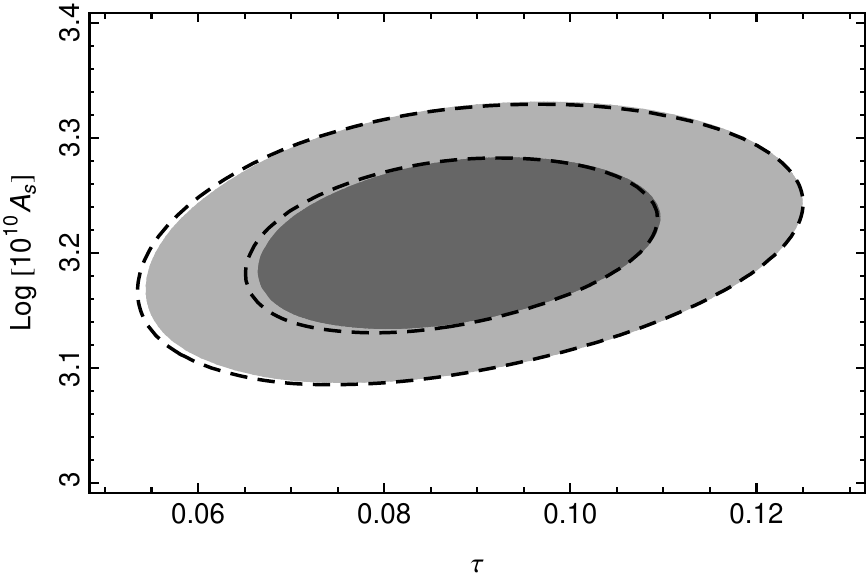}
\par\end{centering}

\caption{Comparison of constraints on $\tau$ and $A_{s}$ using the
  split temperature-polarization \WMAP\ likelihood approach (dashed
  contours; Eqs.~\ref{eq:Qtransf} and \ref{eq:Utransf}) with those
  obtained with the exact brute-force pixel likelihood (shaded
  contours; Eq.~\ref{pbLike}). \label{fig:T+QUvsTQU}}
\end{figure}

In Fig.~\ref{fig:T+QUvsTQU}, we compare constraints on $\tau$ and
$A_{s}$ as derived with this split likelihood with those obtained
through an exact brute-force evaluation of Eq.~\ref{pbLike},
simultaneously including temperature and polarisation measurements at
$N_{{\rm side}}=16$. The two methods produce almost indistinguishible
results.

In Appendix \ref{sec:Dust-cleaning-353} we assess the robustness of
the {\it WMAP} polarisation likelihood with respect to dust contamination,
by replacing the {\it WMAP} polarised dust template with the far more
sensitive HFI 353\,GHz polarisation map. We find that the optical
depth to reionization, $\tau$, is reduced by about
$0.5-1\,\sigma$, depending on the template removal method adopted. However, since the \Planck\ polarisation maps are
excluded from the current data release, we adopt the {\it WMAP} polarisation
likelihood without further changes for now, and will return to this
topic in the next data release.

\subsection{Low-$\ell$ power spectrum -- consistency and robustness}
\label{sec:low_ell_spectrum}

\begin{figure*}
\begin{centering}
\includegraphics[width=1\textwidth]{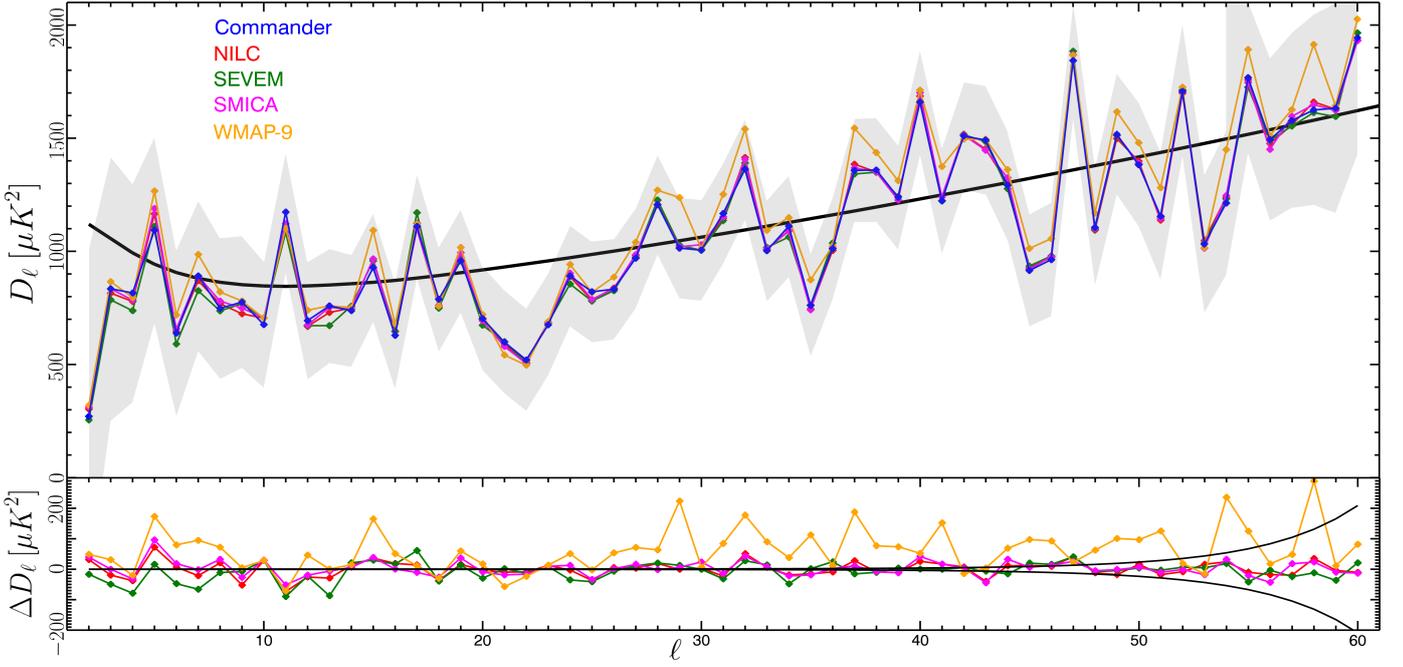}
\par\end{centering}

\caption{Top panel: temperature power spectra evaluated from downgraded \Planck\ maps, estimated with  \commander, \nilc, \sevem, or \smica, and the 9-year \WMAP\ ILC map, using the \texttt{Bolpol} quadratic estimator. The grey shaded area indicates the $1\,\sigma$ Fisher errors while the solid line shows the \Planck\ \LCDM\ best fit model. Bottom panel: Power spectrum differences for each algorithm/data set
  relative to the \commander\ spectrum, estimated from the spectra
  shown in the panel above. The black lines show the
  expected $1\,\sigma$ uncertainty due to (regularization)
  noise.}
\label{fig:TTspectrum} 
\end{figure*}

\begin{figure}
\begin{centering}
\includegraphics[width=88mm]{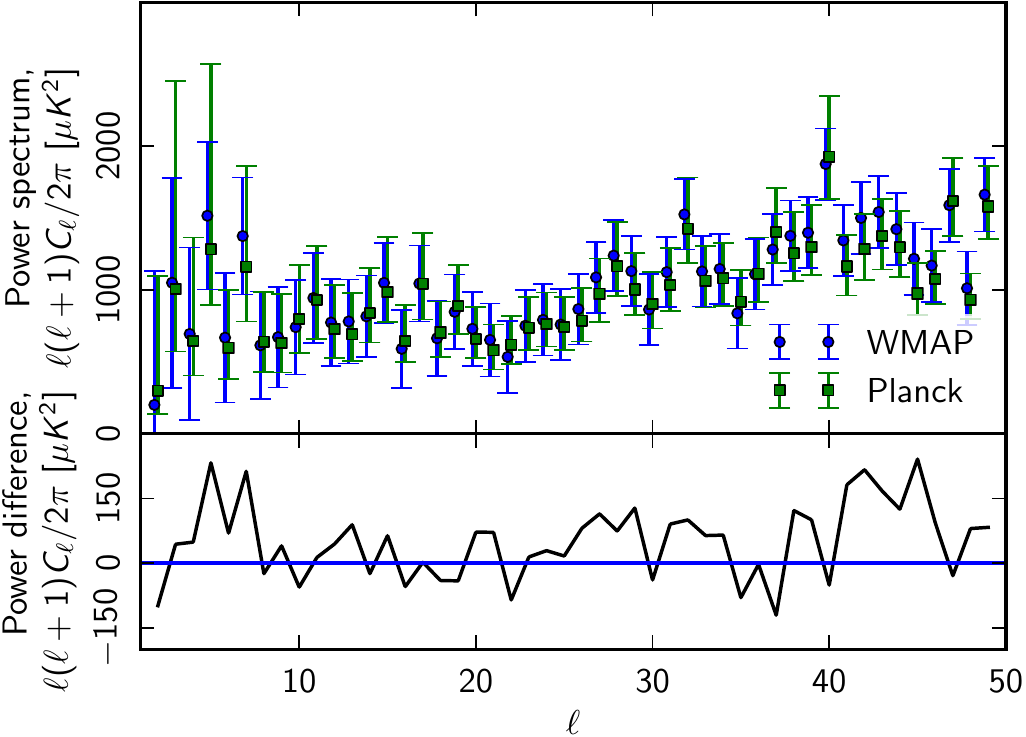}
\caption{\emph{Top}: Comparison between the low-$\ell$ \Planck\ temperature power spectrum estimated by
  \commander\ and the 9-year \WMAP\ spectrum
  \citep{BennettWMAP9}. Error bars indicate 68\% confidence regions. \emph{Bottom}: Difference between the \WMAP\ and \Planck\ low-$\ell$ spectra.}
\label{fig:low-ell-powerspectCommWMAP} 
\par\end{centering}
\end{figure}

\begin{figure}
\includegraphics[width=88mm]{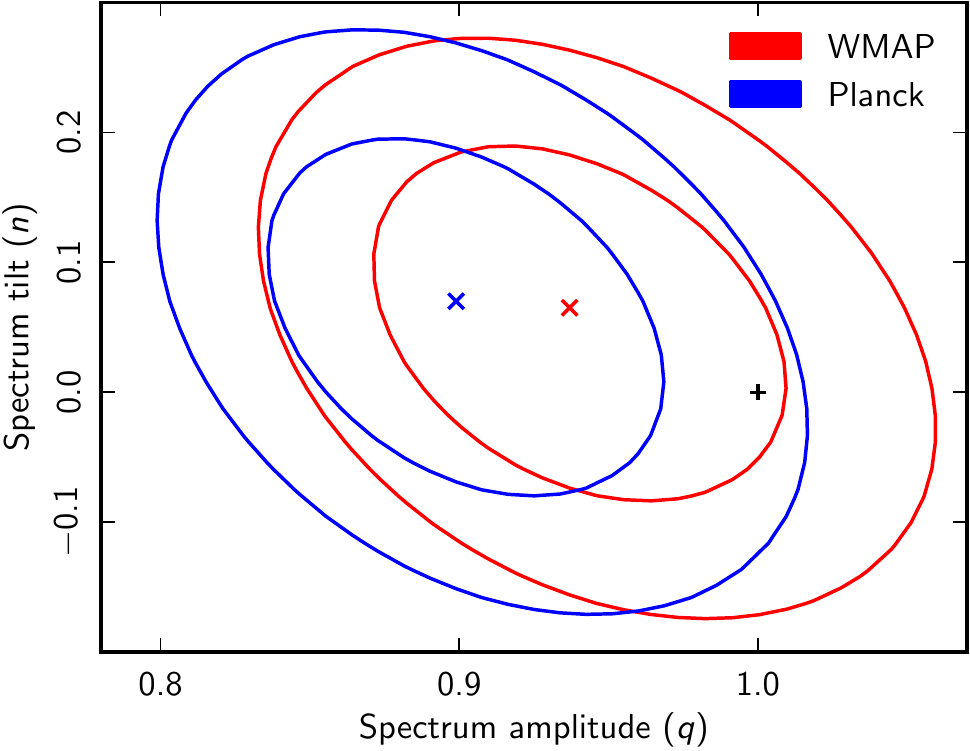} 
\caption{Probability distribution for a two-parameter (amplitude and
  tilt) model derived from the \Planck\ (blue) and \textit{WMAP} (red)
  low-$\ell$ likelihoods, including angular scales between $\ell=2$
  and 30. \label{fig:A_n_gibbs}}
\end{figure}

In this section, we present the low-$\ell$ \Planck\ CMB temperature
power spectrum derived using the \commander\ approach described above,
and assess its robustness through comparisons with three alternative
foreground-cleaned \Planck\ CMB maps (\nilc, \sevem, and \smica;
\citealp{planck2013-p06}), as well as with the 9-year {\it WMAP} ILC
temperature map \citep{BennettWMAP9}.

As a first consistency test, we compute the power spectrum from each
map using \texttt{Bolpol} \citep{Gruppuso2009, Molinari2013}, an
implementation of the quadratic maximum-likelihood power
spectrum estimator \citep{Tegmark1997}. Each map is smoothed to an
effective resolution of $329.81^{\prime}$ FWHM, to suppress aliasing
from high multipoles \citep{Keskitalo2010}, and repixelized on an
$N_{\textrm{side}}=32$ HEALPix grid. Gaussian white noise with a
variance of 4\,$\mu$K$^{2}$ is added to each map to regularize the
noise covariance matrix. 

Here we adopt the U78 common mask, defined in \citet{planck2013-p06}, to
exclude regions of high Galactic emission, leaving 78\% of the sky
for analysis. We remove the observed monopole and dipole in the mask. The resulting power spectra up to $\ell\le64$ are shown
in the top panel of Fig.~\ref{fig:TTspectrum}, while
the bottom panel shows the power spectrum residuals
of each map relative to the \commander\ map. Note that the same noise
realization was added to each map, and the regularization noise
therefore contributes little in this plot. For the different 
internally-derived \Planck\ maps, no
residual spectrum exceeds $\lesssim 100\:\mu\mathrm{K}^{2}$ and is
typically $\lesssim 50\:\mu\mathrm{K}^{2}$ at $\ell \gtrsim 10$. The \WMAP\ spectrum exhibits significantly larger residuals,
and are typically of the order of $\sim$100\,$\mu\mathrm{K}^{2}$ at
$\ell \gtrsim 30$.

Figure~\ref{fig:low-ell-powerspectCommWMAP} shows the \Planck\ and
\WMAP\ temperature power spectra derived directly from the respective
likelihood code, while Fig.~\ref{fig:A_n_gibbs} shows the
corresponding constraints on the two-parameter amplitude-tilt model
employed in Sect.~\ref{sec:commander_dataselection_prep}, including
multipoles between $\ell=2$ and 30. Neglecting the minor differences
in the masks adopted by the two codes, these power spectra and
parameter constraints are largely dominated by cosmic variance, and
one should therefore expect the two distributions to be almost
identical. Instead, from Fig.~\ref{fig:A_n_gibbs} we see that the {\it
  WMAP} low-$\ell$ spectrum is 2.5--3\% higher than the
\Planck\ spectrum. For a detailed discussion of this discrepancy,
including a comparison at higher $\ell$, see \citet{planck2013-p01a}.
Here we only note that the effect is robust with respect to foreground
removal and power spectrum evaluation algorithms, and also point out
that the effect at low multipoles is too large to be explained by
uncertainties in the \Planck\ transfer functions
\citep{planck2013-p02,planck2013-p03c} or calibration
\citep{planck2013-p02b,planck2013-p03f}. Also note that the amplitude
of the low-$\ell$ spectrum relative to the \Planck\ best-fit model,
$(q,n) = (1,0)$, derived including the full multipole range between
$2\le\ell\le2500$, is somewhat low in Fig.~\ref{fig:A_n_gibbs}, with a
best-fit amplitude of $q\sim0.9$.  This observation is discussed and
quantified in greater detail in Sect.~\ref{sub:low-low_l}.

%% file: 9_Planck-CMB-spectrum-likelihood.tex
\section{The \Planck\ CMB spectrum and likelihood\label{sec:Planck-Likelihood}}

\subsection{Hybridisation of low- and high-$\ell$ likelihoods\label{sub:LL-HL-matching}}

The high-$\ell$ and low-$\ell$ likelihoods introduced in
Sects.\ \ref{sec:High-multipoles-likelihood} and
\ref{sec:Low-multipoles-Likelihood} each describe only a part of the
full \planck\ data set. To estimate cosmological parameters from
all the angular scales probed by \planck, they must be combined into a single
likelihood function that describes all multipoles from
$\ell=2$ to 2500.

In principle, it is desirable to include as many multipoles as
possible in the low-$\ell$ likelihood, since it captures
the full non-Gaussian structure of the
likelihood. The Gaussian approximation for the likelihood 
using pseudo-spectra also improves at higher multipole due to 
the increasing number of degrees of freedom \citep{Efstathiou2004}. 
For \planck\ we adopt a transition multipole
of $\ell_{\textrm{trans}} = 50$, a compromise between obtaining robust
convergence properties for the low-$\ell$ likelihood, and ensuring
that the Gaussian approximation holds for the high-$\ell$ likelihood
\citep{HL09}.

To combine the likelihoods, we must account for the weak correlations 
between the low- and high-$\ell$ components. We consider three options:
\begin{enumerate}
\item \emph{Sharp transition}: The low-$\ell$ likelihood ends at
  $\ell_{\mathrm{max}}=49$; the high-$\ell$ likelihood starts at
  $\ell_{\mathrm{min}}=50$; no correlations are accounted for. 
\item \emph{Gap}: The low-$\ell$ likelihood ends at $\ell_{\mathrm{max}}=32$;
  the high-$\ell$ likelihood starts at $\ell_{\mathrm{min}}=50$; no
  correlations are accounted for, but the gap is sufficiently wide
  that any correlations are negligible. 
\item \emph{Overlap with correction}: The low-$\ell$ likelihood ends
  at $\ell_{\mathrm{max}}=70$; the high-$\ell$ likelihood starts at
  $\ell_{\mathrm{max}}=50$; the double-counting of the overlap region
  is accounted for by subtracting from the log-likelihood a
  contribution only including $50\le\ell\le70$ as evaluated by the
  \commander\ estimator. Under the assumption that no correlations
  extend from $\ell\le50$ to $\ell\ge70$, this approach is exact (for
  further details, see \citealp{Gjerlow2013}).
\end{enumerate}

We estimate cosmological parameters using all three methods, and find that
the posterior means typically vary by $<0.1\,\sigma$.  The largest
variation is seen including the running of the spectral index
of scalar perturbations, in which the posterior mean changes by $0.2\,
\sigma$. Further, all deviations at the 0.1--0.2$\sigma$ level are
seen for case 2 above, which excludes data compared to the other two;
case 1 and 3 give nearly indistinguishable results.  Since case 1 is
implementationally simpler, and can be estimated more efficiently
(see Sect.~\ref{sec:Low-multipoles-Likelihood}), we select
this method, adopting a sharp transition at $\ell_{\textrm{max}}=50$.

\subsection{The \planck\ power spectrum and \LCDM\ constraints}

\begin{figure*}
\centering{}
\includegraphics[bb=0bp 0bp 680bp 410bp,clip,width=1\textwidth]{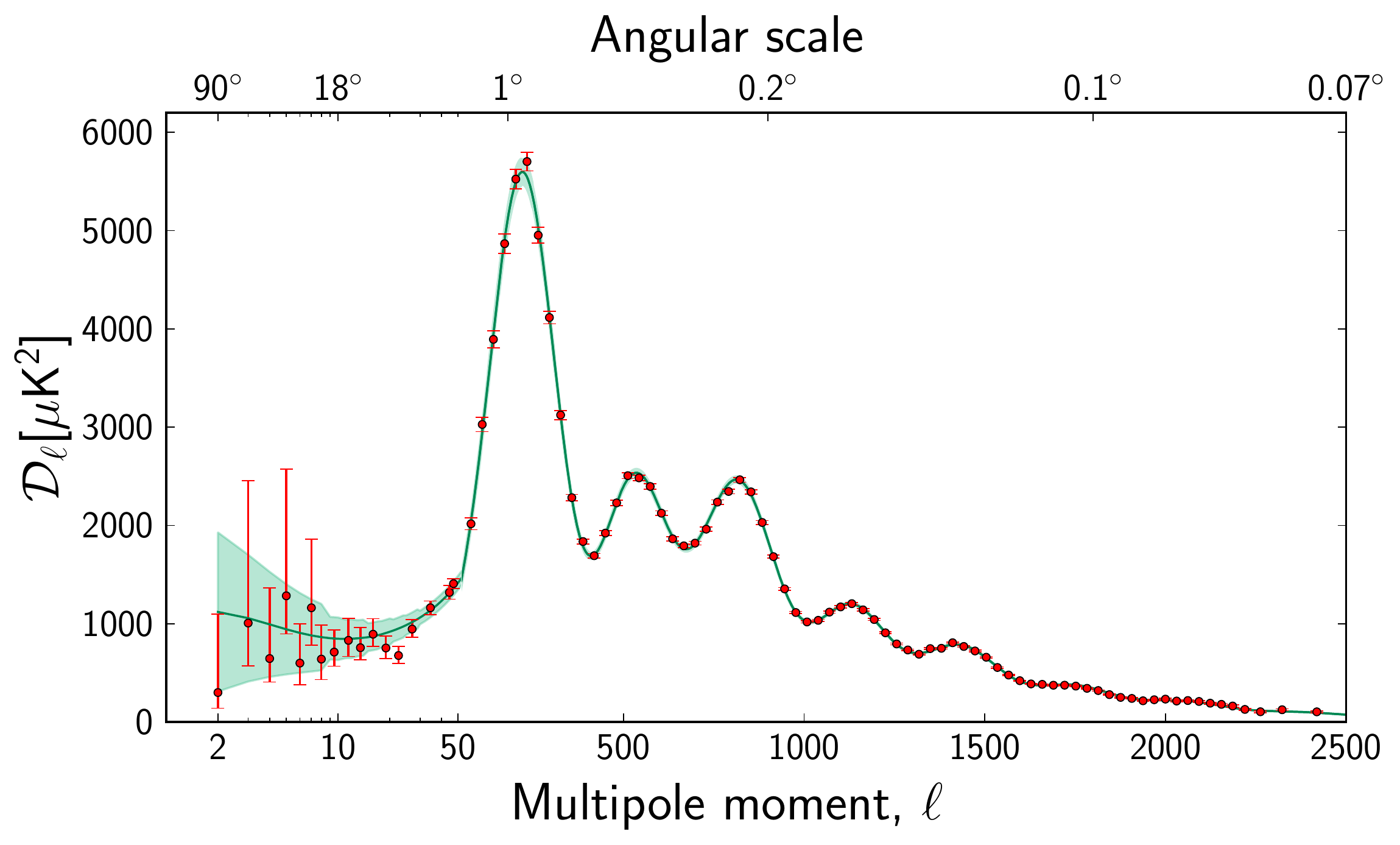}
\caption{The 2013 \Planck\ CMB temperature angular power spectrum. The error bars include cosmic variance, whose magnitude is indicated by the green shaded area around the best fit model. The low-$\ell$ values are plotted at 2, 3, 4, 5, 6, 7, 8, 9.5, 11.5, 13.5, 16,   19, 22.5, 27, 34.5,  and 44.5. } 
\label{fig:planckCMB}
\end{figure*}

Using the full \Planck\ likelihood, we now present the final 2013
\Planck\ CMB power spectrum. For this, we fix all nuisance parameters
to their maximum-likelihood values. The resulting spectrum is shown in
Fig.~\ref{fig:planckCMB} together with the corresponding best-fit
six-parameter \LCDM\ model.  The agreement between the observations
and the model is excellent over most of the multipole range. Only at
low $\ell$s is it possible to see a systematic offset in the form of a
slight power deficit; this will be addressed separately in the next
section.

\begin{table*}
\caption{Constraints on the basic six-parameter $\Lambda$CDM model
  using \Planck\ data. The top section contains constraints on the six
  primary parameters included directly in the estimation process, and
  the bottom section contains constraints on derived parameters.}
\label{tab:params-ref} 
\centering{}

\input{Table-LCDM_cosmo_params.tex} 
\end{table*}

Table~\ref{tab:params-ref} provides a summary of the $\Lambda$CDM
parameters derived using the methodology described in
\citet{planck2013-p11} from the \Planck\ likelihood. Here we use the same prior 
ranges on all parameters as in \cite{planck2013-p11}. These are as in Table~\ref{tab:cosmo-params}, except for $A^{\mathrm{CIB}}_{143}$, $A^{\mathrm{tSZ}}$, and $A^{\mathrm{kSZ}}$, which are modified to $[0,20]$, $[0,10]$, and $[0,10]$ respectively.
Results are given
for \Planck\ alone, and in combination with the low-$\ell$
\WMAP\ polarisation likelihood (\Planck+WP). For each case, we
report both posterior maximum and mean values. Uncertainties denote
68\% confidence limits.

A detailed discussion of these results is presented in
\cite{planck2013-p11}, including an analysis of extended cosmological
models, and their compatibility with other astrophysical data sets.
The bounds derived from \Planck\ alone are significantly tighter than
those from the 9-year \WMAP\ data alone, and comparable or better than
those inferred from \WMAP\ combined with SPT and ACT
observations. These new constraints provide a precision test of the
\LCDM\ model. In general, we find good agreement with results derived
from other astrophysical data sets, although there are a few
exceptions that are in moderate tension with
\Planck\ \citep{planck2013-p11}.

Considering each of the six $\Lambda$CDM parameters in turn, we first
note that \Planck\ constrains the physical baryon density to
$\Omega_{\rm b}h^2=0.02207\pm0.00033$, which is in remarkable agreement with
standard BBN predictions based on a determination of the primordial
abundance of deuterium, $\Omega_{\rm b}h^2=0.021\pm0.001$ \citep{Iocco:2008va}, but
with a fractional uncertainty of 1.5,\%, three times smaller than the
BBN uncertainty. The physical density of dark matter is measured with
a fractional uncertainty of 2.6\%, providing new constraints on
specific dark matter production scenarios. The single most precise
parameter, however, is the angular size of the sound horizon at the
last-scattering surface, $\theta_*$, which is measured with a
fractional uncertainty of 0.065\,\% by \Planck, improving on the
combined \WMAP, ACT, SPT, and SNLS33 constraint by a factor of two.

Next, given that no polarisation data are included in the current data
release, it is remarkable that \Planck\ alone constrains the optical
depth to reionization, $\tau$, with a fractional error of 40\%. This
is made possible by \Planck's high angular resolution and sensitivity,
which allows a high signal-to-noise measurement of lensing in the
small-scale CMB power spectrum. This in turn breaks the well-known
$e^{-2\tau}A_{\textrm{s}}$ degeneracy between $\tau$ and the amplitude
of scalar perturbations, $A_\textrm{s}$. The fractional uncertainty on
$A_{\textrm{s}}$ from Planck alone is 7\,\%.

Having sufficient power to measure $\tau$ from small angular scale
temperature data, \Planck\ naturally also provides very strong
constraints on the spectral index of scalar perturbations,
$n_{\textrm{s}}$, leading to a fractional uncertainty of 0.97\,\%. The
scale-invariant Harrison-Zeldovich spectrum, $n_{\textrm{s}}=1$, is
ruled out at a significance of $4.1\sigma$ from the
\Planck\ temperature spectrum alone. The analyses presented in
\cite{planck2013-p11} and \cite{planck2013-p17} show that the
preference for a (red) tilted primordial spectrum remains very strong
also within most extensions beyond the minimal \LCDM\ model. The
implications of this results for inflationary models are discussed in
\cite{planck2013-p17}.

With our choice of cosmological parameters, the Hubble parameter,
$H_0$, and the fractional density of the cosmological constant,
$\Omega_\Lambda=1-\Omm$, are derived parameters. They are probed by
CMB observations mainly through their impact on
$\theta_{\mathrm{MC}}$, and, to lesser extent, by the impact of
$\Omega_\Lambda$ on the late-time integrated Sachs-Wolfe effect.
Since $\theta_{\mathrm{MC}}$ is accurately measured, a particular
combination of $H_0$ and $\Omega_\Lambda$ is very well constrained by
\Planck, although in a model-dependent way; $\theta_*$ depend on other
cosmological parameters, such as the spatial curvature radius,
neutrino masses, the number of relativistic degrees of freedom, or a
possible dark energy equation of state parameter.

The results reported in Table~\ref{tab:params-ref} rely on the
assumption of a flat \LCDM\ cosmology with three neutrino species, two
of which are assumed to massless and one featuring a small mass
$m_\nu=0.06$~eV, reflecting the lower bound on neutrino masses imposed
by neutrino oscillation experiments. Under these assumptions,
\Planck\ finds preferred ranges for $H_0$ and $\Omega_\Lambda$ that
are lower than previous CMB experiments. For instance, \Planck+WP
gives $H_0=67.3\pm1.2 ~\rm{km}\, \rm{s}^{-1}\Mpc^{-1}$, to be compared
with $70.5\pm1.6 ~\rm{km}\, \rm{s}^{-1}\Mpc^{-1}$ for the combined
WMAP9+eCMB data set presented by \cite{hinshaw2012}. The underlying
cosmology in the two analyses is the same, excepted for the small
neutrino mass introduced in our default \LCDM\ model. However, if we
assume all three neutrino species to be massless, our best-fit and
mean values for $H_0$ increase only by $0.6~\rm{km}\,
\rm{s}^{-1}\Mpc^{-1}$. Thus, the tension is clearly driven by the data
rather than by theoretical assumptions. \cite{planck2013-p11} shows
that our results for $H_0$ and $\Omega_\Lambda$ are in very good
agreement with Baryon Acoustic Oscillation data, but in moderate
tension with other cosmological probes.  For instance, our \Planck+WP
bounds on $H_0$ disagree at the $2.5\sigma$ level with direct
determinations of the Hubble parameter using cepheids and
supernovae~\citep{Riess:98} or quasar time delays~\citep{Suyu:12}, as
well as with the results of the Carnegie Hubble
Program~\citep{Freedman:12}.  Our bounds on $\Omega_\Lambda$
are in a slight $2\sigma$ tension with the results of the SNLS
supernovae collaboration \citep{Conley:11,Sullivan:11}, although in
better agreement with the Union2.1 compilation~\citep{Suzuki:12}.  Our
combined determination of $\sigma_8$ and $\Omm$ shows larger tension
with recent data based on cosmic shear or cluster count techniques. On
the other hand the \Planck\ best-fit \LCDM\ model is in good agreement
with the halo power spectrum derived from the luminous red galaxy
catalogue of the Sloan Digital Sky Survey \citep{Reid:2009xm},
especially when the analysis is restricted to linear scales.

\subsection{Significance of the low-$\ell$ tension with $\Lambda$CDM models\label{sub:low-low_l}}

From the above discussion, it is clear that the $\Lambda$CDM framework
provides an excellent model for most of the \Planck\ data. However, as
noted in Sect.~\ref{sec:low_ell_spectrum}, and seen in
Fig.~\ref{fig:planckCMB}, the low-$\ell$ \Planck\ temperature power
spectrum appears to be in some tension with the best-fit
\Planck\ $\Lambda$CDM model, which for \Planck\ is almost exclusively
determined by the small-scale spectrum. In this section we assess the
significance and impact of this tension between low and high $\ell$s
using three different statistical tests.

\begin{figure}
\begin{centering}
\includegraphics[width=1\columnwidth]{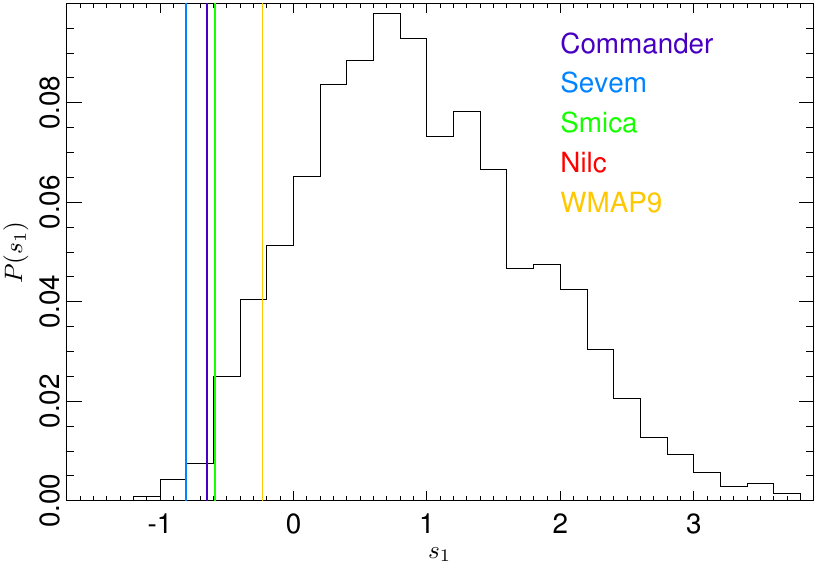}
\par\end{centering}
\caption{Results of the Hausman test applied to the temperature power
  spectrum for $2\le\ell\le 32$. The black histogram shows the
  expected distribution, estimated with simulations, of the
  $s_{1}$ test statistic. The vertical bars
  represent \planck\ CMB maps and the 9-year \WMAP\ ILC map. Note that
  the statistic is indistinguishable for the \nilc\ and \commander\ maps. 
\label{fig:haus_lowlvshighl}}
\end{figure}

We start by applying a modified Hausman test
\citep{Polenta_CrossSpectra, planck2013-p02} to the low-$\ell$ spectra
derived from the four foreground-cleaned \Planck\ maps
\citep{planck2013-p06} and the 9-year \WMAP\ ILC map, using
multipoles up to $\ell_{\textrm{max}}=32$. This test uses the
statistic $s_{1} = \sup_{r}B(\ell_\mathrm{max},r)$, where
\begin{align}
B(\ell_\mathrm{max},r)&=\frac{1}{\sqrt{\ell_\mathrm{max}}}\sum_{\ell=2}^{\textrm{int}(\ell_\mathrm{max}r)}H_{\ell},r\in\left[0,1\right] \label{eq:hausman1}\\
H_{\ell}&=\frac{\hat{C_{\ell}}-C_{\ell}}{\sqrt{\textrm{Var}\,\hat{C_{\ell}}}},
\end{align}
and $\hat{C}_{\ell}$ and $C_{\ell}$ denote the observed and model
power spectra, respectively. Intuitively, this statistic measures the
relative bias between the observed spectrum and model, measured in
units of standard deviations, while taking into account the so-called
``look-elsewhere effect'' by maximizing $s_1$ over multipole
ranges. We use realistic \Planck\ `FFP6' simulations
\citep{planck2013-p01} to derive the empirical distribution of $s_1$
under the null hypothesis. Figure~\ref{fig:haus_lowlvshighl} compares
the results obtained from the data with the simulation distribution,
and Table~\ref{tab:haus_lowlvshighl} lists significances. As measured
by this statistic, we see that a negative bias is found in the
low-$\ell$ \Planck\ power spectrum relative to the $\Lambda$CDM model
at the 99\% confidence level. 

For the \WMAP\ ILC map the significance of the negative bias nominally
decreases to 93\%. This is consistent with the findings in
Sect.~\ref{sec:low_ell_spectrum}, where it was shown that the
\WMAP\ temperature power spectrum is 2.5--3\,\% higher than the
\Planck\ spectrum at low $\ell$'s.  However, as discussed in
\citet{planck2013-p01a}, a similar amplitude difference between the two
experiments is also seen at smaller scales. Since the current test
compares the observed \WMAP\ data with the best-fit
\Planck\ $\Lambda$CDM model, the present test is not optimal for
assessing internal consistency between low and high $\ell$s within the
\WMAP\ data.

\begin{table}
\begingroup 
\newdimen\tblskip \tblskip=5pt
\caption{Results of the Hausman test applied to the temperature power spectrum for $2\le\ell\le 32$.}
\label{tab:haus_lowlvshighl}
\nointerlineskip
\footnotesize 
\setbox\tablebox=\vbox{ %
\newdimen\digitwidth 
\setbox0=\hbox{\rm 0}
\digitwidth=\wd0
\catcode`*=\active
\def*{\kern\digitwidth}
\newdimen\signwidth
\setbox0=\hbox{+}
\signwidth=\wd0
\catcode`!=\active
\def!{\kern\signwidth}
\halign{
\hbox to 1in{#\leaderfil}\tabskip=0em&
\hfil#\hfil\tabskip=0.2cm&
\hfil#\hfil\tabskip=0pt\cr
\noalign{\doubleline}
\omit Data set\hfill  &  $s_1^{\tt obs}$ & $P(s_1 <
s_1^{\tt obs})$ \cr
\omit \hfill & & {[}\%{]} \cr
\noalign{\vskip 3pt\hrule\vskip 3pt}
\commander\ & -0.647& 0.73\cr
\nilc\ & -0.649& 0.73\cr
\sevem\ & -0.804& 0.50\cr
\smica\ & -0.589& 1.33\cr
\WMAP 9 ILC& -0.234& 7.18\cr
\noalign{\vskip 3pt\hrule\vskip 3pt}}}
\endPlancktable 
\endgroup
\end{table}

\begin{figure}
\centering{}\includegraphics[width=88mm]{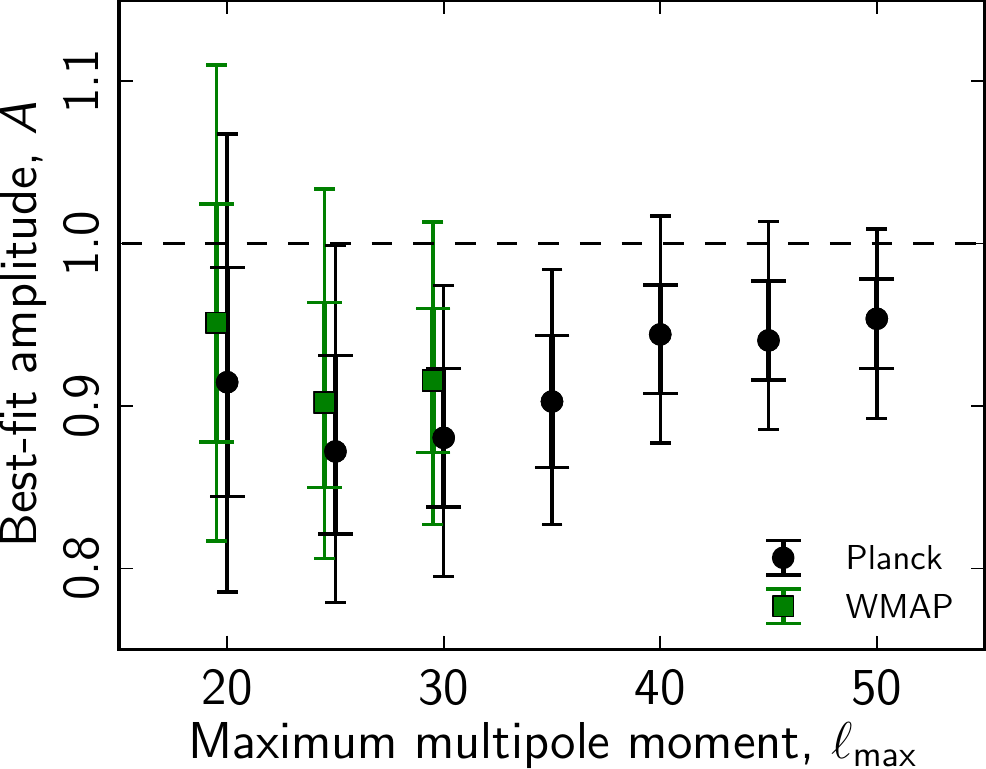}
\caption{
Power spectrum amplitude, $q$, relative to the best-fit \Planck\ model as a function of $\ell_\mathrm{max}$, as measured by the low-$\ell$ \Planck\ and \WMAP\ temperature likelihoods, respectively. Error bars indicate 68 and 95\% confidence regions. 
\label{fig:commander-lowl-tension}
}
\end{figure}

Next, to obtain a quantitative measure of the relative power discrepancy
between low and high $\ell$s, we fit the two-parameter amplitude--tilt
power spectrum model (see
Sect.~\ref{sec:commander_dataselection_prep}) to the \Planck\ data
using the low-$\ell$ likelihood restricted to various multipole ranges
defined by $2\le\ell\le\ell_{\textrm{max}}$, where
$\ell_{\textrm{max}}$ is allowed to vary. Thus, this measures the
amplitude of the low-$\ell$ spectrum relative to the best-fit
\Planck\ $\Lambda$CDM spectrum, which is driven by the smaller
angular scales.  Figure~\ref{fig:commander-lowl-tension} shows the
resulting constraints on the power spectrum amplitude, $q$, as a
function of $\ell_{\textrm{max}}$, after marginalizing over the tilt,
$n$. For comparison, we also show similar constraints derived using
the low-$\ell$ \WMAP\ temperature likelihood up to $\ell=30$. The
best-fit amplitude is $q\sim0.9$ for $\ell_{\textrm{max}}=20$--35,
different from unity at a statistical significance of 2--$2.5\sigma$
by this measure.  The \WMAP\ spectrum shows a consistent
behaviour, up to the same overall scaling factor of 2.5--3\% between
\Planck\ and \WMAP\ discussed above. We have verified that these
results are insensitive to the (well-known) low quadrupole moment by
excluding $C_2$ from the analysis; the large cosmic variance of this
particular mode results in a low overall statistical weight in
the fit.

\begin{figure}[h]
\includegraphics[width=88mm]{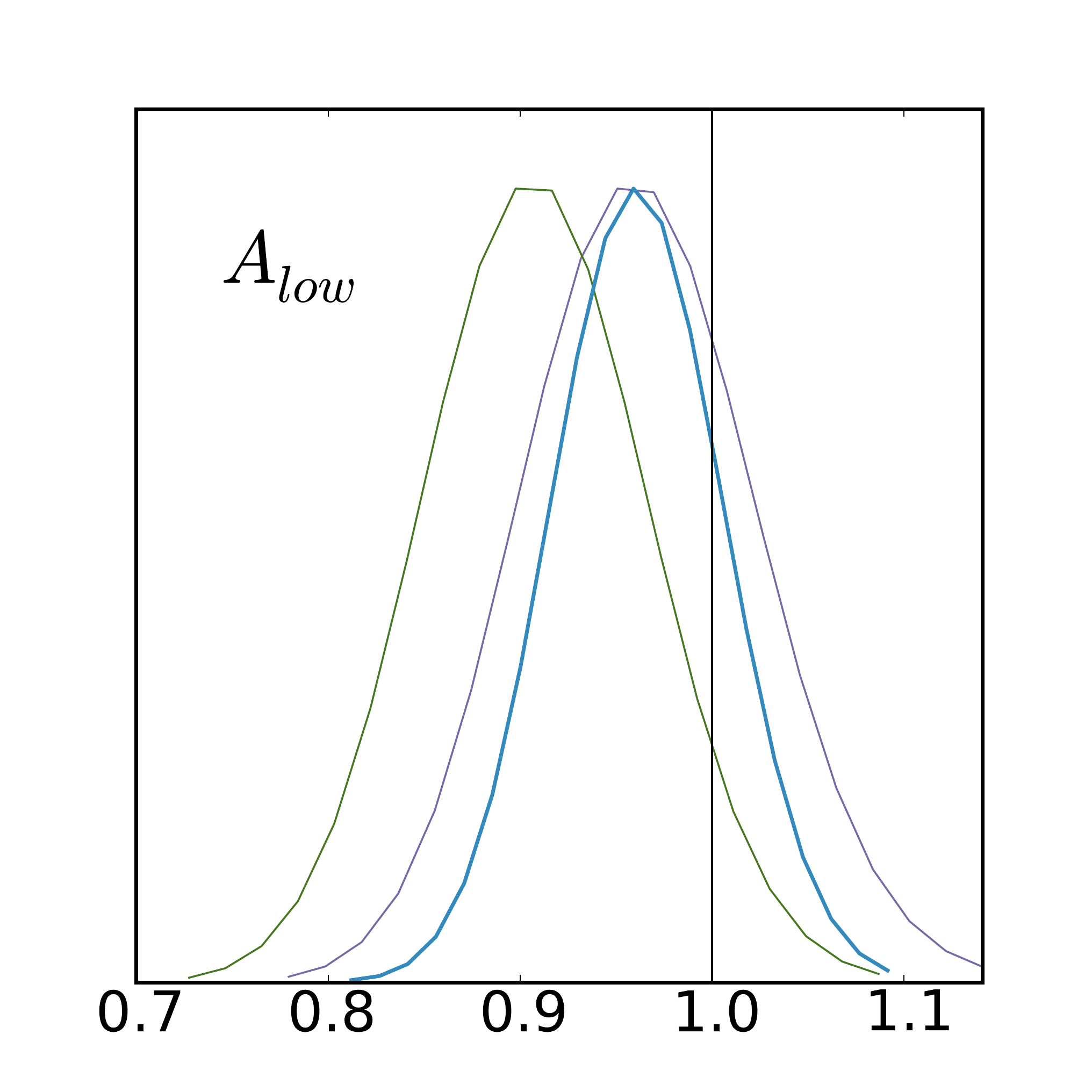}
\caption{Posterior distributions for the low-$\ell$ spectrum
  amplitude, $A_\mathrm{low}$, estimated using the \Planck\ likelihood,
  with $\ell_{\textrm{low}}=32$ (green) and $\ell_\textrm{low}=49$ (blue). 
  The purple line show the distribution derived using
  the \WMAP\ temperature likelihood with $\ell_{\textrm{low}}=32$. 
 \label{fig:commander-wmap-lowl-tension}}
\end{figure}

Finally, we assess the impact of the low-$\ell$ power deficit on the
$\Lambda$CDM model estimated using the \Planck\ likelihood\footnote{We have verified that the following results are
  insensitive to whether \plik\ or \camspec\ are used for the
  high-$\ell$ likelihood.}
(augmented with the \WMAP\ polarisation likelihood). We fit a
low-$\ell$ rescaling amplitude, $A_\mathrm{low}$ for
$\ell<\ell_\mathrm{low}$ jointly with the $\Lambda$CDM parameters,
i.e., $C_{\ell} = A_{\textrm{low}} C_{\ell}^{\Lambda\textrm{CDM}}$ for
$\ell<\ell_\mathrm{low}$ and $C_{\ell} =
C_{\ell}^{\Lambda\textrm{CDM}}$ for
$\ell\ge\ell_{\textrm{low}}$. Figure~\ref{fig:commander-wmap-lowl-tension}
shows the resulting posterior distributions for $A_\mathrm{low}$ for
$\ell_\mathrm{low}=32$ (green) and $\ell_\mathrm{low}=49$ (blue). The
purple line shows the same when replacing the \Planck\ low-$\ell$
likelihood with the \WMAP\ low-$\ell$ likelihood
($\ell_\mathrm{low}=32$). The corresponding best-fit values are
$A_\mathrm{low}=0.899\pm 0.046$ (\Planck; $\ell_\mathrm{low}=32$),
$A_\mathrm{low}=0.953\pm 0.033$ (\Planck; $\ell_\mathrm{low}=49$) and
$A_\mathrm{low}=0.953\pm0.048$ (\WMAP; $\ell_\mathrm{low}=32$),
respectively. As already noted in Sect.~\ref{sec:low_ell_spectrum},
these values are too large to be explained by the $<1\%$ uncertainties
in the \Planck\ transfer functions \citep{planck2013-p02,
  planck2013-p03}.

\begin{figure}
\centering{}\includegraphics[width=1\columnwidth]{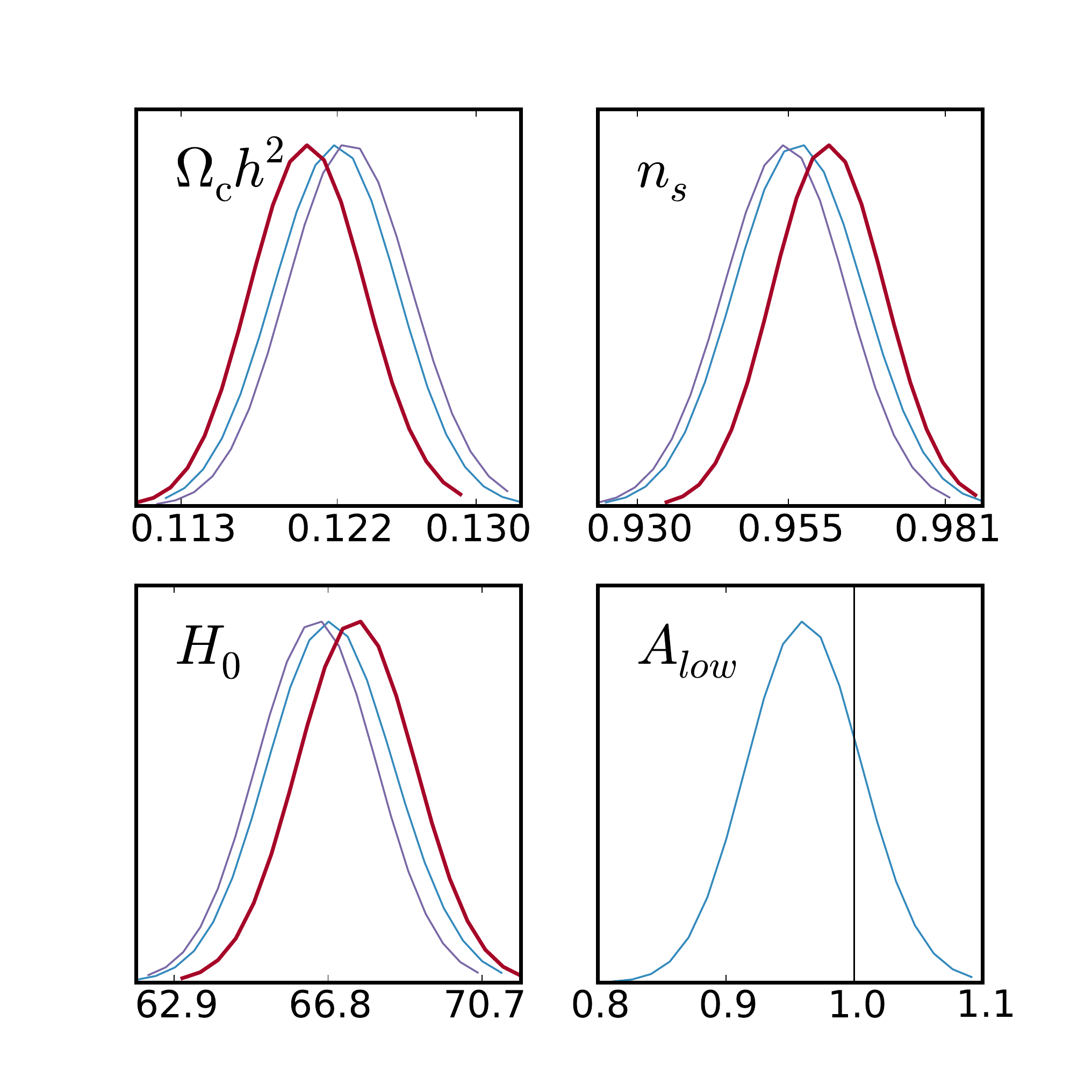}
\caption{Comparison of the posterior distributions for
  $\Omega_\textrm{c}h^2$, $n_{\textrm{s}}$, and $H_0$ for the default
  six-parameter $\Lambda$CDM model  constrained by 
  \Planck\ (\emph{red}); compared to the case when we allow a variable 
  low-$\ell$
  power spectrum amplitude at $\ell\le49$ (\emph{blue}); and when
  replacing the low-$\ell$ temperature likelihood with a Gaussian
  prior on $\tau$, the optical depth of reionization
  (\emph{purple}). The lower right panel shows the posterior
  distribution for the low-$\ell$ amplitude,
  $A_{\textrm{low}}$. \label{fig:lowl-tension}}
\end{figure}

In Fig.~\ref{fig:lowl-tension} we show the posterior distributions for
$\Omega_\textrm{c}h^2$, $n_{\textrm{s}}$ and $H_0$ after marginalizing
over $A_\mathrm{low}$ for $\ell_\textrm{low}=49$. (Adopting
$\ell_\mathrm{low}=32$ results in negligible differences for all
parameters except $A_\mathrm{low}$).  Shifts of 0.6--$1\sigma$ are
observed compared to the reference model, $A_\mathrm{low} = 1$. We note 
that $H_0$, which already has a `low' value (for a detailed discussion, see
\citealp{planck2013-p11}), prefers an even lower value when allowing a
rescaling of the low-$\ell$ spectrum. As a final test, we replace the
entire low-$\ell$ likelihood, both temperature and polarisation, with
a Gaussian prior on the optical depth of reionization, $\tau=0.089\pm
0.014$, matching the \WMAP\ measurement \citep{hinshaw2012}. The
resulting posteriors are shown as purple lines in
Fig.~\ref{fig:lowl-tension}, and agree well with the case including a
low-$\ell$ scaling factor, but are, in fact, slightly further away
from the reference model. Although not very significant in an absolute
sense, these results do indicate that the high-$\ell$ likelihood is
challenged in finding models that also fit the low-$\ell$ power
spectrum.

To summarize, we have phenomenologically quantified a tension between
the low-$\ell$ CMB power spectrum at $\ell \lesssim 40$ and the
\Planck\ best-fit \LCDM\ model. 
Its significance varies between 2.5 and $3\,\sigma$ depending on the
estimator used. The effect is seen in all four \Plancks
foreground-cleaned CMB maps with little variation. It is also present
in the 9-year \WMAP\ data, although an overall amplitude difference of
2.5--3\,\% between the data sets complicate a direct comparison. 
To make further progress, one would seek to establish a
physical model that predicts a low-$\ell$ power deficit compared to
high $\ell$'s, and that may also predict other observable
effects which may be tested with cosmological data. Such a model may 
be related to the tentative detections of violations of statistical
isotropy discussed in \citet{planck2013-p09}, e.g., the low CMB
temperature variance, the hemispherical power asymmetry, or the
alignment between the quadrupole and octopole moments.

%% file: Table-LCDM_cosmo_params.tex
\begingroup
\openup 5pt
\newdimen\tblskip \tblskip=5pt
\nointerlineskip
\vskip -3mm
\footnotesize
\setbox\tablebox=\vbox{
    \newdimen\digitwidth
    \setbox0=\hbox{\rm 0}
    \digitwidth=\wd0
    \catcode`"=\active
    \def"{\kern\digitwidth}
    \newdimen\signwidth
    \setbox0=\hbox{+}
    \signwidth=\wd0
    \catcode`!=\active
    \def!{\kern\signwidth}
\halign{
\hbox to 0.9in{$#$\leaderfil}\tabskip=1.5em&$#$\hfil&$#$\hfil&$#$\hfil&\hfil$#$\hfil\tabskip=0pt\cr
\noalign{\doubleline}
 \multispan1\hfil \hfil & \multispan2\hfil \planckonly\hfil & \multispan2\hfil \Planck+\WP\hfil\cr
\noalign{\vskip -3pt}
\omit&\multispan2\hrulefill&\multispan2\hrulefill\cr
 \omit\hfil Parameter\hfil & \omit\hfil Best fit\hfil & \omit\hfil 68\% limits\hfil & \omit\hfil Best fit\hfil & \omit\hfil 68\% limits\hfil\cr
\noalign{\vskip 3pt\hrule\vskip 5pt}
\Omega_{\mathrm{b}} h^2& 0.022068& 0.02207\pm 0.00033 & 0.022032& 0.02205\pm 0.00028\cr
\Omega_{\mathrm{c}} h^2& 0.12029& 0.1196\pm 0.0031 & 0.12038& 0.1199\pm 0.0027\cr
100\theta_{\mathrm{MC}}& 1.04122& 1.04132\pm 0.00068 & 1.04119& 1.04131\pm 0.00063\cr
\tau& 0.0925& 0.097\pm 0.038 & 0.0925& 0.089^{+0.012}_{-0.014}\cr
n_\mathrm{s}& 0.9624& 0.9616\pm 0.0094  & 0.9619& 0.9603\pm 0.0073\cr
\ln(10^{10} A_\mathrm{s})& 3.098& 3.103\pm 0.072 & 3.0980& 3.089^{+0.024}_{-0.027}\cr
\noalign{\vskip 5pt\hrule\vskip 3pt}
\Omega_\Lambda& 0.6825& 0.686\pm 0.020  & 0.6817& 0.685^{+0.018}_{-0.016}\cr
\Omega_{\mathrm{m}}& 0.3175& 0.314\pm 0.020  & 0.3183& 0.315^{+0.016}_{-0.018}\cr
\sigma_8& 0.8344& 0.834\pm 0.027  & 0.8347& 0.829\pm 0.012\cr
z_{\mathrm{re}}& 11.35& 11.4^{+4.0}_{-2.8}  & 11.37& 11.1\pm 1.1\cr
H_0& 67.11& 67.4\pm 1.4 & 67.04& 67.3\pm 1.2\cr
10^9 A_{\mathrm{s}}& 2.215& 2.23\pm 0.16 & 2.215& 2.196^{+0.051}_{-0.060}\cr
\Omega_{\mathrm{m}} h^2& 0.14300& 0.1423\pm 0.0029 & 0.14305& 0.1426\pm 0.0025\cr
\mathrm{Age}/\mathrm{Gyr}& 13.819& 13.813\pm 0.058  & 13.8242& 13.817\pm 0.048\cr
z_\ast& 1090.43& 1090.37\pm 0.65  & 1090.48& 1090.43\pm 0.54\cr
100\theta_\ast& 1.04139& 1.04148\pm 0.00066  & 1.04136& 1.04147\pm 0.00062\cr
z_{\mathrm{eq}}& 3402& 3386\pm 69 & 3403& 3391\pm 60\cr
\noalign{\vskip 5pt\hrule\vskip 3pt}
} 
} 
\endPlancktable
\endgroup

%% file: A_10_Discussion-and-Conclusions.tex
\section{Discussion \& conclusions\label{sub:conclusions}}

We have presented the \Planck\ likelihood, which provides a detailed
and accurate characterisation of the two-point statistics of the CMB
temperature field, accounting for all significant sources of
uncertainty; statistical, instrumental, and astrophysical. This
likelihood function allows us to present an estimate of the CMB
temperature power spectrum that spans more than three decades in
$\ell$ with unprecedented precision; a spectrum that saturates the
cosmic variance limit at all scales $\gtrsim 0.1^\circ$, nearly
exhausting the information content of the temperature anisotropies,
and, in fact, is becoming limited by uncertainties due to astrophysical
foreground modelling. This is precisely what was originally promised at
the time when \Planck\ was selected by ESA in March 1996.

On large angular scales, $\ell < 50$, the \Planck\ likelihood is based
on a Gibbs sampling approach that allows joint CMB power spectrum and
component separation analysis, while accurately marginalizing over a
physically motivated foreground model constrained by the 30--353~GHz
\Planck\ frequencies. On intermediate and small scales, the
\Planck\ likelihood employs a fine-grained set of cross-spectrum
combinations among the 100, 143, and 217 GHz detector maps to constrain
the high-$\ell$ CMB power spectrum, ensuring that no noise bias can
compromise the results, while at the same time allowing for physical
foreground modelling in terms of power spectrum templates. This
emphasis on physical foreground modelling has made it possible to
combine the full power of the \Planck\ data with observations from
higher-$\ell$ CMB experiments.

We have validated our results through an extensive suite of
consistency and robustness analyses, propagating both instrumental and
astrophysical uncertainties to final parameter estimates.  Further, we
have studied in detail the well-known degeneracies that exist between
the foreground and cosmological parameters at high $\ell$s when only
including \Planck\ observations, and shown that they have only a weak
impact on cosmological conclusions.

On a more detailed level, we draw the following conclusions:
\begin{itemize}
\item The consistency between power spectra measured at different
  frequencies is remarkable. In the signal-dominated regime for single
  detectors, at $\ell \lesssim 1000$, the cross-spectra show an RMS
  dispersion of a few $\mu\textrm{K}^2$ in multipole bands of
  $\delta\ell=31$. This confirms the relative calibration of the 100,
  143, and 217\,GHz detectors to $\sim$0.2\%.

\item The differences, $\Delta D_\ell$, between the $143\times143$,
  $143\times217$, and $217\times217$ cross-spectra averaged over
  multipole bands of $\delta\ell\approx31$ have a dispersion over
  $800\le\ell\le1500$ of 9, 5, and 5\,$\mu\textrm{K}^2$, respectively,
  after subtracting the best-fit foreground model. This dispersion is
  not primarily of instrumental origin, but can be predicted from a
  model of the chance correlations between foregrounds and CMB
  fluctuations.

\item At high $\ell$s, the power spectrum of the four
  foreground-cleaned CMB maps derived through component separation are
  consistent within their uncertainties. The cosmological parameters
  derived from these maps are consistent with those estimated by the
  \Planck\ likelihood for $\ell \lesssim 2000$, despite very different
  foreground models.

\item At low $\ell$s, the power spectrum differences among the four
  foreground-cleaned CMB maps are below $50\,\mu\textrm{K}^2$ for
  nearly every single multipole. Residuals with respect to the 9-year
  \WMAP\ ILC map are slightly larger, typically $100\,\textrm{K}^2$ or
  more.  A detailed comparison between \Planck\ and \WMAP\ reveals a
  systematic power spectrum amplitude difference at the $2\%$ -- $3\%$
  level that cannot be accounted for within the \Planck\ instrumental
  error budget. This is consistent with the findings presented in
  \cite{planck2013-p01a}.

\item Parameters derived from the 70~GHz \Planck\ frequency map are in
  excellent agreement with the reference results derived using the
  \Planck\ likelihood; when the latter is limited to $\ell\le1000$,
  the agreement is even more striking. This confirms the strong
  internal consistency between the LFI and HFI instruments.

\item The best-fit $\Lambda$CDM model derived from the
  \Planck\ likelihood predicts $TE$ and $EE$ spectra in exquisite
  agreement with the measured polarization signature over a broad
  range of frequencies (70 to 217~GHz) and multipoles ($\ell \lesssim
  1000$). At 100, 143, and 217\,GHz, the instrumental noise in the $EE$
  spectrum is at the $\mu\textrm{K}^2$ level for $\ell \lesssim 1000$,
  and the visible differences between the spectra are dominated by
  their different levels of foreground contribution, not by systematic
  effects.

\item We report a tension between the \Planck\ best-fit $\Lambda$CDM
  model and the low-$\ell$ spectrum in the form of a power deficit of
  5--10\% at $\ell\lesssim40$, with a statistical significance of
  2.5--3$\,\sigma$. Thus, while the minimal \LCDM\ model provides an
  outstanding fit for intermediate and small angular scales, this
  tension may suggest that the model is incomplete. In this respect,
  it is worth noting that other, but possibly related, anomalies have
  been reported in a companion paper studying statistical isotropy in
  the \Planck\ sky maps at statistically significant levels.

\end{itemize}

In summary, we find that the majority of the \planck\ data can be
described by a minimal six-parameter \LCDM\ model with a very high
degree of accuracy. Within this model the statistical uncertainties
are dominated by astrophysical foreground modelling by scales of $\ell
\simeq 1500$. At lower $\ell$s, the unprecedented quality of the
\planck\ data is such that the only fundamental limit is 
that we can only observe one CMB sky. In other words, \Planck\ is
cosmic variance dominated at $\ell \lesssim 1500$, extragalactic foreground
dominated at $\ell\gtrsim1500$, and dominated nowhere by instrumental
noise or systematic errors.

Using only \Planck\ data, we report a detection of
$\ns < 1$ at more than $4\,\sigma$ confidence, significantly stronger
than the limit derived from \WMAP, SPT, ACT, and SNLS3
combined. Complementing the \Planck\ observations with the 9-year
\WMAP\ polarization data increases the significance to $5.4\sigma$.
The multipole range above $\ell > 1500$ is crucial for
constraining possible extensions to the minimal $\Lambda$CDM model; for a
detailed exploration of a wide range of such models,
see \cite{planck2013-p11} and \cite{planck2013-p17}. There 
we report some tensions among the CMB damping tail parameters,
including $\Omega_K$, $n_\mathrm{run}$, and $Y_{\mathrm P}$. However, none of
these indicate significant departures from the \LCDM\ framework.

In the near future, we will extend our analysis to produce a cosmic
variance limited likelihood and power spectrum reaching to higher
multipoles. To some extent, this will be achieved through more
sophisticated astrophysical foreground modelling, and by exploiting
additional frequency information. However, the two major steps forward
will be, first, to include the \Planck\ polarization observations in
the likelihood analysis, and, second, to exploit the full data set
generated by the two \Planck\ instruments. The amount of HFI
data available for analysis is nearly double that which is presented here,
and the LFI instrument is still observing at the time of writing.

%% file: App-sec2.tex
\subsection{Power spectra and the coupling matrix \label{app:Power-Spectra}}

We denote the pixel weight function for temperature by $w_{i}^{T}$.
The pseudo-spectra of Eq.~\ref{C1} are constructed using the
following: 
\begin{equation}
\tilde{a}_{\ell m}^{T}=\sum_{s}\Delta T_{s}w_{s}^{T}\Omega_{s}Y_{\ell m}^{*}(\mathbf{\theta}_{s}),\label{A0a}
\end{equation}
where the sum is over the pixels in the map.

The coupling matrix appearing in Eq.~\ref{C2} is given by \cite{Hietal02}:
\begin{eqnarray}
M_{\ell_{1}\ell_{2}}^{TT} & = & \frac{(2\ell_{2}+1)}{4\pi}\sum_{\ell_{3}}(2\ell_{3}+1)\tilde{W}_{\ell_{3}}\left(\begin{array}{ccc}
\ell_{1} & \ell_{2} & \ell_{3}\nonumber\\
0 & 0 & 0
\end{array}\right)^{2},\label{A1a}\\
 & \equiv & (2\ell_{2}+1)\Xi_{TT}(\ell_{1},\ell_{2},\tilde{W})
\end{eqnarray}
where for the cross spectrum $(i,j)$, $\tilde{W}_{\ell}$ is the
power spectrum of the window function 
\begin{equation}
\tilde{W}_{\ell}^{ij}=\frac{1}{(2\ell+1)}\sum_{m}\tilde{w}_{\ell m}^{i}\tilde{w}_{\ell m}^{j*},\label{A2}
\end{equation}

\subsection{Pseudo-$C_\ell$ covariance matrices\label{app:PCL-Covariance-Matrices}}

For the case of narrow window functions and uncorrelated pixel noise
$(\sigma_{i}^{T})^{2}$, the covariance matrices can be approximated
as 
\begin{eqnarray}
\langle\Delta\tilde{C}_{\ell}^{T_{ij}}\Delta\tilde{C}_{\ell^{\prime}}^{T_{pq}}\rangle & \approx & C_{\ell}^{T}C_{\ell^{\prime}}^{T}\left[\Xi_{TT}(\ell,\ell^{\prime},\tilde{W}^{(ip)(jq)})+\Xi_{TT}(\ell,\ell^{\prime},\tilde{W}^{(iq)(jp)})\right]\nonumber \\
 & + & (C_{\ell}^{T}C_{\ell^{\prime}}^{T})^{1/2}\times\nonumber \\
 &  & \left[\:\Xi_{TT}(\ell,\ell^{\prime},\tilde{W}^{2T(ip)(jq)})\right.\nonumber \\
 &  & \left.+\Xi_{TT}(\ell,\ell^{\prime},\tilde{W}^{2T(iq)(jp)})\right.\label{eq:B1}\\
 &  & \left.+\Xi_{TT}(\ell,\ell^{\prime},\tilde{W}^{2T(jp)(iq)})\right.\nonumber \\
 &  & \left.+\Xi_{TT}(\ell,\ell^{\prime},\tilde{W}^{2T(jq)(ip)})\right]\nonumber \\
 & + & \Xi_{TT}(\ell,\ell^{\prime},\tilde{W}^{TT(ip)(jq)}+\Xi_{TT}(\ell,\ell^{\prime},\tilde{W}^{TT(iq)(jp)}),\nonumber 
\end{eqnarray}
where $\Xi$ is the matrix defined in Equation~\ref{A1a}. The window
functions are given by: 
\begin{eqnarray}
\tilde{W}_{\ell}^{(ij)(pq)} & = & \frac{1}{(2\ell+1)}\sum_{m}\tilde{w}_{\ell m}^{(ij)}\tilde{w}_{\ell m}^{(pq)*},\label{B2a}\\
\tilde{W}_{\ell}^{TT(ij)(pq)} & = & \frac{1}{(2\ell+1)}\sum_{m}\tilde{w}_{\ell m}^{T(ij)}\tilde{w}_{\ell m}^{T(pq)*},\label{B2b}\\
\tilde{W}_{\ell}^{2T(ij)(pq)} & = & \frac{1}{(2\ell+1)}\sum_{m}\tilde{w}_{\ell m}^{(ij)}\tilde{w}_{\ell m}^{T(pq)*},\label{B2d}
\end{eqnarray}
where
\begin{eqnarray}
\tilde{w}_{\ell m}^{(ij)} & = & \sum_{s}w_{s}^{i}w_{s}^{j}\Omega_{s}Y_{\ell m}^{*}(\theta_{s}),\\
\tilde{w}_{\ell m}^{T(ij)} & = & \sum_{s}(\sigma_{s}^{T})^{2}w_{s}^{i}w_{s}^{j}\Omega_{s}^{2}Y_{\ell m}^{*}(\theta_{s}).
\end{eqnarray}
To avoid cumbersome notation, we have omitted indices from the theoretical
spectra appearing in Eq.~\ref{eq:B1}. In practice, these spectra
include unresolved foreground contributions and are smoothed by the
appropriate beam transfer functions $b^{ij}$. In addition, these
covariance matrices are corrected for the pixel window
functions $p_{\ell}$ (i.e., covariance matrices such as $\langle\Delta\tilde{C}_{\ell}^{T_{ij}}\Delta\tilde{C}_{\ell^{\prime}}^{T_{pq}}\rangle$
are divided by $p_{\ell}^{2}p_{\ell^{\prime}}^{2}$)


\subsection{Combining intra-frequency cross-spectra \label{app:Combining-intra-frequency}}

For \Planck, the vector containing the power spectra, and its associated covariance
matrix, are both large and so require substantial compression to make the computation of a high-$\ell$ likelihood fast enough for parameter estimation. 
As described in Appendix \ref{app:Calibration-and-beam},
after we correct for the `effective' calibration factors for each individual detector set, the power spectra at each freqency are consistent
to extremely high accuracy. Any remaining residuals
have a negligible impact on the cosmological analysis. Thus, we combine 
the cross-spectra from different detectors within a given 
frequency combination into a single power spectrum. We do 
not average across frequency combinations since the unresolved 
foregrounds depend on frequency. Further compression can be 
accomplished, if desired, only after unresolved foreground parameters have been determined.

We form the linear combination of individual cross-spectra, for each
multipole, in the following way: 
\begin{equation}
\hat{C}_{\ell}^{T_{k}}=\sum_{ij\subset k,\, i\ne j}\frac{\alpha_{\ell}^{TT_{ij}}y_{i}y_{j}{\hat{C}_{\ell}^{T_{ij}}}}{(b_{\ell}^{Tij})^{2}(p_{\ell}^{T})^{2}}.\label{C6}
\end{equation}
Here the index $k$ denotes the particular frequency cross-spectrum
combination (e.g., $100\times100$, $143\times217$), the
coefficients $y_{i}$ denote the multiplicative factors for each map,
$b_{\ell}^{Tij}$ is the (isotropised) beam transfer function for
the map combination $ij$, and $p_{\ell}$ is the isotropised pixel
window function%
\footnote{Note that for the masks used here, the isotropised pixel window function
provided by HEALPIX is sufficiently accurate.%
}. The coefficients $\alpha_{ij}$ are normalized so that 
\begin{equation}
\sum_{ij\subset k,\, i\ne j}\alpha_{\ell}^{TT_{ij}}=1,\qquad\alpha_{\ell}^{TT_{ii}}=0.
\end{equation}

How can we determine the coefficients $\alpha_{ij}$? A near optimal
combination, $\hat{X}_{\ell}^{k}$, is given by solving 
\begin{equation}
\sum_{pq}\hat{{\cal {M}}}_{pq}^{-1}\hat{X}_{\ell}^{k}=\sum_{pq}\hat{{\cal {M}}}_{pq}^{-1}\hat{X}_{\ell}^{pq},\label{C7}
\end{equation}
where $\hat{{\cal {M}}}_{pq}^{-1}$ is the block of the inverse covariance
matrix appropriate to the spectrum combination $k$. If the covariance
matrix $\hat{{\cal {M}}}$ accurately describes the data, the solution
of Eq.~\ref{C7} properly accounts for the correlations between
the cross-spectra. Solving Eq.~\ref{C7} requires the inversion
of a large matrix, so we adopt a simpler solution by weighting each
estimate by the diagonal component of the relevant covariance matrix, e.g., 
\begin{equation}
\alpha_{\ell}^{TT_{ij}}\propto1/{\rm Cov}(\hat{C}_{\ell}^{T_{ij}}\hat{C}_{\ell}^{T_{ij}}).\label{C8}
\end{equation}
This has the effect of assigning each cross-spectrum equal weight
in the signal dominated regime and the correct inverse variance weighting
in the noise dominated regime. This is the correct solution in the
noise dominated regime. The analysis of intra-frequency residuals
presented in Appendix~\ref{app:Calibration-and-beam} shows that in
the signal dominated regime we see excess variance (with no obvious
dependence on the detector/detector set combination) compared to what
we expect from instrument noise alone. This excess variance is small
compared to the signal and is caused by residual beam errors, consistent
with the beam eigenmode amplitudes discussed in Appendix~\ref{app:Calibration-and-beam},
that are not included in the covariance matrices. This is our justification
for assigning roughly equal weight to the spectra in the signal dominated
regime.

When we construct a likelihood from the combined estimates we construct
the full covariance matrix including cross-correlations between the
various spectra. In this matrix, the cross-correlations in the signal
dominated regime are dominated by cosmic variance if different masks
are used for different frequencies. If identical masks are used for
all frequencies, the cross-correlations in the signal-dominated regime
are dominated by the cross correlations between the CMB and unresolved
foregrounds, which are included in the analytic covariance matrices
and act as a regularizing contribution (see Appendix~\ref{app:ChanceCorrealtions}).


\subsection{Covariance matrix of combined spectra \label{app:Covariance-matrix-of-combined-estimates}}

\begin{figure*}
\begin{centering}
\includegraphics[bb=35bp 30bp 595bp 580bp,clip,width=0.48\textwidth]{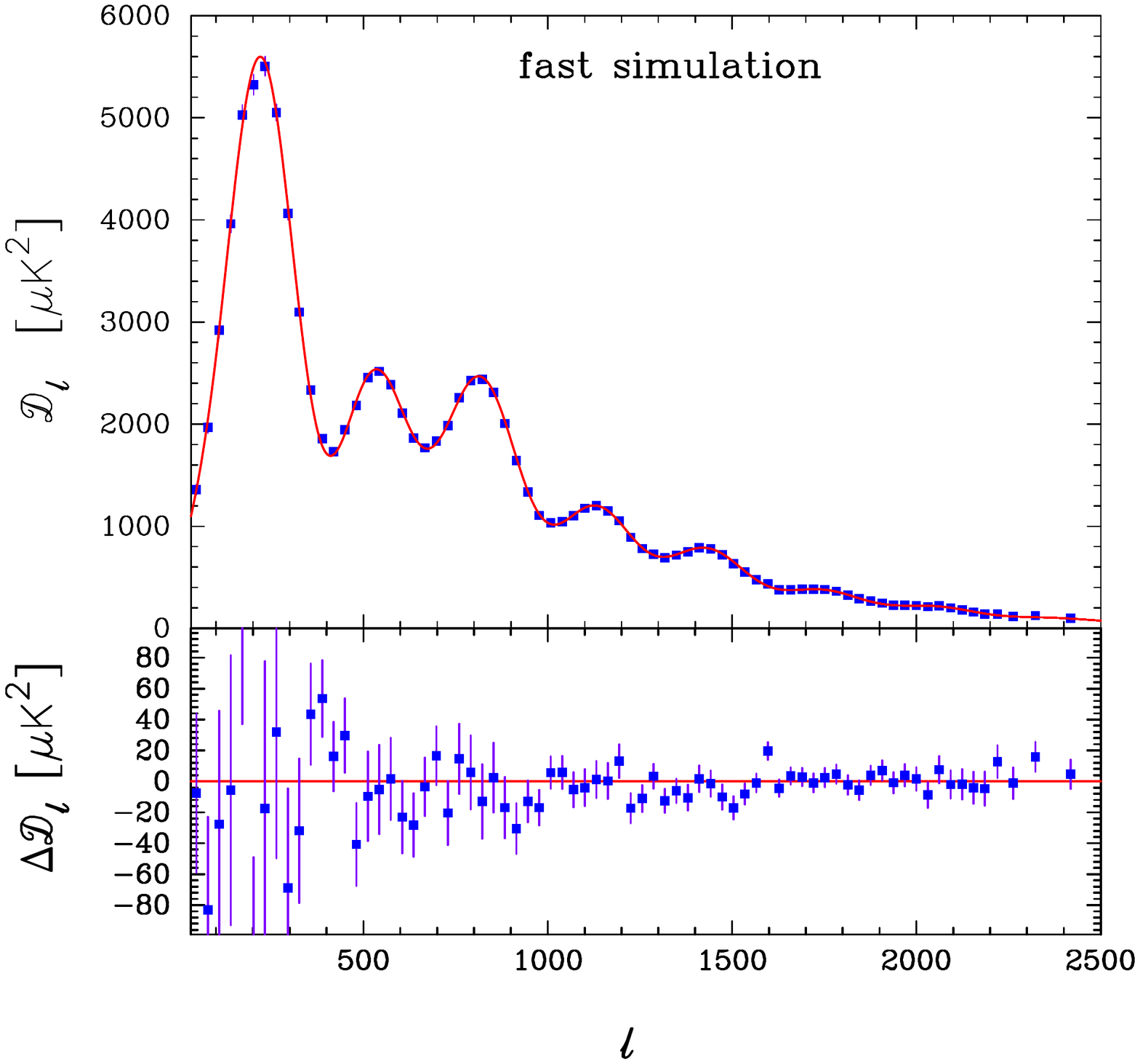}
\includegraphics[bb=35bp 30bp 595bp 580bp,clip,width=0.48\textwidth]{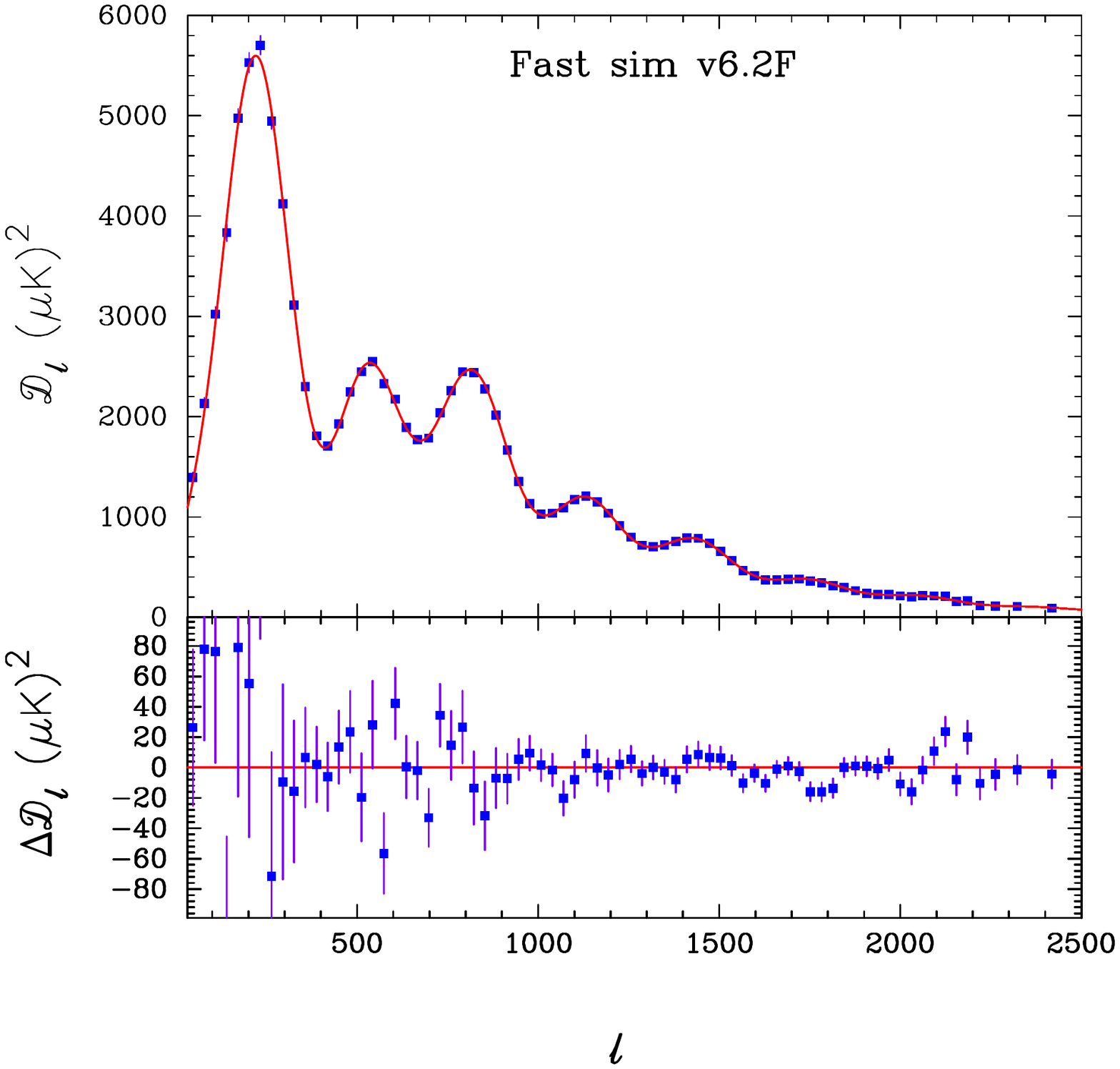}
\par\end{centering}

\centering{}\caption{Toy power spectra drawn from a Gaussian
  distribution with mean given by a $\Lambda$CDM power spectrum and
  a covariance given by the band-averaged $C_{\ell}$
  covariance matrix,  (\emph{top}), and the corresponding differences
  with respect to the input $\Lambda$CDM model. Note the apparent
  presence of ``coherent oscillatory features'' in the difference
  spectra. This which are fully described by the power spectrum covariance
  matrix. To assess the statistical significance of apparently
  ``unexpected features'' in the power spectrum, it is critical to
  include all sources of systematic errors in the evaluation, as these
  can, and do, introduce significant correlations among different
  $C_{\ell}$'s.
\label{fig:cspec_sims}}
\end{figure*}

The estimates of the \Planck\ cross-spectra are linear combinations of the pseudo-$C_{\ell}$
estimates, so their covariance matrices are given by e.g., 
\begin{equation}
{\rm Cov}(\hat{C}_{\ell}^{T_{k}}\hat{C}_{\ell^{\prime}}^{T_{k'}})=\sum\alpha_{\ell}^{TT_{ij}}\alpha_{\ell^{\prime}}^{TT_{pq}}{\rm Cov}(\hat{C}_{\ell}^{T_{ij}}\hat{C}_{\ell^{\prime}}^{T_{pq}}).\label{C9}
\end{equation}
Analytic expressions for these covariance matrices have been given
in \cite{Efstathiou2004,Efstathiou2006,HL08}, and are described in Appendix \ref{app:PCL-Covariance-Matrices}.
The covariance matrices are computed assuming a \textit{fixed} fiducial
theoretical model including an unresolved foreground model for each
frequency combination. Typically, the unresolved foregrounds introduce
corrections to the covariance matrices of a few percent in the transition
region between signal and noise domination. In addition, we compute
the fiducial model by applying appropriate beam functions $b^{T_{ij}}$
for each detector combination.

As discussed above, the number of coupling matrices required to compute
expressions such as Eq.~\ref{C9} scales as $N_{\mathrm{map}}^{4}$ and so becomes
prohibitively expensive as the number of cross-spectra becomes large.
However, most of these coupling matrices are similar, differing
primarily in the amplitude of the noise levels and in minor respects
such as a small number of missing pixels. We can therefore adopt the
same masks and weightings for groups of cross-spectra and compute
coupling matrices only for distinct combinations. This dramatically
reduces the computational burden. A similar approach was adopted by
\cite{L08} to analyse the \WMAP\ 5-year temperature maps.

It is also straightforward to calculate covariance matrices for differences
between different averages. If we form two spectra averaging over
different detector combinations 
\begin{eqnarray}
\hat{C}_{\ell}^{T_{a}} & = & \sum_{i,j,i\ne j}\alpha^{ij}\hat{C}_{\ell}^{T_{ij}},\label{C10a}\\
\hat{C}_{\ell}^{T_{b}} & = & \sum_{i,j,i\ne j}\beta^{ij}\hat{C}_{\ell}^{T_{ij}},\label{C10b}
\end{eqnarray}
then the covariance matrix of the difference $\hat{C}_{\ell}^{T_{a}}-\hat{C}_{\ell}^{T_{b}}$
is simply 
\begin{eqnarray}
{\rm Cov}(\hat{C}_{\ell}^{T_{a}}-\hat{C}_{\ell}^{T_{b}}) & = & \sum_{i\ne j,p\ne q}(\alpha^{ij}\alpha^{pq}+\beta^{ij}\beta^{pq}-\alpha^{ij}\beta^{pq}-\beta^{ij}\alpha^{pq})\nonumber \\
 &  & \times\langle\Delta\hat{C}_{\ell}^{T_{ij}}\Delta\hat{C}_{\ell}^{T_{pq}}\rangle.\label{eq:C11}
\end{eqnarray}

As a pedagogical illustration of the importance of these correlations,
we show in Fig.~\ref{fig:cspec_sims} two toy power spectra drawn from
a Gaussian distribution with mean given by a $\Lambda$CDM spectrum and
covariance given by ${\rm
  Cov}(\hat{C}_{\ell}^{T_{a}}-\hat{C}_{\ell}^{T_{b}})$. That is, these
spectra are not computed from a real sky map, but simply drawn
directly from the $C_{\ell}$ error model, and therefore exclusively
contain correlations modelled by the covariance matrix. The apparent
``coherent oscillatory features'' seen in the difference spectra
(bottom panels) are therefore fully described by the
\camspec\ covariance matrix, accounting for correlated beam and
foreground uncertainties, mask-induced coupling etc. The left panel
shows a typical realisation, while the case in the right panel is
selected as, visually speaking, one of the most ``peculiar'' within a
relatively small set of simulatoins. When assessing the statistical
significance of ``unexpected features'' in the real CMB spectrum,
e.g., similar to those seen in Figs.~\ref{fig:ps_CMB_ref} and
\ref{fig:ps_CMB_ref-zoom}, it is critical to account for these
correlations.


\subsection{The ``fiducial Gaussian'' approximation \label{app:The-fiducial-gaussian}}

We use a likelihood based on the so-called ``fiducial Gaussian''
approximation. Here we present our justification for this choice,
based on an ``expansion in covariance'' of the exact likelihood
in the exact full-sky, isotropic noise case.

Assuming the CMB, noise, and foregrounds are Gaussian, then the probability
distribution for the $a_{lm}$ coefficients of a collection of maps,
given a model, is $p=e^{-S}$, where 
\begin{equation}
S=\sum_{\ell}(\ell+1/2)\left(\mathrm{tr}\left[{\bf C_{\ell}}^{-1}{\bf \hat{C}_{\ell}}\right]+\ln\left|{\bf C_{\ell}}\right|\right),\label{eq:action}
\end{equation}
up to a model-independent normalization.
Here ${\bf \hat{C_{\ell}}}$ is the matrix of empirical spectra at a
given multipole, and ${\bf C_{\ell}}$ are their expectation values for
the model in question.

Now, a key point to note is that theoretical power spectra typically
differ from each other at each $\ell$ by less than they differ from
the observed ${\bf \hat{C_{\ell}}}$, because of cosmic variance and
noise in the latter. So we are justified in expanding Eq.~\ref{eq:action}
about a reasonable fiducial model. Considering a single value of $'ell$
for simplicity, writing 
\begin{equation}
{\bf C}={\bf C_{f}}+{\bf \Delta},\label{eq:cexp}
\end{equation}
we obtain 
\begin{eqnarray}
S & = & S_{f}+(l+1/2)\mathrm{tr}\left(-{\bf C_{f}}^{-1}\Delta{\bf C_{f}}^{-1}{\bf \hat{C}_{l}}{\bf C_{f}}^{-1}\Delta\right.\nonumber \\
 &  & \left.+{\bf C_{f}}^{-1}\Delta{\bf C_{f}}^{-1}\Delta{\bf C_{f}}^{-1}{\bf \hat{C}_{l}}\right.\nonumber \\
 &  & \left.-\frac{1}{2}{\bf C_{f}}^{-1}\Delta{\bf C_{f}}^{-1}\Delta+\ldots\right)
\end{eqnarray}
to second order in $\Delta$. We may now complete the square in $\Delta$
after extracting a term that is small if the fiducial model is accurate.
Up to terms independent of $\Delta$, we have, to second order in
$\Delta$: 
\begin{eqnarray}
S_{2} & = & (\ell+1/2)\mathrm{tr}\left({\bf C_{f}}^{-1}\Delta{\bf C_{f}}^{-1}\Delta({\bf C_{f}}^{-1}{\bf \hat{C}}-1)\right) \nonumber\\
 &  & +\frac{1}{2}(\ell+1/2)\mathrm{tr}\left(({\bf C}-{\bf \hat{C}}){\bf C_{f}}^{-1}({\bf C}-{\bf \hat{C}}){\bf C_{f}}^{-1}\right).
\end{eqnarray}
Here we have recombined the perturbation and the fiducial model back
together, using Eq.~\ref{eq:cexp}, in the second term to obtain
exactly the ``fiducial Gaussian'' likelihood. The first term is
a correction to the fiducial Gaussian likelihood that is typically
small if the fiducial model is accurate.

One can motivate neglecting this term by noticing that in its absence
the approximate likelihood is unbiased (as the exact one is). One
trades getting second derivatives exactly right in the vicinity of
the fiducial model with getting the position, though not the depth,
of the minimum right.

Vectorizing the distinct elements of ${\bf C}-{\bf \hat{C}}$ \citep[following
Appendix A of][]{HL09}, and recognizing the coefficients as the
inverse covariance matrix elements of the spectra under the fiducial
model, we obtain 
\begin{equation}
S_{\mathrm{fid}}=\frac{1}{2}(\hat{C}^{1T}-C^{1T},\hat{C}^{2T}-C^{2T},\ldots)\hat{{\cal M}}^{-1}(\hat{C}^{1T}-C^{1T},\hat{C}^{2T}-C^{2T},\ldots)^{T},
\end{equation}
where $\hat{\mathcal{M}}$ is the fiducial covariance matrix of the
spectra, and the upper indices run on the different pairs of frequencies.

This suggests an easy generalization to the coupled cut sky pseudo-spectra,
given our calculation of their covariances in Appendix~\ref{app:PCL-Covariance-Matrices}.
We now replace the power spectra above with corresponding appropriate
averages of detector cross-spectra. With bold face now denoting spectra
laid out as vectors, and ${\bf \hat{{\cal M}}}$ the grand fiducial
covariance, our final action is: 
\begin{equation}
S_{\mathrm{fid}}=\frac{1}{2}({\bf \hat{C}}^{1T}-{\bf C}^{1T},{\bf \hat{C}}^{2T}-{\bf C}^{2T},\ldots){\bf \mathcal{\hat{M}}}^{-1}({\bf \hat{C}}^{1T}-{\bf C}^{1T},{\bf \hat{C}}^{2T}-{\bf C}^{2T},\ldots)^{T}.\label{eq:sfinal}
\end{equation}
Another advantage of the ``fiducial Gaussian'' approximation is
that instrumental uncertainties (calibration errors, beam errors,
etc.) do not appear in the inverse covariance, but only in the expression
of the theoretical spectra ${\bf C}^{T}$ in Eq.~\ref{eq:sfinal}
above.

Note that if we fix the foreground model $C^{Fk}$ for each spectrum
$k$, together with the calibration coefficients and beam parameters,
we can minimize the likelihood (Eq.~\ref{eq:sfinal}) with respect to
a `best-fit' \textit{primary CMB spectrum}. This `best-fit' spectrum
is given by the solution of 
\begin{equation}
\sum_{kk^{\prime}\ell^{\prime}}(\hat{{\cal M}}_{\ell\ell^{\prime}}^{-1})^{kk^{\prime}}\hat{C}_{\ell'}^{\mathrm{CMB}}=\sum_{kk^{\prime}\ell^{\prime}}(\hat{{\cal M}}_{\ell\ell^{\prime}}^{-1})^{kk^{\prime}}(c^{k^{\prime}}\hat{C}_{\ell^{\prime}}^{k^{\prime}}-\hat{C}_{\ell^{\prime}}^{Fk^{\prime}}),\label{CSL3}
\end{equation}
where the $c^{k}$ are \textit{spectrum} effective calibration factors
(see Appendix \ref{app:Calibration-and-beam}). The covariance matrix
of the estimates $\hat{C}_{\ell}^{\mathrm{CMB}}$ is given by the inverse of
the Fisher matrix:\foreignlanguage{english}{ 
\begin{equation}
\langle\Delta\hat{C}_{\ell}^{\mathrm{CMB}}\,\Delta\hat{C}_{\ell^{\prime}}^{\mathrm{CMB}}\rangle=\left(\sum_{kk^{\prime}}(\hat{{\cal M}}_{\ell\ell^{\prime}}^{-1})^{kk^{\prime}}\right)^{-1}.\label{CSL4}
\end{equation}
}


\subsection{Uncertainties on individual detector sets beams and calibrations\label{app:Det-beam-errors}}

Let us consider two detectors (or detector sets) $X$ and $Y$. Neglecting 
instrumental noise, the cross-spectrum $C^{\rm{XY,obs}}_\ell$ is related to the true one, $C^{\rm{XY,sky}}_\ell$, through 
\begin{equation}
C^{\rm{ XY, obs}}_\ell=C^{\rm {XY, sky}}_\ell W^{\rm{ XY, eff,true}}_\ell,
\end{equation}
where $W^{\rm{XY,eff,true}}_\ell$ is the effective beam window function.
Note that because of the optical beam non-circularity and the \Planck\ scanning
strategy, $W^{\rm{XY}}_\ell \ne\left(W^{\rm{XX}}_\ell W^{\rm{YY}}_\ell \right)^{1/2}$
when $X \ne Y$, while $W^{\rm{XY}}=W^{\rm{YX}}$ for any $X$ and $Y$. In the $\ell$ range of interest, $W^{\rm{XY}}_\ell\ge0$,
so we denote $W^{\rm{XY}}=\left(B^{{\rm XY}}\right)^{2}$, following the usual prescription for simple (circular) beam models.
In what follows, we will drop the $XY$ pair superscript except when
they are required for clarity. 

Our analyses use the best estimated $C^{{\rm est}}_\ell$ of the sky
power spectrum, where the measured $C^{{\rm obs}}_\ell$ is corrected
by a nominal effective window $W^{{\rm eff,nom}}_\ell$:
\begin{align}
C^{{\rm est}}_\ell & =C^{{\rm obs}}_\ell /W^{{\rm eff,nom}}_\ell,\nonumber \\
 & =C^{{\rm sky}}_\ell W^{{\rm eff,true}}_\ell /W^{{\rm eff,nom}}_\ell,\nonumber \\
 & =C^{{\rm sky}}_\ell \left(B^{{\rm eff,true}}_\ell /B^{{\rm eff,nom}}_\ell\right)^{2}.
\end{align}
The ratio $B^{{\rm eff,true}}_\ell /B^{{\rm eff,nom}}_\ell$, which
determines the uncertainty on the angular power spectrum due to the beam, is 
estimated using Monte-Carlo simulations of planet transits.

We estimate $B^{\rm{mean}}_\ell$ and $W^{\rm{mean}}_\ell$ from the Monte Carlo 
simulations as 
\begin{align}
B^{\rm{mean}}_\ell & =\sum_{i=1}^{\nmc}(W^{i}_\ell)^{1/2}/\nmc, \label{eq:bmean}\\
W^{\rm{mean}}_\ell & =\sum_{i=1}^{\nmc}W^{i}_\ell/\nmc\label{eq:wmean},
\end{align}
and compute the deviations around the mean 
\begin{equation}
\Delta^{i}_\ell=\ln\left(B^{i}_\ell/B^{\rm{mean}}_\ell\right).\label{eq:delta_beam}
\end{equation}
Since the relative dispersion of the simulated $W^{i}_\ell$ 
is small (less than 1\%), the deviations are well approximated by
\begin{equation}
\Delta^{i}_\ell \simeq\frac{1}{2}\ln\left(W^{i}_\ell /W^{\rm{mean}}_\ell \right).
\end{equation}
The matrix ${\bf \Delta}$ then has $\nmc$ rows and $\lmax+1$
columns.
Its Singular Value Decomposition
(SVD) is given by
\begin{equation}
{\bf \Delta}={\bf M}{\bf D}{\bf V}^{T}
\end{equation}
where ${\bf M}$ is an orthogonal $\nmc\times\nmc$ matrix (i.e.,
${\bf M}^{T}{\bf M}={\bf M}{\bf M}^{T}={\bf I}_{\nmc}$), ${\bf D}$
is a diagonal matrix with $\nmc$ non-negative eigenvalues, and ${\bf V}$
is a matrix with $\lmax+1$ rows whose $\nmc$ columns are orthonormal
vectors (i.e., ${\bf V}^{T}{\bf V}={\bf I}_{\nmc}$). Here ${\bf I}_{\nmc}$
is the identity matrix. 

The covariance matrix of the beam deviations is defined as 
\begin{align}
{\bf C} & \equiv{\bf \Delta}^{T}{\bf \Delta}/(\nmc-1)\nonumber \\
 & ={\bf V}{\bf D}^{2}{\bf V}^{T}/(\nmc-1),
\end{align}
from which we compute the eigenmode matrix 
\begin{equation}
{\bf E}\equiv{\bf D}{\bf V}^{T}/(\nmc-1)^{1/2}
\end{equation}
using the SVD of ${\bf \Delta}$. 
Most of the statistical content of ${\bf \Delta}$
or ${\bf C}$ is limited to the first few modes $\nmodes$ with
the largest eigenvalues. We therefore keep only the largest $\nmodes=5$ of the $E$ matrix. The beam uncertainty for a given spectrum is then given by
\begin{align}
B_\ell & =B^{\rm{mean}}_\ell\exp\left({\bf g}^{T}{\bf E}\right)_{\ell}\nonumber\\
 & =B^{\rm {mean}}_\ell\exp\left(\sum_{k=1}^{\nmodes}g^{k}E^{k}_\ell\right)
\end{align}
where ${\bf g}$ is a vector of independant Gaussian
variates of unit variance with $\nmodes$ elements, and 
$E_{k}(\ell)$ is the $k$-th row of ${\bf E}$.

This can be generalized to a set
of spectra.
Taking three
pairs of detector sets $a=\{UV\}$, $b=\{XY\}$ and $c=\{ZT\}$,
one can write
\begin{equation}
\left({\bf \Delta}^{a}\ {\bf \Delta}^{b}\ {\bf \Delta}^{c}\right)=\left({\bf M}^{a}\ {\bf M}^{b}\ {\bf M}^{c}\right).\matthreethree{{\bf E}^{a}}000{{\bf E}^{b}}000{{\bf E}^{c}}.
\end{equation}
and the covariance matrix is given by
\begin{align}
{\bf C}^{abc} & =\left({\bf \Delta}^{a}\ {\bf \Delta}^{b}\ {\bf \Delta}^{c}\right)^{T}.\left({\bf \Delta}^{a}\ {\bf \Delta}^{b}\ {\bf \Delta}^{c}\right)\nonumber \\
 & =\matthreethree{{\bf E}^{a}}000{{\bf E}^{b}}000{{\bf E}^{c}}^{T}.\matthreethree{{\bf I}}{{\bf M}^{aT}{\bf M}^{b}}{{\bf M}^{aT}{\bf M}^{c}}{{\bf M}^{bT}{\bf M}^{a}}{{\bf I}}{{\bf M}^{bT}{\bf M}^{c}}{{\bf M}^{cT}{\bf M}^{a}}{{\bf M}^{cT}{\bf M}^{b}}{{\bf I}}.\matthreethree{{\bf E}^{a}}000{{\bf E}^{b}}000{{\bf E}^{c}},
\end{align}
The beam errors
can therefore be correlated \citep[and in fact are strongly so, see][]{planck2013-p03c}.
In the next Appendix, this general covariance matrix is used to
derive the beam error eigenmodes of combined spectra for the \camspecs
likelihood.


\subsection{Calibration and beam uncertainties for the \camspec\ likelihood \label{app:Calibration-and-beam}}

\begin{figure}
\centering{}\includegraphics[bb=20bp 50bp 305bp 280bp,clip,width=1\columnwidth]{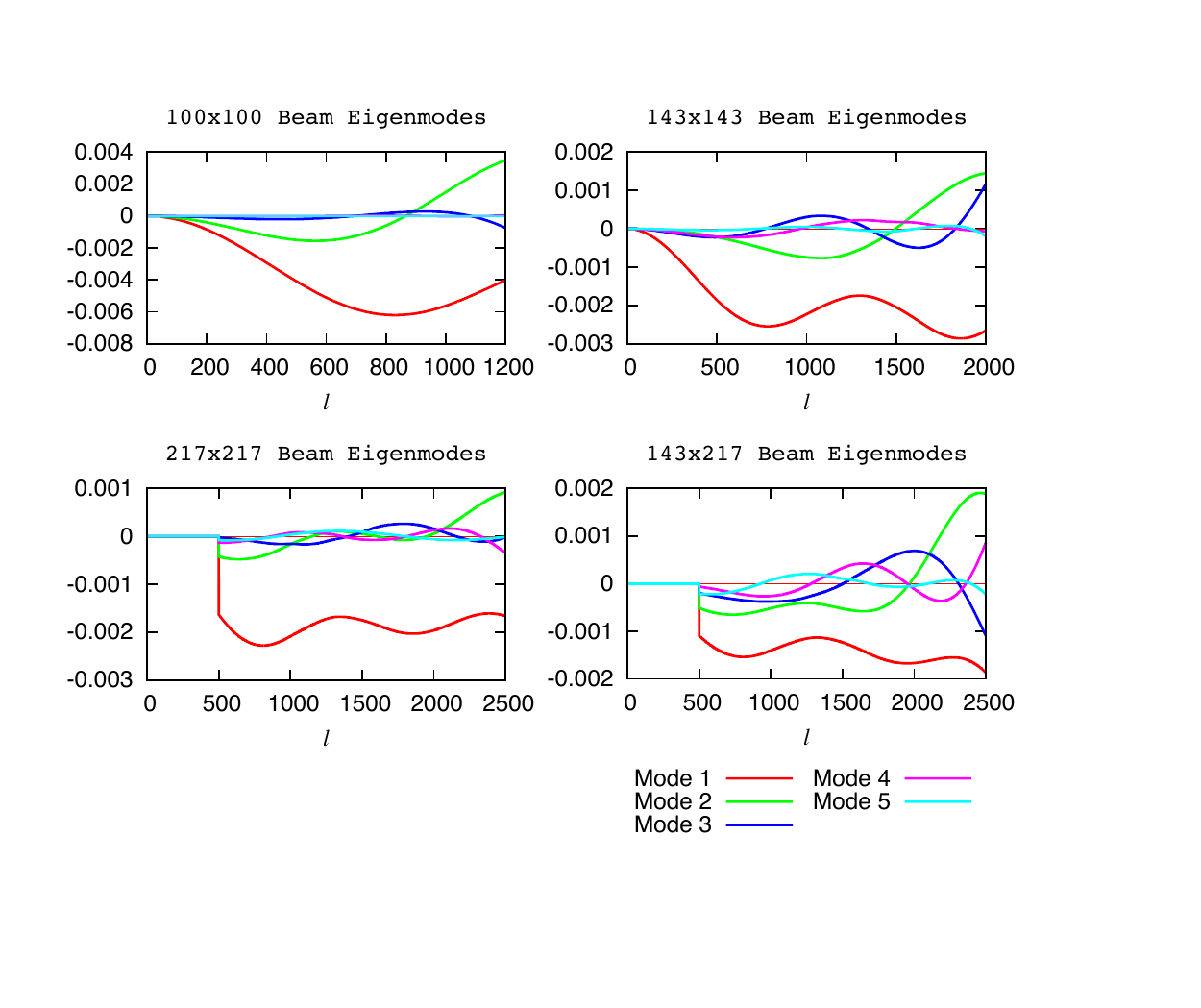}
\caption{\label{fig:effeigmodes}The eigenmodes of the beam covariance matrix,
shown for each cross-spectrum used in the analysis. The largest five
modes are shown for each cross-spectrum.}
\end{figure}

Four effective cross-spectra are used in the \camspec\ likelihood, each
using an individually-prescribed $\ell$-range. For each
of the effective power spectra, all eligible mask- and beam-deconvolved
cross-spectra $\hat{C}_{l}^{XY}$ are used, weighted according to
\begin{equation}
\hat{C}_{\ell}^{p}=\sum_{XY}\alpha_{\ell}^{XY,p}\hat{\, C}_{\ell}^{XY},
\end{equation}
with $p$ labelling the effective spectrum. As describe in Appendix~\ref{app:Det-beam-errors}, uncertainties in the determination of the HFI effective
beams are described in terms of beam eigenmodes, $E_{k}$, and distributions
of the coresponding eigenvalues. To propagate beam errors
into the likelihood, we start by using the eigenvalues, along with
the $\alpha_{\ell}^{XY}$ weights, to construct an appropriate covariance
matrix for the effects of beam errors in $\langle\langle\hat{C}_{\ell}^{p}\hat{C}_{\ell'}^{p}\rangle\rangle_{\text{beam}}$:
\begin{equation}
\langle\langle\hat{C}_{\ell}^{p}\,\hat{C}_{\ell'}^{p}\rangle\rangle_{\text{beam}}\approx4\sum_{ij}\sum_{XY}\sum_{ZW}\alpha_{\ell}^{XY,p}\,\alpha_{\ell'}^{ZW,p}\, E_{i}^{XY}(\ell)\, E_{j}^{XY}(\ell')\, R_{ij}^{XY,ZW},
\end{equation}
where $R_{ij}^{XY,ZW}$ is the correlation between the $i-$th eigenmode
of the $XY$ cross-spectrum with the $j$-th eigenmode of the $ZW$ cross-spectrum.
The portion of this matrix corresponding to the $\ell$-range used
in the likelihood is then extracted, and itself singular-value-decomposed.
We keep the first $n_{\text{effmodes}}$ (typically five) eigenmodes
$E_{i}^{p}(\ell)$, $i=1,\ldots,n_{\text{effmodes}}$, orthogonal
over the $\ell$-range and normalized such that the sum of their outer
product directly approximates the covariance: 
\begin{equation}
\langle\langle\hat{C}_{\ell}^{p}\,\hat{C}_{\ell'}^{p}\rangle\rangle_{\text{beam}}\approx\sum_{i}E_{i}^{p}(\ell)\, E_{i}^{p}(\ell').
\end{equation}
The eigenmodes are illustrated in Fig. \ref{fig:effeigmodes}.

Next we calculate a suitable covariance matrix between the eigenmodes.
This requires the (non-diagonal in $\ell$) inter-effective-spectrum
covariance matrices $\langle\langle\hat{C}_{\ell}^{p}\,\hat{C}_{\ell'}^{q}\rangle\rangle_{\text{beam}}$,
given by: 
\begin{equation}
\langle\langle\hat{C}_{\ell}^{p}\,\hat{C}_{\ell'}^{q}\rangle\rangle_{\text{beam}}\approx4\sum_{ij}\sum_{XY}\sum_{ZW}\alpha_{\ell}^{XY,p}\,\alpha_{\ell'}^{ZW,q}\, E_{i}^{XY}(\ell)\, E_{j}^{ZW}(\ell')\, R_{ij}^{XY,ZW}.
\end{equation}

Given these matrices, we can ``stack'' the effective spectra to
form a data vector $X$ and form a grand beam-covariance matrix $\langle\langle X\, X^{\text{T}}\rangle\rangle_{\text{beam}}$.
$X$ has length $n_{X}=\sum_{p}(\ell_{\text{max}}^{p}-\ell_{\text{min}}^{p}+1)$.
Zero-extending each eigenmode at both ends, and arranging these into
a matrix, we can form an $n_{X}$ by $n_{\text{effmodes}}\cdot n_{\text{eff}}$
matrix of extended eigenmodes, $E_{\text{eff}}$, where $n_{\rm {eff}}=4$. Now we imagine approximating
$\langle\langle X\, X^{\text{T}}\rangle\rangle_{\text{beam}}$ as
a correlated outer product of $E_{\text{eff}}$, 
\begin{equation}
\langle\langle X\, X^{\text{T}}\rangle\rangle_{\text{beam}}\approx E_{\text{eff}}M_{\text{eff}}E_{\text{eff}}^{\text{T}}.
\end{equation}
Requiring that the covariance be chosen to minimize the summed squared-difference
between elements on the two sides yields: 
\begin{equation}
M_{\text{eff}}=E_{\text{eff}}^{\text{T}}\,\langle\langle XX^{\text{T}}\rangle\rangle_{\text{beam}}\, E_{\text{eff}}.
\end{equation}
The $E_{i}^{p}(\ell)$'s and the associated covariance $M_{\text{eff}}$
are then passed to the likelihood.

\subsection{Noise model of HFI detector sets \label{app:Detector-noise-model}}

For strictly uncorrelated pixel noise $(\sigma_{i}^{T})^{2}$, and pixel
weights $w_{i}$, the noise power spectra for the temperature maps
are: 
\begin{equation}
\tilde{N}^{T}=\frac{1}{4\pi}\sum_{i}(\sigma_{i}^{T})^{2}w_{i}^{2}\Omega_{i}^{2},\label{N1}
\end{equation}
with contribution to the pseudo-C$_\ell$ estimates, for uncorrelated
noise, of
\begin{equation}
\tilde{N}^{T}=\frac{1}{4\pi}\sum_{i}(\sigma_{i}^{T})^{2}w_{i}^{2}\Omega_{i}^{2}.
\end{equation}
Values for $\tilde{N}^{T}$ are listed in Table~\ref{tab:Noise-estimates}.
There is a significant dispersion in the noise properties of the two
maps at 100\,GHz. At 143 and 217\,GHz, the PSB maps have significantly
lower noise than the SWB maps by a factor of two, as expected.

$\;$ 
\begin{table}
\begin{centering}
\begin{tabular}{lcc}
\hline 
\hline
Map  & Mask  & $\tilde{N}^{T}$ \tabularnewline
\hline 
100-ds1  & 3  & $2.717\times10^{-4}$ \tabularnewline
100-ds2  & 3  & $1.144\times10^{-4}$ \tabularnewline
 &  & \tabularnewline
143-5  & 1  & $6.165\times10^{-5}$ \tabularnewline
143-6  & 1  & $6.881\times10^{-5}$ \tabularnewline
143-7  & 1  & $5.089\times10^{-5}$ \tabularnewline
143-ds1  & 1  & $2.824\times10^{-5}$ \tabularnewline
143-ds2  & 1  & $2.720\times10^{-5}$ \tabularnewline
 &  & \tabularnewline
217-1  & 1  & $1.159\times10^{-4}$ \tabularnewline
217-2  & 1  & $1.249\times10^{-4}$ \tabularnewline
217-3  & 1  & $1.056\times10^{-4}$ \tabularnewline
217-4  & 1  & $9.604\times10^{-5}$ \tabularnewline
217-ds1  & 1  & $6.485\times10^{-5}$ \tabularnewline
217-ds2  & 1  & $7.420\times10^{-5}$ \tabularnewline
\hline 
\end{tabular}
\par\end{centering}

\caption{Noise estimates for the detector maps, applying the Galactic masks
used in the \camspecs likelihood (`mask$\_$3' for 100\,GHz, retaining
58\,\% of the sky, and `mask$\_1$' for 143 and 217\,GHz, retaining
37\,\% of the sky, combined with an extragalactic point source mask).
\label{tab:Noise-estimates}}
\end{table}

\begin{figure}
\begin{centering}
\includegraphics[bb=90bp 120bp 590bp 465bp,clip,width=0.5\columnwidth]{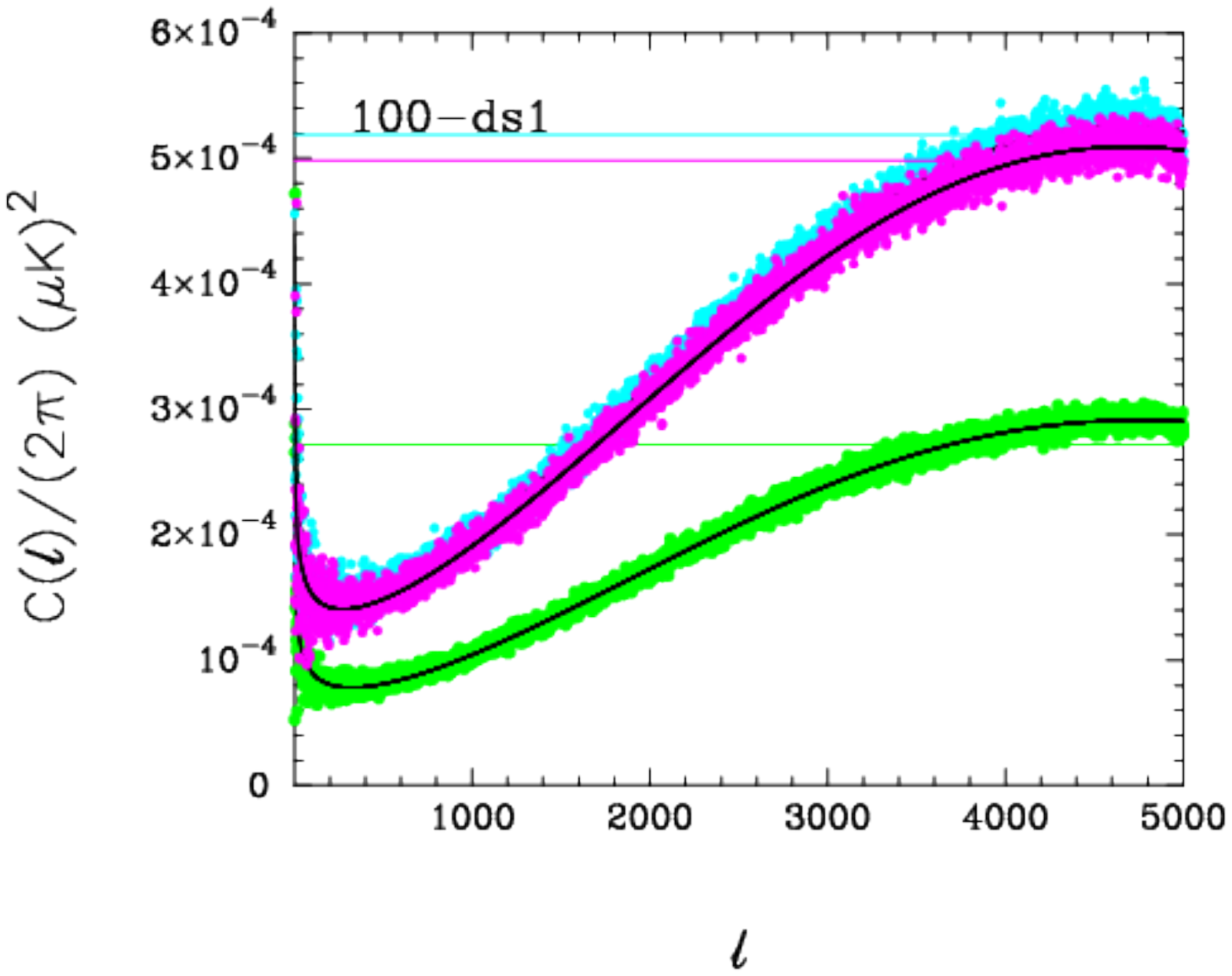}\includegraphics[bb=90bp 120bp 590bp 465bp,clip,width=0.5\columnwidth]{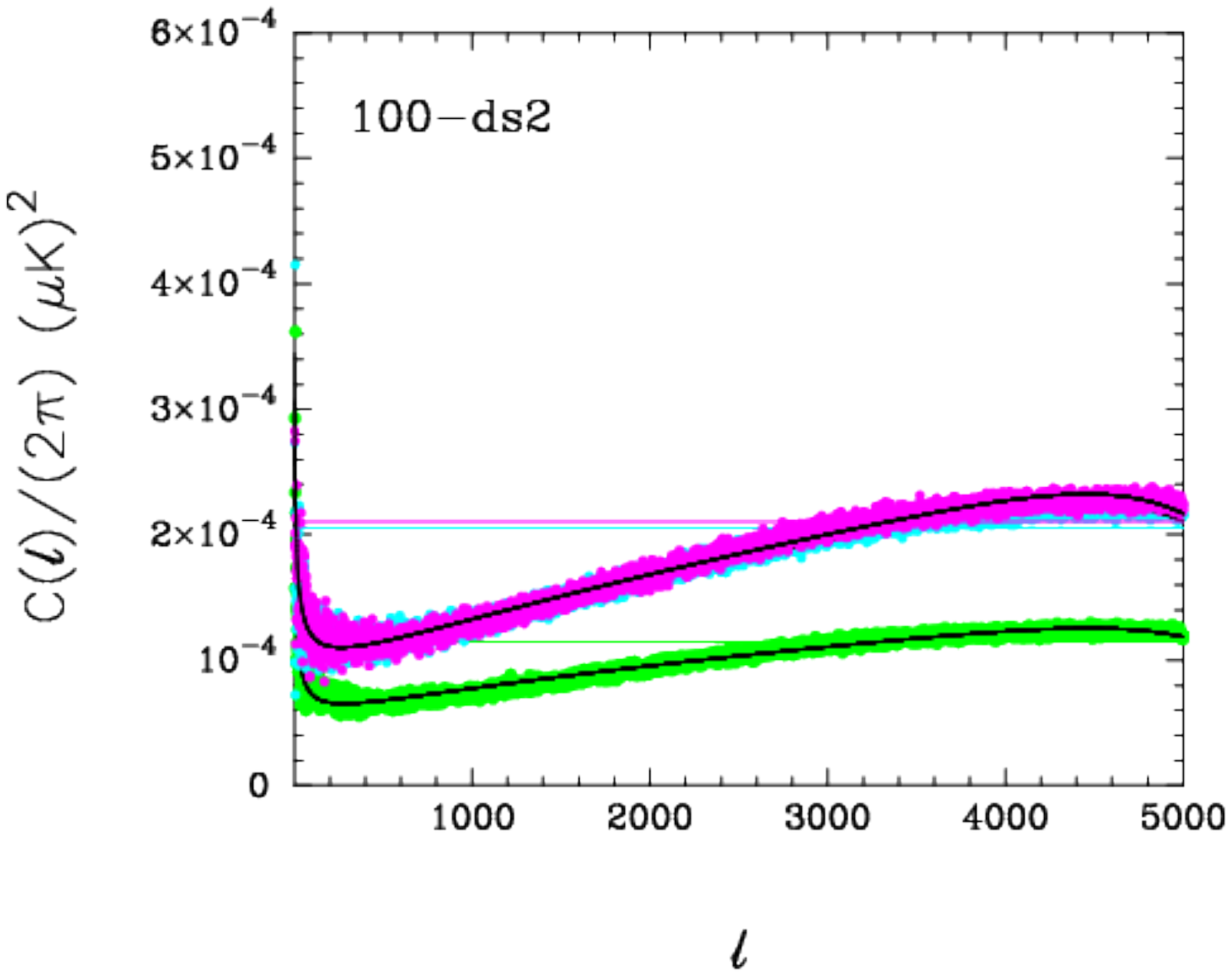}
\par\end{centering}

\begin{centering}
\includegraphics[bb=90bp 120bp 590bp 465bp,clip,width=0.5\columnwidth]{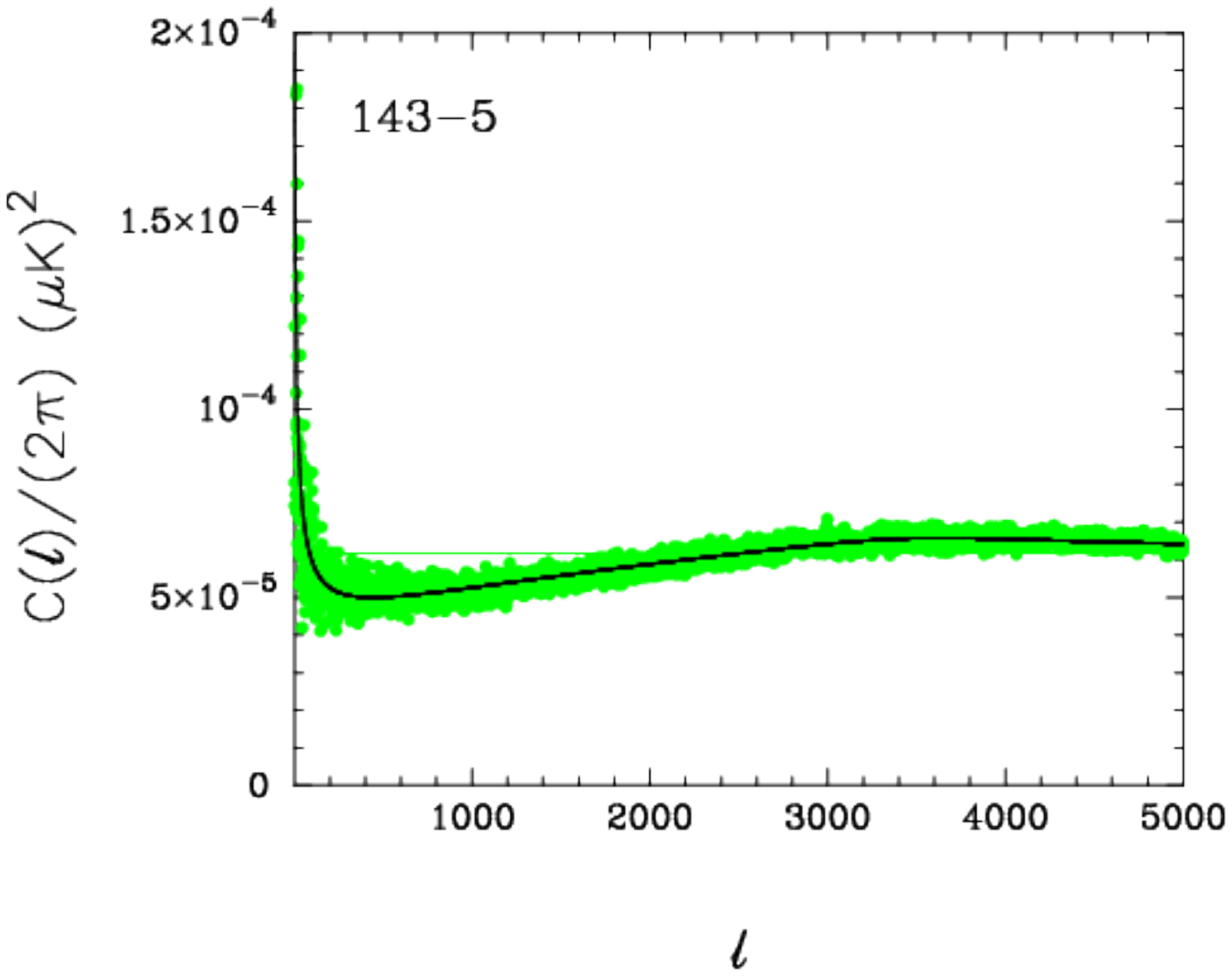}\includegraphics[bb=90bp 120bp 590bp 465bp,clip,width=0.5\columnwidth]{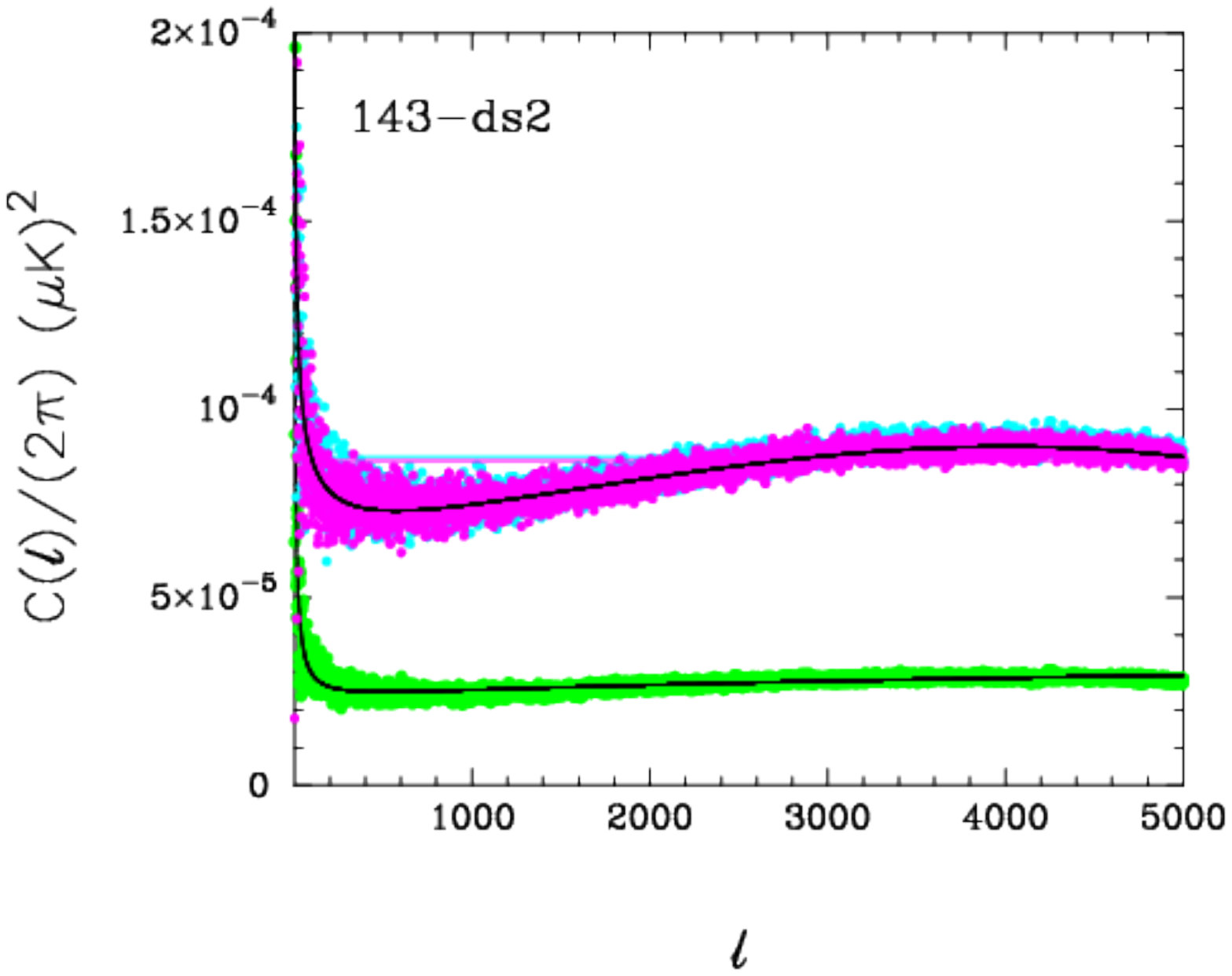}
\par\end{centering}

\begin{centering}
\includegraphics[bb=90bp 70bp 590bp 465bp,clip,width=0.5\columnwidth]{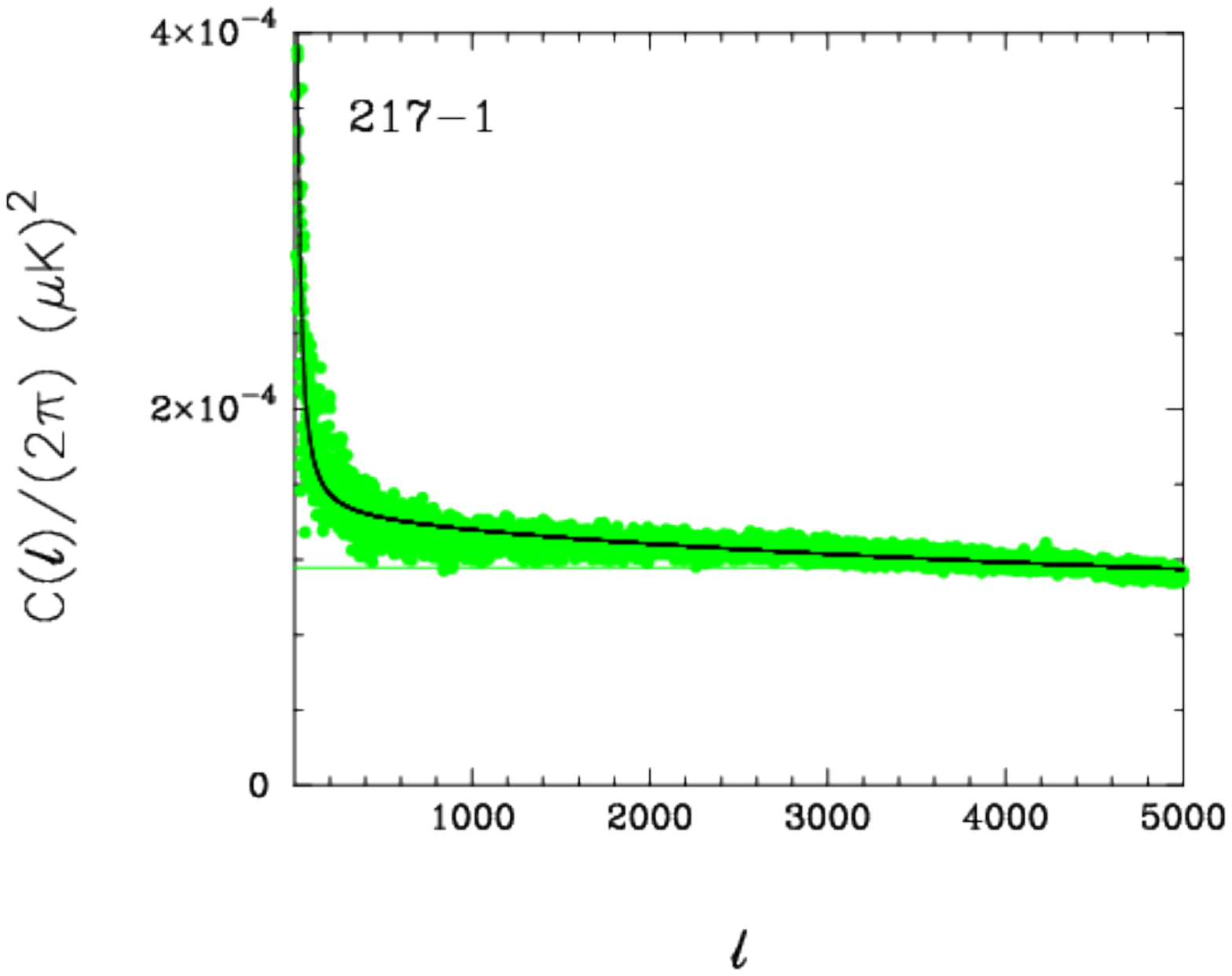}\includegraphics[bb=90bp 70bp 590bp 465bp,clip,width=0.5\columnwidth]{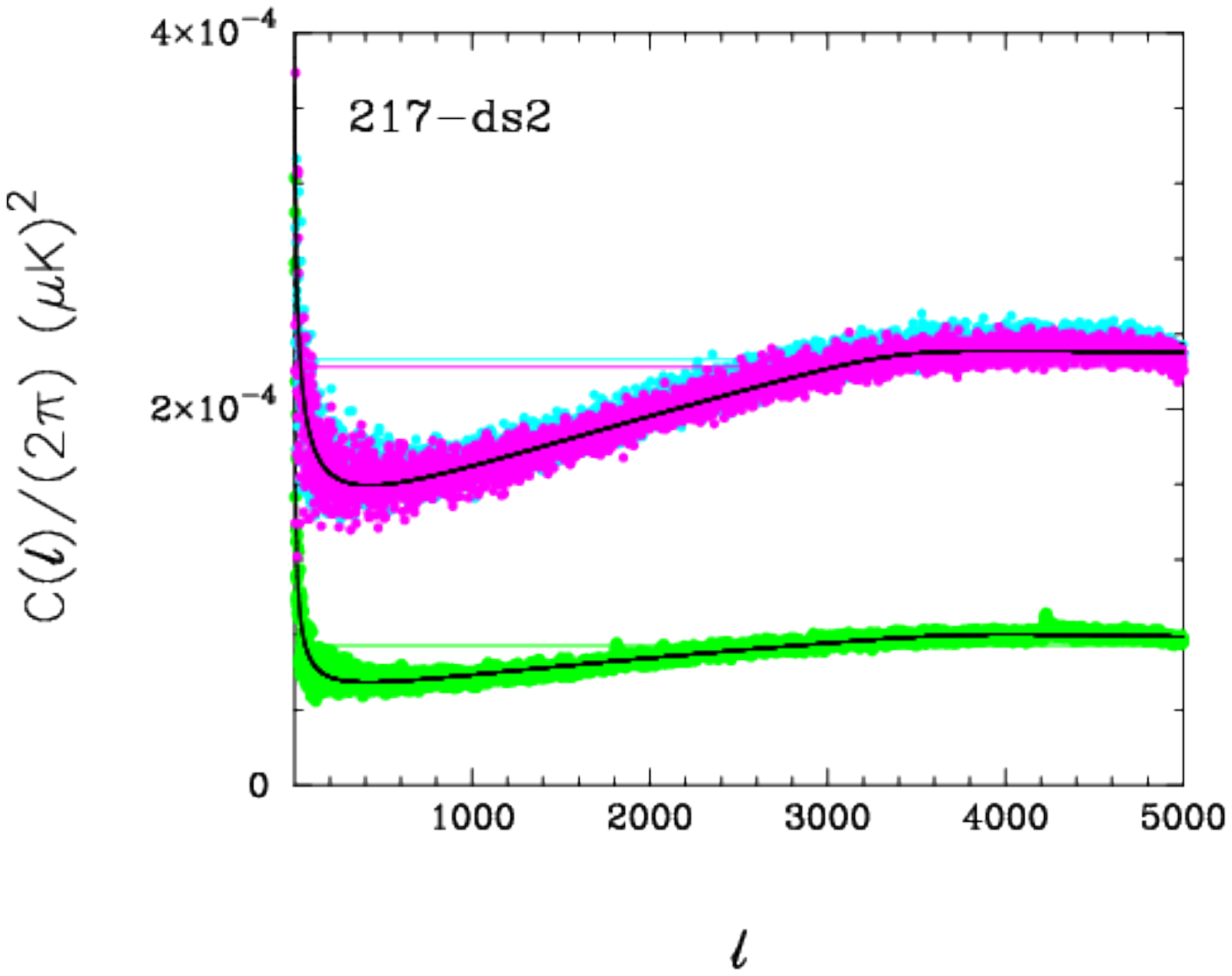}
\par\end{centering}

\caption{Noise spectra computed from difference maps for a selection of detector
sets: $100\;{\rm GHz}$ (top), $143\;{\rm GHz}$ (middle), and $217\;{\rm GHz}$
(bottom). Green lines show the spectra for $T$, purple for $Q$ and
magenta points for $U$. The solid black lines show the modeled
fits to the spectra (Eq.~\ref{N3}), and the coloured horizonal lines
show the white-noise levels of Eq.~\ref{N1}. These are computed using
the same masks as used in Table \ref{tab:Noise-estimates}. \label{fig:noisespec} }
\end{figure}

The noise spectra for the \Plancks HFI maps are non-white. The following
$7$-parameter function 
\begin{equation}
\tilde{N}_{\ell}^{{\rm fit}}=A\left(\frac{100}{\ell}\right)^{\alpha}+\frac{B(\ell/1000)^{\beta}}{(1+(\ell/\ell_{c})^{\gamma})^{\delta}},\label{N3}
\end{equation}
is a flexible parameterization that provides accurate fits for all
of the HFI channels. The first term on the left models the excess
`1/f'-like noise while the second term models the `bell shaped' noise
spectrum at high multipoles introduced by time constant deconvolution
applied to the time-ordered data, and the low-pass filter designed to remove high-frequency
noise due to demodulation. Estimates of the noise spectra can be computed
from difference maps constructed from different half-ring surveys
\footnote{As described in the HFI Data Processing paper \cite{planck2013-p03}, these
difference maps provide an estimate of the noise level in the sum
maps with an accuracy of about $2\%$.%
}. Examples of fits to noise spectra for the $143\;{\rm GHz}$ and $217\;{\rm GHz}$
channels are shown in Figure \ref{fig:noisespec}. Note that the $100\;{\rm GHz}$
noise spectra are significantly non-white. At $143$ and $217\;{\rm GHz}$,
the deviations from white noise are smaller. Since these cross spectra
contribute almost all of the weight in the likelihood at high multipoles,
the modelling of non-white noise is not a critical factor in forming
an accurate likelihood. 

\noindent $\;$$\;$ 

We adopt a heuristic approach to fold departures from white noise
into the power spectrum covariance estimates. We define a set of noise
weight functions, e.g., 
\begin{equation}
\nu_{\ell}^{m}=\frac{\tilde{N}_{\ell}^{{\rm fit,m}}}{\tilde{N}_{\ell}^{m}}.\label{N4}
\end{equation}
Wherever a $\sigma^{2}$ term appears in a covariance matrix, we multiply
the appropriate coupling matrix by a factor 
\begin{equation}
(\nu_{\ell}^{m}\nu_{\ell^{\prime}}^{m})^{1/2}.\label{N5}
\end{equation}
This heuristic approach can be partially justified by noting that
for \textit{isotropic} Gaussian noise over the full sky, the distribution
of $\hat{C}_{\ell}$ is given by the inverse Wishart distribution:
\begin{equation}
dP({\hat{C}_{\ell}}\vert{\bf C_{\ell}})\propto\vert{\bf W_{\ell}}\vert^{-\left(\frac{2\ell+1}{2}\right)}{\rm exp\left(-\frac{1}{2}Tr{\bf W_{\ell}}^{-1}{\bf \hat{C}_{\ell}}\right),\label{W1}}
\end{equation}
where 
\begin{equation}
{\bf \hat{W}_{\ell}}=\frac{1}{(2\ell+1)}{\left(\begin{array}{c}
C_{\ell}^{TT}+N_{\ell}^{TT}\end{array}\right)},\label{W3}
\end{equation}
e.g., \cite{PB06}. In this special case, our heuristic correction
is exact. Further justification of the accuracy of this heuristic
approach comes from direct comparisons with numerical simulations
incorporating non-white noise (see Sect.~\ref{sec:HL-accuracy})
and from the accurate agreement of covariance matrices with the $\ell$-by-$\ell$
scatter measured in all of the cross-spectra used to form the likelihood.

%% file: App-Sky-masks.tex
\section{Sky masks\label{App:Sky-masks}}

\begin{figure*}
\begin{centering}
\includegraphics[bb=0bp 0bp 483bp 243bp,clip,width=1\columnwidth]{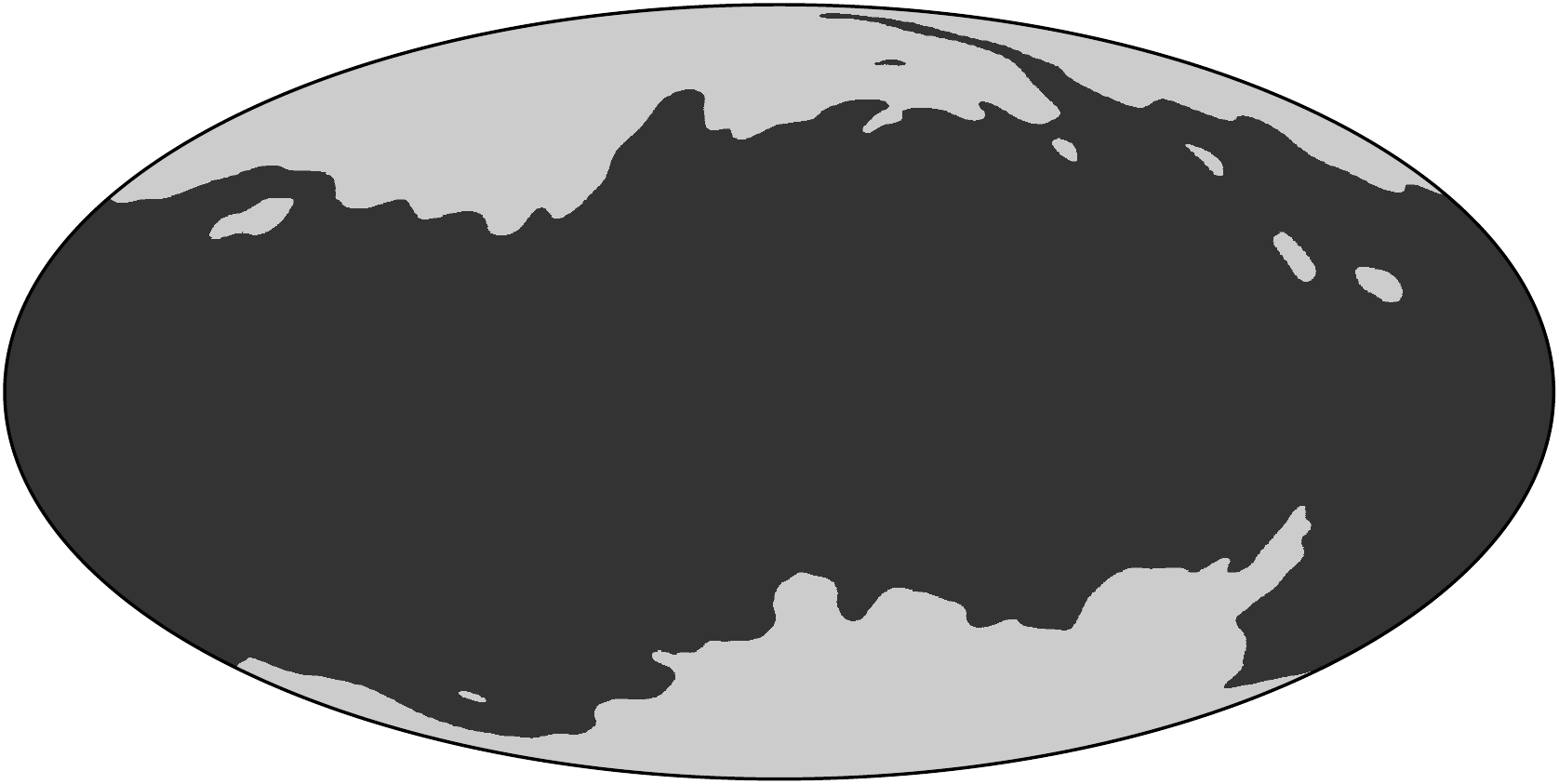}
\includegraphics[bb=0bp 0bp 483bp 243bp,clip,width=1\columnwidth]{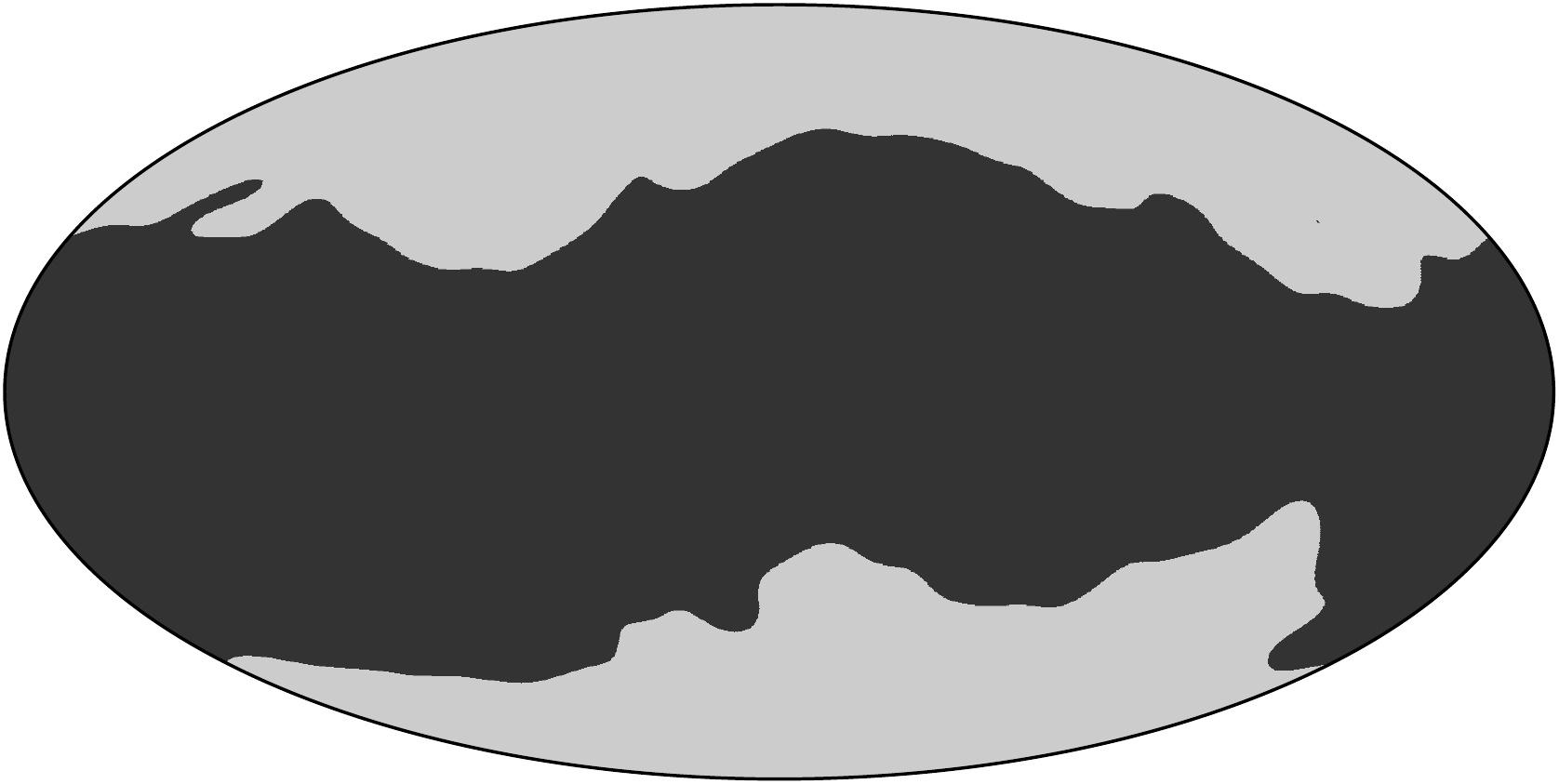} 
\includegraphics[bb=0bp 0bp 483bp 243bp,clip,width=1\columnwidth]{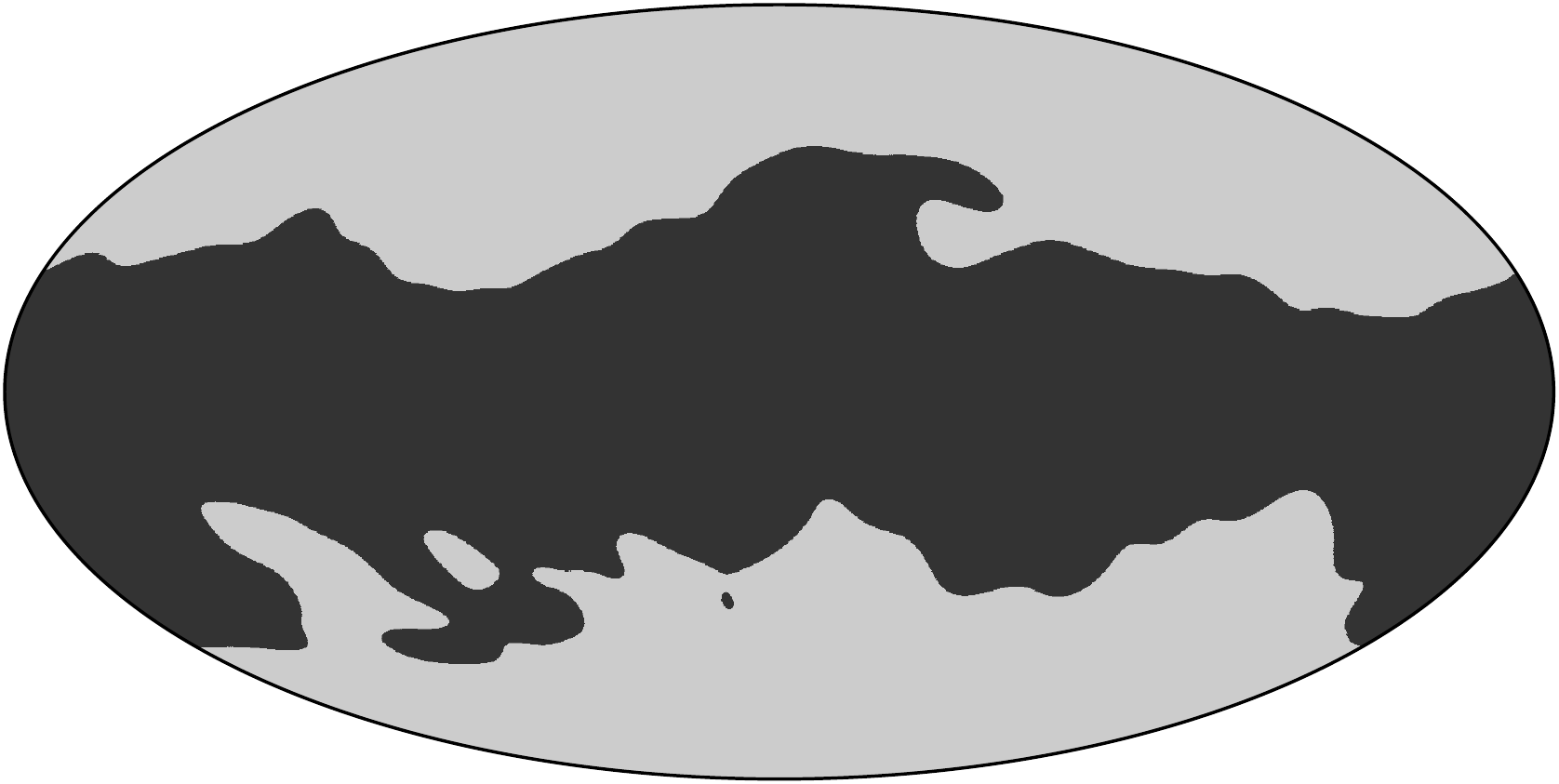}
\includegraphics[bb=0bp 0bp 483bp 243bp,clip,width=1\columnwidth]{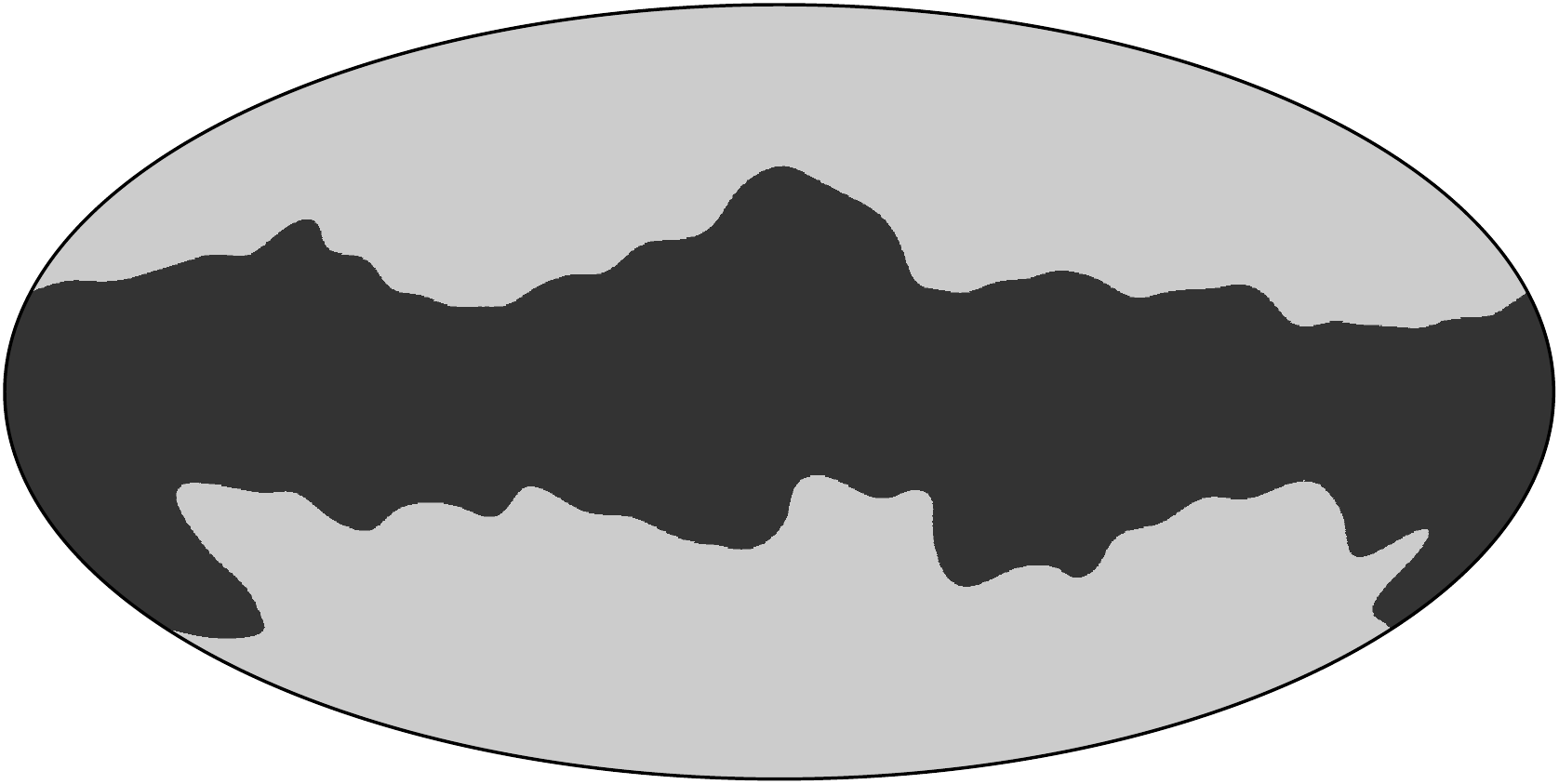} 
\includegraphics[bb=0bp 0bp 483bp 243bp,clip,width=1\columnwidth]{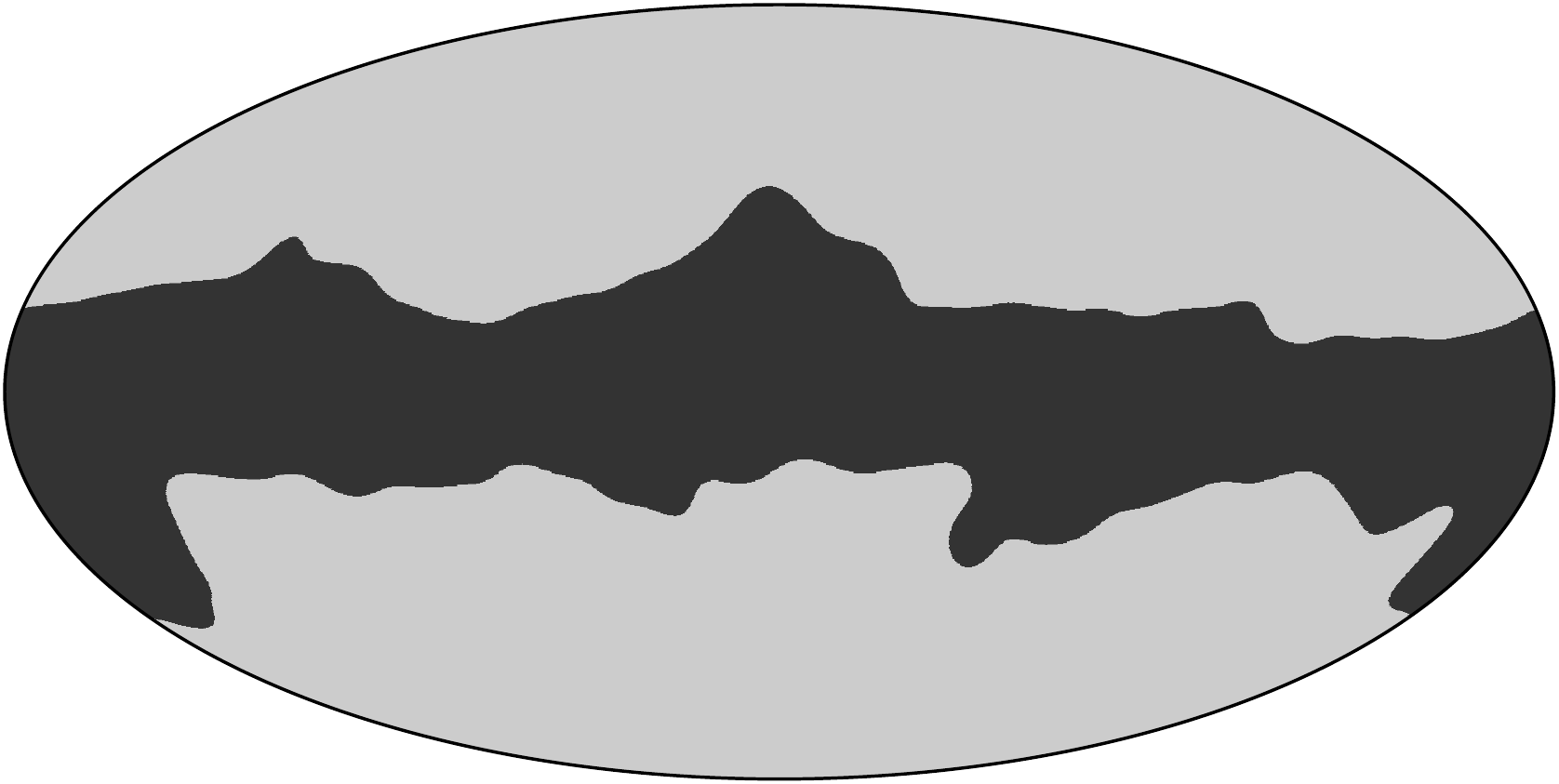}
\includegraphics[bb=0bp 0bp 483bp 243bp,clip,width=1\columnwidth]{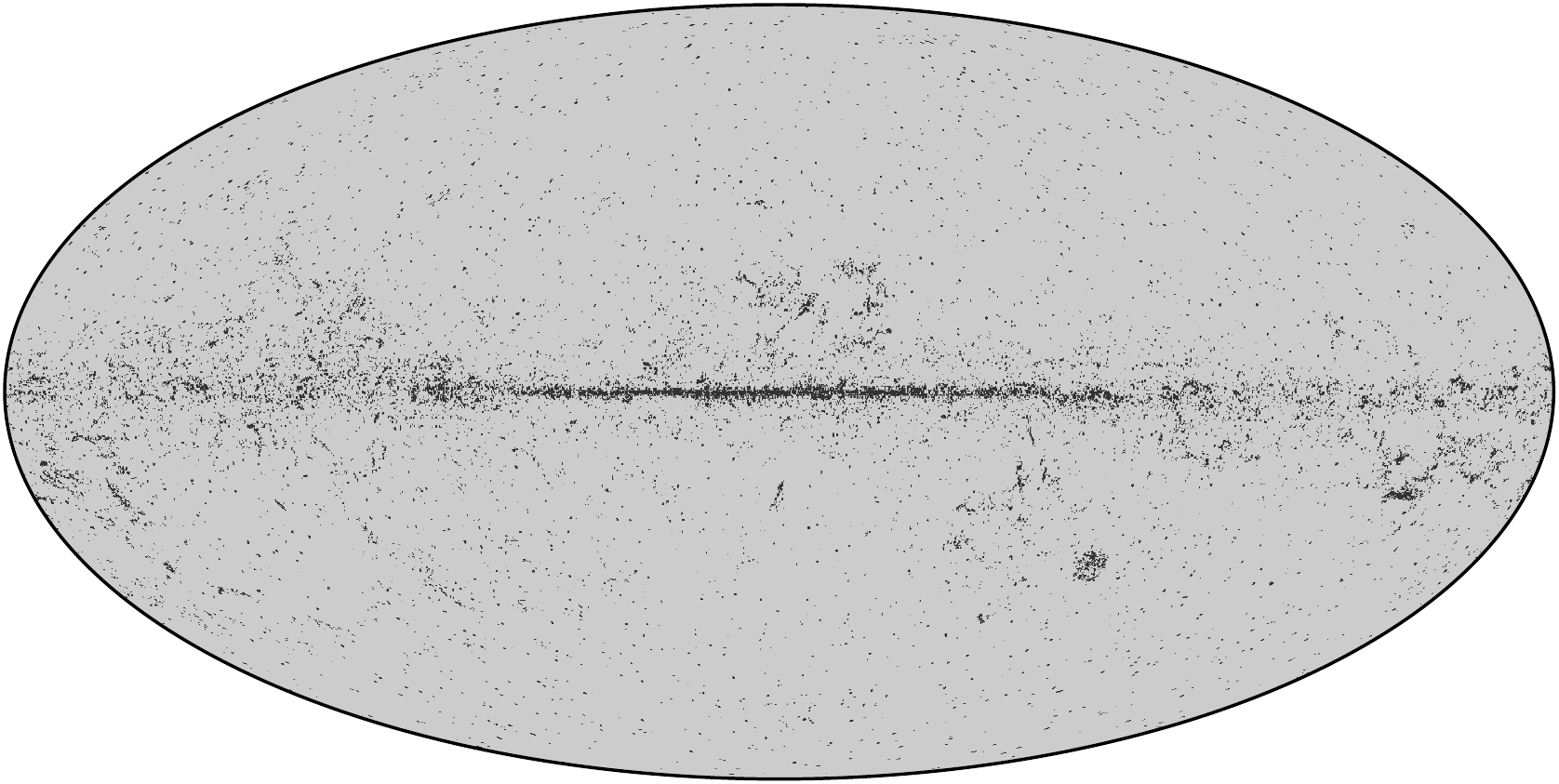} 
\end{centering}
\caption{The set of unapodised foreground masks, G22, G35, G45, G56, G65, PS96, which, once apodised, are used
for the likelihood analyses. These Galactic masks are defined using
a threshold of the 353\,GHz \Planck\ temperature map.}
\label{fig:masks_noapod} 
\end{figure*}

We apply a threshold to the $353$\GHz\ temperature map to
define a set of diffuse Galactic masks shown in Fig.~\ref{fig:masks_noapod}.
We refer to them using the percentage of the sky retained: G22, G35, G45, G56, G65. 
We also use a point source mask, labeled PS96, 
which is based on the union of the point sources
detected from the channels in the range $100$ to $353$\,GHz

\begin{figure*}
\begin{centering}
\includegraphics[bb=0bp 0bp 483bp 243bp,clip,width=1\columnwidth]{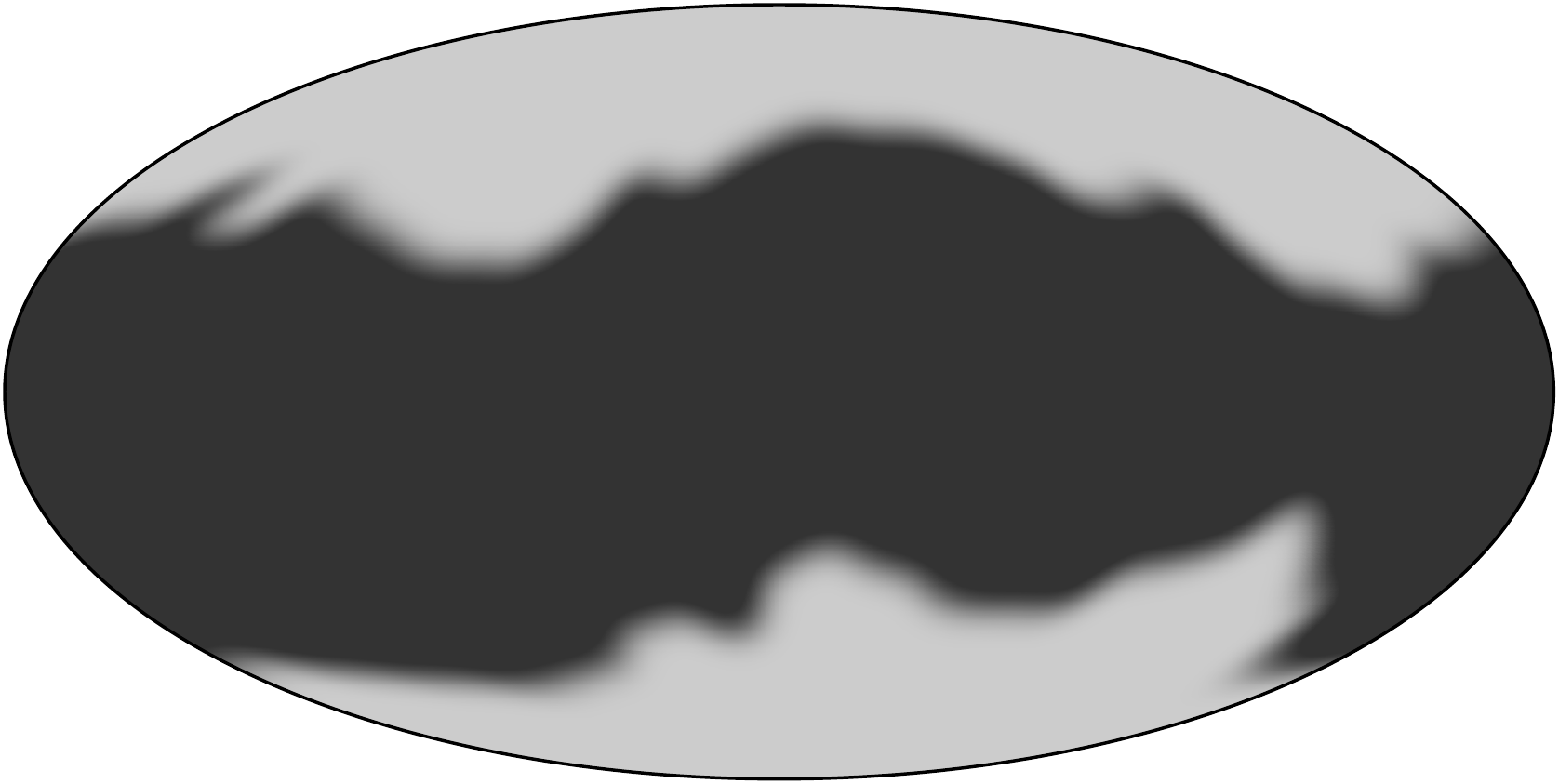} 
\includegraphics[bb=0bp 0bp 483bp 243bp,clip,width=1\columnwidth]{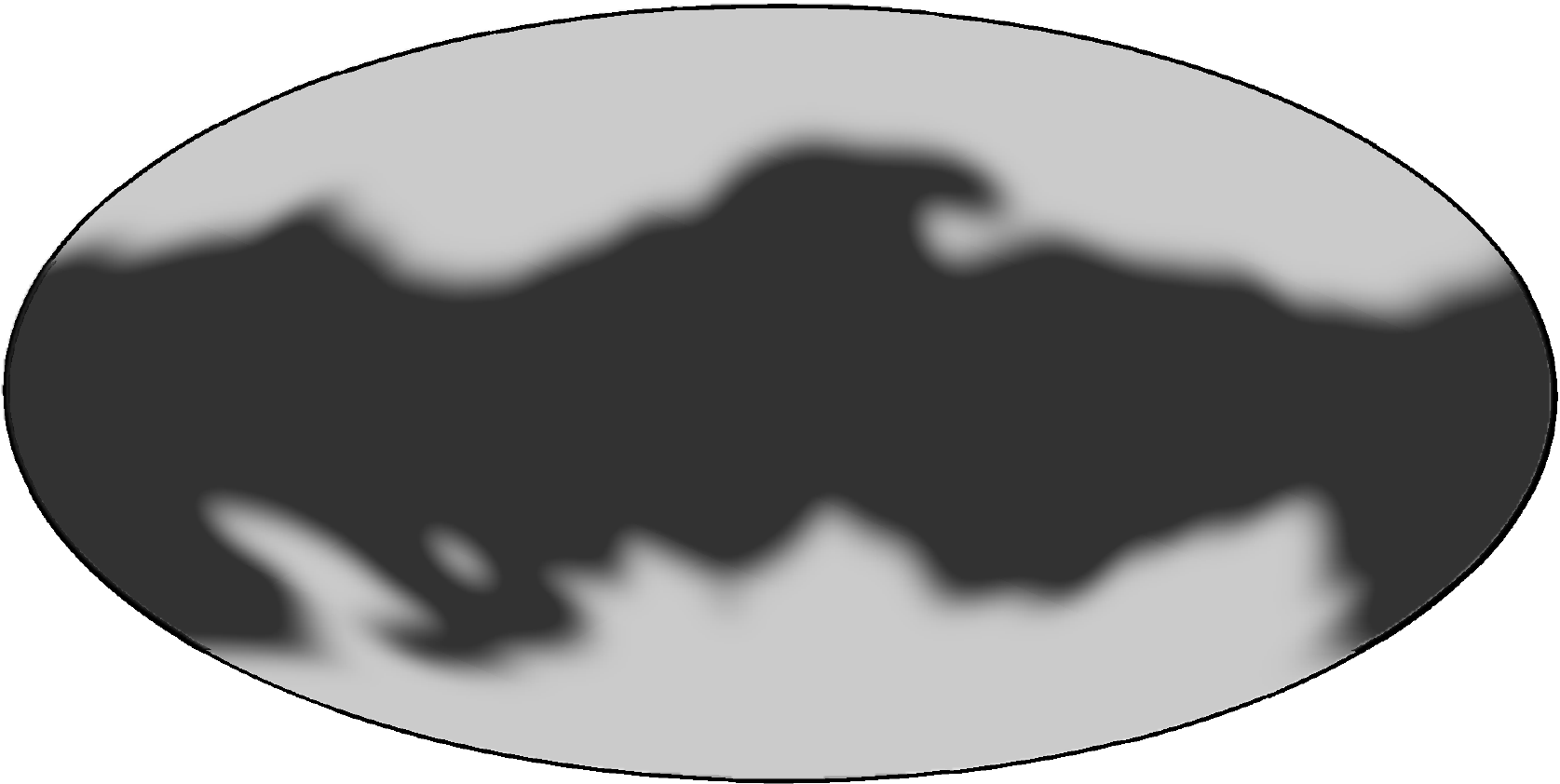}
\includegraphics[bb=0bp 0bp 483bp 243bp,clip,width=1\columnwidth]{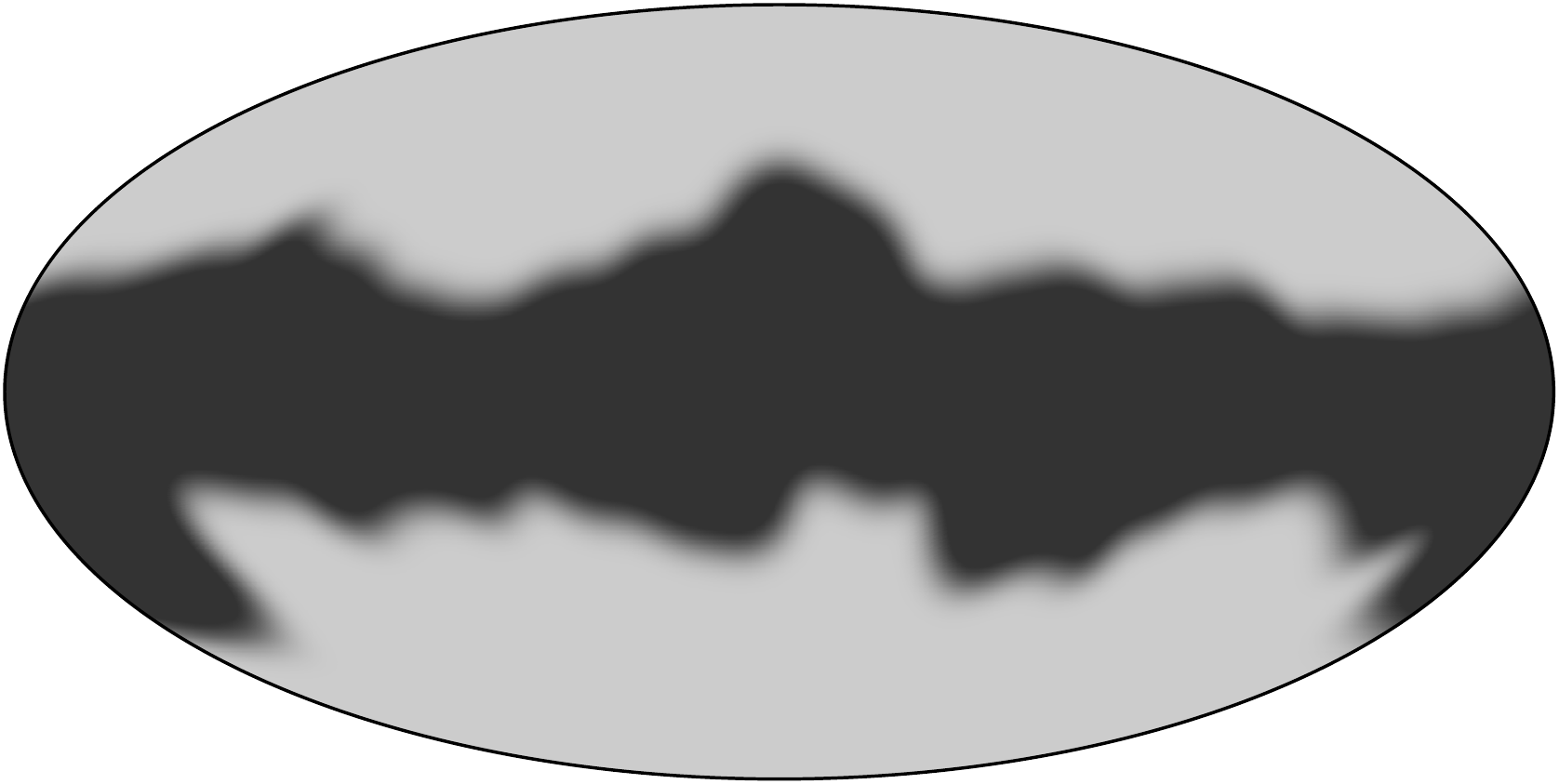}
\includegraphics[bb=0bp 0bp 483bp 243bp,clip,width=1\columnwidth]{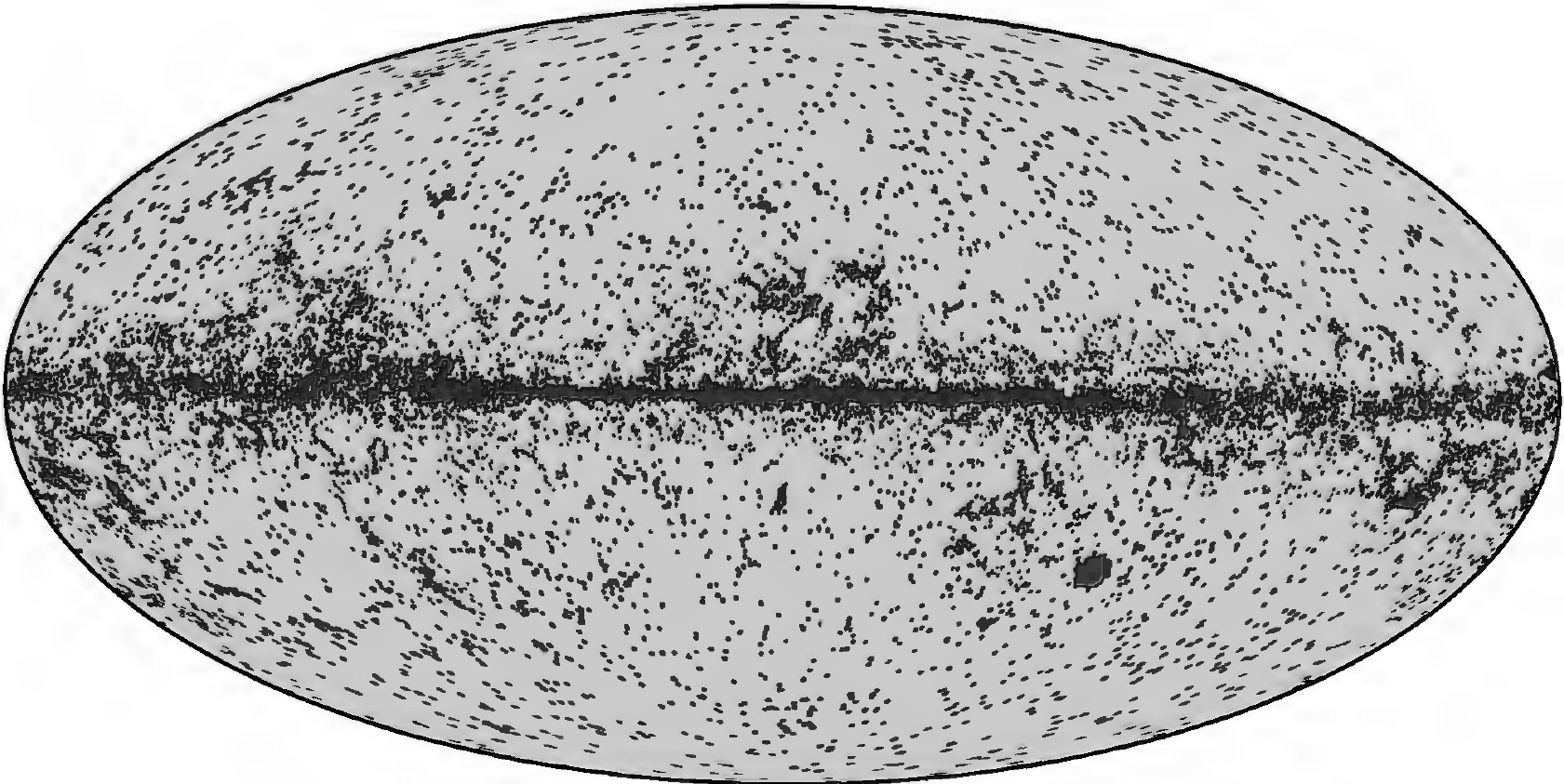}
\end{centering}
\caption{Apodised Galactic and point source masks which we combine to use in
the likelihood analysis. From left to right and top to bottom, the
panels show the GA34, GA38, GA54 and PSA82 masks.}
\label{fig:masks_apod} 
\end{figure*}

In order to avoid power leakage, we also derive a series of apodised masks. 
For the Galactic masks, we proceed as follows.
First, we smooth each mask with a five-degree Gaussian
beam, and zero any pixels below a threshold of $0.15$. 
We then subtract $0.15$ from the remaining pixels, and rescale the resulting 
map by $1/(1-0.15)$. The resulting masks are shown
in Fig.~\ref{fig:masks_apod}.  In order to retain sufficient sky
area for the most conservative sky mask, a slightly less aggressive
version of mask G22 was used to seed the apodization process for that
case. Each point source is apodised to
30\arcmin\ FWHM, resulting in the PSA82 point source mask. A set of the resulting masks are shown in Fig.~\ref{fig:masks_apod}.

For all cosmology analyses, we use three of the apodised galactic and point sources masks: CL31, CL39, and CL49, which are shown in Fig.~\ref{fig:CLmasks}. Table~\ref{tab:masks} summarizes the various masks. For a limited set of tests in \S\ref{sub:Galactic-emission} we also used `mask0' and `mask1' which combine the non-apodised Galactic masks G22 and G35 with the apodised PSA82 point source mask. 

\begin{table}[tmb] 
\begingroup 
\newdimen\tblskip \tblskip=5pt
\caption{Series of Galactic and cosmology masks.\label{tab:masks}}
\vskip -6mm
\footnotesize 
\setbox\tablebox=\vbox{ %
\newdimen\digitwidth 
\setbox0=\hbox{\rm 0}
\digitwidth=\wd0
\catcode`*=\active
\def*{\kern\digitwidth}
\newdimen\signwidth
\setbox0=\hbox{+}
\signwidth=\wd0
\catcode`!=\active
\def!{\kern\signwidth}
\newdimen\decimalwidth
\setbox0=\hbox{.}
\decimalwidth=\wd0
\catcode`@=\active
\def@{\kern\signwidth}
\halign{ 
    \hfil#\hfil\tabskip=1em& 
    \hfil#\hfil\tabskip=1em& 
    \hfil#\hfil\tabskip=1em& 
    \hfil#\hfil \tabskip=0pt\cr
\noalign{\doubleline}
Series& Galactic Mask& Apodised Galactic masks& Cosmology Mask\cr
\noalign{\vskip 2pt\hrule\vskip 2pt}
0&G22&GA21&-\cr
1&G35&GA34&CL31\cr
2&G45&GA38&CL39\cr
3&G56&GA54&CL49\cr
4&G65&GA60&-\cr
\noalign{\vskip 2pt\hrule\vskip 2pt}
}}
\endPlancktable 
\endgroup
\end{table}

%% file: App-Chance-correlations.tex
\section{Chance correlations and inter-frequency consistency tests\label{app:ChanceCorrealtions}}

Here we explicitly show that, even if the foreground contamination
is much smaller than the CMB, chance cross-correlations can produce
scatter in the inter-frequency power spectra that is large in the
signal dominated regime. To see this, consider the case of two frequencies.
Frequency 1 provides a faithful map of the CMB fluctuations. Frequency
2 contains a foreground component F. We therefore write the maps at
the two frequencies as: \beglet 
\begin{eqnarray}
{\bf X}_{1} & = & {\bf S},\label{P1a}\\
{\bf X}_{2} & = & {\bf S}+{\bf F},\label{P1b}
\end{eqnarray}
\endlet with spherical transforms \beglet 
\begin{eqnarray}
a_{\ell m}^{1} & = & S_{\ell m},\\
a_{\ell m}^{2} & = & S_{\ell m}+F_{\ell m}.
\end{eqnarray}
\endlet The power spectra of the two maps are therefore: \beglet
\begin{eqnarray}
C_{\ell}^{1} & = & \frac{1}{(2\ell+1)}\sum_{m}S_{\ell m}S_{\ell m}^{*}=C_{\ell}^{{\rm CMB}},\\
C_{\ell}^{2} & = & \frac{1}{(2\ell+1)}\sum_{m}(S_{\ell m}+F_{\ell m})(S_{\ell m}^{*}+F_{\ell m}^{*})\nonumber \\
 & = & C_{\ell}^{{\rm CMB}}+2C_{\ell}^{{\rm CMB}\times{\rm F}}+C_{\ell}^{{\rm F}},
\end{eqnarray}
\endlet and the difference between the power spectra is 
\begin{equation}
C_{\ell}^{2}-C_{\ell}^{1}=2C_{\ell}^{{\rm CMB}\times{\rm F}}+C_{\ell}^{{\rm F}}.\label{IFS1}
\end{equation}

If the CMB is uncorrelated with the foreground, the first term will
average to zero over a large number of CMB realizations. But we observe
only one realization of the CMB, and so the cross-term will dominate
the inter-frequency residuals even if the foreground contamination
is much lower than the CMB ($C^{{\rm F}}\ll C^{{\rm CMB}}$). This
is the origin of the excess scatter between the $143$ and $217$\,GHz
power spectra at low multipoles shown in Fig.~\ref{fig:interband_residues}.

\begin{figure}[h]
\begin{centering}
\includegraphics[bb=5bp 25bp 795bp 500bp,clip,width=1\columnwidth]{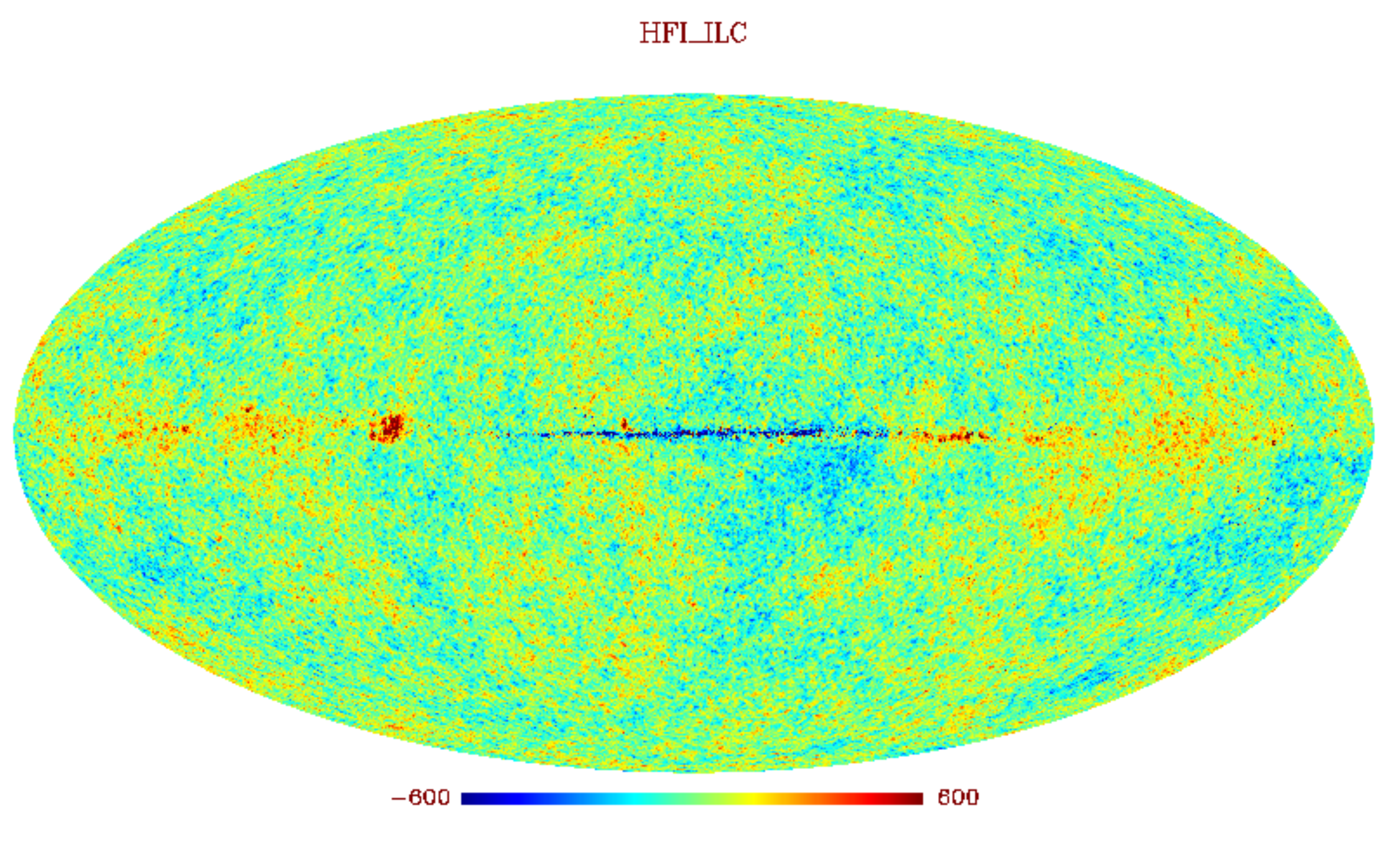}
\par\end{centering}

\begin{centering}
\includegraphics[bb=5bp 25bp 795bp 500bp,clip,width=1\columnwidth]{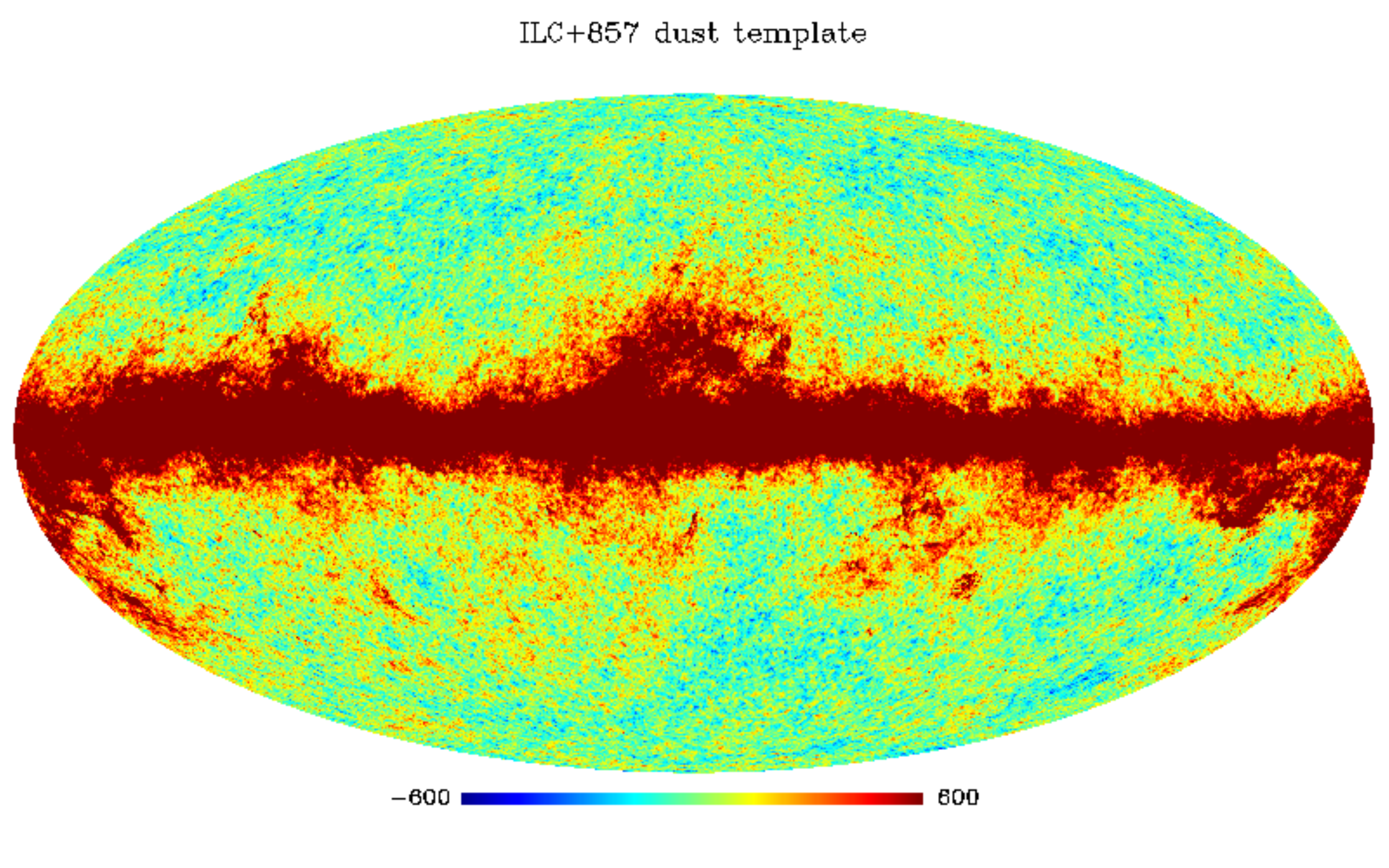}
\par\end{centering}

\begin{centering}
\includegraphics[bb=5bp 25bp 795bp 500bp,clip,width=1\columnwidth]{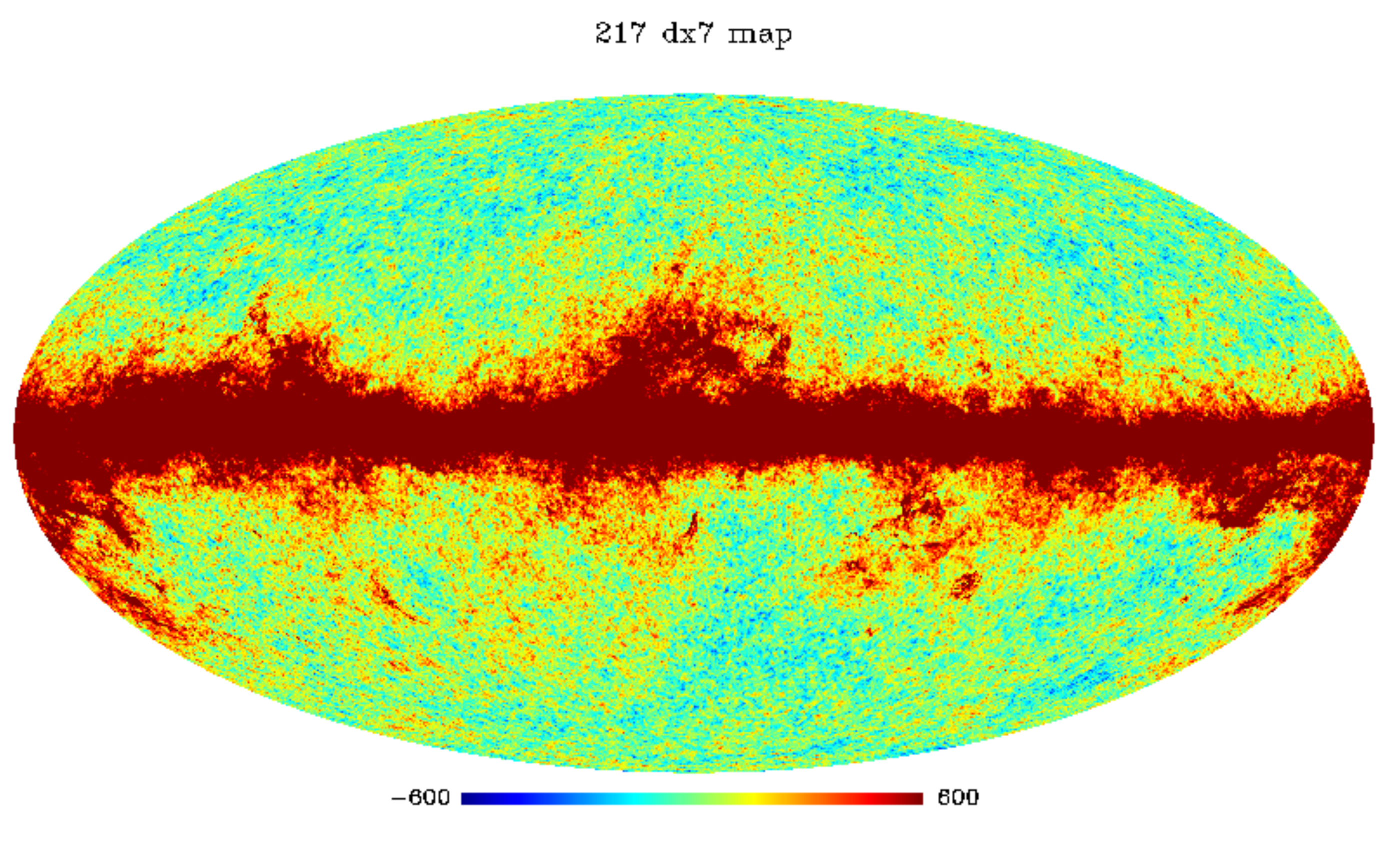}
\par\end{centering}

\centering{}\caption{{\it Top}: An ILC CMB map constructed from $100$-$353$\,GHz
maps. {\it Middle}: The ILC map added to the 857\,GHz
map, scaled to match the diffuse dust emission at 217\,GHz. This map
(and an equivalent at 143\,GHz) is used as the CMB+dust template
to assess CMB/foreground cross correlations. {\it Bottom}: The real 217\,GHz map.\label{217maps}}
\end{figure}

We construct a specific example of this. The upper map in Fig.~\ref{217maps}
shows an ILC map, estimating the CMB, generated from the $100$, $143$, $217$, and $353$\,GHz
maps. The map in the middle panel shows a `fake' 217\,GHz map, \textit{i.e.}
the sum of the ILC map and the $857$\,GHz map scaled in amplitude
to match dust emission at 217\,GHz. The real $217$\,GHz map is
shown in the lower panel of Figure \ref{217maps}. The `fake' 217\,GHz
map is evidently quite a good match to the real $217$\,GHz. By rescaling
the $857$\,GHz map to estimate the dust emission at $143$\,GHz, we can generate
a `fake' 143\,GHz map in an analogous way. These `fake' maps each
contain two components by construction, and so the inter-frequency
residuals from these maps will be dominated by the CMB-foreground
cross term in Eq.~\ref{IFS1}.

\begin{figure}
\begin{centering}
\includegraphics[bb=40bp 170bp 600bp 600bp,clip,width=1\columnwidth]{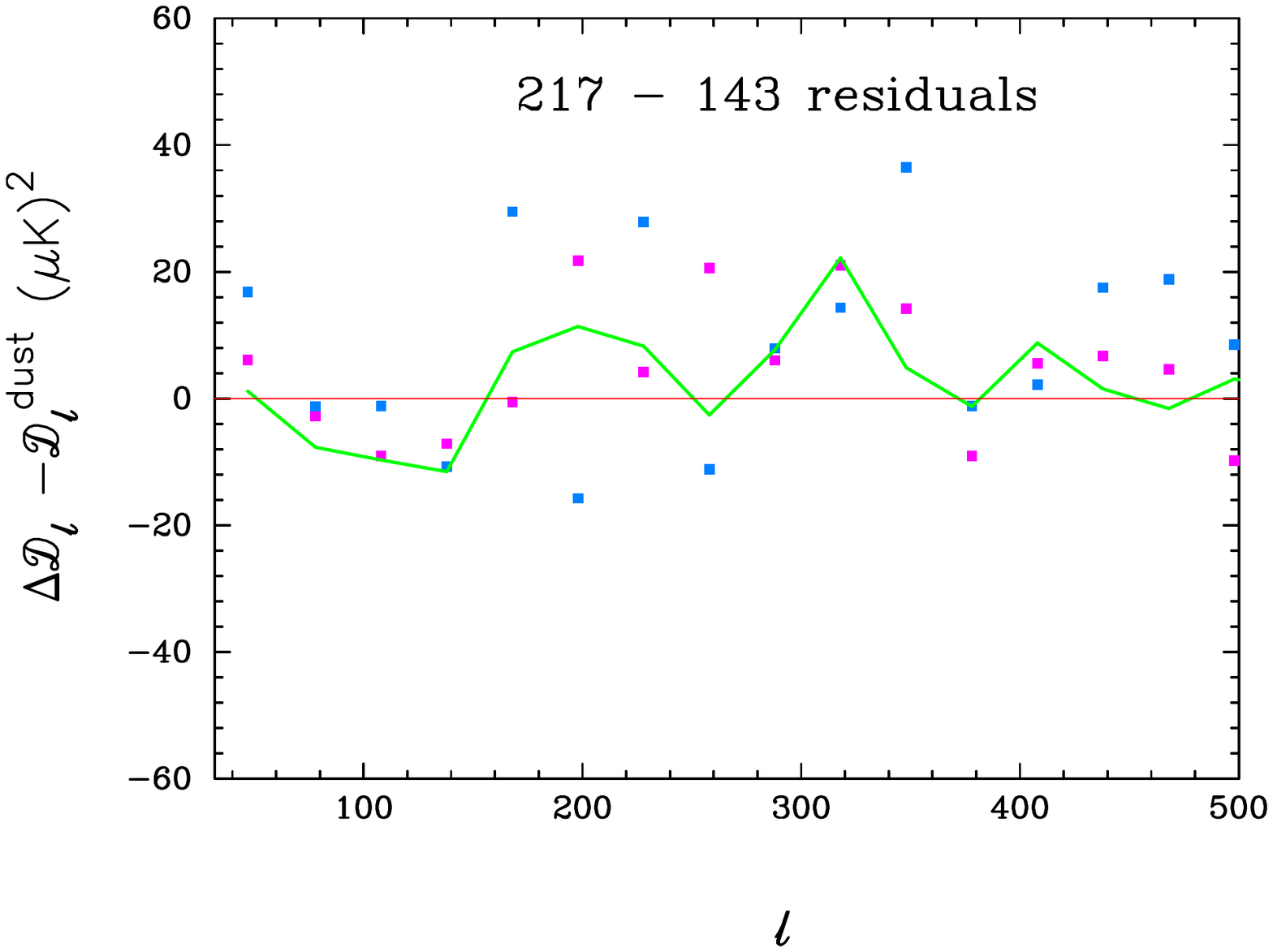}
\par\end{centering}

\caption{The magenta points show the double difference power spectrum (as in
Figure \ref{fig:doublediff}. The green line shows the same double
difference spectrum computed from the ILC + 857 dust template maps
described in the text. The blue points show the difference of the
$217$ and $143$ power spectra for mask$\_$1. In all cases, the
smoothed dust power spectrum model of equation (\ref{SSD1}) has been
subtracted.}

\label{doublediffsim} 
\end{figure}

The $143$ and $217$~GHz power spectrum difference from these fake maps
are compared to the $143-217$ residuals of the real data in Fig.~\ref{doublediffsim}. The magenta points show the same mask1-mask0
double-difference power spectrum between $217$ and $143$\,GHz as
shown in Fig.~\ref{fig:doublediff}. The only difference here is
that the smoothed dust fit of Eq.~\ref{SSD1} has been subtracted
from the spectra so that the points scatter around zero. There are
advantages to using the double difference because: (a) it is insensitive
to calibration differences between frequencies; (b) the contrast between
dust emission and other foregrounds (point sources/SZ) is stronger
in the area of sky defined by mask1 - mask0 and so the double
difference power spectrum should be closer to the results from the
fake maps, which use only a dust template%
\footnote{Actually, as demonstrated in Fig.~\ref{857fit}, the CIB dominates
over Galactic dust emission over most of the area of mask0, but
this will not be a precise template for the CIB emission at cosmological
channels: (a) because the spectrum of the CIB differs slightly from
Galactic dust; (b) the CIB emission decorrelates from high to low
frequencies because lower frequencies probe galaxies at higher redshifts.%
} . The solid green line shows the double difference power spectrum
computed from the fake maps. The amplitude of the scatter from the
fake maps and the real data are very similar. In fact, there is almost
point-by-point agreement between the results from the real data and
the fake maps. This provides compelling evidence that the observed
inter-frequency scatter at low multipoles is dominated by the CMB-foreground
cross term in Eq.~\ref{IFS1} rather than some mysterious systematic
effect in the data.

The blue points in Figure \ref{doublediffsim} show the difference
of the $217$ and $143$\,GHz power spectra for mask1. The scatter
at multipoles $\lesssim100$ is almost identical to the scatter of
the purple points, but increases slightly at higher multipoles. This
behaviour is expected and is caused by the additional foreground components
(CIB/point sources/SZ) which become comparable in amplitude to Galactic
dust at multipoles greater than a few hundred.

\begin{figure}[h]
\begin{centering}
\includegraphics[bb=40bp 30bp 640bp 590bp,clip,width=1\columnwidth]{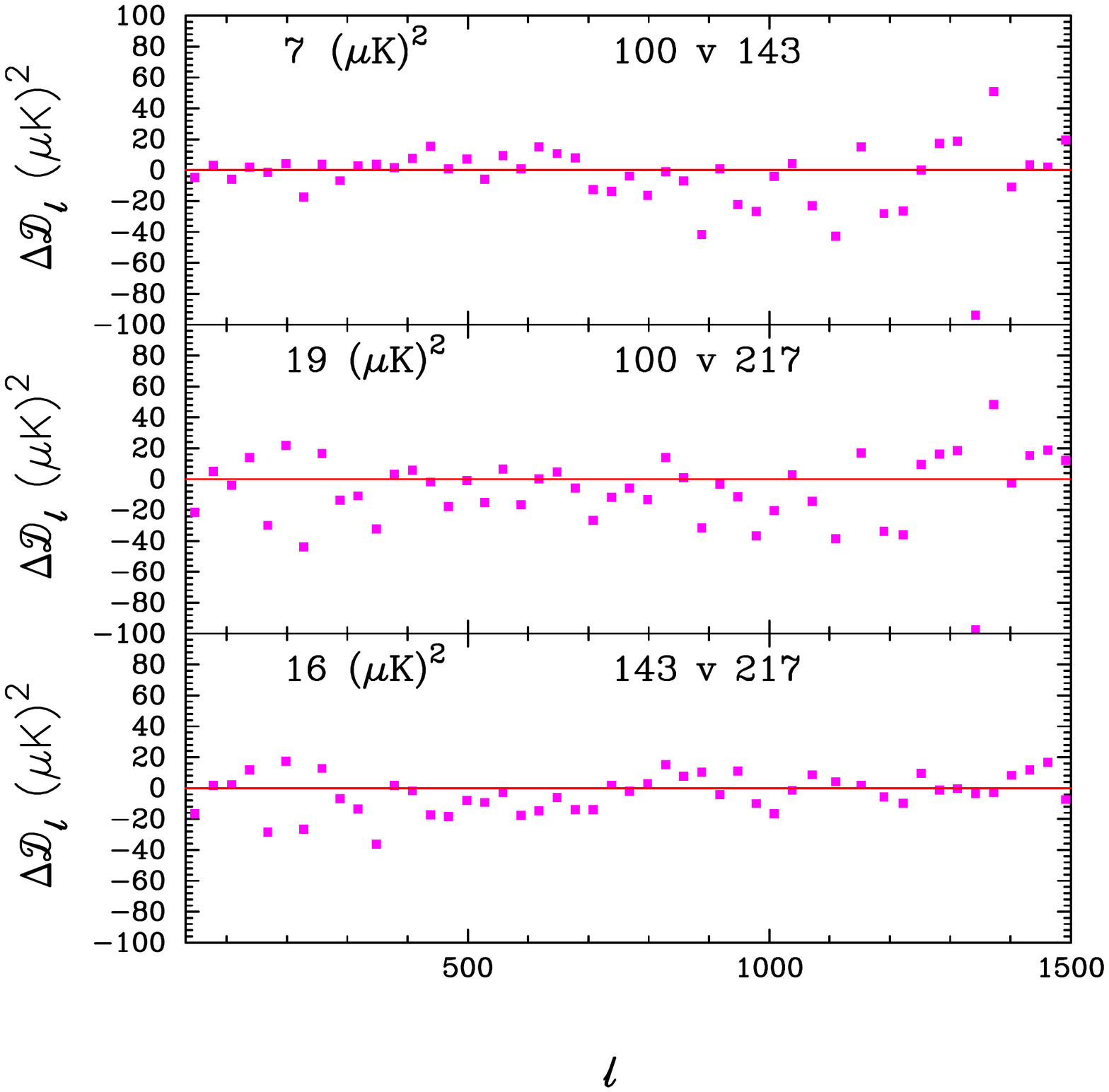}
\par\end{centering}

\caption{Power spectrum residuals between the summed cross-spectra at three
HFI frequencies. The top panel shows $100\times100$ vs $143\times143$,
the middle panel shows $100\times100$ vs $217\times217$ and the
bottom panels shows $143\times143$ vs $217\times217$. A `best fit'
model for unresolved foregrounds has been subtracted from the power
spectrum at each frequency (\textcolor{red}{see Section xxx for further
details)}. The scatter in the multipole range $50\le\ell\le500$ is
listed in each panel.}

\label{100-217res} 
\end{figure}

We would also expect a strong dependence of the inter-frequency residuals
with frequency at low multipoles. Diffuse Galactic emission rises
steadily in amplitude from $100$\,GHz to $217$\,GHz and hence
we would expect the inter-frequency scatter to rise as we go up in
frequency. This is what we see in the real data (shown in Fig.~\ref{100-217res}).
Since diffuse Galactic emission is well approximated by the 857\,GHz
map at all frequencies, we can predict the scatter seen in this figure
by scaling 857\,GHz to lower frequencies. For $143-217$~GHz we observe
a scatter of $16$ \muk over the multipole range $50\le\ell\le500$.
So, from the $857$~GHz scalings to lower frequencies we predict the scatter given in the Table, which is in excellent agreement with the scatter seen in Figure \ref{100-217res}.

\begin{center}
\begin{tabular}{ccc}
\hline 
\hline
 & Predicted scatter & Observed scatter\tabularnewline
\hline 
100 - 143 & \selectlanguage{english}%
$7\;(\mu{\rm K})^{2}$\selectlanguage{british}%
 & \selectlanguage{english}%
$7\;(\mu{\rm K})^{2}$\selectlanguage{british}%
\tabularnewline
100 - 217 & \selectlanguage{english}%
$18\;(\mu{\rm K})^{2}$\selectlanguage{british}%
 & \selectlanguage{english}%
$19\;(\mu{\rm K})^{2}$\selectlanguage{british}%
\tabularnewline
\hline 
\end{tabular}
\label{tab:scatter}
\par\end{center}

%% file: App-sec7.tex
\subsection{Detailed Validity checks\label{App:Checks}}

In this appendix we show the full set of figures for the distribution of the cosmological and foreround model parameters for the suite of tests described in Sect.~\ref{sec:consistency}. We also show the correlation matrix between these parameters and the calibration coefficents of each detector.

\begin{figure*}[h]
\begin{centering}
\includegraphics[bb=55bp 45bp 535bp 530bp,clip,width=1\textwidth]{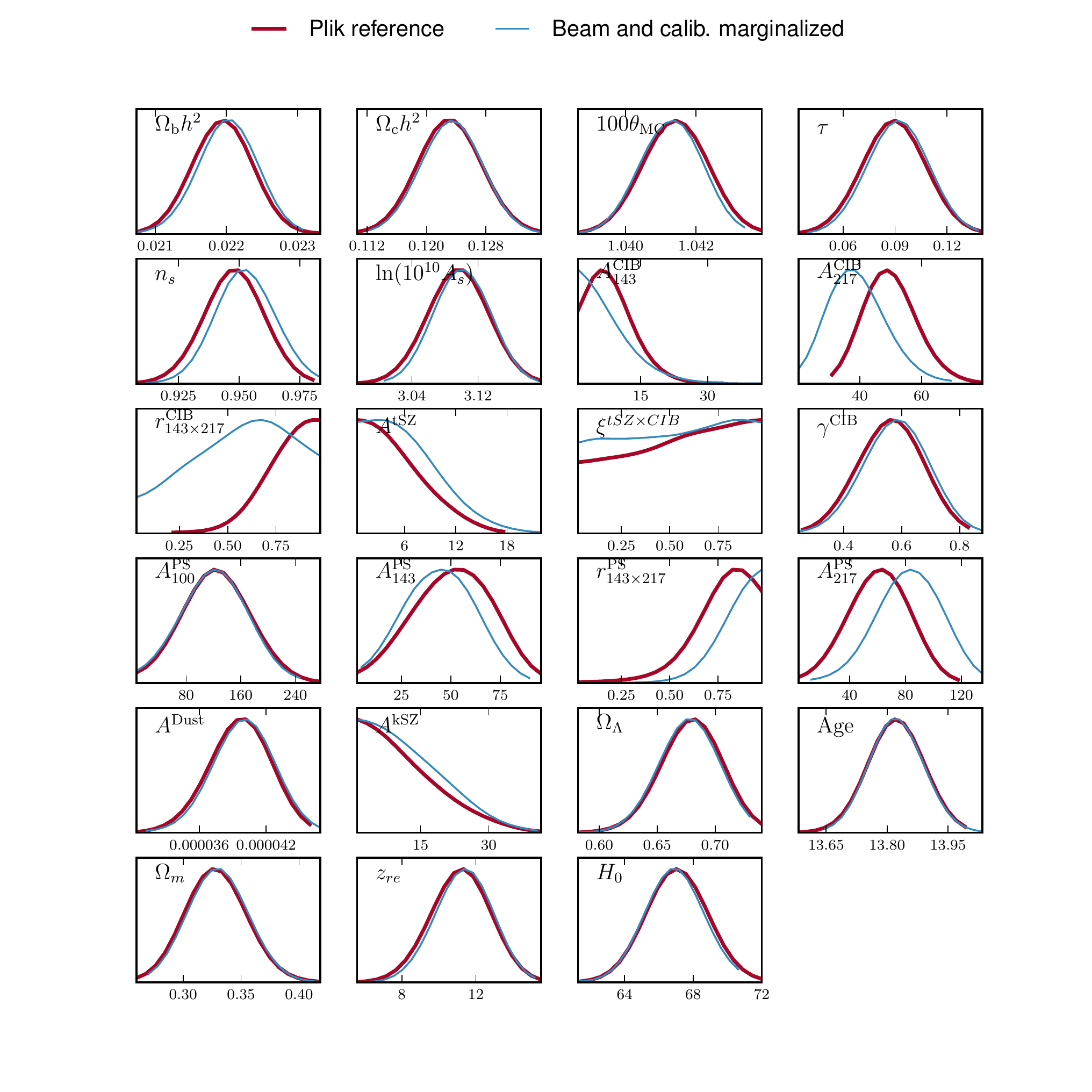}
\par\end{centering}
\centering{}\caption{
Impact on cosmological and foreground parameters of fixing the
calibration and beam coefficients at their maximum posterior value
(red), compared to marginalizing over these nuisance parameters (blue).
\label{fig:beamcalib-marginals}
}
\end{figure*}

\begin{figure*}[h]
\begin{centering}
\includegraphics[bb=100bp 280bp 1330bp 1350bp,clip,width=1\textwidth]{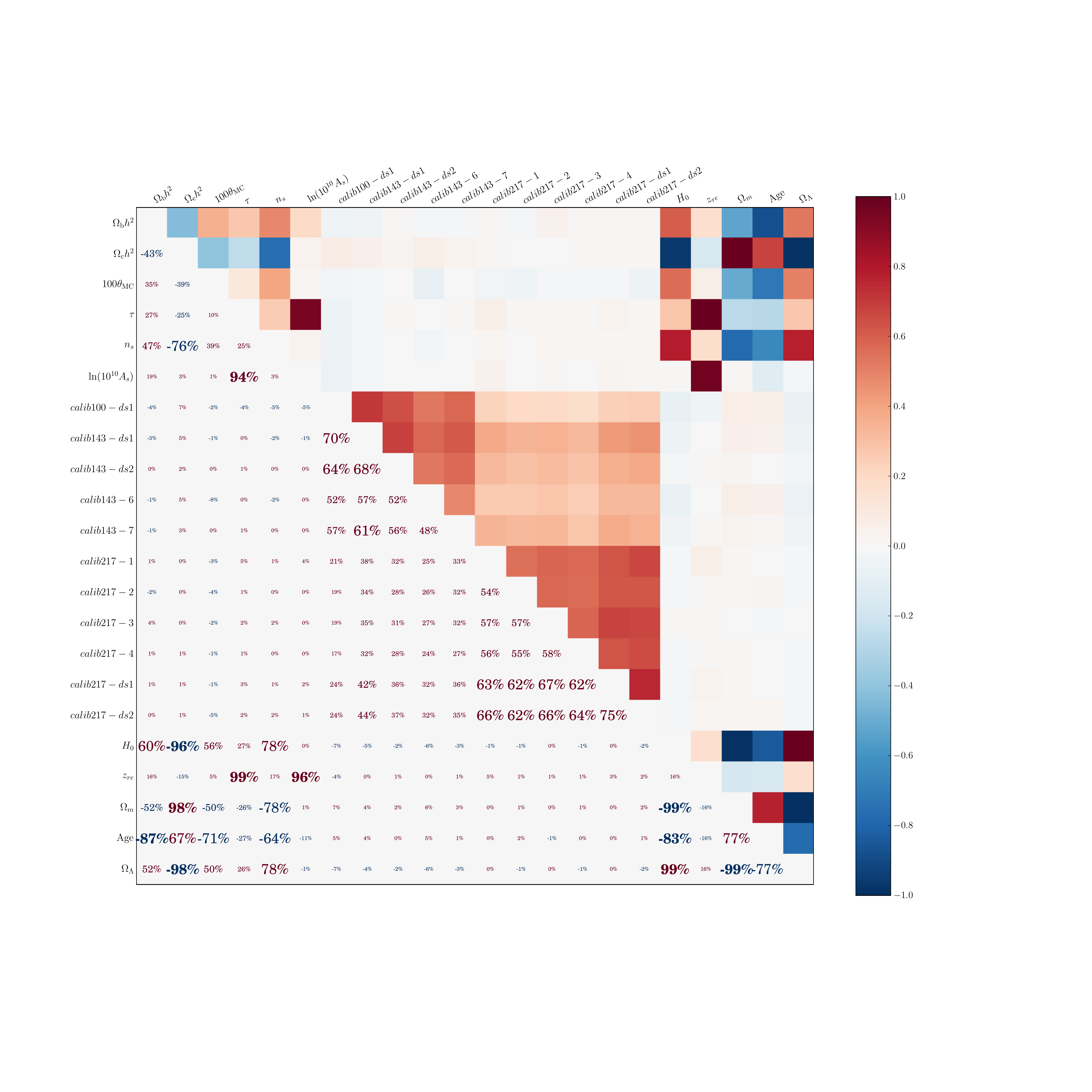}
\par\end{centering}
\caption{Covariance matrix between cosmological parameters and
detector cross-calibration coefficients (the calibration of the 143-5 detector
is set to $1$ to avoid an overall degeneracy with the total signal
amplitude).
\label{fig:beamcalib-cosmocalib-covariance}
}
\end{figure*}

\begin{figure*}[h]
\begin{centering}
\includegraphics[bb=100bp 290bp 1390bp 1400bp,clip,width=1\textwidth]{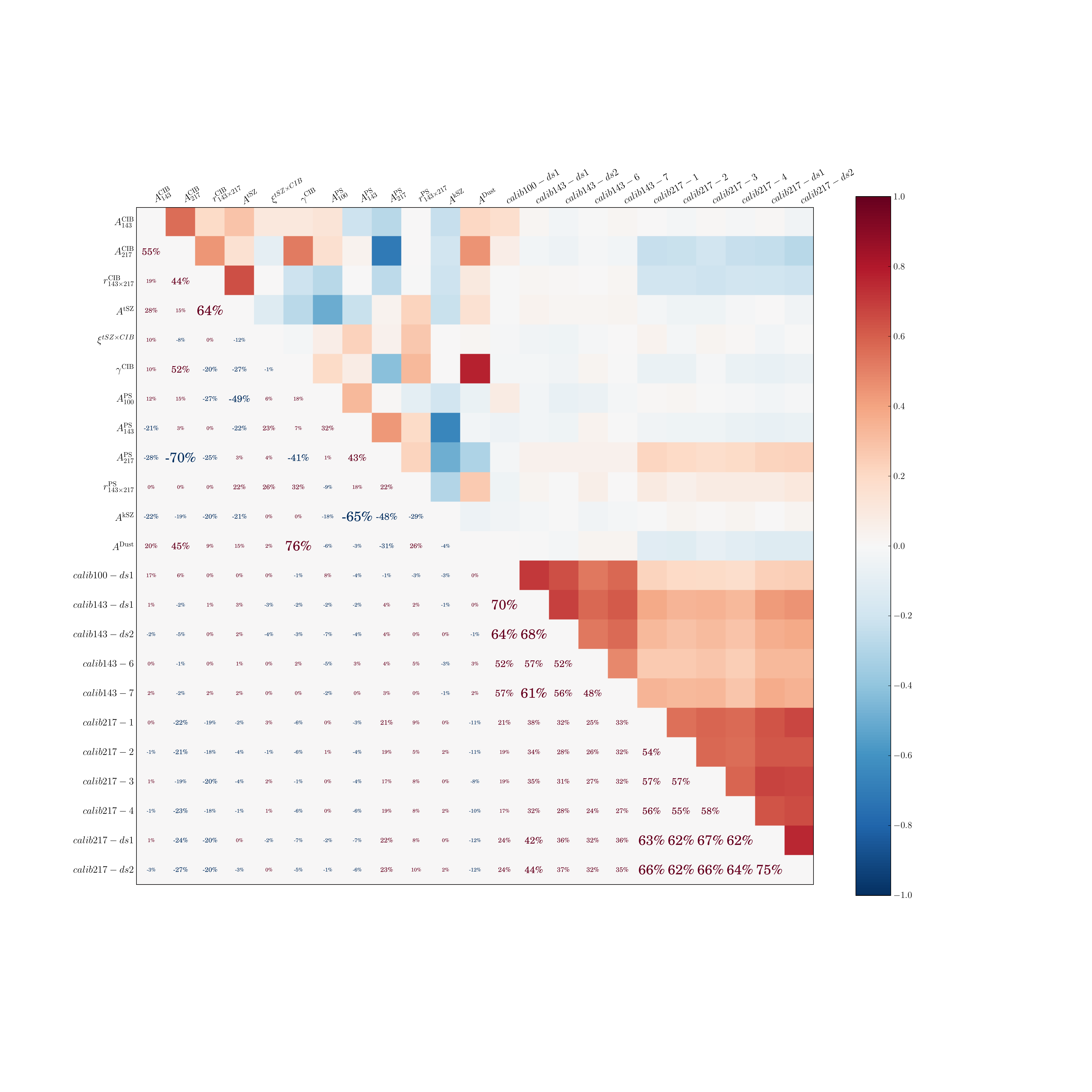}
\par\end{centering}
\caption{Covariance matrix between foreground parameters and detector
cross-calibration coefficients, as in Fig.~\ref{fig:beamcalib-cosmocalib-covariance}.
\label{fig:beamcalib-fgcalib-covariance}
}
\end{figure*}

\begin{figure*}[h]
\begin{centering}
\includegraphics[bb=55bp 45bp 535bp 530bp,clip,width=1\textwidth]{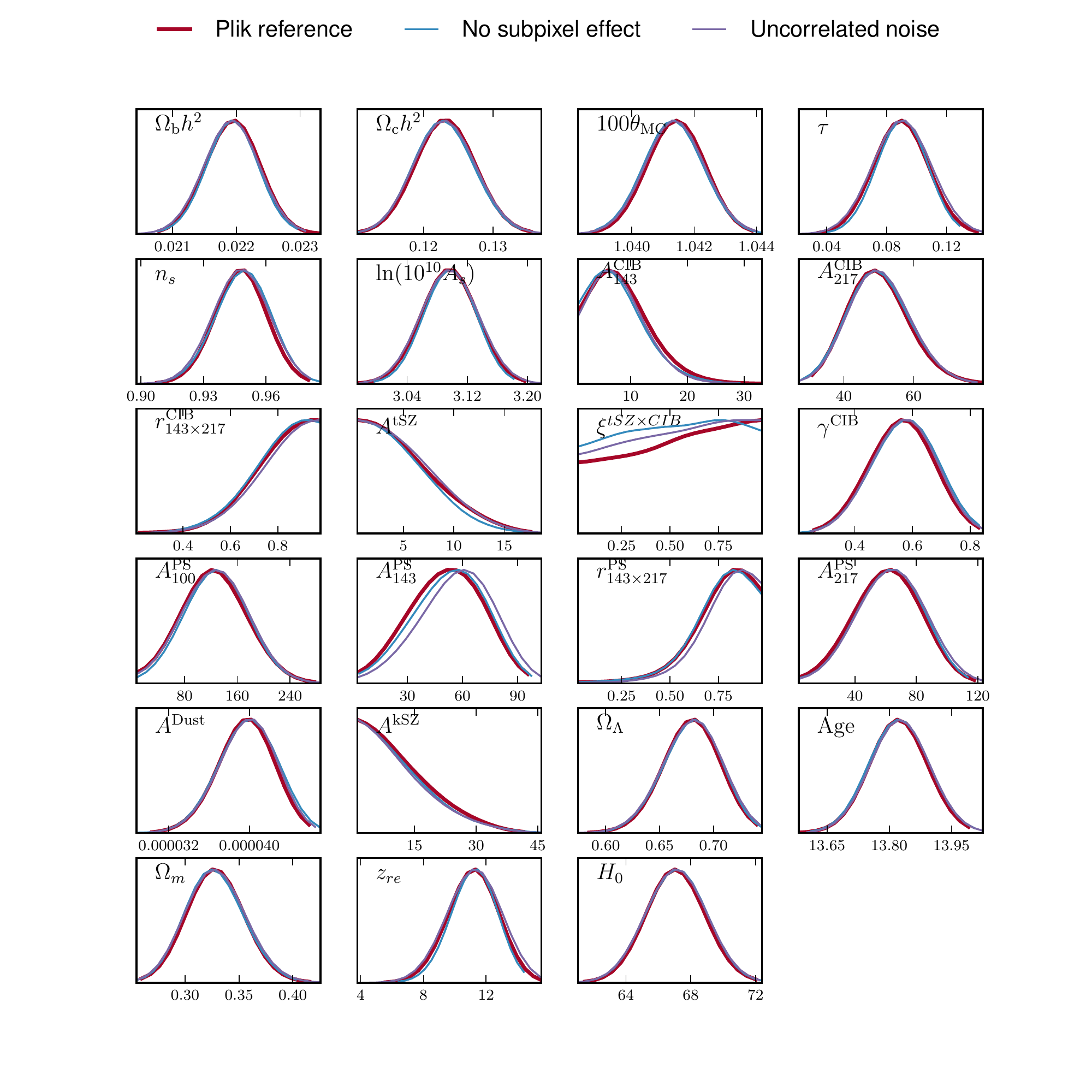}
\par\end{centering}
\caption{Impact of removing the sub-pixel effect (blue) or the
correlated noise between detector sets (purple), compared to
the reference case (red).
\label{fig: plik-validation-faint_test}
}
\end{figure*}

\begin{figure*}[h]
\begin{centering}
\includegraphics[bb=55bp 45bp 535bp 530bp,clip,width=1\textwidth]{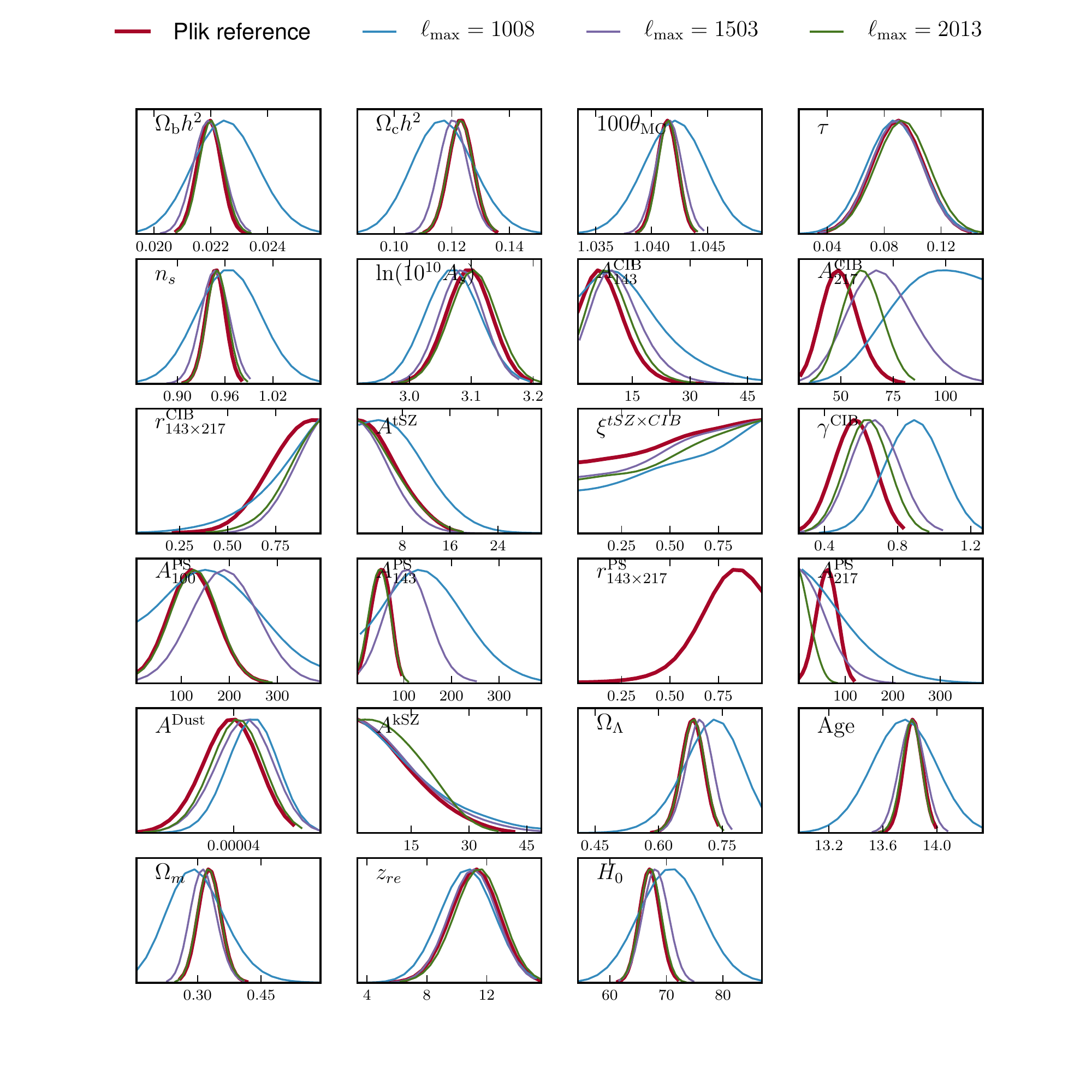}
\par\end{centering}
\caption{
Impact on the cosmological and foreground parameters of varying the maximum multipole, $\ell_{\mathrm{max}}$. We consider $\ell_{\mathrm{max}}=1008$ (blue), $1503$ (purple), and $2013$ (green); compared to the reference $\ell_{max}=2508$ (red).
\label{fig:varying-lmax}
}
\end{figure*}

\begin{figure*}[h]
\begin{centering}
\includegraphics[bb=55bp 45bp 535bp 530bp,clip,width=1\textwidth]{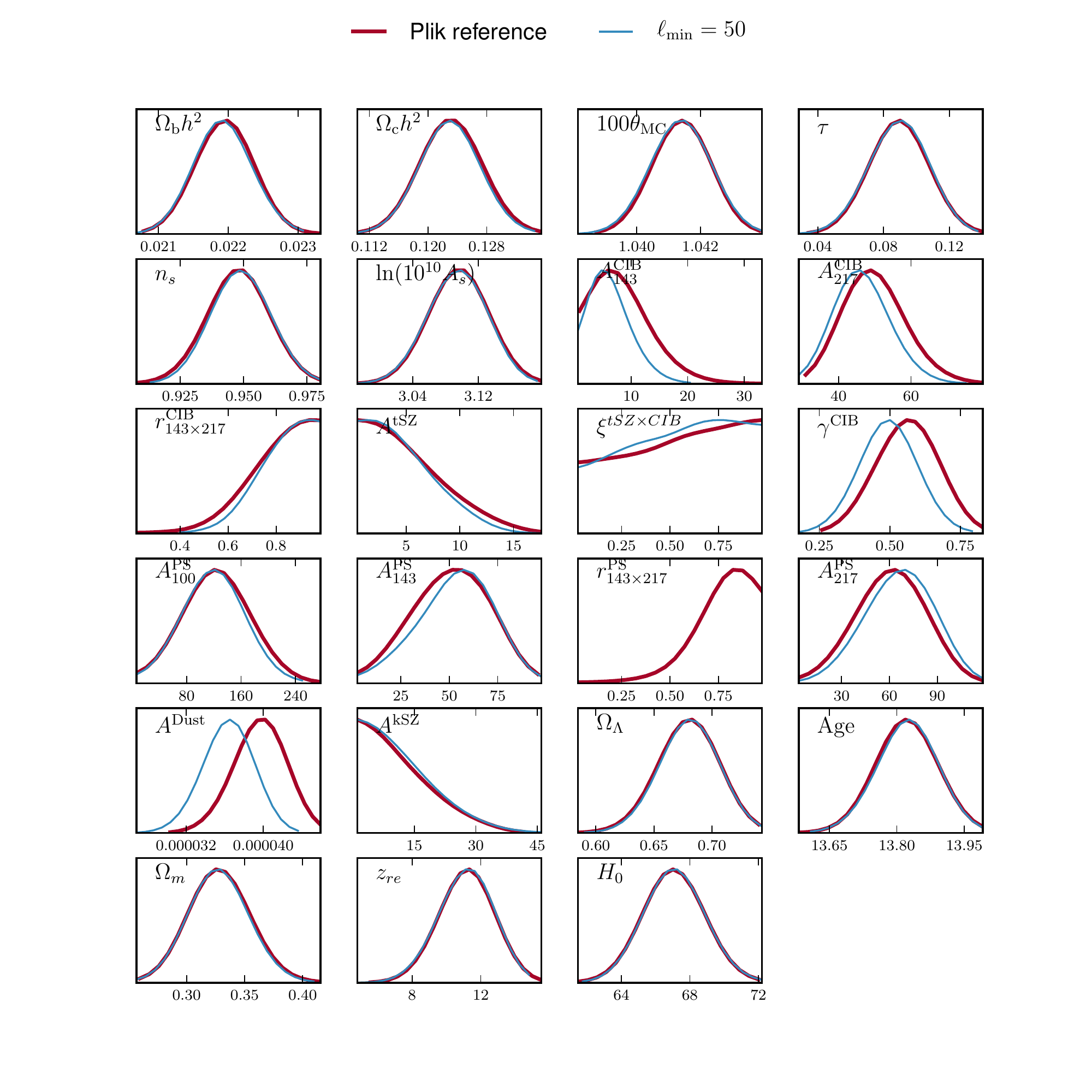}
\par\end{centering}
\caption{Impact of changing the minimum multipole from $\ell_{\mathrm{min}}=100$ (red, as in reference case) to $\ell_{\mathrm{min}}=50$ (blue). 
\label{fig:varying-lmin}
}
\end{figure*}

\begin{figure*}[h]
\begin{centering}
\includegraphics[bb=55bp 45bp 535bp 530bp,clip,width=1\textwidth]{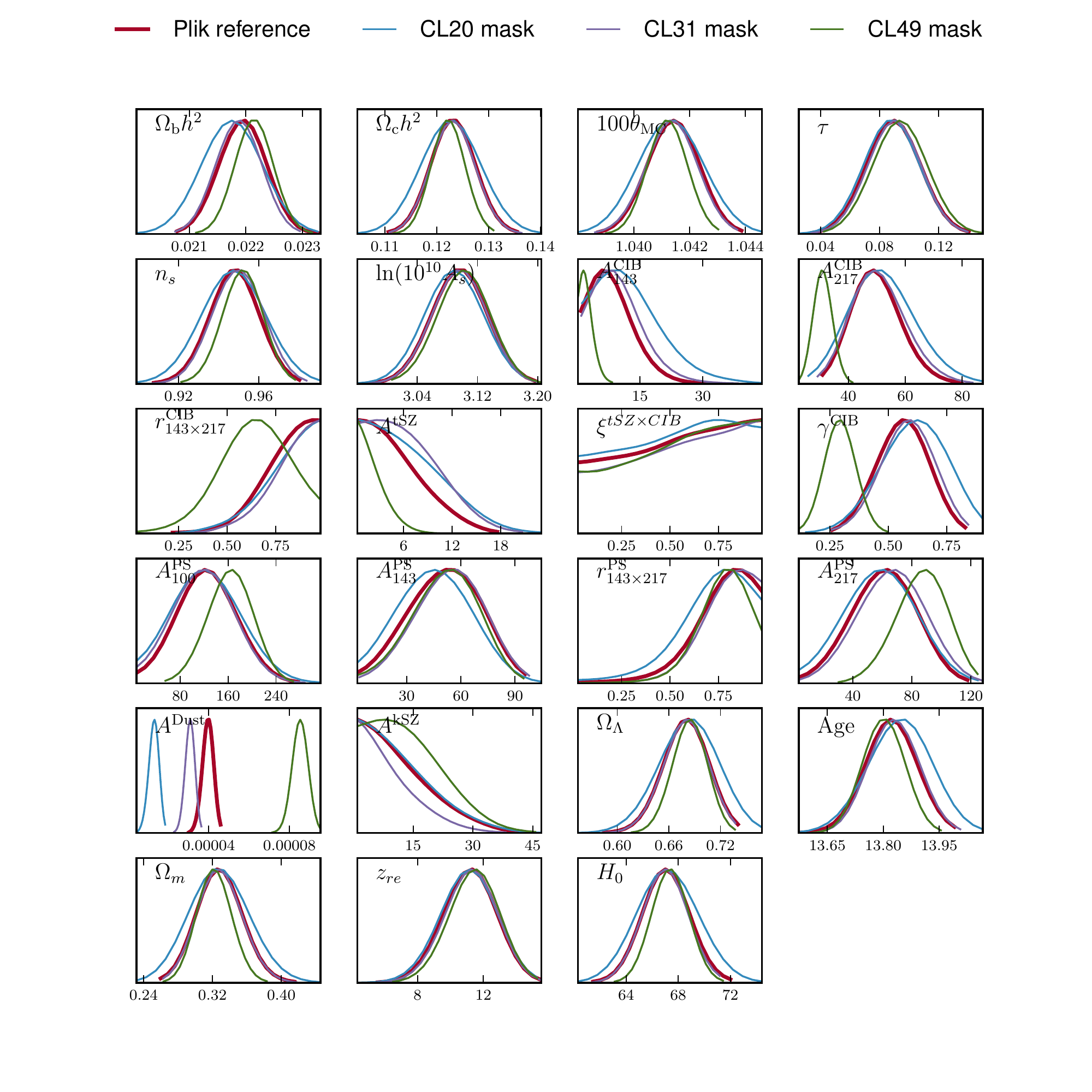}
\par\end{centering}
\caption{Impact of changing the Galactic mask, increasing the sky area used from G22 (blue), G35 (purple), G45 (red, reference), to the least conservative G56 (green). All results use the $100\leq\ell\leq2508$ range. Note that the 
\camspec\ likelihood uses the G35 mask for the $143$ and $217$~GHz channels, and 
G56 at $100$~GHz, but with a restricted, composite multipole range.
\label{fig:varying-mask}
}
\end{figure*}

\begin{figure*}[h]
\begin{centering}
\includegraphics[bb=55bp 45bp 535bp 530bp,clip,width=1\textwidth]{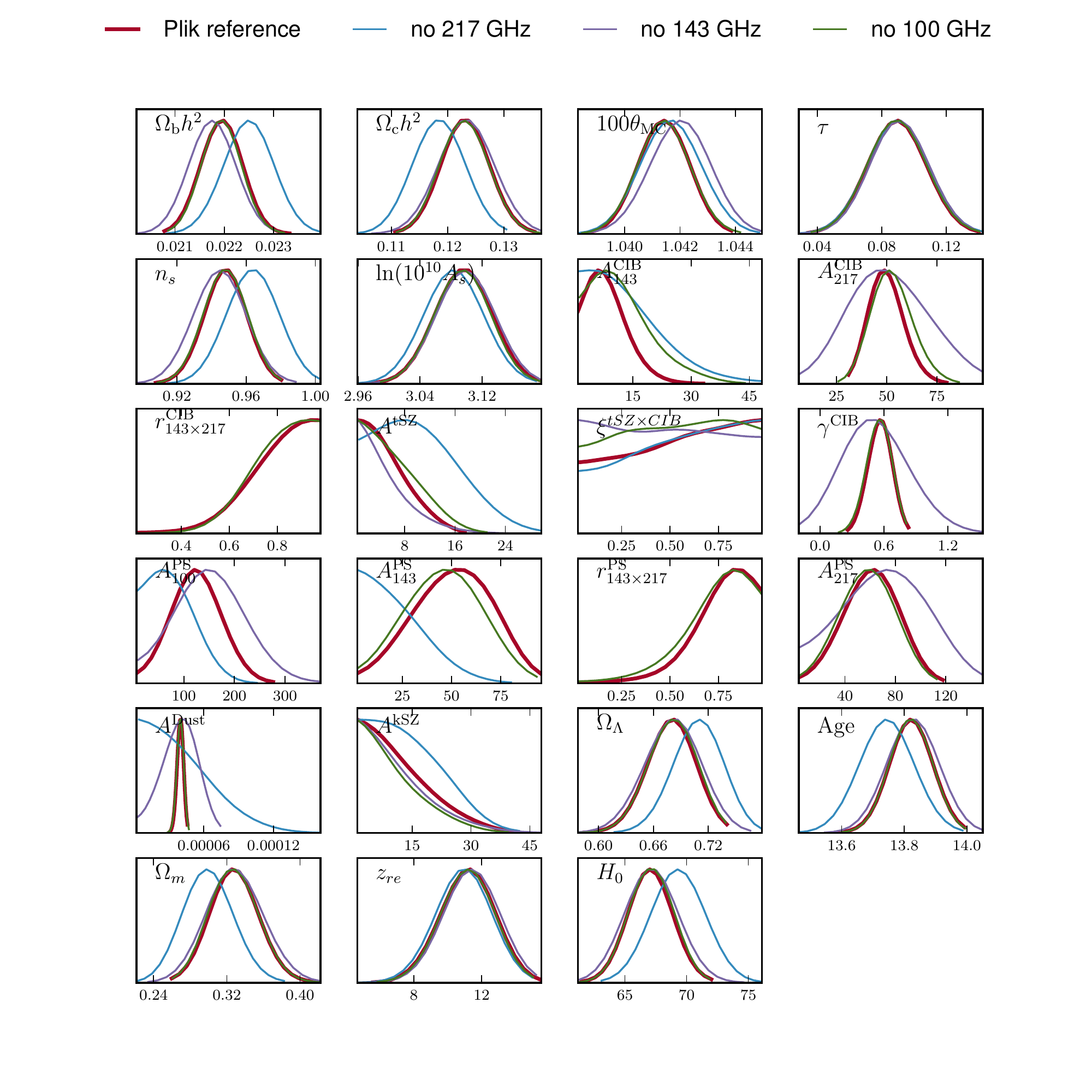}
\par\end{centering}
\caption{Impact on parameters of removing one single frequency channel (i.e., \emph{all} spectra with at least one frequency in the removed
channel). Results are shown removing the $100$~GHz (green), $143$~GHz (purple), or $217$~GHz (blue) channels, compared to the reference case (red). Where the $217$~GHz channel is removed, the CIB spectral index is held fixed at $\gamma^{\rm CIB}=0.6$. 
\label{fig:removing-one-channel}}
\end{figure*}


\subsection{Cosmological parameters from Planck 70\,GHz data\label{App:70GHz-Cosmo}}

For this analysis we implement the pseudo-$C_{\ell}$ method described in \cite{Hietal02}
extended to derive both auto- and cross-power spectra from the 70\,GHz maps \citep[see, e.g.][for a comparison between the two estimators]{Polenta_CrossSpectra}.
The noise power spectrum and the covariance matrix are computed using
1000 realistic \Plancks simulations \citep[FFP6,][]{planck2013-p01} of both signal and noise maps. The beam
window functions are presented in \cite{planck2013-p02d}, and mode-coupling
kernels to correct for incomplete sky coverage are computed from formulae
analogous to those in Appendix.~\ref{app:Power-Spectra}.

\begin{figure}[h]
\centering{}\includegraphics[width=1\columnwidth]{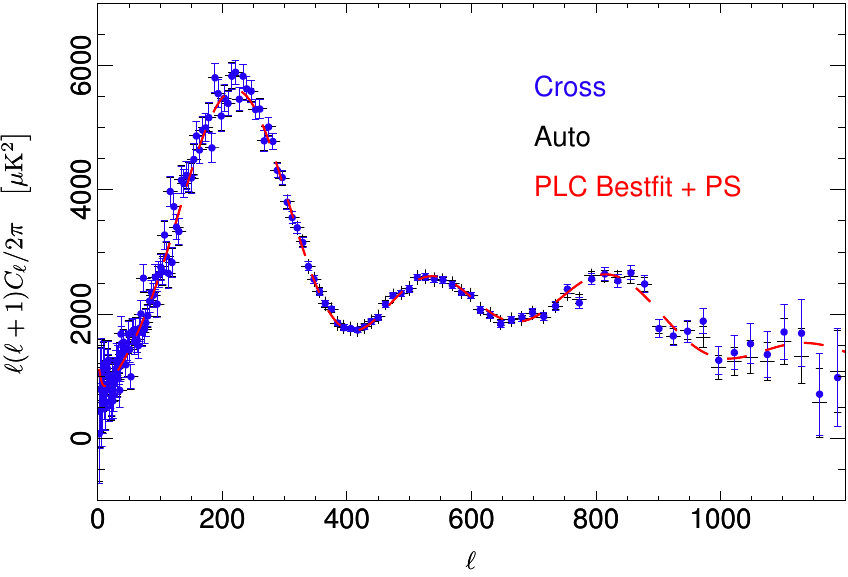}\caption{The power spectrum from the \Plancks 70\,GHz channel, before removal
of unresolved sources. Both the auto-spectrum from the 70\,GHz maps,
and the weighted cross spectra from maps of the three 70\,GHz horn
pairs are shown. The best-fitting cosmological model from \camspec,
shown for comparison with the best-fitting source power added, gives
a good fit to the data. \foreignlanguage{english}{\label{fig:Planck-70-GHz}}}
\end{figure}

\begin{figure*}[h]
\begin{centering}\includegraphics[width=1\textwidth]{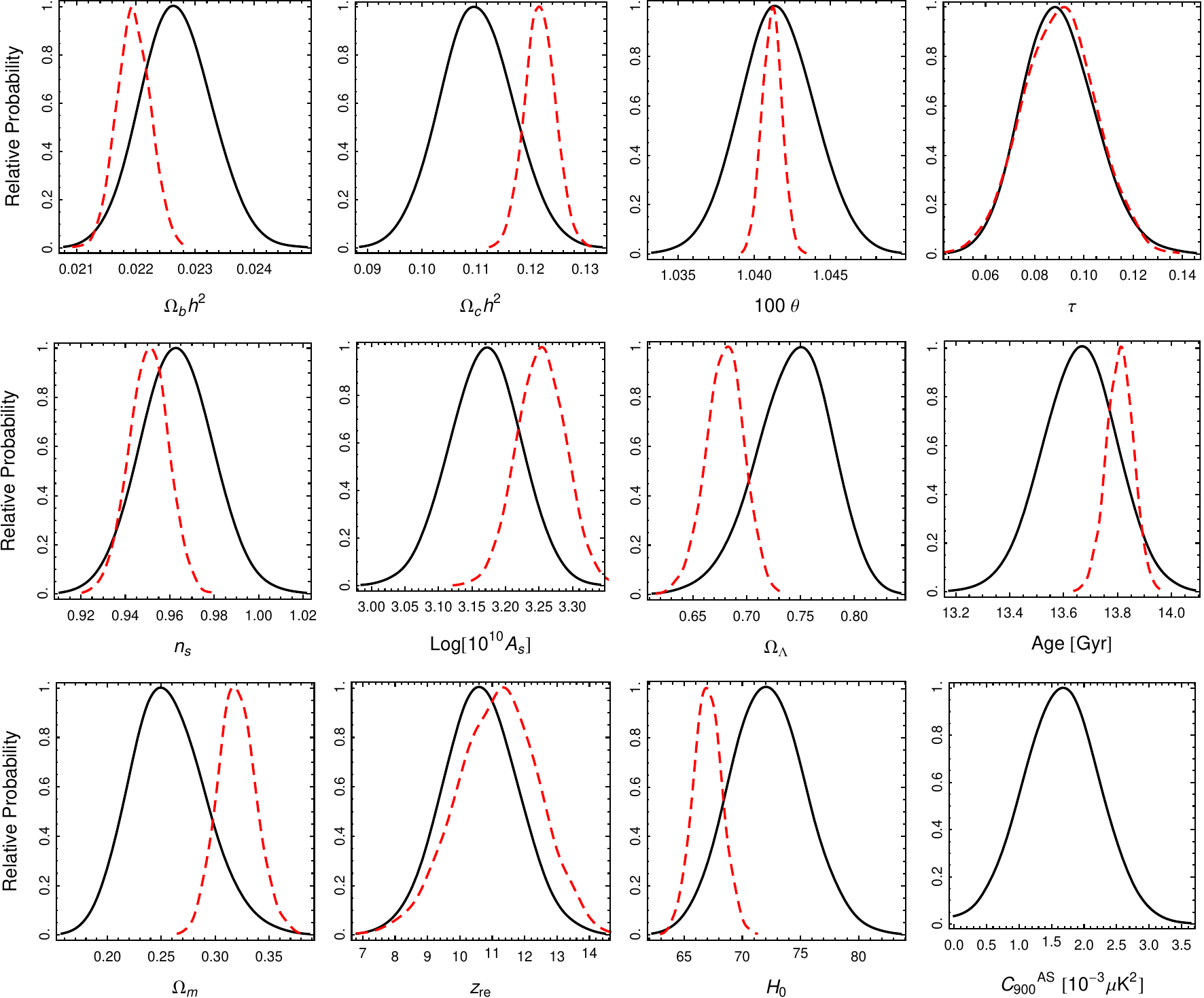}
\par\end{centering}
\centering{}\caption{Cosmological parameters derived from the 70\,GHz maps (solid black) are compared to \camspec\ results (red dashed). \label{fig:70Ghz_parameters_LFIvsHFI} }
\end{figure*}

\begin{figure*}[h]
\begin{centering}\includegraphics[width=1\textwidth]{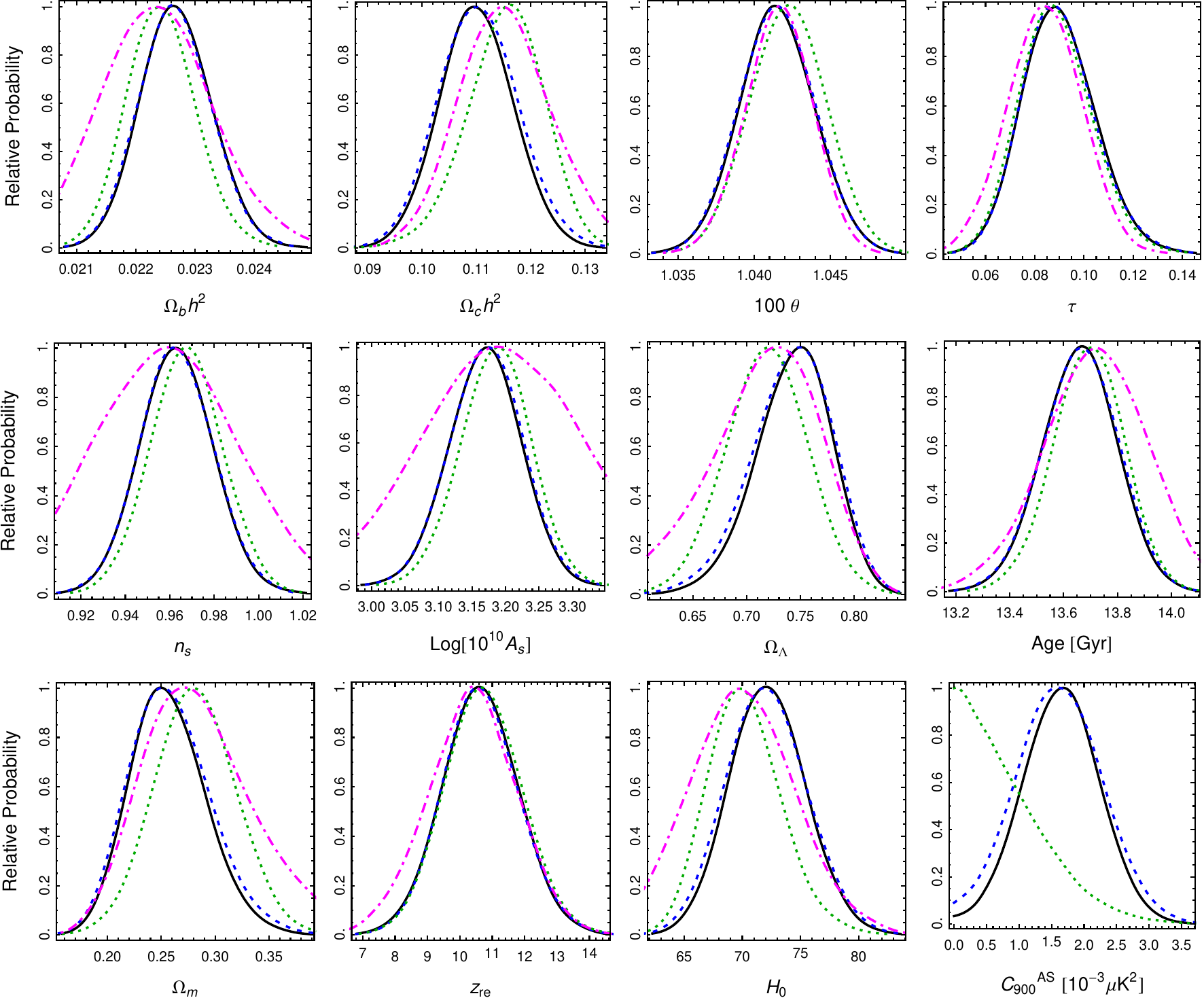}
\par\end{centering}
\centering{}\caption{Cosmological parameters derived from the 70\,GHz maps for different values of the maximum multipole -- $\ell_{max}=800$ (green dotted), $\ell_{max}=1000$ (blue dashed), and $\ell_{max} = 1200$ (solid black) -- are compared to \plik\ at  $\ell_{max}=1008$  (pink dot-dashed). \label{fig:70Ghz_parameters_LFIvsHFI-lmax} }
\end{figure*}

In Fig.~\ref{fig:Planck-70-GHz} we show the auto- and cross-power
spectra computed from the 70\,GHz maps, where cross-spectra are obtained
by cross-correlating maps from different pairs of horns (there are three such pairs in total). We use these
to construct a likelihood at $\ell>49$ by assuming a Gaussian distribution
for the band-powers, and include the covariance matrix estimated from
simulations. To estimate cosmological parameters we use this likelihood
in combination with the \Plancks low-$\ell$ likelihood. We marginalize over a single extragalactic foreground
parameter, which is a Poisson term $C_{900}^{AS}$ modeling unresolved
residual point sources. Fig.~\ref{fig:70Ghz_parameters_LFIvsHFI} shows
the resulting parameters, compared to those from \camspec. Considering the different $\ell$ range contributing to the two analysis, the parameter distributions are consistent.  In Fig.~\ref{fig:70Ghz_parameters_LFIvsHFI-lmax}, we show 70\,GHz parameters for three choices of the maximum $\ell$ considered: $\ell_{max}=800$, $1000$, and $1200$. The latter two are consistent. Minor discrepancies are displayed at $\ell_{max}=800$, which can be explained since there is, in this range, no detection of the point source component. In the same figure, we show results using the \plik\ likelihood for $\ell_{max}=1008$, which is consistent with \Planck\ 70\,GHz over the same $\ell$ range.

\subsection{Consistency of the \Planck\ low resolution CMB maps \label{app:comp-low-res-maps}}

\begin{figure}[h]
\begin{centering}
\includegraphics[width=1\columnwidth]{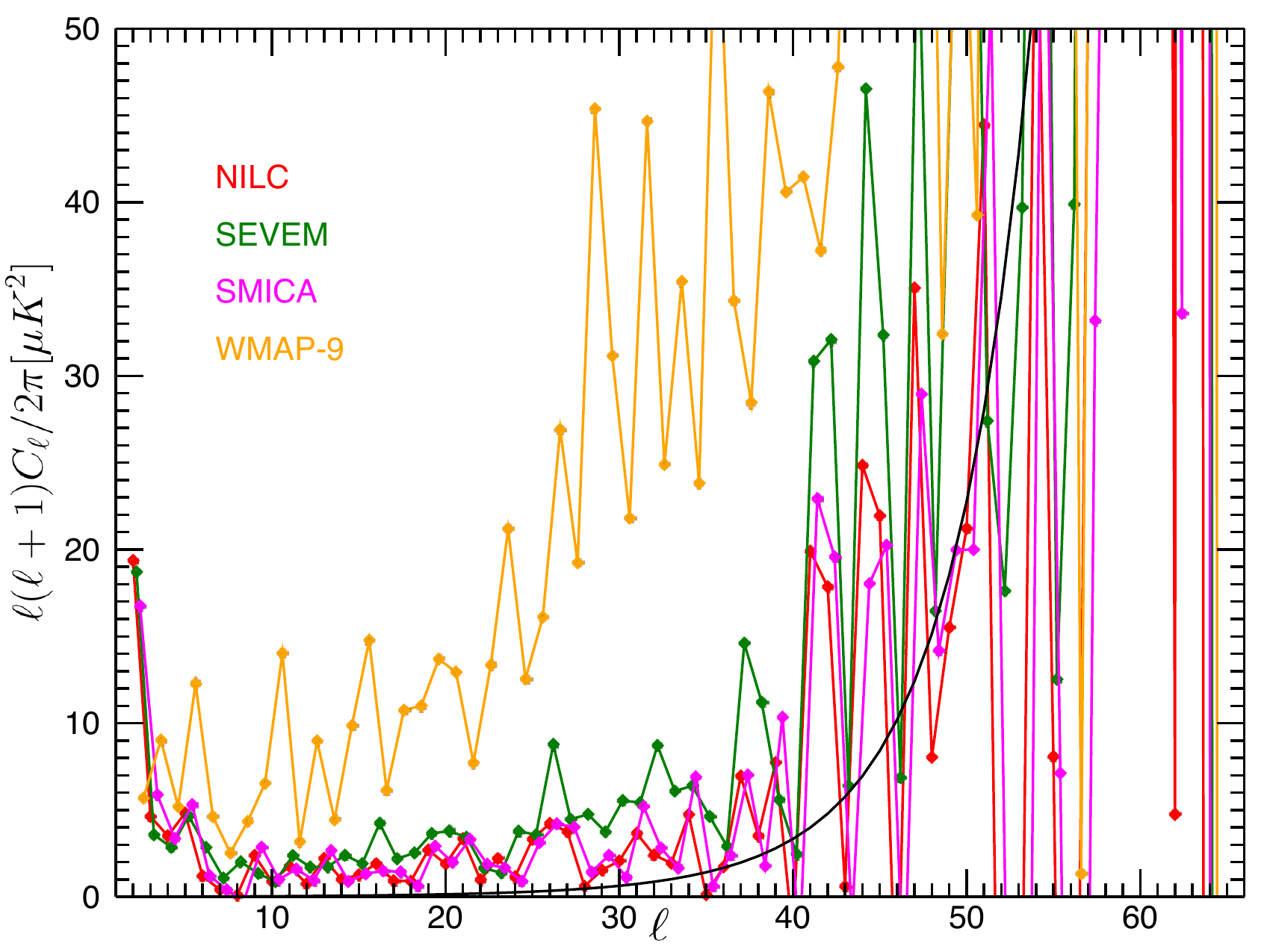}
\par\end{centering}
\caption{The power spectrum of the residual CMB maps, relative to \commander, for the \nilc, \sevem, and \smica\ methods. The $1\sigma$ expected noise level is shown in black.}
\label{fig:TTspectrumdifferencemaps} 
\end{figure}

Here we extend the discussion presented in Sect.~\ref{sec:low_ell_spectrum}. In Fig. \ref{fig:TTspectrumdifferencemaps} we show the power spectrum of the residual maps, relative to \commander, for \nilc, \sevem, and \smica. The maximum discrepancy in the range $\ell\lesssim40$ (the multipole where noise begins to become non-negligible) is localized at the quadrupole and is less than $20\:\mu\mathrm{K}^{2}$,  whereas for the range $3 \le \ell \lesssim40$ the differences are of order $\approx 5\:\mu\mathrm{K}^{2}$.  Overall, the \Plancks maps are  more in agreement among themselves than with \WMAP, except perhaps at the quadrupole. The residual map between \WMAP\ and \Plancks shows power spectrum residuals from $\approx 10\:\mu\mathrm{K}^{2}$ up to $\approx 40\:\mu\mathrm{K}^{2}$ to  $\ell\lesssim 40$. These figures should be compared to the residual estimated from simulated foreground maps, shown to be $\lesssim 10 \:\mu\mathrm{K}^{2}$ at $\ell \lesssim 70$ in \cite{planck2013-p06}.

We complement our results with those obtained from the `FFP6' simulations
described in \cite{planck2013-p06}. They consist of $1000$ signal plus noise maps
processed through each of the four component separation pipelines.

For each Monte Carlo realization, we follow the same
procedure as in the previous section, i.e. smoothing (${\rm FWHM}=329.81^{\prime}$)
and binning the maps to $\ensuremath{N_{{\rm side}}=32}$. We apply
this procedure to both the CMB input maps and the output
maps derived by the four component separation algorithms. Again, a Gaussian white noise with a variance of $4\:\mu\mathrm{K}^2$ is added, 
and the noise covariance matrix is corrected accordingly. 
Note that the additional white noise is taken into account not only for numerical regularization 
(to this extent its amplitude may well be lower), but principally because the output instrumental noise processed 
through component separation and downgraded to low resolution is far from being white. 
The additional white noise makes the detailed knowledge of the full noise covariance matrix unimportant.

For each realization and for each component separation code, we
compute the power spectrum of the processed map and of the input signal
map, using a common mask.
When estimating the power spectrum, the input CMB maps are regularized by adding a negligible Gaussian white noise with $0.1\:\mu\mathrm{K}^{2}$ variance. 
We have checked explicitly that when the white noise is added to the component separated maps, we are able to recover 
the input power spectrum without bias up to $\ell \sim 60$ for all the four component separation methods.

Moreover, we compute the power spectrum of the difference maps (output processed map minus input CMB) for each  realization,
in order to evaluate the total amount of residual noise. Note that this is not only given by the added regularization noise, but also from the intrinsic noise, albeit small, that is present in the maps. 

\begin{figure}[h]
\begin{centering}
\includegraphics[width=1\columnwidth]{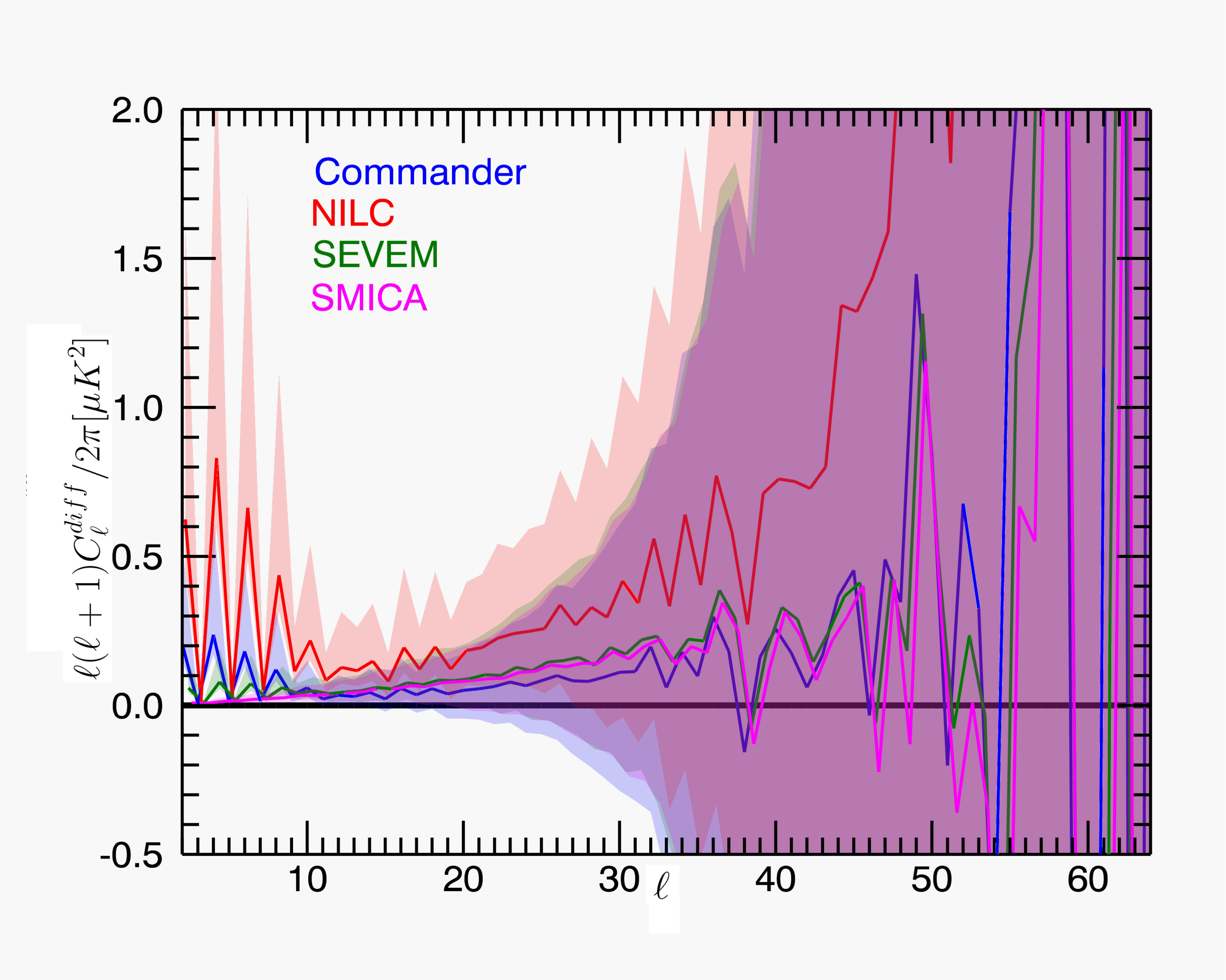}

\par\end{centering}

\caption{Estimated total residual noise (intrinsic and regularizing white noise) levels for each of the four \Plancks
CMB maps: \texttt{Commander}, \texttt{Nilc}, \texttt{Sevem}, and \texttt{Smica}. Solid lines show the average and the hatched regions show the 68\% CL.}

\label{fig:ffp6spectradifferencemaps}
\end{figure}

Fig. \ref{fig:ffp6spectradifferencemaps} shows the average and the $1\sigma$ levels of 
such noise residuals for each of the component separation solutions. We thus see,
that the level of such total noise residuals is well below the difference plotted in Fig. \ref{fig:TTspectrumdifferencemaps}. Therefore, we argue that the existing differences between the codes are due to genuine foreground separation residuals.

%% file: App-Dust-353GHz.tex
\section{Dust cleaning using \Planck\ 353\,GHz\label{sec:Dust-cleaning-353}}

The \textit{WMAP} polarization products are weighted combinations
of Ka, Q, and V band 9 year maps. The \textit{WMAP} analysis mitigates polarized
foreground emission using template fitting, with synchrotron and dust foregrounds. 
As a template for synchrotron emission, the
{\it WMAP} K band channel is used, and for dust a polarization model is used to create a template map \citep{WMAP-3yrsPol}.
Here we assess the impact on the \textit{WMAP}
polarization signal when this dust template is replaced by the \planck\ 353\,GHz map, which can be assumed to trace the dust better.
This is only for comparison; we continue to use the \textit{WMAP} polarization 
products as released by the
\textit{WMAP} team (except using the \planck\ $a_{\ell m}^{TT}$ map as discussed
in Sect.~\ref{sec:Low-multipoles-Likelihood}). 

Foreground cleaned maps can be written as 
\[
\mathbf{m}_{\rm {clean}}=\mathbf{m}_{i}-\alpha_{i}\mathbf{m}_{\mathrm {synch}}-\beta_{i}\mathbf{m}_{\mathrm {dust}}
\]
where $\mathbf{m}=(Q,U)$ are linear polarization Stokes parameter maps, and
the index $i$ is for Ka, Q, and V bands. Here $\mathbf{m}_{\mathrm{synch}}$
is the {\it WMAP} 9-year K band map and for $\mathbf{m}_{\mathrm{dust}}$ we
use either the {\it WMAP} dust template or the \planck\ 353\,GHz maps. For each frequency band,
the scaling coefficients $\alpha_{i}$ and $\beta_{i}$ are estimated
by minimizing the $\chi^{2}$:
\begin{equation}
\chi^{2}(\alpha,\,\beta)=\mathbf{m}_{\mathrm {clean}}^{t}\boldsymbol{C^{-1}}\mathbf{m}_{\mathrm {clean}}\label{eq:353-chi2}
\end{equation}
where $\boldsymbol{C}$ is the covariance matrix. Following the \textit{WMAP} analysis, we do not include the signal contribution in $\boldsymbol{C}$ and we only use the diagonal elements of the noise covariance matrix to estimate $\chi^2$. 
Scaling coefficients are computed using two different {\it WMAP} masks: the `processing mask' that masks a narrow region in the plane of the Galaxy, and the more conservative `P06' mask used for power spectrum estimation and cosmological analysis.

\begin{figure*}
\begin{centering}
\includegraphics[width=0.33\textwidth]{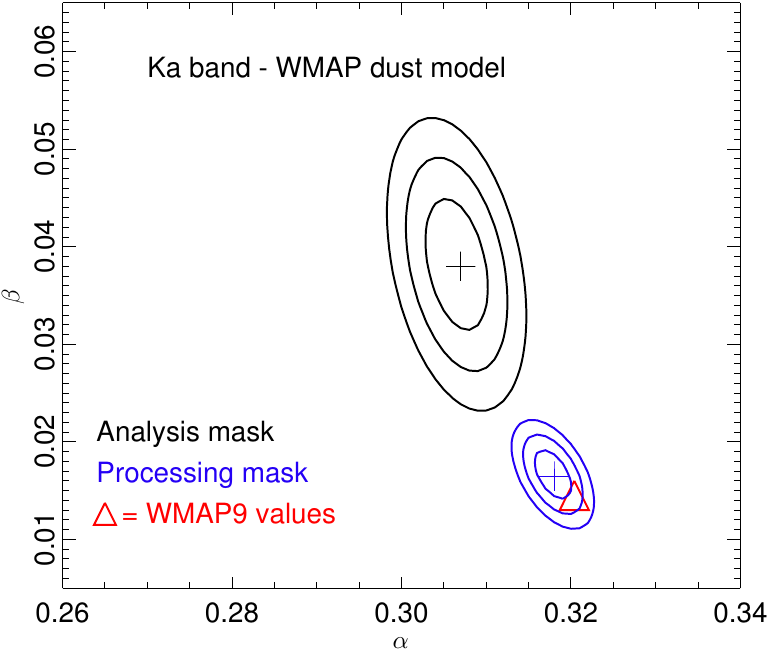}\includegraphics[width=0.33\textwidth]{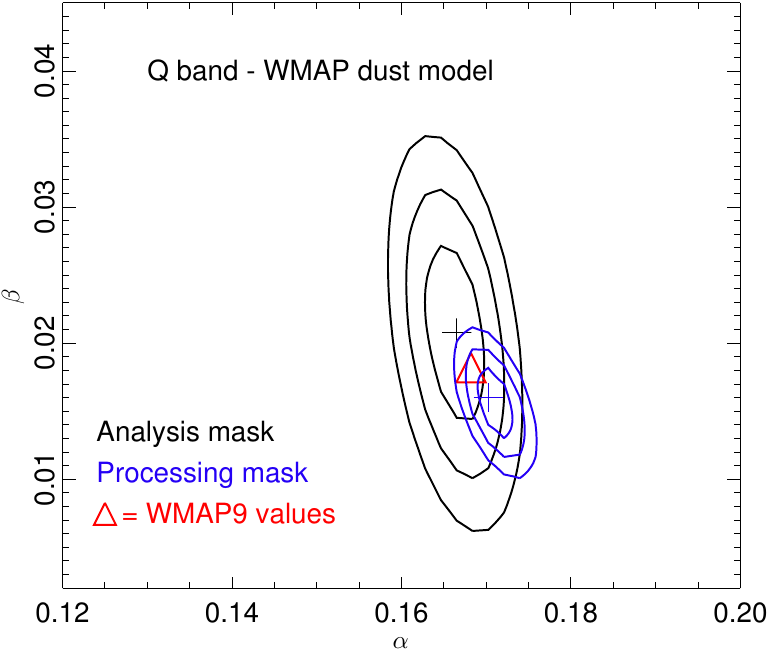}\includegraphics[width=0.33\textwidth]{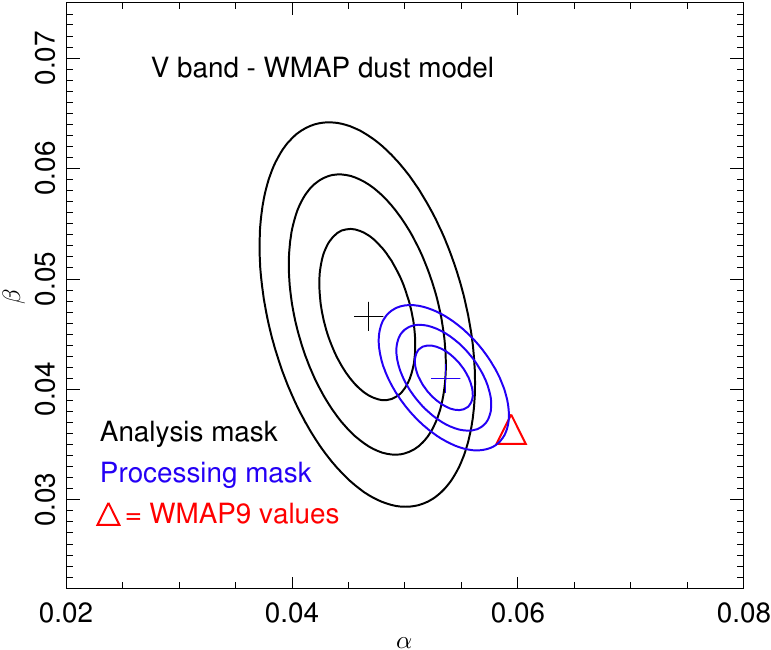}
\par\end{centering}

\begin{centering}
\includegraphics[width=0.33\textwidth]{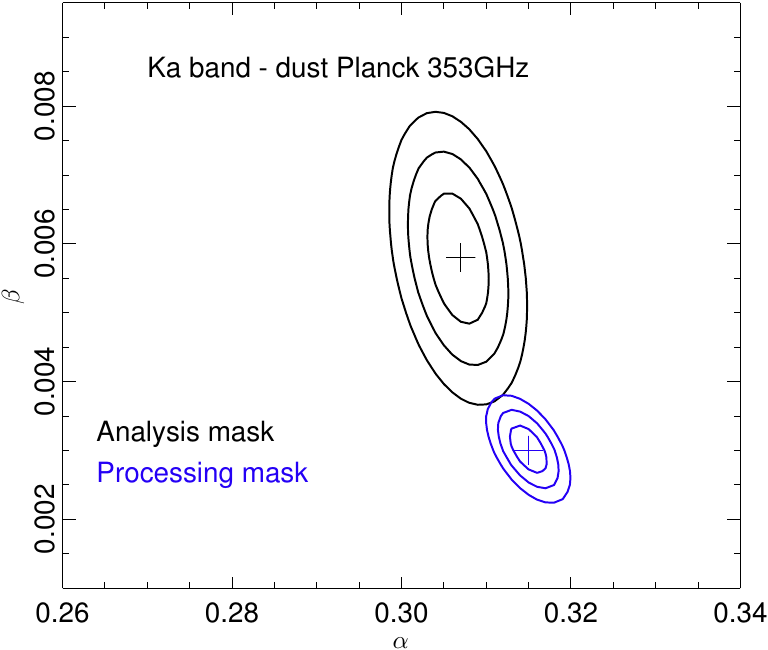}\includegraphics[width=0.33\textwidth]{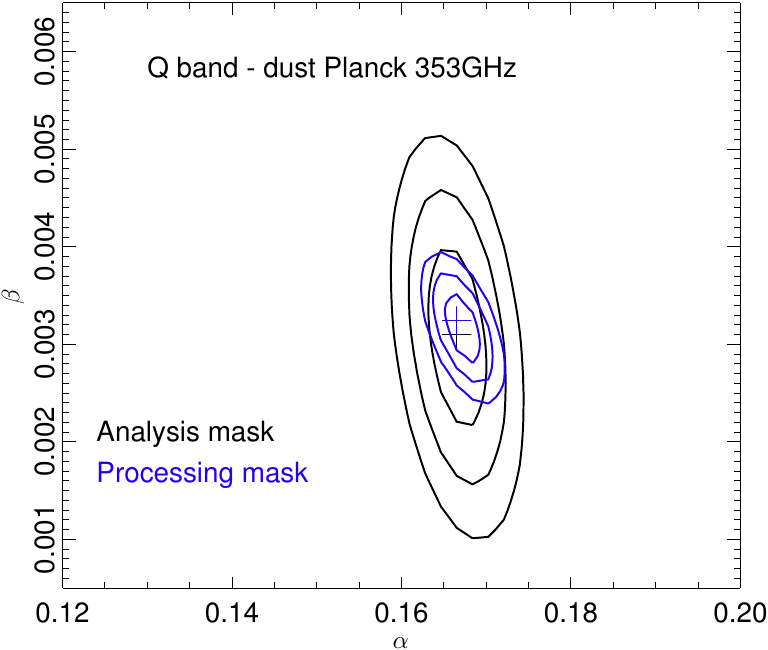}\includegraphics[width=0.33\textwidth]{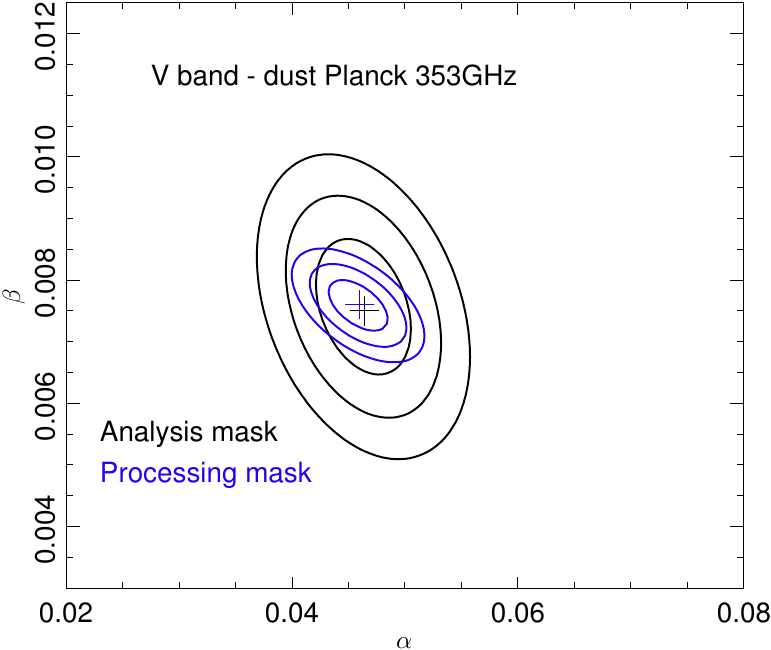}
\par\end{centering}

\begin{centering}
\caption{$1\sigma$ and $2\sigma$ contours for the template coefficient scalings estimated
using the \textit{WMAP} dust template ({\it top}) and the \planck\ 353\,GHz map as a dust template ({\it bottom}), for the Ka, Q, and V bands. We compare the coefficients estimated using the \textit{WMAP} `P06' mask, to this with the smaller \textit{WMAP} `processing
mask'. We also indicate the template values quoted in the \textit{WMAP} paper \cite{BennettWMAP9}. }
\par\end{centering}
\label{fig:scalings} 
\end{figure*}

The template coefficients at each channel are shown in Fig.~\ref{fig:scalings}, estimated using the two different Galactic masks. 
At Q and V band the estimated coefficients are consistent for the two masks; at Ka band the 353~GHz map gives more consistent results than the {\it WMAP} dust template. The coefficients using the P06 mask are more uncertain however, as the residual dust signal outside the mask is low, especially for Ka band. We find that the preferred synchrotron coefficient, $\alpha$, is slightly lower using the 353~GHz map, and the overall $\chi^2$, shown in Table~\ref{tab:scalings}, is slightly improved using the \Planck\ dust map.

\begin{table}
\begin{centering}
\caption{Reduced $\chi^{2}$ values obtained from Eq.~\ref{eq:353-chi2} for map pixels outside the \textit{WMAP} 9-year processing mask. The number of d.o.f. is 5742.}
\begin{tabular}{lccc}
\hline 
\hline
 & Ka & Q & V \\
\hline 
\planck\ 353\,GHz  & 1.127  & 1.132 & 0.991\\
WMAP dust model  & 1.135  & 1.149  & 1.030\\
\hline 
\end{tabular}
\par\end{centering}

\label{tab:scalings} 
\end{table}

\begin{figure}
\begin{centering}
\includegraphics[width=1\columnwidth]{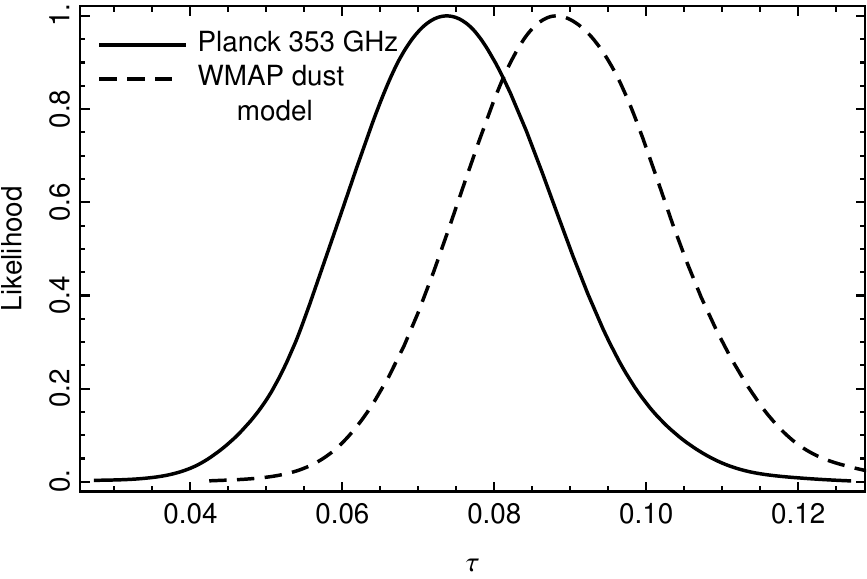}
\par\end{centering}
\centering{}\includegraphics[width=1\columnwidth]{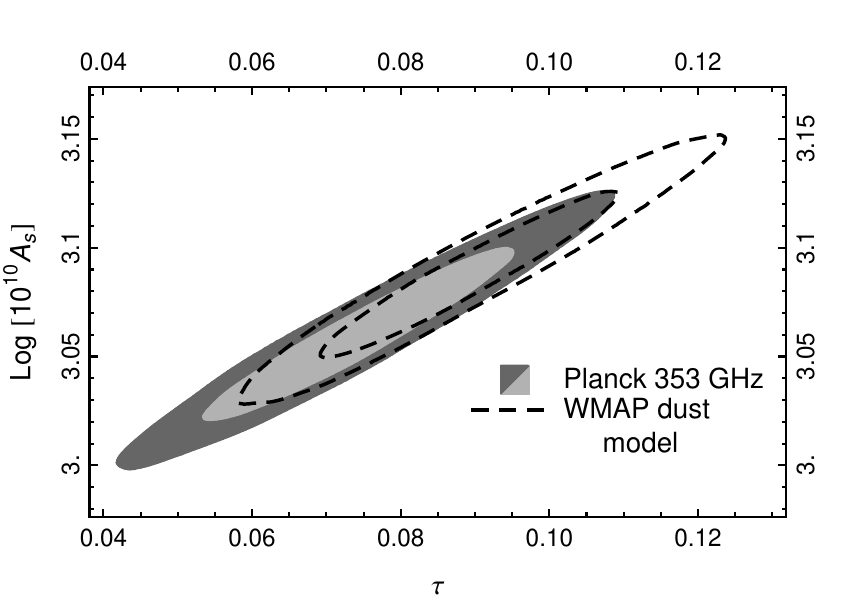}\caption{
1D ({\it top}) and 2D ({\it bottom}) posterior probability for $\tau$
and the combination $\tau - A_{\rm s}$, for two different Galactic dust templates. 
These are computed using dust template coefficients estimated with the {\it WMAP} `processing' mask. The difference in $\tau$ 
is reduced to $\sim0.5\sigma$ if template coefficients are estimated outside the `P06' Galactic mask. 
\label{fig:tau-HFI353}
}
\end{figure}

We test the effect on cosmological parameters, in particular the optical depth to reionization, using the two different templates. Using the \planck\ 353\,GHz channel as the dust template, with coefficients estimated using the processing mask, lowers the best fit value of $\tau$ by about 1$\sigma$ (see Fig.~\ref{fig:tau-HFI353}). we find $\tau=0.075\pm0.013$, compared with $\tau=0.089\pm0.013$ using the \textit{WMAP} dust model.
This also has the effect of lowering $A_{\rm s}$, from $3.088\pm0.025$ to $3.061\pm0.025$, but other $\Lambda${CDM} parameters are not affected.
We note though that using template coefficients estimated outside the P06 Galactic mask, the optical depth using the \planck\ template is lowered by only $0.5\sigma$ compared to the {\it WMAP} template, indicating some spatial dependence.
We conclude that the impact on cosmological parameters from the choice of dust template is not significant, but anticipate more extensive analysis with the full \planck\ polarization data set.

